\documentclass[a4paper,12pt,times,numbered,print,index]{Classes/PhDThesisPSnPDF}

\ifsetCustomFont
\fi

\RequirePackage[labelsep=space,tableposition=top]{caption}
\usepackage{subcaption}

\usepackage{natbib}
\setcitestyle{authoryear,sort&compress, open={[},close={]}} 
\usepackage{hyperref}
\hypersetup{
  colorlinks,
  citecolor=Blue,
  linkcolor=Black}

\usepackage{graphicx,epsfig}
\usepackage{amsmath}
\usepackage{xcolor}
\usepackage{multirow}
\usepackage[normalem]{ulem}
\usepackage{booktabs}
\usepackage{physics}
\usepackage{dsfont}
\usepackage{amssymb}
\usepackage{mathtools}
\usepackage{float}
\usepackage{makecell}
\usepackage{MnSymbol}
\usepackage[ruled,vlined]{algorithm2e}

\let\printindex\relax
\usepackage[nopostdot,% remove dot after description
  indexonlyfirst,% only index first use
  toc]{glossaries} %, style=indexgroup
\usepackage{tocloft}

\makeglossaries

\usepackage[colorinlistoftodos]{todonotes}
\usepackage{indentfirst}
\usepackage{verbatim}
\usepackage{textcomp}
\usepackage{gensymb}

\usepackage{relsize}

\usepackage{lipsum}% http://ctan.org/pkg/lipsum
\usepackage{xcolor}% http://ctan.org/pkg/xcolor
\usepackage{xparse}% http://ctan.org/pkg/xparse
\NewDocumentCommand{\myrule}{O{1pt} O{2pt} O{black}}{%
  \par\nobreak % don't break a page here
  \kern\the\prevdepth % don't take into account the depth of the preceding line
  \kern#2 % space before the rule
  {\color{#3}\hrule height #1 width\hsize} % the rule
  \kern#2 % space after the rule
  \nointerlineskip % no additional space after the rule
}
\usepackage[section]{placeins}

\usepackage{afterpage}

\newcommand\blankpage{%
    \null
    \thispagestyle{empty}%
    \addtocounter{page}{-1}%
    \newpage}
    
\renewcommand{\hbar}{\mathchar'26\mkern-9mu h}

\usepackage{setspace}
\usepackage[skip=6pt plus1pt, indent=15pt]{parskip}

\newtheorem{definition}{Definition}

\ifuseCustomBib
   \RequirePackage[square, sort, numbers, authoryear]{natbib} % CustomBib

\fi

\definecolor{quotemark}{gray}{0.7}
\makeatletter
\def\fquote{%
	\@ifnextchar[{\fquote@i}{\fquote@i[]}%]
}%
\def\fquote@i[#1]{%
	\def\tempa{#1}%
	\@ifnextchar[{\fquote@ii}{\fquote@ii[]}%]
}%
\def\fquote@ii[#1]{%
	\def\tempb{#1}%
	\@ifnextchar[{\fquote@iii}{\fquote@iii[]}%]
}%
\def\fquote@iii[#1]{%
	\def\tempc{#1}%
	\vspace{1em}%
	\noindent%
	\begin{list}{}{%
			\setlength{\leftmargin}{0.1\textwidth}%
			\setlength{\rightmargin}{0.1\textwidth}%
		}%
		\item[]%
		\begin{picture}(0,0)%
		\put(-15,-5){\makebox(0,0){\scalebox{3}{\textcolor{quotemark}{``}}}}%
		\end{picture}%
		\begingroup\itshape}%
	\def\endfquote{%
		\endgroup\par%
		\makebox[0pt][l]{%
			\hspace{0.8\textwidth}%
			\begin{picture}(0,0)(0,0)%
			\put(15,15){\makebox(0,0){%
					\scalebox{3}{\color{quotemark}''}}}%
			\end{picture}}%
		\ifx\tempa\empty%
		\else%
		\ifx\tempc\empty%
		\hfill\rule{100pt}{0.5pt}\\\mbox{}\hfill\tempa,\ \emph{\tempb}%
		\else%
		\hfill\rule{100pt}{0.5pt}\\\mbox{}\hfill\tempa,\ \emph{\tempb},\ \tempc%
		\fi\fi\par%
		\vspace{0.5em}%
	\end{list}%
}%
\makeatother
\usepackage{multirow, makecell}
\title{Neural networks and reservoir computing }
\author{Laia Domingo Colomer}

\dept{ESCUELA TÉCNICA SUPERIOR DE INGENIERÍA AGRONÓMICA, ALIMENTARIA Y
DE BIOSISTEMAS}

\university{Universidad Politécnica de Madrid}
\degreetitle{Doctor of Philosophy}

\subject{LaTeX} \keywords{{LaTeX} {PhD Thesis} {Engineering} {University of
Cambridge}}

\ifdefineAbstract
\fi

\ifdefineChapter
\fi

\begin{document}

\frontmatter

%\maketitle

\begin{titlepage}
\begin{center}

%%%%%%%%%%%%%%%%%%%%%%%%%%%%%%%%%%%%%%%%%%%%%%%%%%%%%
% PAGE 1
%%%%%%%%%%%%%%%%%%%%%%%%%%%%%%%%%%%%%%%%%%%%%%%%%%%%%
\vspace{5pt}\includegraphics[width=11cm]{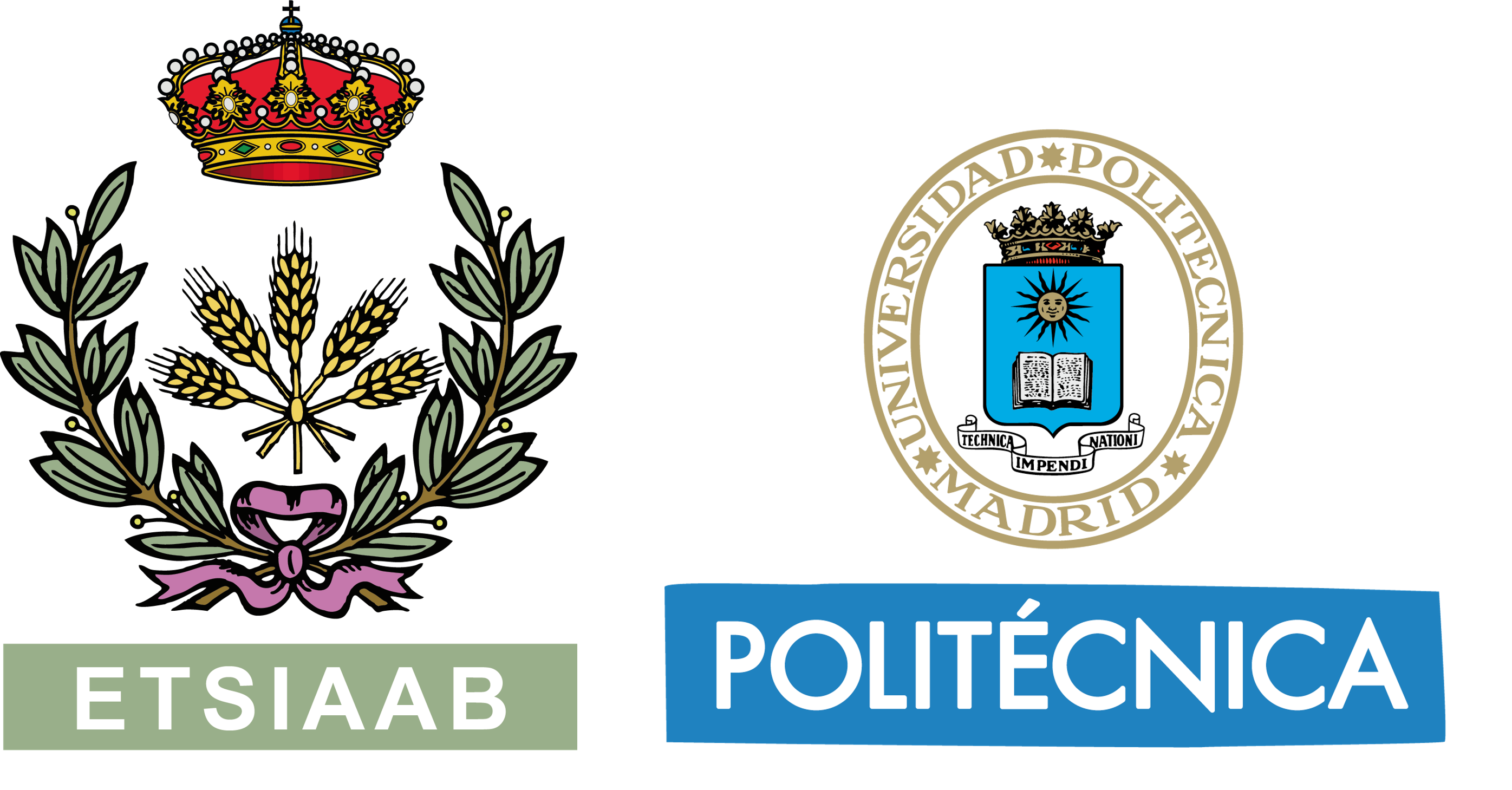}\\
\par
\vspace{20pt} 
{\LARGE UNIVERSIDAD POLITÉCNICA DE MADRID}\\[0.5cm] 
ESCUELA TÉCNICA SUPERIOR DE INGENIERÍA AGRONÓMICA,\\ ALIMENTARIA Y
DE BIOSISTEMAS\\[0.1cm]
%--- COMPLEX SYSTEMS GROUP ---
\vspace{40pt}
\myrule[1pt][7pt]
\textbf{\LARGE  Classical and quantum reservoir computing: development and applications in machine learning}\\
%\vspace{15pt}
%\textbf{\large Reporte de operaciones con matrices}\\
\myrule[1pt][7pt]
\vspace{55pt}
\textit{Tesis doctoral  }\\%A dissertation presented by}\\
\vspace{10pt}
\font\myfontb=cmr12 at 20pt
{\myfontb Laia Domingo Colomer} \\[0.2cm] 
Máster en Fundamental Principles of Data Science

\vspace{60pt}
%Madrid, 2023
2023

%%%%%%%%%%%%%%%%%%%%%%%%%%%%%%%%%%%%%%%%%%%%%%%%%%%%%
% PAGE 2
%%%%%%%%%%%%%%%%%%%%%%%%%%%%%%%%%%%%%%%%%%%%%%%%%%%%%

\afterpage{\blankpage}
\thispagestyle{empty}
\vspace{5pt}\includegraphics[width=11cm]{logos.png}\\
\par
\vspace{20pt} 
{\small GRUPO DE SISTEMAS COMPLEJOS}\\[0.5cm] 
ESCUELA TÉCNICA SUPERIOR DE INGENIERÍA AGRONÓMICA,\\ ALIMENTARIA Y
DE BIOSISTEMAS\\[0.1cm]

\vspace{20pt}
\myrule[1pt][7pt]
\textbf{\LARGE  Classical and quantum reservoir computing: development and applications in machine learning}\\
%\vspace{15pt}
%\textbf{\large Reporte de operaciones con matrices}\\
\myrule[1pt][7pt]
\vspace{35pt}
\font\myfontb=cmr12 at 20pt
{\myfontb Laia Domingo Colomer} \\[0.1cm]  
Máster en Fundamental Principles of Data Science

\vspace{40pt}  
\textit {Directores}\\
\font\myfont=cmr12 at 14pt
\begin{table}[!ht]
    \centering
    \begin{tabular}{c@{\hskip 1.2in}c}
       { \large Florentino Borondo Rodr\'{i}guez}  &  { \large Javier Borondo Benito}\\
       Doctor en Química Cuántica  & Doctor en Sistemas Complejos
    \end{tabular}
\end{table}

\vspace{60pt}
2023
\thispagestyle{empty}

\end{center}

\afterpage{\blankpage}
\par

\end{titlepage}

% ******************************* Thesis Dedidcation ********************************

\begin{dedication} 

A la meva mare, per ser la meva font d'inspiració.

\end{dedication}
% ************************** Thesis Acknowledgements **************************

\begin{acknowledgements}

Mi primer agradecimiento es para mi director, el Prof. Florentino Borondo, por ser una constante fuente de inspiración y motivación a lo largo de la tesis. Muchas gracias por tu apoyo, tu ayuda y extensos conocimientos, que me han llevado a explorar nuevas ideas y forjar mi propio camino. Es muy motivador trabajar con una persona tan versátil y con una visión muy amplia de la ciencia. También quería dar las gracias a mi otro director, el Prof. Javier Borondo, por confiar en mí para desarrollar esta tesis, y por ser un gran mentor tanto en el mundo académico como en el mundo empresarial. 

Muchas gracias a todos los miembros del Grupo de Sistemas Complejos de la Universidad Politécnica de Madird, en particular a la Prof. Rosa María Benito y al Prof. Juan Carlos Losada, por guiarme siempre durante este proceso y por vuestros valiosos consejos, tanto a nivel científico como a nivel personal. Gracias también a mis queridos compañeros del grupo -Javi, Alberto, Julia, Leticia y Tomás- por estar ahí siempre. Gracias Mar, por ser una gran amiga y compañera, estoy encantada de colaborar con alguien tan inteligente y trabajadora como tú. Gracias también al Prof. Fabio Revuelta y el Prof. Javier Villalba, por vuestro rigor científico que ha sido una gran inspiración para mí. 

Muchas gracias también a Fer y Joel, por ser los mejores compañeros en este viaje, por todas las risas y experiencias que han hecho este camino mucho más divertido y llevadero. 

Quería expresar mi profunda gratitud al Prof. Gabriel G. Carlo. Es un verdadero placer trabajar con una persona tan brillante y afable, y con una intuición deslumbrante. Espero que este sea el principio de muchas colaboraciones futuras juntos. 

Very special thanks to my mentors and colleagues at Ingenii, who gave me the opportunity to develop a fruitful research internship with them, which lead to a wide range of future possibilities. Christine Johnson, thank you for teaching me so much, for your firm values and unconditional support. Marko Djukic, thank you for your strong guidance, for challenging me and being a source of inspiration. Thank you Greg Atkins, for being always so nice and thoughtful and for always having the best advice. And thank you, Mahdi, Rina and Maria, for having such brilliant minds and lovely hearts. It is a pleasure working with you all. Together we are going to change the world! 

I would also like to thank Prof. Mark Daniel Ward for his warm welcome to Purdue University and for letting me be part of the Data Mine program. It has been such a rewarding experience working with Roy, Shiva, Lawal, Suriya, Sambit and Shawn.  

I would like to express my deepest gratitude to La Caixa Foundation, which provided me with the incredible opportunity to embark on my PhD journey through their prestigious fellowship. I am immensely grateful for all the assistance and support they have offered, particularly to Gisela, who has consistently been kind-hearted and understanding.

I no podia faltar donar les gràcies a la meva família, amigues i parella. Gràcies mama, pel teu optimisme i entusiasme que fa que la vida sigui més senzilla i emocionant. Em queda molt per aprendre de tu, però em fa feliç pensar que m'acompanyaràs cada dia amb un somriure. Moltes gràcies papa, per ensenyar-me el valor de l'esforç i la dedicació. Gràcies a tu he pogut arribar fins aquí, sabent que dedicaré gran part del meu dia a fer el que més m'agrada, i podré aportar el meu granet de sorra perquè el mon sigui una mica millor. 

Moltes gràcies a les Melonas -Queralt, Georgina, Júlia i Mar- per estar sempre al meu costat, per compartir riures i plors, per ser les germanes que mai no he tingut. I gràcies Laura i Núria, per les converses eternes que curen i em fan sentir acompanyada. Moltes gràcies Nicetu, per haver compartit aquest viatge amb mi durant tants anys i haver-me acompanyat en tots els alts i baixos. Per ser una persona tant brillant en la ciència i en la vida. I moltes gràcies Marta per acompanyar-lo sempre i saber treure el millor d'ell en tot moment. Moltíssimes gràcies Víctor, Clàudia i Bernat per formar part de la meva vida en els petits i grans moments.

I finalment, moltes gràcies Marc, per ser el meu company de vida, per estar al meu costat sempre, en els bons i mal moments, per no rendir-te mai. Gràcies per escoltar-me una i altra vegada, pels ànims i els bons consells. Perquè has fet aquest camí molt més lleuger, i la vida més bonica.

While I could continue expressing my gratitude endlessly, the time has come for me to begin presenting my research. Therefore, I express my heartfelt appreciation to everyone who has supported me throughout my thesis. Thank you!

\end{acknowledgements}

%\newgeometry{top=0cm, bottom=2cm, left=3cm, right=3cm}
% ************************** Thesis Abstract *****************************
% Use `abstract' as an option in the document class to print only the titlepage and the abstract.
\begin{resumen}
\emph{Reservoir computing} es un novedoso algoritmo de aprendizaje automático que utiliza un sistema dinámico no lineal para procesar y aprender eficientemente patrones temporales complejos en los datos, proporcionando una herramienta eficaz y potente para propagar sistemas dinámicos en el tiempo.

El objetivo de esta tesis es investigar los principios de \emph{reservoir computing} y desarrollar variantes originales del algoritmo capaces de abordar diversas aplicaciones en el aprendizaje automático. La investigación demuestra la robustez y adaptabilidad del algoritmo en dominios muy diferentes, incluyendo la predicción de series temporales agrícolas y la propagación temporal de sistemas cuánticos. Además, la tesis explora el algoritmo de \emph{reservoir computing} cuántico, una variante de computación cuántica del método, que tiene el potencial de manejar conjuntos de datos de mayor dimensión y generar representaciones cuánticas de los datos que serían difíciles de simular eficientemente utilizando métodos clásicos.

La primera contribución consiste en desarrollar una novedosa metodología basada de \emph{reservoir computing} a la predicción de precios de productos agrícolas, lo cual es crucial para garantizar la sostenibilidad del mercado alimentario. Los resultados muestran que \emph{reservoir computing}, junto con una inteligente descomposición de la serie en componentes de tendencia, estacionalidad y residuos, supera a los métodos tradicionales de predicción de series temporales.

La siguiente contribución de la tesis está dedicada a resolver la ecuación de Schrödinger para sistemas cuánticos complejos. En primer lugar, se diseña un modelo de redes neuronales, mediante la selección óptima de datos de entrenamiento y la incorporación de una función de pérdida adicional, que permite reproducir con éxito estados de alta energía para Hamiltonianos complejos. Además, se propone un nuevo método de \emph{reservoir computing} para propagar eficientemente las funciones de onda cuánticas en el tiempo, permitiendo el cálculo de todos los estados propios de un sistema cuántico dentro de un rango de energía específico. El método resultante se utiliza para estudiar sistemas prominentes en el campo de la química cuántica y el caos cuántico.

La última contribucióm de la tesis se centra en la el diseño óptimo del algoritmo de \emph{reservoir computing} cuántico, tanto para tareas temporales como no temporales. Los resultados demuestran que las familias de circuitos cuánticos con mayor complejidad, según el criterio de mayorización, ofrecen un rendimiento superior en tareas de aprendizaje automático.  Además, se evalúa el impacto del ruido cuántico en el la calidad de las predicciones del algoritmo, revelando que el ruido de \emph{amplitude damping} puede ser beneficioso para el algoritmo, mientras que la corrección de los ruidos \emph{depolarizing} y  \emph{phase damping} debe ser prioritaria. Además, este estudio se emplea para construir una red neuronal híbrida cuántico-clásica que aborda un problema fundamental en el diseño de nuevos fármacos, destacando los beneficios del \emph{reservoir computing} cuántico para resolver problemas punteros.

\end{resumen}

% ************************** Thesis Abstract *****************************
% Use `abstract' as an option in the document class to print only the titlepage and the abstract.
\begin{abstract}
Reservoir computing is a novel machine learning algorithm that uses a nonlinear dynamical system to efficiently process and learn complex temporal patterns from data, providing an effective and powerful tool for propagating dynamical systems over time. 

The objective of this thesis is to investigate the principles of reservoir computing and develop state-of-the-art variants of the algorithm capable of addressing diverse applications in machine learning. The research demonstrates the algorithm's robustness and adaptability across very different domains, including agricultural time series forecasting and the time propagation of quantum systems. Additionally, the thesis explores quantum reservoir computing, a quantum computing variant of the method, which has the potential to handle higher-dimensional datasets and generate quantum representations of data that would be challenging to simulate efficiently using classical methods.

The first contribution of this thesis consists in developing a reservoir computing-based methodology to predict future agricultural product prices, which is crucial for ensuring the sustainability of the food market. The results show that reservoir computing, together with a clever data decomposition in trend, seasonal and residual components, outperforms traditional forecasting methods.

The next contribution of the thesis is devoted to solving the Schrödinger equation for complex quantum systems. First, a neural network model is designed, by selecting suitable training data samples and incorporating an additional loss function, leading to the successful reproduction of high-energy states for complex Hamiltonians. Additionally, a novel reservoir computing framework is proposed to efficiently propagate quantum wavefunctions in time, enabling the computation of all eigenstates of a quantum system within a specific energy range.  This approach is used to study prominent systems in the field of quantum chemistry and quantum chaos.

The last contribution of this thesis focuses on optimizing algorithm designs for quantum reservoir computing, both for temporal and non-temporal tasks. The results demonstrate that families of quantum circuits with higher complexity, according to the majorization criterion, yield superior performance in quantum machine learning.  Moreover, the impact of quantum noise on the algorithm performance is evaluated, revealing that the amplitude damping noise can actually be beneficial for the performance of quantum reservoir computing, while the depolarizing and phase damping noise should be prioritized for correction. Furthermore, the optimal design of quantum reservoirs is employed to construct a hybrid quantum-classical neural network that tackles a fundamental problem in drug design, highlighting the benefits of quantum reservoir computing in solving cutting-edge problems.

\end{abstract}

%\restoregeometry
% *********************** Adding TOC and List of Figures ***********************

\tableofcontents
\newpage

\listoffigures
\newpage

\listoftables
\newpage

% GLOSSARY
%Term definitions
\newglossaryentry{NN}{name=NN, description={Neural Network}}
\newglossaryentry{RC}{name=RC, description={Reservoir Computing}}
\newglossaryentry{RNN}{name=RNN, description={Recurrent Neural Network}}
\newglossaryentry{ML}{name=ML, description={Machine Learning}}
\newglossaryentry{NISQ}{name=NISQ, description={Noisy Intermediate-Scale Quantum}}
\newglossaryentry{ReLU}{name=ReLU, description={Rectified Linear Unit}}
\newglossaryentry{CNN}{name=CNN, description={Convolutional Neural Network}}
\newglossaryentry{LSTM}{name=LSTM, description={Long Short-Term Memory}}
\newglossaryentry{PDE}{name=PDE, description={Partial Differential Equation}}
\newglossaryentry{QRC}{name=QRC, description={Quantum Reservoir Computing}}
\newglossaryentry{PQC}{name=PQC, description={Parameterized Quantum Circuits}}
\newglossaryentry{MSE}{name=MSE, description={Mean Squared Error}}
\newglossaryentry{FFT}{name=FFT, description={Fast Fourier Transform}}
\newglossaryentry{QR}{name=QR, description={Quantum Reservoir}}
\newglossaryentry{POVM}{name=POVM, description={Positive Operator-Valued Measurement}}
\newglossaryentry{MG}{name=MG, description={Matchgates}}
\newglossaryentry{D2}{name=$D_2$, description={Diagonal-gate circuits applied to 2 qubits}}
\newglossaryentry{D3}{name=$D_3$, description={Diagonal-gate circuits applied to 3 qubits}}
\newglossaryentry{Dn}{name=$D_n$, description={Diagonal-gate circuits applied to $n$ qubits}}
\newglossaryentry{NOON}{name=NOON, description={Natural Orbital Occupation Number}}
\newglossaryentry{QML}{name=QML, description={Quantum Machine Learning}}
\newglossaryentry{UMAP}{name=UMAP, description={Uniform Manifold Approximation and Projection}}
\newglossaryentry{GPU}{name=GPU, description={Graphics Processing Unit}}
\newglossaryentry{FRQI}{name=FRQI, description={Flexible Representation of Quantum Images }}
\newglossaryentry{DREM}{name=DREM, description={Data Regression Error Mitigation }}
\newglossaryentry{RMSE}{name=RMSE, description={Root Mean Squared Error }}
\newglossaryentry{MAE}{name=MAE, description={Mean Absolute Error }}
\newglossaryentry{R2}{name=R2, description={Coefficient of determination R squared}}
\newglossaryentry{SARIMA}{name=SARIMA, description={Seasonal Autoregressive Integrated Moving Average }}
\newglossaryentry{MDA}{name=MDA, description={Market Direction Accuracy }}
\newglossaryentry{E-RC}{name=E-RC, description={Ensemble Reservoir Computing}}
\newglossaryentry{D-RC}{name=D-RC, description={Trend-seasonality decomposition Reservoir Computing}}
\newglossaryentry{NG-RC}{name=NG-RC, description={Next Generation Reservoir Computing }}
\newglossaryentry{ARNN-RC}{name=ARNN-RC, description={Auto-Reservoir Neural Network  }}
\newglossaryentry{CR}{name=CR, description={California Rojo}}
\newglossaryentry{CA}{name=CA, description={California Amarillo}}
\newglossaryentry{CV}{name=CV, description={California Verde}}
\newglossaryentry{IV}{name=IV, description={Italiano Verde}}
\newglossaryentry{IR}{name=IR, description={Italiano Rojo}}
\newglossaryentry{LR}{name=LR, description={Lamuyo Rojo}}
\newglossaryentry{LV}{name=LV, description={Lamuyo Verde}}
\newglossaryentry{CCF}{name=CCF, description={Cross Correlation Function}}
\newglossaryentry{PINC}{name=PINC, description={Prediction Interval Nominal Confidence}}
\newglossaryentry{NIG}{name=NIG, description={Normal Inverse Gaussian }}
\newglossaryentry{ACE}{name=ACE, description={Average Coverage Error }}
\newglossaryentry{PICP}{name=PICP, description={Prediction Interval Coverage Probability}}
\newglossaryentry{PIAW}{name=PIAW, description={Prediction Interval Average Width}}

%Print the glossary
\printglossaries

% \printnomenclature[space] space can be set as 2em between symbol and description
%\printnomenclature[3em]

\printnomenclature

% ******************************** Main Matter *********************************
\mainmatter

%!TEX root = ../thesis.tex

%*********************************** First Chapter *****************************

\chapter{Introduction}  %Title of the First Chapter
\label{chapter1}

\begin{fquote}[Isaac Newton][1642-1727]
    The more I learn, the more I realize how much I don't know.
\end{fquote}

Machine learning (\gls{ML})~\citep{ML}, a transformative field at the intersection of computer science and mathematics, has revolutionized the way we approach complex problems and harness the power of data. ML aims to give computers the ability to learn from examples and experience, without being explicitly programmed for those tasks. That is, instead of giving explicit instructions to solve problems, ML algorithms have the ability to analyze vast datasets, identify patterns, and generate smart predictions. By continuously adapting and improving their performance through experience, these algorithms learn how to solve problems autonomously, generating groundbreaking advancements across various industries and domains.  ML is present in so many segments of technology nowadays that any technology user today has benefited from its applications. For instance, natural language processing allows the creation of powerful language models that generate human-like responses~\citep{openai}, facial recognition software~\citep{face-recognition} allows social media platforms to help users tag and share photos of friends, effective web search~\citep{RLAppGame} eases the acquirement of information and self-driving cars~\citep{self-driving-cars} will soon be available to customers. One of the most paradigmatic ML algorithms, which serves as the core of most current large-scale ML applications, is the Neural Network~(\gls{NN}). 

\section{Neural Networks}
%NEURAL NETWORKS
NNs are inspired by the structure and function of the human brain and are designed to process complex, non-linear relationships in large datasets. Because of their ability to automatically unveil and model hidden patterns in data, without relying on predefined hand-curated features, they can be applied to a wide range of problems, such as image and speech recognition, natural language processing, and time series forecasting~\citep{NNBook}.

NNs have their roots in the 1940s, with the development of early models of artificial intelligence and cognitive psychology. The concept of using simple processing elements that simulate information processing in the brain was first proposed by McCulloch and Pitts in 1943~\citep{mccullochPitts}. In their work, they modelled a simple NN using electrical circuits to describe how neurons in the brain might work. This idea laid the foundation for the development of the first artificial NNs. The first trainable NN, called Perceptron, was designed by Rosenblatt, a Cornell University psychologist, in 1957~\citep{rosenblatt1957}. He described it as “a machine which senses, recognizes, remembers, and responds like the human mind”. Perceptron was built in hardware and is the oldest NN still in use today. 

Until the late 1960s, the field of artificial intelligence experienced a period of rapid growth, and NNs became a popular area of research. However, a critical book written in 1969 by Minsky and Papert, called \emph{Perceptrons}~\citep{minsky1969perceptrons}, showed that Rosenblatt’s original system had severe limitations since it could only learn to classify linearly separable datasets. Therefore, the simple but non-linear XOR (exclusive-or) circuit represented an insurmountable barrier for the Perceptron. As a result of this severe criticism, interest and enthusiasm for NNs among researchers began to decline. What had become known as the field of NNs disappeared for the next decade, which is now known as the first NN winter. 

This long winter stopped in 1982 at the National Academy of Sciences when Hopfield presented an approach to construct more useful networks by using bidirectional lines~\citep{hopfield1982}. Moreover, the development of the backpropagation algorithm in 1986 by three independent groups of researchers, which included Rumelhart, Hinton and Williams~\citep{Rumelhart}, also increased the interest in the field. This was the first practical method for training multi-layered NNs, which led the way to the development of new algorithms such as Convolutional Neural Networks (\gls{CNN}s)~\citep{original_CNN} and Long Short-Term Memory networks (\gls{LSTM}s)~\citep{LSTM_original} which are still widely used today. 

Around the turn of the 21st century, the development of support vector machines, a novel ML method with strong mathematics foundations, increased the skepticism around NNs. The design and training algorithm of NNs seemed intuition-based, and computers were still barely able to meet their computational needs, which caused the field to fall into eclipse again. The interest strongly rose in the second decade of this century, mainly fueled by the increased processing power of graphics processing units (\gls{GPU}s). The advent of modern GPUs has allowed for the development of deep learning networks, with tens to hundreds of layers, from the limited one-layer networks available in the 1960s and two to three-layer networks in the 1980s. As a result, deep learning is currently responsible for the best-performing systems in almost every area of artificial intelligence research. 

% RECURRENT NEURAL NETWORKS
The original design of NNs assumes that data samples are independent of one another. Although this assumption is valid in many scenarios, it is crucial to consider temporal dependencies when dealing with sequential data, such as texts and time series. To overcome this challenge, Recurrent Neural Networks (\gls{RNN}s) were designed in the 1980s~\citep{RNN_original1, RNN_original2}. Their feedback mechanism allows them to capture information from previous inputs and maintain it in a hidden state that can be used for future predictions. While their loop-like architecture allows learning sequential patterns, the training of the original RNN suffers from the vanishing and exploding gradients problem~\citep{vanishing}, which refers to the gradients computed during backpropagation becoming too small or too large to allow for effective learning. This issue causes the network to be unable to capture long-term relationships in data. In 1997, Hochreiter and Schmidhuber proposed the LSTM architecture to overcome this problem~\citep{LSTM_original}. This was a major breakthrough that opened the door to more complex RNN models and led to more practical applications in fields such as natural language processing and time series forecasting. 

More recently, a novel deep learning algorithm called Transformers~\citep{Transformers} has revolutionized natural language processing tasks, allowing the development of large language models, such as the ChatGPT by OpenAI~\citep{openai}. The key feature of Transformers is their self-attention mechanism, which allows the model to weigh the importance of different input elements in making a prediction. This allows Transformers to model long-range dependencies in sequential data, making them well-suited for tasks such as language translation, text classification, and text generation. Additionally, Transformers are parallelizable, allowing for much faster training than traditional RNNs. 

While these models have proven to provide impressive results in processing text, they require access to vast amounts of data during the training process. This is inconvenient for time series forecasting or for the study of dynamical systems, where only small datasets are available to train ML methods. For this purpose, a novel algorithm with lower learning complexity was proposed to overcome the challenges of RNNs: Reservoir Computing (\gls{RC}).

\section{Reservoir Computing}
% RESERVOIR COMPUTING
The RC algorithm was originally designed to compute the time evolution of dynamical systems. The first published work on RC appeared in 2001 by Jaeger~\citep{ENS}, where the author proposed the concept of RC and showed how the corresponding algorithm could be used to integrate dynamical systems with high accuracy. 

At its core, RC uses a fixed, randomly initialized reservoir to generate many high-dimensional, complex representations of the input data. The reservoir is typically a large collection of interconnected recurrent units, such as artificial neurons or even physical systems (as in quantum RC, described below). These units interact with each other, creating complex and rich dynamics within the reservoir. Once the input data has been transformed by the reservoir, creating a vast set of internal trajectories, a linear readout layer is trained to make predictions based on such trajectories. This layer learns to read the relevant information extracted by the reservoir and transform it into a desired output. Such readout layer is typically a simple linear regression model, which learns the best combination of reservoir trajectories to predict the outputs. 

In this way, the reservoir must be able to produce a diverse set of trajectories, enabling the output to be predicted through the computation of a linear combination of these trajectories. As a result, the RC algorithm learns to reproduce the input-output dynamics, allowing the internal trajectories of the reservoir to preserve information from previous inputs and outputs. For this reason, Jaeger referred to this property as \emph{echo states}~\citep{ENS}, since the dynamics of the reservoir nodes can be thought of as an echo of their past. 

The biggest advantage of RC over NNs is that the recurrent weights of the reservoir are randomly generated and remain fixed during training, in such a way that they do not need to be learned. This fact greatly simplifies the training process and reduces the risk of overfitting, a common problem when training ML models~\citep{Domingo_adaptingRC}. When a ML overfits the training data, it performs very well on the training set but poorly when applied to new, unseen data. It occurs when a model becomes too complex and starts to memorize the training data rather than learning general patterns and relationships that can be applied to new data. 

In the case of RC, the training process requires learning only the weights of the readout layer, reducing the complexity of the algorithm to just performing a linear regression. The simplicity of this learning strategy reduces the chances of overfitting and also overcomes the problem of vanishing and exploding gradients present in RNNs, allowing the algorithm to be trained with much smaller datasets and to efficiently process time-series data with long-range temporal dependencies. 

The concept of RC was also simultaneously introduced in the field of computational neuroscience as liquid state machines~\citep{LSM}, which were inspired by the behavior of liquid substances, where information is distributed and processed by the collective behavior of the constituent molecules. 

The interest in RC has greatly grown over time due to the advent of big data and the increasing need for real-time processing. More recently, the RC gained a lot of popularity thanks to the work of the chaos theorist Ott and collaborators~\citep{chaos1, chaos2}, which showed how RC could propagate chaotic dynamical systems to an impressive distant horizon. Chaotic systems are characterized by a high dependence on initial conditions, meaning that even small changes in those conditions can lead to large differences in the behavior of the system over time. For this reason, chaotic systems are extremely difficult to predict and control, as small perturbations can lead to significant changes in the system's evolution. This behavior is commonly known as the \emph{butterfly effect}, the term named after the famous talk from Lorentz in 1995~\citep{lorenz1995essence}. Chaotic systems are responsible, for example, for the difficulties in predicting weather changes or the evolution of financial markets. The ability of RC to accurately propagate such systems proved its potential in predicting highly complex and volatile dynamics and opened the door to the development of relevant applications in multiple fields. 

The ML methods presented so far are highly versatile since they can be adapted to solve very different problems. Even though ML solutions are capable of learning systems of relatively moderate sizes, doing so in more complex patterns often requires an enormous number of parameters and long training times. For instance, tasks such as protein folding or learning language models require training deep learning models with millions or even billions of weights~\citep{LanguageModels, Alphafold}. Because of the high complexity of these models, even larger datasets are required to learn the hidden patterns of the data. Therefore, the success of these models is conditioned to having access to high-performance computational resources and vast datasets. 

Even though in this thesis we provide a method for RC to prevent overfitting in high-dimensional datasets, the current exponential growth and complexity of available data remains a clear bottleneck for ML methods. For this reason, tremendous interest has recently arisen in a technological field with the potential to dramatically improve some of these algorithms: quantum computing.

% JUMP TO QUANTUM COMPUTING
\section{Quantum computing}
The foundations of quantum computing were formalized in the 1980s~\citep{Feynman,Feynman2}, where it was shown that the quantum units of information, called \emph{qubits}, can carry more information than classical bits. For this reason, quantum computers were expected to be able to solve certain tasks faster than any classical device. In the following decade, two algorithms demonstrated that quantum computers could reach quantum supremacy for certain difficult problems when run on ideal quantum computers. The first one was Grover search~\citep{Grover, Grover2}, which provided a quantum solution to perform unstructured searches quadratically faster than classical computers. The second one, Shor's algorithm~\citep{Shor}, showed an exponential speedup for large integer factorization, which implied that the common cryptography schemes would be no longer secure when fault-tolerant quantum computers existed.

Since then, researchers have put efforts into building practical quantum computers that can provide the promised quantum advantage. For example, Shor's algorithm was first implemented by a physical quantum computer in 2001 by IBM~\citep{IBMShor}. Currently, companies such as IBM, Google, Microsoft, Rigetti, IonQ, or D-Wave provide quantum computing platforms that allow to implementation of small-scale quantum algorithms for both research and industrial applications. Recently, Google Quantum AI proved the scalability of error-correcting techniques~\citep{google2023suppressing}, which is the first step towards fault-tolerant computation. However, there is still a long way until large-scale fault-tolerant quantum computing becomes a reality.  

With the existing proofs of concepts and claims that quantum supremacy has been achieved~\citep{quantumsupremacy}, not without controversy~\citep{IBM19, no-supremacy}, quantum computing is very likely to provide practical applications in the field of ML, commonly known as Quantum Machine Learning (\gls{QML}). In a QML setting, quantum and classical methods are combined to generate hybrid algorithms that enhance the performance of the original classical ML methods. Even though quantum computers are not yet able to unravel their full potential, a new alternative has attracted much attention recently. This new trend, called \emph{Noisy Intermediate Scale Quantum} (\gls{NISQ}) era~\citep{preskill1998quantum}, is devoted to developing quantum algorithms to reach quantum advantage using the (small) quantum computers available today~\citep{NISQ}. A highly relevant quantum algorithm for the NISQ era, which will be widely studied in this thesis, is Quantum Reservoir Computing (\gls{QRC})~\citep{QRC2, DynamicalIsing,  QRC,  reviewQRC, NaturalQRC}. 

QRC represents a promising advancement in the field of ML, building upon the foundations of classical RC. While classical RC relies on the dynamics of a reservoir of interconnected neurons to perform computations, QRC leverages the unique properties of quantum mechanics such as superposition and entanglement to offer novel capabilities. In QRC, the reservoir is implemented using a quantum system, such as a quantum circuit, which consists of a set of interacting qubits. In this way, QRC exploits the dynamics and properties of quantum systems as computational resources to perform ML tasks. 

The main idea of QRC is to use quantum operations to capture complex patterns and dependencies in the input data. The first step of the algorithm is to encode the input data as a quantum state, which is then passed through a quantum circuit that acts as the reservoir. The reservoir consists of a set of quantum operations -also known as quantum gates- that entangle the input data in a way that amplifies its features and makes it easier to extract useful information~\citep{Domingo_hybrid}. Then, quantum measurements are used to build the final features, which are passed through a classical readout layer, which uses a classical ML algorithm (usually a linear model) to predict the desired output.

The advantage of QRC over classical RC is that the quantum-entangling operations potentially allow for a more efficient and powerful feature space transformation. On the one hand, the entangling operations can exploit the use of quantum operations to create complex, non-linear transformations of the input data that are difficult to achieve with classical methods. On the other hand, the exponential scaling of the Hilbert space (the space where qubits live) with the number of qubits allows us to encode and process larger datasets with significantly shorter training times. A quantum system consisting of $N$ qubits has a Hilbert space of size $2^N$, enabling quantum states to encode significantly larger amounts of data compared to classical systems. Quantum reservoirs are particularly beneficial when working with quantum data, such as molecular data. In such cases, since the data itself describes a quantum-mechanical phenomenon, it is logical to use quantum operations to identify its hidden patterns. Moreover, although quantum computers are susceptible to noise and decoherence, the simplicity of the training method for QRC makes it suitable for NISQ devices.

QRC has been employed to solve both temporal and non-temporal tasks. Regarding temporal tasks, QRC has mostly been used for time-series forecasting~\citep{time_series1,time_series2,OptQRC,time_Series4,time_series5, DynamicalIsing}. Regarding non-temporal tasks in the quantum domain, QRC is used in Ref.~\citep{QRC2} for the detection of entanglement and computation of associated quantities, which are challenging to measure accurately in experimental setups. Additionally, in Ref.~\citep{QuantumTomography}, quantum state tomography using a quantum reservoir has been developed. This method enables the reconstruction of the unknown density matrix of the input quantum state with only a single measurement on local observables of the reservoir nodes, without requiring correlation detection. Quantum reservoirs have also been used to compute the preparation of desired quantum states, such as anti-bunched and cat states in Ref.~\citep{StatePreparation1} or maximally entangled states, NOON, W, cluster, and discorded states in Ref.~\citep{StatePreparation2}. Finally, QRC has been used in the field of quantum chemistry to predict the first two excited molecular energies and transition dipole moments from the ground-state wavefunction of the molecule~\citep{quantumchemQRC, Domingo_optimalQRC, Domingo_QRCNoise}.

\section{Thesis contributions}
\label{sect:thesis}
The main objective of this thesis is to explore the fundamental principles of RC and create advanced variations of the algorithm that can effectively handle a wide range of applications in the field of ML. Through this research, the thesis showcases the algorithm's resilience and flexibility in adapting to highly diverse domains and tasks.

%%%%%%%%%%%%%%%%%%%%%%%%%%%%%%%%%%%%%%%%%%%%%%%%%%%%%%%%%%%
% RC for time series forecasting
%%%%%%%%%%%%%%%%%%%%%%%%%%%%%%%%%%%%%%%%%%%%%%%%%%%%%%%%%%%
The first contribution of this thesis is devoted to studying a critical agro-industrial problem: forecasting the prices of agricultural products~\citep{FW, scirep1,nfood,nfood2}. Forecasting changes in the agri-food market is crucial for ensuring food security and sustainability of the food system~\citep{dash}. However, this task is challenging as it involves analyzing highly volatile and small time series that are susceptible to external influences. The study presented here uses RC to predict the price dynamics of three agricultural products -zucchini, tomato, and aubergine- in the southeastern region of Spain, one of the largest suppliers to the European market~\citep{agri-food-Spain}. The results show that RC, in combination with a clever data decomposition in trend, seasonality and residuals, outperforms traditional time series forecasting methods~\citep{Domingo_zucchini}.

Moreover, in this thesis we discuss the best strategies to predict confidence intervals for the forecasted prices, which provides useful information about the uncertainty associated with the forecasts, as well as the inherent variability present in real-world situations. Then, we investigate how to generate multivariate RC models, that take into account the correlation between the prices of different agricultural products to enhance the predictions of the RC models.

%%%%%%%%%%%%%%%%%%%%%%%%%%%%%%%%%%%%%%%%%%%%%%%%%%%%%%%%%%%
% RC for quantum systems
%%%%%%%%%%%%%%%%%%%%%%%%%%%%%%%%%%%%%%%%%%%%%%%%%%%%%%%%%%%
In addition to exploring the benefits of RC in forecasting relevant time series, this thesis also demonstrates the versatility of the RC algorithm by addressing a fundamental problem in quantum mechanics: solving the Schrödinger equation, which completely determines the dynamics and properties of the quantum system.

In quantum chemistry, solving the time-independent Schrödinger equation is of central importance, as it allows calculating the energy levels and wavefunctions of molecules, which in turn provides insights into their electronic and geometric properties~\citep{molecular_book}. This information is critical for understanding and predicting the behavior of molecules in a wide range of applications, such as drug design~\citep{molecularsim1, molecularsim4, quantummechanical}, materials science~\citep{materials}, and catalysis~\citep{catalysis2, catalysis}. For example, the energy levels of a molecule provide information about its chemical reactivity, and the eigenfunctions provide information about its geometry, such as the bond lengths, bond angles, and dihedral angles~\citep{molecular_book2}.

 The most widespread method for solving the Schrödinger equation is based on the variational principle, whose resulting numerical method ~\citep{variational_original} is reduced to an eigenvalue problem. As a result, in order to approximate the $N$-th energy level, all the lower-lying $N-1$ levels also need to be calculated, which makes the task particularly difficult when one is interested in finding highly excited states. Moreover, this method needs to be adapted to the quantum system in the study. On the other hand, data-based methods such as ML do not rely on the underlying physical model of the system, so the same architecture can be used to study multiple Hamiltonians, thus providing a more versatile approach. 

 In the first project on this topic, we begin by studying how NNs, the precursors of RC, can be designed to predict the low and high energy states of multiple quantum systems~\citep{Domingo_DL}. The training set of the NN contains Hamiltonian with analytical solution, so that the training data generation is computationally efficient and convenient. Then, the network is asked to generalize to more complex systems by predicting the eigenfunctions of new Hamiltonians which do not have an analytical solution. The data generation process and learning algorithm are crucial for guaranteeing the good performance of the algorithm. By choosing the appropriate training data samples and introducing an additional loss function that ensures that the predicted wavefunction follows the time-independent Schrödinger equation, the resulting method can reproduce high-energy states of complex molecular Hamiltonians. 

While this strategy overcomes the problem present in the variational method, where the computation of the $N$-th excited state requires calculating the lower-lying $N-1$ states, this approach requires training one NN for every energy level. As a result, to obtain all the excited states within a certain energy range, a separate NN needs to be trained for each state within that range. To overcome this limitation and unravel the full potential of RC, we introduce a novel RC-based approach~\citep{Domingo_adaptingRC} that allows computing all the eigenfunctions in a certain energy range. The complexity of the problem is reduced to propagating initial wavepackets with time by solving the time-dependent Schrödinger equation. 

The proposed method uses a novel variant of the RC algorithm, called \emph{multi-step RC}, to propagate such state in time. This approach is computationally more efficient than using a classical numerical integrator, such as the widespread method proposed by Kosloff and Kosloff~\citep{kosloff}, or a NN approach such as an LSTM~\citep{LSTM_original}. Moreover, since RC is model agnostic, it is easily adaptable to multiple quantum systems. The proposed adaptation of RC to quantum systems reduces the chances of overfitting, thus providing a better generalization capacity. 

With the multi-step RC method, we are able to successfully propagate quantum states in time, recovering the energies and wavefunctions with high accuracy. The method is first evaluated with simple one-dimensional (1D) and two-dimensional (2D) systems, to show its advantages over the standard RC method and traditional approaches~\citep{Domingo_adaptingRC}. Then, the method is used to integrate more complex systems. The first one is described by the coupled Morse Hamiltonian~\citep{Domingo_Morse}, which has been extensively used to model the vibrational stretching of the H$_2$O molecule~\citep{H2O2,H2O, H20_NN}. The second one~\citep{Domingo_scars} is the coupled quartic oscillator~\citep{quartic1,quartic2}, which has been of great interest in the field of quantum chaos~\citep{Fabio2, Fabio1,Fabio3}.

 Notice that propagating quantum states in time is a very different task than forecasting agri-commodity prices. While the former requires dealing with complex-valued, high-dimensional data, the latter studies low-dimensional, real-valued, volatile time series, which are in turn highly influenced by external factors. Therefore, the fact that the RC algorithm can be adapted to solve both problems proves the vast power and versatility of the algorithm.

%%%%%%%%%%%%%%%%%%%%%%%%%%%%%%%%%%%%%%%%%%%%%%%%%%%%%%%%%%%
% Quantum reservoir computing
%%%%%%%%%%%%%%%%%%%%%%%%%%%%%%%%%%%%%%%%%%%%%%%%%%%%%%%%%%%

The last part of this thesis is devoted to the study of QRC, the quantum variant of the standard RC, for both temporal and non-temporal tasks. QRC uses a quantum circuit as a reservoir, instead of a NN, which allows the use of the properties of quantum mechanics to produce enhanced embeddings of the input data and accelerate the computations with high-dimensional data. This is of special importance in the advent of large and complex datasets, which poses a challenge for classical ML methods. 

 To achieve optimal performance in QML, the design of the quantum reservoir is crucial. Therefore, this thesis studies the relationship between the complexity of a family of quantum circuits and its performance in QRC. Our results show that the families with higher complexity, which is measured with the majorization indicator, are the most suitable to act as quantum reservoirs~\citep{Domingo_optimalQRC}. Thus, we propose an optimal family of quantum circuits, the G3 family, that serve as reservoirs. Such family requires fewer quantum gates than other commonly used reservoirs, such as the transverse-field Ising model~\citep{Ising}, and are easily implemented in NISQ devices. We also examine the effect of noise on the performance of QRC and show that the \emph{amplitude damping} noise can improve the performance of quantum reservoirs for small error rates, while the \emph{depolarizing} and \emph{phase damping} should be prioritized for correction~\citep{Domingo_QRCNoise}. 

Finally, we apply QRC to solve a significant drug design problem regarding target validation, which consists of predicting the binding affinity of a new biomolecule and its target proteins~\citep{Domingo_hybrid}. By predicting such binding affinity, it is possible to identify the most promising candidates from a large pool of compounds, reducing the number of molecules that need to be experimentally tested. The approach presented in this thesis consists of using a hybrid quantum-classical CNN for binding affinity prediction, where the first layer of the classical CNN is replaced by a layer of quantum reservoirs. The resulting model is able to reduce by 20\% the complexity of the classical network while maintaining optimal performance in the predictions. Additionally, this results in significant time savings of up to 40\% in the training process, providing meaningful speed up of the drug design process.

\section{Thesis organization}
The organization of this thesis is as follows. First, Chapter~\ref{chapter2} presents the theoretical concepts and ideas involving NNs, RC, quantum computing and QRC. 

Chapter~\ref{chapter4} is devoted to applying and adapting the RC algorithm to forecast the prices of agricultural products. In particular, in Sect.~\ref{paper:RC_calabacines} we show that RC, together with a clever decomposition of the data in components, can outperform traditional time series forecasting approaches. Section~\ref{sect:discussion_agro} presents the remaining challenges for achieving accurate and reliable price forecasting. In particular, we discuss multiple methodologies to perform confidence interval prediction instead of deterministic forecasting. Moreover, we discuss the importance of multivariate time series prediction, which considers the relationships between multiple products to enhance price forecasting models. 

In Chapter~\ref{chapter3}, we develop NN and, more importantly, RC methods to integrate the time-independent and the time-dependent Schrödinger equation. In particular, Sect.~\ref{sect:paper1} introduces a NN method to integrate the time-independent Schrödinger equation, which allows for the efficient calculation of high-energy states for multiple Hamiltonians. In Sect.~\ref{sect:paper2}, the RC algorithm is adapted to propagate an initial wavefunction through time, thus solving the time-dependent Schrödinger equation. This method is used and evaluated to study, in Sect.~\ref{sect:paper3}, the coupled Morse oscillator describing the stretching dynamics of the H$_2$O molecule and the coupled quartic oscillator, in Sect.~\ref{sect:paper4}, which is a prominent system in quantum chaos. 

In Chapter~\ref{chapter5}, the design and performance of QRC is evaluated. The first part of this chapter (Sect.~\ref{sect:optimalQRC}) is devoted to studying the optimal design of quantum reservoirs for non-temporal QML tasks. Then,  Sect.~\ref{sect:noise_QRC} provides the study of the effect of noise on the performance of QRC. The analysis of the optimal reservoir designs is extended to temporal tasks in Sect.~\ref{sect:QRC_forecasting}. Section~\ref{sect:hybrid_CNN} presents an application of quantum reservoirs to solve a drug design problem.  

Chapter~\ref{chapter6} outlines the conclusions and final remarks. Additionally, this thesis contains three appendices. In Appendix~\ref{appendix1}, the variational method used to compare the performance of the NN and RC algorithms in Chapter~\ref{chapter3} is described. Appendix~\ref{appendix2} introduces the canonical transformations used to rewrite the coupled Morse Hamiltonian in generalized coordinates so that the coupling appears in the spatial coordinates instead of in the momentum coordinates. Finally, Appendix~\ref{appendix3} provides the research publications associated with this thesis.
%!TEX root = ../thesis.tex
%*******************************************************************************
%****************************** Second Chapter *********************************
%*******************************************************************************
\renewcommand{\hbar}{\mathchar'26\mkern-9mu h}

\chapter{Theoretical concepts}
\label{chapter2}
\begin{fquote}[Robert Orben][1927-2023]
    To err is human — to blame it on a computer is even more so.
\end{fquote}

This chapter presents the theoretical basis used to develop this thesis and is divided into three main sections. First, in Sect.~\ref{sect:MLNN} we introduce the basic concepts of ML (Machine Learning) and NNs (Neural Networks). Second, Sec.~\ref{sect:RC} presents the details of RC (Reservoir Computing), a novel algorithm based on NNs used to propagate dynamical systems. Finally, Sect.~\ref{sect:QC} introduces the fundamentals of QML (Quantum Machine Learning), a novel technology that has the potential to improve classical ML algorithms.

  %%%%%%%%%%%%%%%%%%%%%%%%%%%%%%%%%%%%%%%%%%%%%%%%%%%%%%%%%%%%%
 % NEURAL NETWORKS
 %%%%%%%%%%%%%%%%%%%%%%%%%%%%%%%%%%%%%%%%%%%%%%%%%%%%%%%%%%%%%
 
\section{Types of machine learning tasks}
\label{sect:MLNN}
 ML is a field of artificial intelligence that enables computers to learn from data and improve their performance on specific tasks without being explicitly programmed. ML tasks are usually classified into three categories, depending on how learning is received and what feedback is given to such learning:

\begin{enumerate}
    \item \textbf{Supervised learning:} The algorithm is given data samples labelled with their desired output. The goal is to infer a relationship between the input and the associated output so that the ML model can learn to predict the labels of unseen data~\citep{MLbook2}. 
    \item \textbf{Unsupervised learning:} The algorithm is given data without labels, and the goal is to identify patterns and common trends between the samples~\citep{ML}.
    \item \textbf{Reinforcement learning: } The algorithm interacts with a dynamic environment in which it must achieve a particular goal. The program is given rewards and punishments so that the ML model adapts its strategy to improve its performance~\citep{RLBook}. 
\end{enumerate}

In this thesis, we will focus on studying supervised ML methods, since in all cases there will be a clear output to predict. Let us now introduce and describe a prominent algorithm in ML, the Neural Network (NN).

\section{Neural networks}
\label{sect:NNs}
 A NN~\citep{NN} is an information processing paradigm that is inspired by the way biological nervous systems process information. It is composed of a large number of interconnected \emph{neurons}, which work in unison to solve specific problems. A NN is considered to be \textit{deep} if it is formed by a large number of neuron layers. To use a NN for function approximation, it has to go through a training process, where examples are fed to the model, and the parameters are tuned to provide accurate predictions.

\begin{figure}[!ht]
    \centering
    \includegraphics[width=1.00\textwidth]{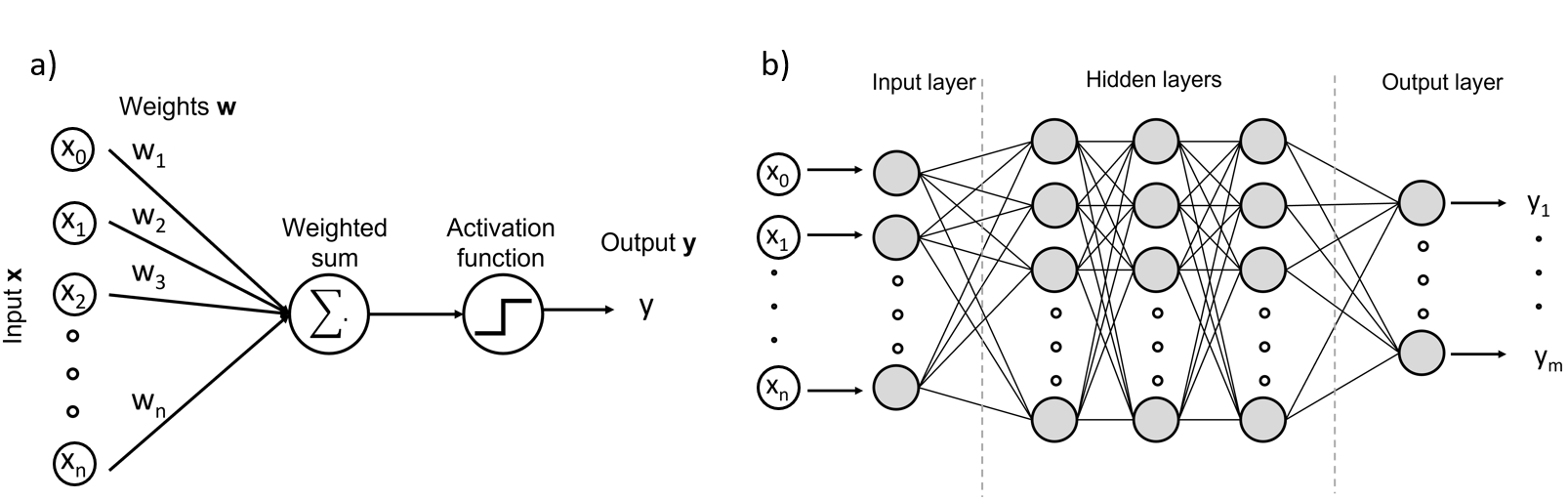}
    \caption{Representation of the architecture of a single-neuron network a) and a feed-forward neural network b).}
    \label{fig:NN}
\end{figure}

Figure~\ref{fig:NN} a) shows the elements of a NN with a single neuron. Such neuron can be described as a computational unit receiving input $\vec{x} = (x_0, x_1, x_2, \cdots, x_n)$, whose output is a real number $y$. Mathematically, a neuron $\Sigma$ can be described in the following way:
\begin{equation}
    \begin{split}
        & \Sigma(\vec{x}) = W \vec{x} + \vec{b}\\
        & y = \varphi(\Sigma(\vec{x})),
    \end{split}
    \label{eq:NN}
\end{equation}
where $\varphi$ is the activation function, $W$ is the matrix of weights, $\vec{b}$ is the bias vector, and $\vec{x}$ is the input vector. The weights $W$ determine the importance of each component $x_i$ of the input. In this way, a large value of $w_i$ means that the component $x_i$ contributes more significantly to the output of the neuron. After making a linear combination of the inputs, the resulting value is passed through an activation function $\varphi$, which provides the final output. The activation function decides whether the node is "activated" or not. The most common activation function is the Rectified Linear Unit (\gls{ReLU}), $\varphi(z) = \max(0,z)$. In this way, if $\Sigma(\vec{x})$ exceeds a given threshold (0 in this case), the function $\varphi$ activates the node, providing a non-zero output. 

Let us show a simple example of the training of a NN with only one neuron, that is, a Perceptron. In this example, the task of the NN is to convert from Celsius to Fahrenheit degrees. That is, given an input $x$ which represents the temperature in Celsius degrees, the network will learn to predict the temperature in Fahrenheit $y$. Notice that in this case, the dimension of both $x$ and $y$ is 1. In order to train the network, we need a training dataset, containing the input-output pairs. Consider the following training set: $\{(-40,-40), (-10,14), (0,32), (8,46), (15,59), (22,72), (38,100)\}$. The goal is to learn the best values of $W$ and $b$ in Eq.~\ref{eq:NN} that correctly reproduce the input-output relationship. For simplicity, a linear activation function $\varphi(x)=x$ is chosen. The first step is to randomly initialize the weight and bias values, for example with $W=0.1$ and $b=1$. Now, we can use the Perceptron update rule to learn the optimal values of the weight and bias in an iterative process. Given a number of epochs, the Perceptron learning algorithm is described in Algorithm~\ref{algorithm}, where $\eta$ is the learning rate, which controls the velocity of change of the weight and bias values.

\begin{algorithm}[!ht]
 \KwData{Training set $\{(x_i,y_i)\}_i^n$}
 \KwResult{Weight $W$ and bias $b$}
 Initialize $W\leftarrow 0.1$, $b\leftarrow 1$ \\
 \For{$j=0$ to num\_epochs}{
  \For{$i=0$ to $n$}{
    $\hat{y_i} = Wx_i + b$ \\
    Error =$y_i - \hat{y_i}$ \\
    $W \leftarrow W + \eta \times$ Error$ \times x_i$\\
    $b \leftarrow b + \eta \times$ Error
    }
 }
 \caption{Perceptron}
 \label{algorithm}
 \end{algorithm}
\begin{figure}[!ht]
    \centering
      \centering
      \includegraphics[width=0.75\textwidth]{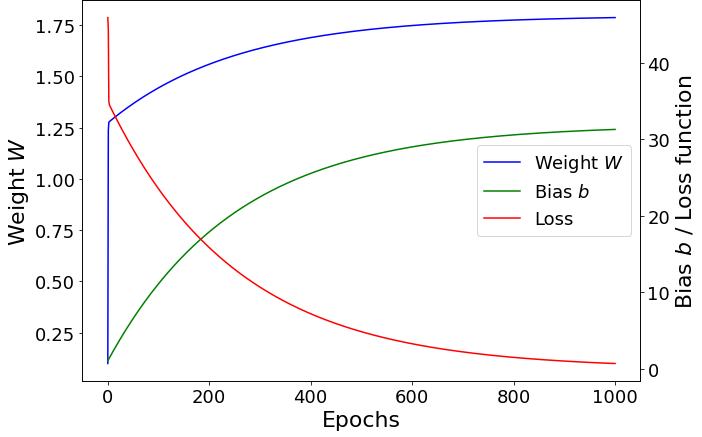}
      \caption{Training process of a Perceptron learning the conversion from Celsius to  Fahrenheit degrees.}
\label{fig:perceptron_learning}
\end{figure}

Figure~\ref{fig:perceptron_learning} shows the evolution of the weight and bias values for 1000 epochs and $\eta=0.0005$. As can be seen, the values of both $W$ and $b$ converge to a constant value as the number of epochs increases. Moreover, the loss function $\mathcal{L} = \sum_{i} |\hat{y}_i -y_i|$, which controls the error made in the predictions, decreases to zero as the training goes on. At the end of the training phase, the values have converged to values $W=1.79$ and $b=31.3$, which are quite similar to the actual relationship $W_\text{true} =1.8$, $b_\text{true} = 32$. In this way, the Perceptron network has learned to approximately change from Celsius to Fahrenheit degrees. If more precise results were needed, the number of epochs and the size of the training set should be increased.

The previous example shows how to solve a simple task using a NN with only one neuron. Obviously, in order to solve meaningful tasks, NNs with more than one neuron are needed. The way neurons are connected within the network determines the type of NN. In the following sections, we introduce the NN architectures used in this thesis.

\subsection{Feed-forward neural networks}
The most common NNs are \emph{feed-forward} networks or \emph{fully-connected} networks. In this case, neurons are organized in layers with connections between two adjacent layers. That is, all neurons in two adjacent layers are fully pairwise connected but neurons within the same layer share no connections. A representation of this network is shown in Fig.~\ref{fig:NN} b). Each of the neurons in Fig.~\ref{fig:NN} b) provides an output $y_j$ according to Eq.~\ref{eq:NN}. This output, together with other neurons' output $\{y_i\}$, will be passed to the next layer in the network. In this way, the output of one node becomes the input of another node in the next layer of neurons. 

The output layer behaves slightly differently from the other layers of the network. While the input and hidden layers aim to decide what information is passed to the next layer, the output neurons aim to provide the final prediction. Thus, their activation function depends on the task that the network aims to solve. For regression tasks, a linear activation function $\varphi(z) = z$ is usually used. For binary classification problems, a sigmoid activation function $\displaystyle\varphi(z) = \frac{1}{1 + e^{-z}}$ is usually chosen, which predicts the probability of belonging to a certain class. Finally, for a multi-class classification a softmax function $\displaystyle\varphi(z)_i = \frac{e^{z_i}}{\sum_{j=1}^K e^{z_j}}, \ i \in \{1, \cdots , K\}$ is used, where $K$ is the number of classes. Notice that in this case, $K$ neurons are needed, and each neuron will predict the probability of belonging to class $i$, for $i \in \{1, \cdots, K\}$. 

The number of layers and neurons in each layer are hyperparameters that must be tuned. If the NN is too shallow it may not be capable of adequately solving the problem, whereas too deep networks may not be able to generalize to new data due to their too high complexity.

\subsection{Convolutional neural networks}
\label{sect:CNN}
\begin{figure}[!ht]
    \centering
    \includegraphics[width=0.95\textwidth]{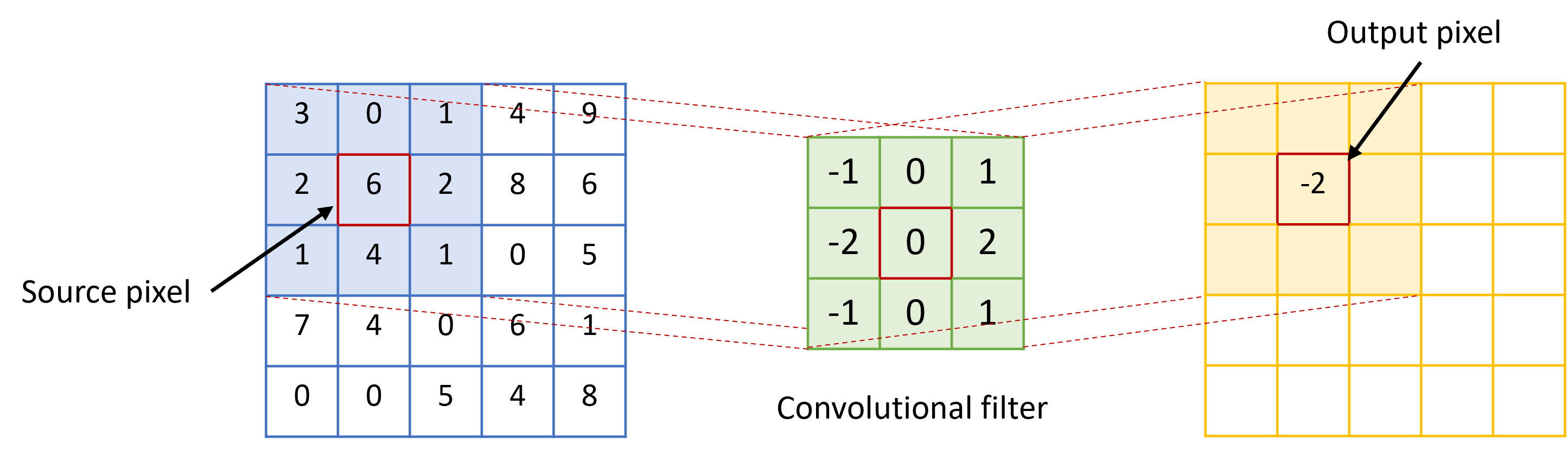}
    \caption{Example of a convolutional layer.}
    \label{fig:convolutional}
\end{figure}

A type of deep network that has proven to produce very successful results in deep learning is the Convolutional Neural Network (CNN)~\citep{CNN}. This type of network is specialized in processing high-dimensional data in the form of spatial arrays, such as time series (in 1D), images (in 2D) or volumes (in 3D). The name stems from the fact that instead of general matrix multiplication, it employs a mathematical convolution in at least one of its layers. Mathematically, given a convolution \emph{kernel} $K$, also called \emph{filter}, represented by a $(M \times N)$ array, the convolution of an array $A$ with $K$ is:
\begin{equation}
   h(x,y) = (A \otimes K)(x,y) = \sum_{m=0}^{M-1} \sum_{n=1}^{N-1} K(m,n) A(x-n, y-m).
   \label{eq:CNN}
\end{equation}
The output of the convolution is another array that might represent some kind of information that was present in the initial array in a very subtle way. In this way, a filter of a CNN is responsible for detecting one feature of the input. The kernel matrix $K$ contains free parameters that must be learned to perform the optimal feature extraction. An example of convolution is shown in Figure~\ref{fig:convolutional}.

\begin{figure}[H]
    \centering
    \includegraphics[width=0.6\textwidth]{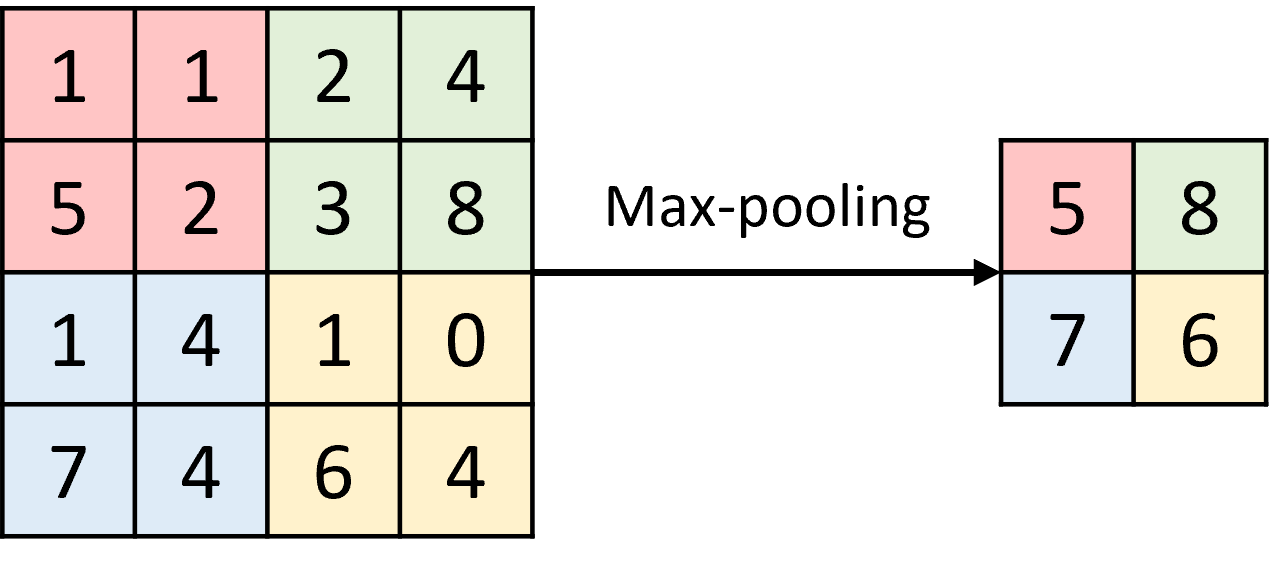}
    \caption{Example of a max-pooling layer.}
    \label{fig:max-pooling}
\end{figure}

Another hyperparameter of the network is the stride size with which the kernel slides over the input. Typically the convolutional operation is followed by a nonlinear activation function which adds non-linearity to the system. In the last step, a pooling layer is added to progressively reduce the spatial size of the array. The usual pooling functions are max pooling and average pooling, which are applied to the output of the filter convolution. An example of a max-pooling layer is shown in Figure~\ref{fig:max-pooling}. After a series of convolutional and pooling layers, a flattening layer and some fully-connected layers are used to combine the extracted features and predict the final output. Notice that Eq.~\ref{eq:CNN} can easily be generalized to 3D or 1D data so that CNNs can also be used to analyze volumes or series. 

\subsection{Recurrent neural networks}
\label{sect:RNN}

The two types of NNs that we have seen so far assume that all data samples are independent of each other. However, this is not true for sequential data, such as time series or text, where you require information from past data to make accurate future predictions. For example, when reading this text, one understands the meaning of each word based on the understanding of the previous words. Traditional networks do not have the memory of past samples and thus are not suitable to deal with this kind of data. Recurrent Neural Networks (RNN)~\citep{RNN} address this issue by introducing loops within the network, allowing information from prior inputs to influence the current output. 
\begin{figure}[H]
    \centering
    \includegraphics[width=0.8\textwidth]{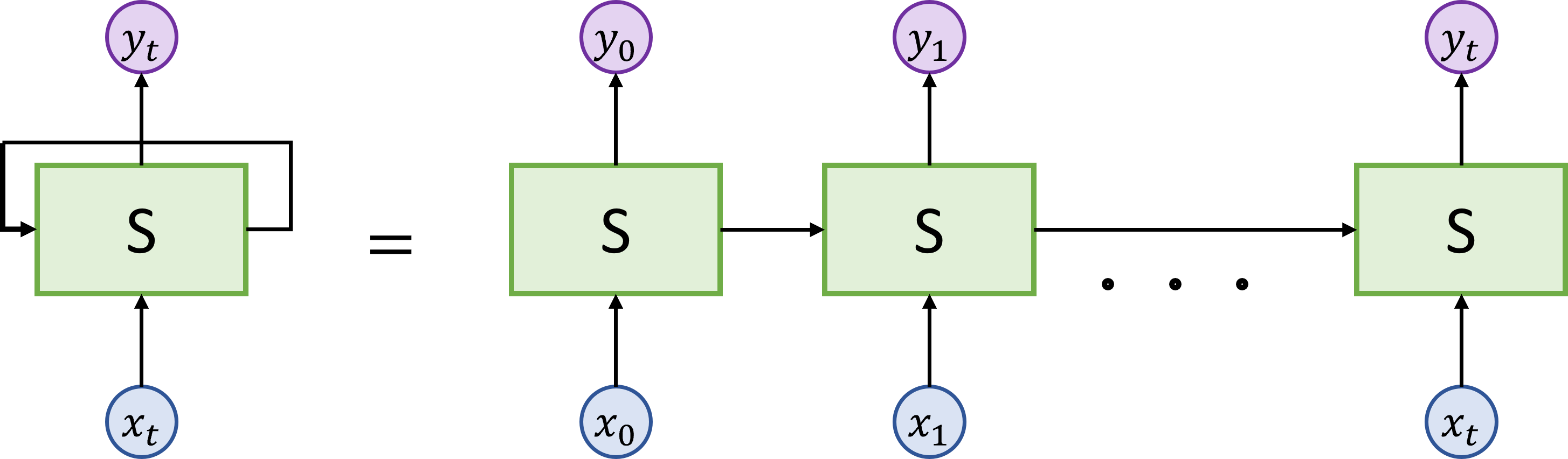}
    \caption{Schematic representation of a Recurrent Neural Network.}
    \label{fig:RNN}
\end{figure}
A representation of the architecture of a RNN is shown in Fig.~\ref{fig:RNN}. The network S takes the input $x_t$ and outputs the prediction $y_t$. The loop within the network allows passing information from $x_t$ to make predictions for $y_{t+1}, y_{t+2}, \cdots$. Even though it may seem hard to implement a loop within a NN, in reality it is nothing more than several copies of the same network, each passing a message to its successor. Thus, the RNN becomes a chain of networks, which is a very natural architecture for analyzing sequential data. Notice that each network in the chain is a copy of the previous one, so they all share the same architecture and training parameters. Mathematically,
\begin{equation}
\begin{split}
    & \vec{s}_t = \tanh(W_s \vec{s}_{t-1} + W_x  \vec{x}_t + \vec{b}_s)\\
    & \vec{y}_t = W_y \vec{s}_t,
\end{split}
    \label{eq:vanilla_RNN}
\end{equation}
where $\vec{s_t}$ is the internal state of the network and $W_s, \vec{b_s}, W_x$ and $W_y$ are the training matrices.

There are many types of RNNs, that slightly modify the original architecture. One of the most popular types of RNNs is the Long-Short Term Memory (LSTM) network~\citep{LSTM_original}. LSTMs aim to solve the problem of long-term dependencies. In theory, the original architecture of RNNs should allow the network to remember information from the distant past. However, in practice, the networks depicted in Fig.~\ref{fig:RNN} fail to capture long-term relationships. For example, imagine we wanted to predict the italicized word in the following sentence: "I grew up in Spain. Therefore, I speak fluent \emph{Spanish}". The context of growing up in Spain allows us to infer that the spoken language is Spanish. However, if that context was a few sentences prior, it would be difficult, or even impossible, for the RNN in Eq.~\ref{eq:vanilla_RNN} to connect the information. LSTMs, on the other hand, are designed to keep knowledge from the distant past, overcoming this problem.

\begin{figure}[H]
     \centering
     \includegraphics[width=1.0\textwidth]{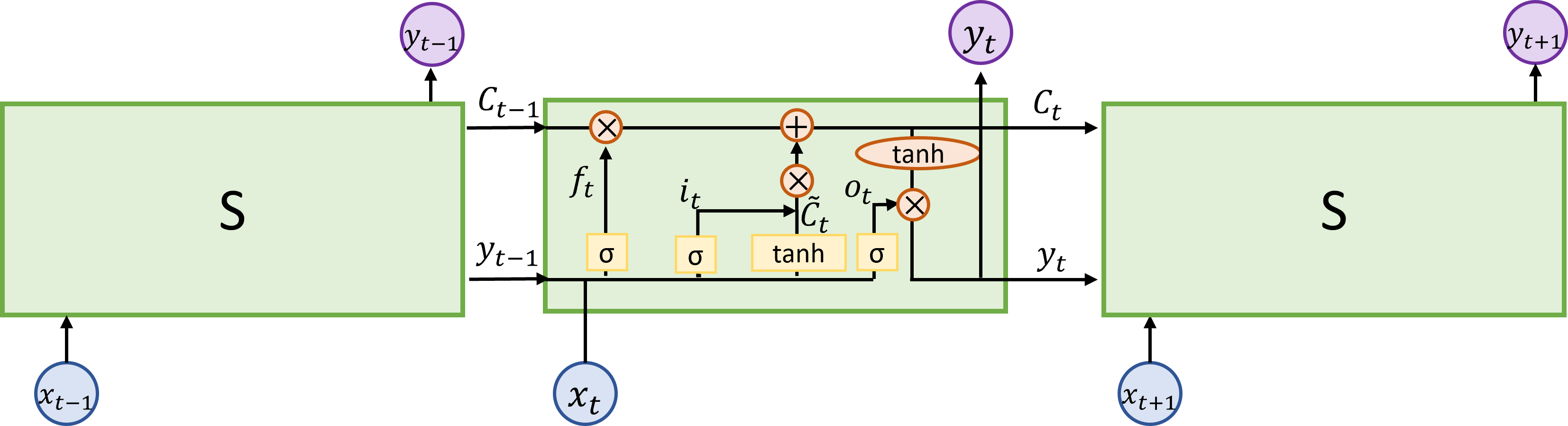}
     \caption{Architecture of a Long Short-Term Memory network.}
     \label{fig:LSTM}
 \end{figure}

The architecture of a LSTM is depicted in Fig.~\ref{fig:LSTM}.
In this figure, each line represents a vector which carries information from the output of one layer to the input of the next one. The orange circles represent pointwise operations and the yellow boxes are network layers. Merging lines denote concatenation while line forking denotes making copies of the vector representing that line. 
Let us explain in detail the architecture of a LSTM. The most important part of the network is the \emph{cell state} $C_t$, which corresponds to the horizontal line at the top of the diagram. This vector represents the memory gained from past inputs. When new input arrives, the cell state can be modified by using three gates: the \textit{forget gate}, the \textit{input gate} and the \textit{output gate}. The forget gate decides which information is going to be erased from the cell state. It analyzes the previous output $ \vec{y}_{t-1}$ and the current input $ \vec{x}_t$ and outputs a vector $ \vec{f}_t$ with the same dimension as $ \vec{C}_t$ with numbers between 0 and 1. A 0 in the $i$-th position of $ \vec{f}_t$ means that the information in the $i$-th position of $ \vec{C}_{t-1}$ should be completely forgotten, and a 1 means that it should be entirely kept. Mathematically,
 \begin{equation}
     \vec{f_t} = \sigma(W_f  [\vec{y}_{t-1}, \vec{x}_t] +  \vec{b}_f),
 \end{equation}
 where $\sigma$ is the sigmoid function which scales the output between 0 and 1, and $W_f$ and $ \vec{b}_f$ are the weights and biases of the forget gate. 
 
 The next step of the algorithm is the input gate, which decides what information from the new input $ \vec{x}_t$ should be stored in the cell state. The vector $ \vec{i}_t$ determines which components of the input are relevant to be stored in $ \vec{C}_t$, by using a sigmoid function just as the forget gate
 \begin{equation}
      \vec{i}_t = \sigma(W_i[ \vec{y}_{t-1},  \vec{x}_t] +  \vec{b}_i).
 \end{equation}
 Then, a new vector $ \vec{\tilde{C}}_t$ is calculated, which contains the information from each of the input components that will be added to the new state $ \vec{C}_t$
 \begin{equation}
      \vec{\tilde{C}}_t = \sigma(W_C [ \vec{y}_{t-1},  \vec{x}_t] +  \vec{b}_C).
 \end{equation} 
To obtain the final state $ \vec{C}_t$, we multiply the forget gate by the previous state $ \vec{C}_{t-1}$ and add the information we want to remember from the current input $ \vec{i}_t *  \vec{C}_t$
\begin{equation}
     \vec{C}_t =  \vec{f}_t *  \vec{C}_{t-1} +  \vec{i}_t *  \vec{\tilde{C}}_t,
\end{equation}
where $*$ denotes pointwise multiplication. The last gate of the algorithm is the \emph{output} gate, which decides how to use the information in $\vec{C}_t$ to make a prediction $\vec{y}_t$. First, a sigmoid layer decides which parts of the cell state are going to be used as an output. Then, the cell state is passed through a $\tanh$ function to map the values to $[-1,1]$. Finally, we multiply it by the sigmoid layer to output only the chosen information
\begin{equation}
\begin{split}
    & \vec{o}_t = \sigma(W_o [ \vec{y}_{t-1}, x_t] +  \vec{b}_o), \\
    &  \vec{y}_t =  \vec{o}_t * \tanh( \vec{C}_t).
\end{split}
\end{equation}
This architecture allows us to make useful long-term predictions. However, it requires learning many parameters during training, which usually requires large datasets and long training times. The steps used to train NNs are presented in the following section.

\subsection{Training neural networks}
\label{sect:train_NN}
 Finding the optimal parameters of a NN can be formulated as an optimization problem, which tries to minimize a loss function. A loss function $\mathcal{L}(y, f(\vec{x}))$ $= \frac{1}{n} \sum_i \ell(y_i, f(x_i))$ represents the price paid for the inaccuracy of predictions in solving a specific problem. Common loss functions are Mean Squared Error (\gls{MSE}) or Mean Absolute Error (\gls{MAE}) for regression tasks, and logistic loss and cross-entropy loss for classification tasks. For more complex tasks, a customized loss function can be used. For example, in Chapter~\ref{chapter3}, we will design a loss function that ensures that the output of the network is a solution to a quantum-mechanical problem.
 
In order to minimize a loss function $\mathcal{L}$, many approaches have been proposed. The most popular method is called gradient descent~\citep{ruder2016overview}, which updates the weight parameters in the direction of the negative gradient of the loss function. This direction is the one with the highest decrease in the loss function. Starting from a random point $\vec{x}_0$, the gradient descent method iteratively updates the point $\vec{x}_t$ by:
\begin{equation}
    \vec{x}_t = \vec{x}_{t-1} - \eta \nabla \mathcal{L}(\vec{x}_{t-1}),
    \label{eq:gradient_decent}
\end{equation}
where $\eta$ is the learning rate, which controls the step size of the learning algorithm. If $\eta$ is too small, the algorithm will take too long to converge to the minimum. On the other hand, if $\eta$ is too big, the algorithm may jump around the optimal point or even diverge completely. Equation~\ref{eq:gradient_decent} is repeated until the termination condition is met, which is either a prefixed number of iterations or that $|\vec{x}_t - \vec{x}_{t-1}|$ is smaller than some chosen tolerance. Notice that the performance of the algorithm is only guaranteed for convex functions. For non-convex functions, the algorithm may converge to a local minimum, which is not the global solution to the problem. Variations of this algorithm have been proposed, the most popular being the Adam optimization~\citep{Adam}, which takes into account the gradient of previous steps and adapts the learning rate independently to every dimension in space.

The algorithm for calculating the gradients of the network is called backpropagation. The backpropagation algorithm~\citep{backprop} starts from the output layer calculating the gradients of the previous layer until the input
layer. In general, the gradients are calculated based on the multivariate chain rule throughout the computational graph. The computation of derivatives is done by using \textit{Automatic differentiation}~\citep{ baydin2015automatic}, which systematically applies the chain rule at the elementary operator level. Automatic differentiation relies on the fact that all numerical computations are ultimately compositions of a finite set of elementary operations for which derivatives are known. For this reason, given a library of derivatives of all elementary functions in a deep network, it is possible to compute the derivatives of the network with respect to all parameters at machine precision and apply gradient descent methods to its training. Without automatic differentiation, the design and debugging of optimization processes for complex NNs with millions of parameters would be impossible. 

For RNNs, the backpropagation algorithm is applied to the unrolled network (which is depicted in the right part of Fig.~\ref{fig:RNN}). That is, for a given sequence of inputs and outputs, the gradient is calculated for each input-output pair. Then, the gradients at each time step are aggregated to provide the total gradient. Notice that this step is not necessary for other types of networks, since they do not share weights among layers. This variation of the algorithm is called backpropagation through time. 

The gradient computation for RNNs requires a recurrent multiplication by the weight matrices of the network $W_s,W_x$ and $W_y$ (see Eq.~\ref{eq:vanilla_RNN}). If the norm of the weight matrices is smaller than~1, continuous multiplication by such matrices may cause the gradient to approach 0, which is known as the \emph{vanishing gradient} problem~\citep{vanishing}. On the other hand, if the norm of the derivatives is larger than~1, gradients may diverge, leading to the opposite effect, the \emph{exploding gradients} problem. The LSTM design mitigates this inconvenience by using the cell state and the three gates that control the flow of information throughout the network.

 %%%%%%%%%%%%%%%%%%%%%%%%%%%%%%%%%%%%%%%%%%%%%%%%%%%%%%%%%%%%%
 % RESERVOIR COMPUTING
 %%%%%%%%%%%%%%%%%%%%%%%%%%%%%%%%%%%%%%%%%%%%%%%%%%%%%%%%%%%%%

\section{Reservoir computing}
\label{sect:RC}
So far, we have discussed how RNNs can be used to predict sequential data, such as text or time series. Within this context, RNNs have proven to give optimal results, due to their ability to keep the knowledge learned from past data. However, RNNs are known to be computationally expensive during training, and such training may be complicated due to the problem of exploding or vanishing gradients during backpropagation. For this reason, a novel technique was proposed, called Reservoir Computing (RC), devoted to solving these problems. 

The concept of RC was introduced simultaneously as Echo State Networks~\citep{ENS} in the field of ML, and as Liquid State Machines~\citep{LSM} in computational neuroscience. In the RC framework, the learning complexity of the problem is reduced to performing a linear regression. The key is to design a dynamical system (the internal network) which learns to reproduce the input-output dynamics. Jaeger~\citep{ENS, ESN2} introduced the notation of \emph{echo states}, which refers to the network dynamics being unequivocally determined by the input and output. In this case, the nodes of the network can be thought of as an \emph{echo} of their past. 

In this section, we present the details of the RC algorithm based on echo state formalism. We begin by presenting the topology of the algorithm. Then, the echo state properties are introduced, together with an illustrative example. Next, the vanilla training process is explained, followed by some popular variations on the standard method. We end this section with a comparison with NNs.

\subsection{Network topology}

\begin{figure}[!ht]
    \centering
    \includegraphics[width=0.99\textwidth]{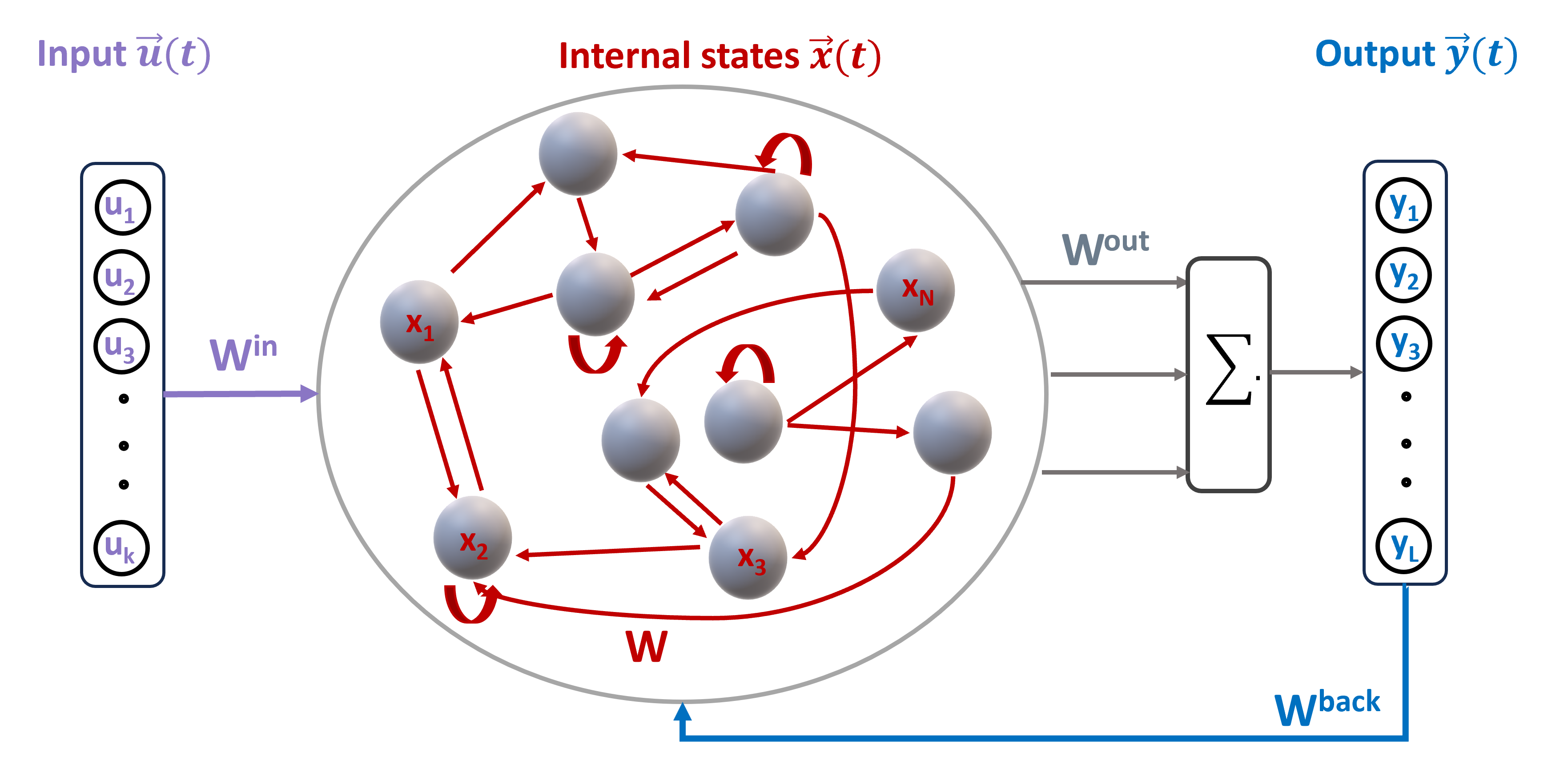}
    \caption{Architecture of a reservoir computing model.}
    \label{fig:reservoir}
\end{figure}

The network associated with the RC algorithm is composed of a $K$-dimensional input $ \vec{u}$, a $N$-dimensional internal network $\vec{x}$ and a $L$-dimensional output $\vec{y}$ (see Fig.~\ref{fig:reservoir}). The nodes of the internal network are called \emph{internal states} or simply \emph{states}. These states are updated at each training step, according to the input and output values, as we will see shortly. That is,

\begin{itemize}
    \item $\vec{u}(t)=(u_1(t),u_2(t),\ldots,u_K(t))$ is a $K$-dimensional vector giving the inputs 
    at time $t$,
    \item $\vec{x}(t)=(x_1(t),x_2(t),\ldots,x_N(t))$ is an $N$-dimensional vector giving the internal 
    states at time $t$, and
    \item $\vec{y}(t)=(y_1(t),y_2(t),\ldots,y_L(t))$ is an $L$-dimensional vector giving the outputs at time $t$.
\end{itemize}
Notice that $t$ takes discrete values. Also, the series $\vec{y}(t)$ refers to the predicted output of the internal network, which is not necessarily equal to the expected output of the system, which is denoted by $\vec{y}_\text{teach}(t)$.  Figure~\ref{fig:reservoir} shows a schematic representation of the reservoir architecture, which contains the following elements  
\begin{itemize}
    \item\textbf{The input connections}: $W^{\text{in}} \in M(\mathbb{R})_{N\times K}$ gives the weights from the input units to the internal states.
    \item \textbf{The internal connections}: $W \in M(\mathbb{R})_{N\times N}$ gives the weights between the different internal states. This network is usually referred to as the \emph{reservoir}.
    \item \textbf{The feedback connections}: $W^{\text{back}}\in M(\mathbb{R})_{N\times L}$ gives the weights from the output units to the internal states.
    \item \textbf{The output connections}: $W^{\text{out}}\in M(\mathbb{R})_{L\times K+N+L}$ gives the wights from the inputs, internal states and outputs to the output units:
        \begin{itemize}
            \item $W_{i,j}^\text{out}$ for $1 \leq j \leq K $ is the weight from the $j$-th input to the $i$-th output.
            \item $W_{i,j}^\text{out}$ for $K \leq j \leq K +N$ is the weight from the $j$-th internal state to the $i$-th output.
            \item $W_{i,j}^\text{out}$ for $K + N \leq j \leq K +N + L$ is the weight from the $j$-th output  to the $i$-th output.
        \end{itemize}
\end{itemize}
With this definition, all types of connections between units are allowed. In particular, the network can have connections from the input units to the output units and from some output units to other output units. Even though all these connections are possible, it is usually considered in the literature that the output units are only connected to the internal states. In this thesis, we will consider this approach unless stated otherwise, as depicted in Fig.~\ref{fig:reservoir}. Notice that the output connections matrix is then $W^{\text{out}}\in M(\mathbb{R})_{L\times N}$. Matrices $W^{\text{in}}$, $W$ and $W^{\text{back}}$ are fixed, so that they do not change during training. The only learnable parameters are the weights $W^{\text{out}}$. 

Once we have defined the elements of the network, we can introduce the dynamics of the internal states, which evolve according to the following expression 
\begin{equation}
    \vec{x}(t) = f(W^{\text{in}} \vec{u}(t)+W\vec{x}(t-1)+W^{\text{back}}\vec{y}(t-1)),
    \label{eq:update_x}
\end{equation}
where $f$ is the activation function, usually $f(\cdot)=\tanh(\cdot)$, which is applied component-wise. Since the evolution of the internal states is a time series, we usually talk about the dynamics of the nodes referring to the evolution of the internal states. The output of the network is calculated by
\begin{equation}
        \vec{y}(t) = f^{\text{out}}\Big[W^{\text{out}} \vec{x}(t)\Big],
        \label{eq:RC_out}
\end{equation}
with $f^{\text{out}}$ being an activation function, usually $f^{\text{out}}(\cdot) = id(\cdot)$. To apply Eq.~\ref{eq:RC_out} we need to know the value of $W^{\text{out}}$, which is determined during the training process. Before introducing the steps of the training algorithm, let us consider the properties that the nodes' dynamics must follow to achieve accurate predictions, i.e. the echo state properties.

\subsection{Echo states properties}
\label{sect:echo_states}
The performance of RC depends on the ability of the reservoir network to learn the input-output dynamics. Jaeger~\citep{ENS} introduced a series of properties that the internal network must follow: the existence of echo states. Formal definitions and proofs for these concepts are provided in Ref.~\citep{ENS}. In this section, we provide an informal explanation of the echo state properties, illustrated with some examples to give an intuitive view of how the reservoir network should behave.

\begin{definition}[Echo states]
\label{def:ES}
Assume that the inputs, outputs and internal states belong to compact sets ($\{\vec{u}(t)\}_t \in U^\text{in}$, $\{\vec{y}(t)\}_t \in U^\text{out}$, $\{\vec{x}(t)\}_t \in A$ with $U^\text{in}$, $U^\text{out}$ and $A$ compact sets). Then, the network has echo states if the network internal states $\vec{x}(t)$ are uniquely determined by the input and output sequences. That is, if there exists a function $E:U^\text{in} \times U^\text{out} \longrightarrow \mathbb{R}^N$ such that
\begin{equation}
    \vec{x}(t) = E\Big(\cdots, \vec{u}(t-1), \vec{u}(t), \ \cdots, \vec{y}(t-1), \vec{y}(t)\Big).
\end{equation}
\end{definition}
This definition requires that the internal states' dynamics are completely determined by the input-output evolution. Therefore, the internal states are an \emph{echo} of their past and are thus named \emph{echo states}. Notice that, according to this definition, all the units of the network are bounded. Otherwise, information would be lost when applying the activation function, since the hyperbolic tangent function collapses large values to $\pm 1$. Moreover, the series needs to belong to closed sets, since, otherwise, the network prediction may converge to values that are outside the set and are thus unattainable. In practice, given a reservoir network, it is very hard to find the function $E$ needed to fulfill Def.~\ref{def:ES}. Luckily, there exist some properties that guarantee the existence of echo states.
\begin{itemize}
    \item \textbf{State forgetting property:} For large enough $t$, the values of the internal states $\vec{x}(t)$ do not depend on the initial values $\vec{x}(0)$. 
    \item \textbf{Input/Output forgetting property:} For $t$ large enough, the values of the internal states $\vec{x}(t)$ do not depend on the initial values of the input $\vec{u}(0)$ or output $\vec{y}(0)$. 
\end{itemize}
These properties imply that regardless of the initial value of the internal states, inputs, or outputs, the reservoir's trajectories will eventually converge to the same dynamics.  As a result, the RC algorithm demonstrates robustness to the system's initial conditions and learns solely from the input-output relationships, offering a valuable collection of diverse trajectories used to predict the behavior of the target system. Moreover, these two properties are \emph{equivalent} to Def.~\ref{def:ES}, that is, if the network has echo states then the state forgetting and input forgetting properties are fulfilled. Fulfilling these two properties also ensures the existence of echo states.

\begin{figure}[!ht]
    \centering
    \includegraphics[width=1.0\textwidth]{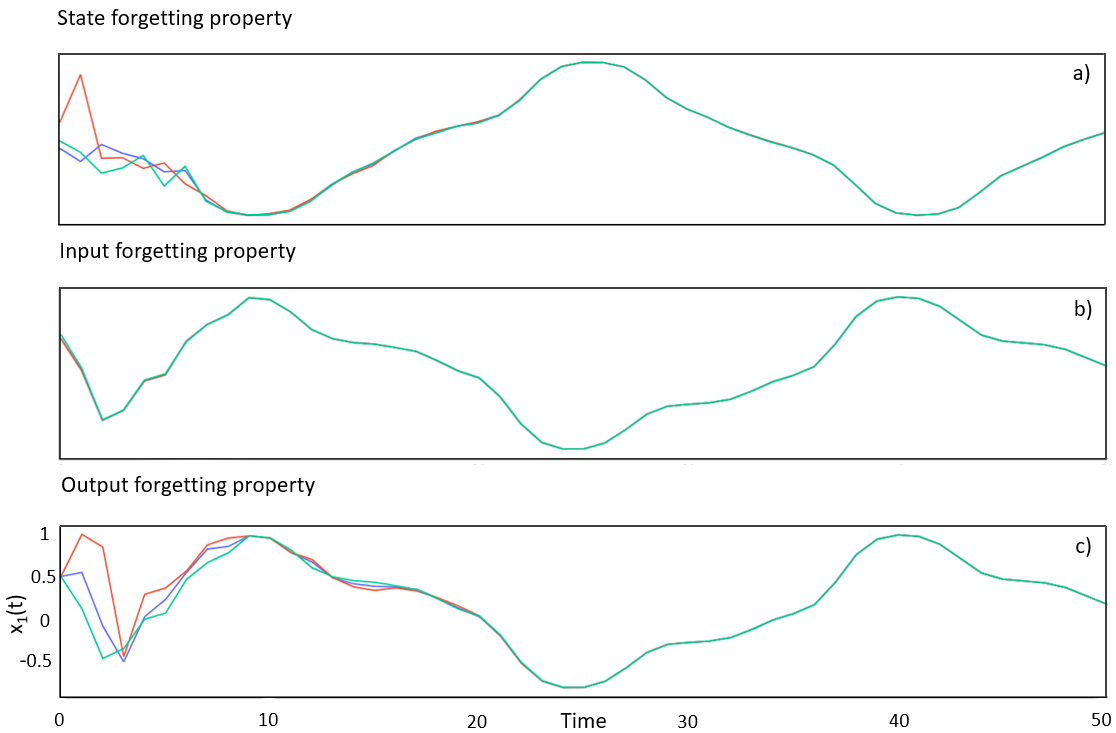}
    \caption[Example of echo state network following the state, input and output forgetting properties.]{Example of echo state network following the state, input and output forgetting properties. The plots show an example of the time evolution for one of the internal states $x_1(t)$ for different initial conditions (see main text) of the nodes (panel a)), inputs (panel b)) and outputs (panel c)).  }
    \label{fig:RC_forgetting}
\end{figure}

Let us illustrate these properties by evolving the internal states with a toy example. The input series corresponds to a periodic time series with period $T=\pi/10$
\begin{equation}
    u(t) = \sin(t/5),
\end{equation}
and we wish to learn the output
\begin{equation}
    y_\text{teach}(t) = \frac{1}{2}\sin^7(t/5).
\end{equation}
The state-forgetting property implies that if we perform two different initializations of the network but evolve the states with the same input and output, both sequences of internal states will end up converging. The influence of the initial states fades out with time, thus the reservoir dynamics do not depend on the initialization of the network. In our example, we initialize the internal states in three different ways 
\begin{equation}
    1. \ \ \vec{x}(0) = (0,\cdots, 0) \quad 2. \ \ \vec{x}(0) = (1,\cdots, 1) \quad 3. \ \ \vec{x}(0) \sim U(-0.5,0.5)^N.
\end{equation}
Figure~\ref{fig:RC_forgetting} a) shows the dynamics of the one internal state for the three cases. Even though the beginning of the internal dynamics is different for the three internal states, their dynamics converge with time. 
The input/output forgetting property states that recent values of the input (and output) will have more influence on the internal state dynamics than older inputs and outputs. Therefore, two different inputs (outputs) that coincide after some steps will produce the same sequence of internal states. To test this property in our example we evolve the internal states initializing the input in three different ways 
\begin{equation}
    1. \ \ \vec{u}(0) = \sin(0/5) = 0 \quad 2. \ \ \vec{u}(0) = 1 \quad 3. \ \ \vec{u}(0) \sim U(-0.5,0.5).
\end{equation}
Notice that the first initial input is the true value of the input. Fig.~\ref{fig:RC_forgetting} b) shows the evolution of the internal states with different values of inputs. We see that the nodes' dynamics converge even faster than in Fig.~\ref{fig:RC_forgetting} a).

Finally, we test the output forgetting property, by selecting three different initial outputs 
\begin{equation}
    1. \ \ y_\text{teach}(0) = 1/2\sin^7(0/5) = 0 \quad 2. \ \ y_\text{teach}(0) = 1 \quad 3. \ \ y_\text{teach}(0) \sim U(-0.5,0.5).
\end{equation}
The node dynamics are shown in Fig.~\ref{fig:RC_forgetting} c). Again, we see that the echo state network follows the output forgetting property since all the nodes converge to the same dynamics. We conclude that the evolution of the internal states after a certain time solely depends on the input-output dynamics.

Visualizing the state forgetting and input forgetting properties helps us guess if the reservoir network has echo states. However, it is still hard to verify these properties exactly. Predicting whether a given network architecture will provide echo states is still an open question. However, some sufficient conditions guarantee the existence or non-existence of echo states 

\begin{enumerate}
    \item If the largest singular value of $W$ is $\sigma_\text{max} <1$, then for all inputs $\vec{u}(t)$, outputs $\vec{y}(t)$ and states $\vec{x}(t) \in [-1,1]^N$, the network contains echo states.
    \item If the spectral radius of $W$ is $|\lambda_\text{max}|>1$, the network has no echo states for any input and output sets $U^\text{in} \times U^\text{out}$ containing 0 and internal state set $A = [-1,1]^N$. 
\end{enumerate}
These propositions give us an idea of how to design the internal network $W$. However, they are too restrictive, since there exist networks which have echo states when $W$ has singular values greater than 1. In these cases, the input and outputs are not 0 at any time step, thus the proposition is not contradicted. The spectral radius of $W$ is related to the memory of the reservoir. For large values of $\lambda_\text{max}$, the reservoir has a longer memory. However, if $|\lambda_\text{max}|>>1$, due to the previous proposition, the reservoir may not have echo states. 

We have defined the echo state properties and discussed how we can test if the internal network follows these properties. A natural question to ask is how these properties help the model predict future outputs. Let us assume we have a time series $\vec{y}_\text{teach}$ that we want our network to predict. That is, the network prediction series $\vec{y}$ should be as close as possible to $\vec{y}_\text{teach}$. After defining the matrices $W^\text{in}$, $W^\text{back}$ and $W$, the internal states of the network would evolve obtaining $\{\vec{x}(t\_\min+1 ), \cdots \vec{x}(T)\}$, regardless of the initialization of the states, inputs and outputs. Notice that the first $t\_\min$ values of $\vec{x}$, $\vec{u}$ and $\vec{y}$ need to be discarded, since they can be influenced by the initial values of the network. 

\begin{figure}[!ht]
    \centering
    \includegraphics[width=1.0\textwidth]{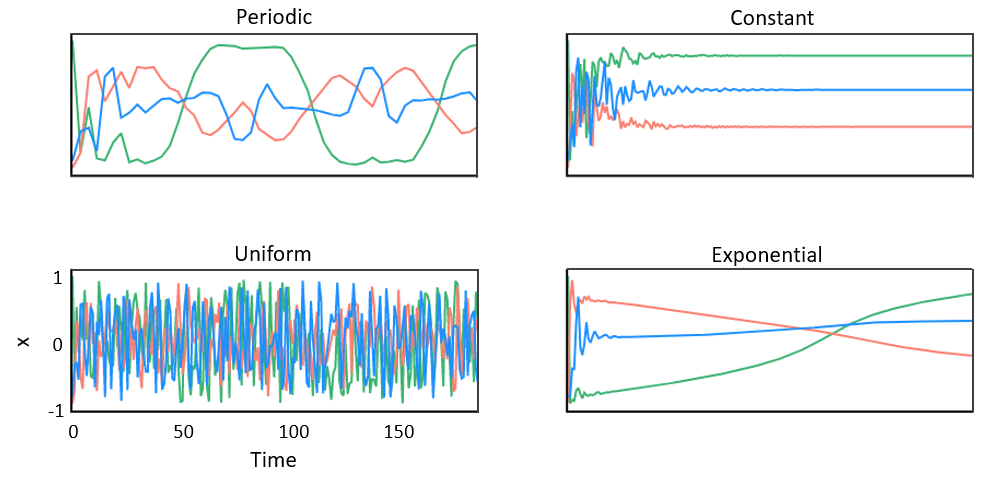}
    \caption{Example of the evolution of the internal state for four different input signals: periodic input, constant input, uniformly distributed input and exponential input. The curves show the trajectories of different internal states (see Eq.~\ref{eq:update_x}) for each reservoir.  }
    \label{fig:RC_inputs}
\end{figure}

The goal is to obtain a wide variety of trajectories $\vec{x}$ that span a subspace that contains the desired trajectory $\vec{y}_\text{teach}$. In this way, making linear combinations of such trajectories will be enough to provide an accurate prediction $\vec{y}(t)$. The matrices $W^\text{in}$, $W^\text{back}$ and $W$ need to be chosen appropriately so that the internal states' trajectories span the desired subspace. As an example, Fig.~\ref{fig:RC_inputs} shows the evolution of the network states for different series $\vec{y}_\text{teach}$. As can be seen, the nodes' evolution resembles the behavior of the input series. For example, when the input series is a periodic signal, the internal states of the reservoir also have a periodic behavior. On the other hand, when the input signal has an exponential behavior, the internal states also present such exponential evolution. In this way, the internal states are learning the possible dynamics of the input so that the network can make future predictions just by combining its internal trajectories. This is the key to learning the dynamics of $\vec{y}_\text{teach}$: obtaining a wide variety of trajectories so that the internal states are a \emph{reservoir of dynamics}. 

 \subsection{Training process}
 \label{sect:training_RC}
The echo state properties provide intuition on how the network should evolve to learn the input-output dynamics. In this section, we present the training and prediction steps for RC. Equations~\ref{eq:update_x} and~\ref{eq:RC_out} already introduce the formulas to calculate the dynamics of the network. However, these formulas need to be slightly modified during the training phase. The steps to train a RC model are the following

\begin{enumerate}
    \item \textbf{Generate the reservoir matrices $W, W^{\text{in}}, W^{\text{back}}$ randomly} \\
    The matrix $W$ should be sparse and in-homogeneous so that the internal states contain a diverse set of trajectories. $W$ is usually a sparse matrix whose non-zero entries take values $\{a, -a\}$ with the same probability.  The value of $a$ is chosen to achieve the desired value of the spectral radius $|\lambda_\text{max}|$, which controls the memory of the reservoir. That is, learning dynamics which require remembering long-term relationships requires higher values of $|\lambda_\text{max}|$. However, as we have seen in the previous section, too large values of $|\lambda_\text{max}|$ may cause the network to not have echo states. It is common to use large reservoirs (large $N$) to ensure rich reservoir dynamics. The exact size depends on the amount and dimensionality of the training data. Larger reservoirs tend to produce better results but at the expense of requiring more computational resources and increasing the chances of overfitting the training data. Matrices $W^{\text{in}}, W^{\text{back}}$ are constructed similarly as $W$ but they are \emph{dense} matrices to ensure that all inputs and outputs are used to update the dynamics of the reservoir. 
    
    \item \textbf{Update the internal states by \emph{teacher forcing}}
    
    Let $T$ be the length of the training series $\vec{y}_\text{teach}$. The internal states $\vec{x}(t)$ for $t=1, \cdots, T$ are calculated by 
    \begin{equation}
    \vec{x}(t) = f(W^{\text{in}} \vec{u}(t)+W\vec{x}(t-1)+W^{\text{back}}\vec{y}_{\text{teach}}(t-1)).
    \label{xn_train}
    \end{equation}
    Recall that the initial state $\vec{x}(0)$ can be initialized randomly since it does not affect its long-term dynamics because of the state forgetting property. Once this step finishes we have $N$ time series of dimension $T$, which can be stored in a $T \times N$ matrix, where the $j$-th column contains the trajectory of the $i$-th node. Notice that this step is called teacher forcing because we only use the known series $\vec{y}_\text{teach}$ instead of the predicted series $\vec{y}$. This step does not require learning any matrix parameters, since the values of $W, W^{\text{in}}, W^{\text{back}}, f$ are all fixed, and the series $\vec{u}(t), \vec{y}_\text{teach}(t)$ are known.

    \item \textbf{Discard a transient of $t_\text{min}$ states to guarantee the convergence of the reservoir dynamics}\\
    \\
    To ensure that the echo state properties are satisfied, the first $t_\text{min}$ time steps should be discarded, since the first values of $\vec{x}(t)$ are influenced by the initial values of the input, output and states.

    \item \textbf{Find the readout matrix $W^\text{out}$ by minimizing the mean squared error}
    %Equation 2
    \begin{equation}
        MSE(\vec{y}, \vec{y}_{\text{teach}}) =   \frac{1}{T - t_{\text{min}}} \sum_{t=t_{\min}}^T \Big[f^{{\text{out}}^{-1}}(\vec{y}_{\text{teach}}(t))  - W^{\text{out}} \vec{x}(t)\Big]^2 .
        \label{eq:MSE}
    \end{equation}
    Notice that this step only requires performing a linear regression to learn $W^{\text{out}}$. Remember that, according to Eq.~\ref{eq:RC_out}, the prediction of the reservoir is a linear combination of the internal states of the network. If the reservoir is designed appropriately, the internal state dynamics provide enough variety of trajectories to span a subspace containing the desired output.
\end{enumerate}

Once the reservoir network has been trained, it evolves autonomously to make future predictions $\vec{y}(t)$.  The steps to make such predictions are the following: 
\begin{enumerate}
    \item  \textbf{Given an input $\vec{u}(t)$, update the state of the reservoir using Eq.~\ref{eq:update_x}:}
    \begin{equation*}
        \vec{x}(t) = f[(W^{\text{in}} \vec{u}(t)+W\vec{x}(t-1)+W{^\text{back}}\vec{y}(t-1)],
    \end{equation*}
    where $\vec{y}(t-1)$ is the prediction of the output at time $t-1$.
    \item \textbf{Compute the prediction $\vec{y}(t)$ using Eq.~\ref{eq:RC_out}:}
    \begin{equation*}
        \vec{y}(t) = f^{\text{out}}\Big[W^{\text{out}} \vec{x}(t)\Big].
    \end{equation*}
\end{enumerate}

 \subsection{Training improvements}
 \label{sect_training_improvements}
 The previous section presented the original RC training, proposed in Ref.~\citep{ENS}. However, multiple improvements have been proposed in the literature to make training more efficient in certain situations. In this section, we review some of the most popular variations of the original algorithm. Then, we train the RC algorithm with two simple examples and explore how the presented improvements affect the learning process in those cases.
 
 \subsubsection{Leaking rate}
 There are occasions when it is important to keep the knowledge obtained in the distant past. We have seen that we cannot increase too much the spectral radius of the internal network, since the reservoir may stop having echo states. Another possibility to increase the memory capacity of the reservoir is changing the formula to update the internal state. In this case, Eq.~\ref{eq:update_x} becomes
\begin{eqnarray}
        \vec{\tilde{x}}(t) = f[(W^{\text{in}} u(t)+W\vec{x}(t-1)+W{^\text{back}}\vec{y}(t-1)],\\ \nonumber
        \vec{x}(t) =(1 - \alpha) \vec{x}(t-1) + \alpha \vec{\tilde{x}}(t),
        \label{eq:update_x_leaking}
\end{eqnarray}
where $\alpha \in (0,1]$ is the leaking rate. Similarly, Eq.~\ref{xn_train} becomes
\begin{eqnarray}
        \vec{\tilde{x}}(t) = f[(W^{\text{in}} \vec{u}(t)+W\vec{x}(t-1)+W{^\text{back}}\vec{y}_{\text{teach}}(t-1)],\\ \nonumber
        \vec{x}(t) =(1 - \alpha) \vec{x}(t-1) + \alpha \vec{\tilde{x}}(t),
        \label{eq:update_x_leaking_train}
\end{eqnarray}
 When $\alpha$ is close to 1, the internal states at time $t$ are closer to the internal states at time $t-1$. Therefore, such internal states contain more information about the distant past. For $\alpha=0$ we recover Eqs.~\ref{eq:update_x} and~\ref{xn_train}.

 \subsubsection{Regularized linear regression}
 
 One of the main challenges in ML is reducing the chances of overfitting the training data. Overfitting happens when the model learns to predict the training data but fails to generalize to unseen data. In the RC algorithm, the learning algorithm is just a linear model $W^\text{out}$. A strategy to prevent overfitting is adding a regularization term when training such model. In this way, instead of minimizing the MSE in Eq.~\ref{eq:MSE}, we would minimize the following expression
\begin{equation}
 MSE_r(\vec{y}, \vec{y}_{\text{teach}}) = \frac{1}{T - t_{\text{min}}} \sum_{t=t_{\min}}^T \Big(f^{{\text{out}}^{-1}}(\vec{y}_{\text{teach}}(t))
  - W^{\text{out}} \vec{x}(t)\Big)^2 + \gamma ||W^{\text{out}}||^2 ,
 \label{eq:ridge}
 \end{equation}
 where $\gamma$ is the regularization parameter. This linear model is called \emph{ridge regression}~\citep{ridge}. It prevents the algorithm from learning too big coefficients $W^\text{out}$, which usually leads to unstable training and poor generalization capacity. The value of $\gamma$ is a hyperparameter which needs to be tuned depending on the RC task. When $\gamma$ is too large, the model will learn very small values of $W^\text{out}$, resulting in predicting constant values. On the other hand, if $\gamma$ is too small we recover Eq.~\ref{eq:MSE}, increasing the chances of overfitting.
 
 \subsubsection{Adding noise to $\vec{y}_{\text{teach}}$}
 
Another way to reduce the problem of having an ill-conditioned matrix for the linear regression is to add some noise $\vec{\nu}(t)$ to the output $\vec{y}_{\text{teach}}$ during training. That is,
\begin{equation}
    \vec{y}_{\text{teach}}(t) \rightarrow \vec{y}_{\text{teach}}(t) + \vec{\nu} (t).
\end{equation}

Training the model with a noisy output resembles the situation in the prediction phase, thus increasing the stability of the predictions. The noise is uniformly sampled from the interval $[-b,b]$, where $b$ is a small number that depends on each problem. If the noise bound $b$ is too small, adding noise makes no effect. Otherwise, if $b$ is too big, it may give misleading information to the network which decreases the performance of the model.

\subsubsection{Adding a constant input}

 RC can also be used without the input layer when we aim to predict a time series without explanatory variables.  In this case, the term $W^{\text{in}}\vec{u}(t)$ is removed from Eqs.~\ref{eq:update_x_leaking_train} and~\ref{eq:update_x_leaking}. In these situations, it is sometimes useful to add a constant input $\vec{u}(t) = \vec{c} \ \forall t, \ \vec{c} \in \mathbb{R}^k$ to introduce more variability within the internal states.
 
 \subsubsection{Quadratic model}
Until now, the output of the network has always been computed as a linear combination of the internal states of the reservoir. However, sometimes the subspace generated by the internal states does not contain the desired output. In that case, other relationships between the states $\vec{x}$ and the output $\vec{y}$ may be considered. In Refs.~\citep{chaos1, chaos2} the authors consider a quadratic relationship to predict the evolution of the dynamical system. In this case, Eq.~\ref{eq:RC_out} becomes
\begin{equation}
    \vec{y}(t) = W^\text{out}\vec{x}_2(t),
\end{equation}
where $\vec{x}_2(t):=
    \begin{pmatrix}
    \vec{x}(t)\\
    \vec{x}(t)^2
    \end{pmatrix}$ is the concatenation of $\vec{x}(t)$ and $\vec{x}(t)^2$. 

\subsection{Learning a periodic series}
After presenting the training and testing steps of the RC algorithm, we are going to train two simple models and evaluate their performances. Let us start with the example that we used before: predicting a periodic series. Remember that, in this case, the input and output are given by
\begin{equation*}
    u(t) = \sin(t/5), \qquad y_\text{teach}(t) = \frac{1}{2} u(t)^7.
\end{equation*}

\begin{figure}[!ht]
    \centering
    \includegraphics[width=0.85\textwidth]{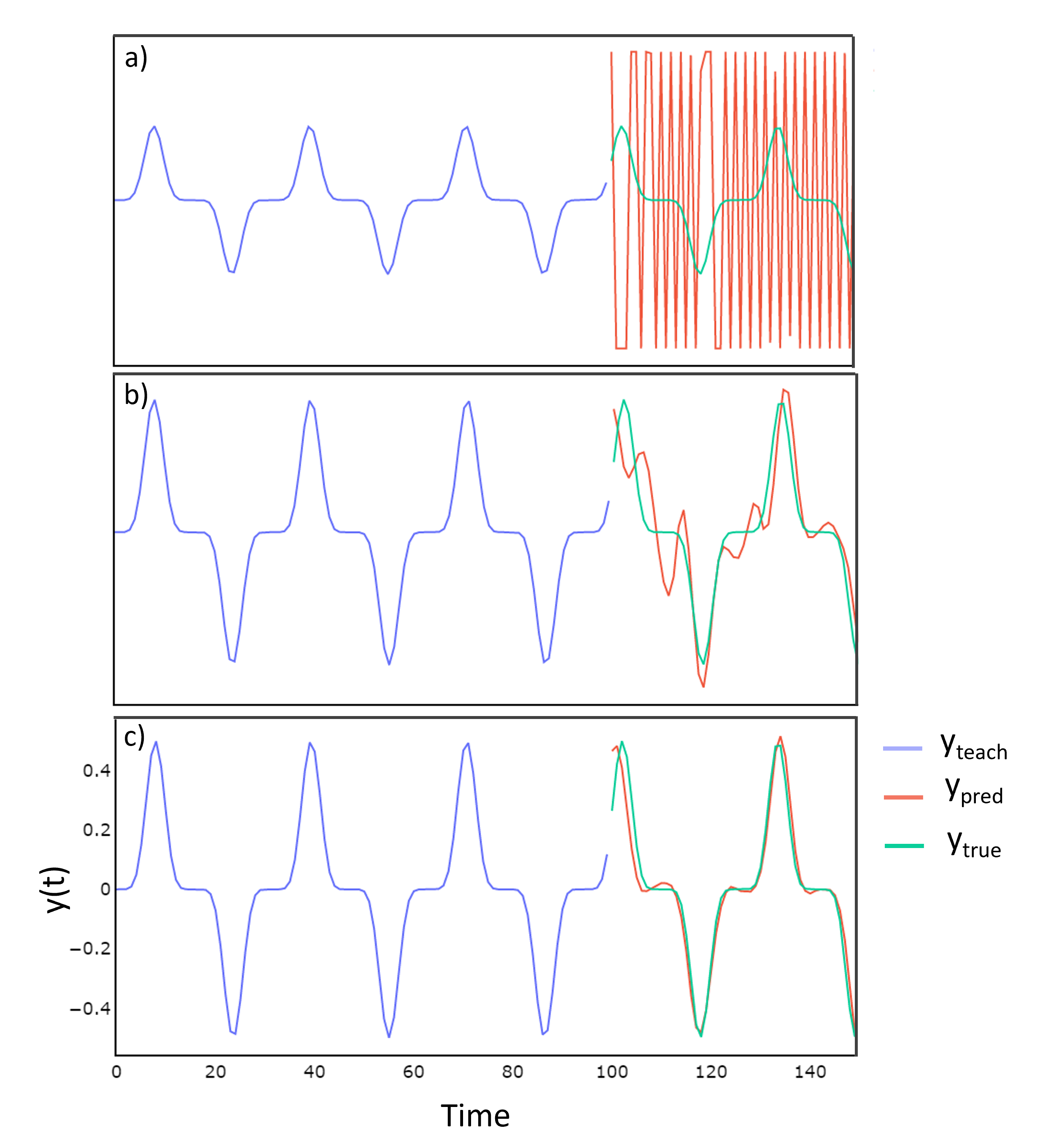}
    \caption{Prediction of the reservoir computing algorithm with a) vanilla training with spectral radius 0.3, b) vanilla training with spectral radius 1.3 and c) training with noisy outputs and ridge regression.}
    \label{fig:RC_predict_periodic}
\end{figure}

We begin by training the model with the original training strategy presented in Sect.~\ref{sect:training_RC} and then test whether the strategies presented in Sect.~\ref{sect_training_improvements} improve the performance of the model. The first step is to define the parameters of the reservoir.  The reservoir consists of a network with N = 200 nodes. Only 5\% of the nodes are connected. The values of the weights are chosen so that the matrix $W$ has a spectral radius of 0.3. The feedback matrix $W^\text{back}$ is chosen to have values $\pm 1$ with equal probability, and the input matrix $W^\text{in}$ is chosen to take values $0, +1, -1$ with probabilities $0.5, 0.25, 0.25$ respectively. We choose $f = f^\text{out} = \tanh$, $t_\text{step} = 100$ and $t_\text{min} = 10$. Once the reservoir is trained, it evolves autonomously for 50 epochs for testing purposes. The results of training the reservoir with this setting are shown in Fig.~\ref{fig:RC_predict_periodic} a). As can be seen, the network is not able to correctly predict future outputs, due to the instability of the linear regression.

\begin{figure}[!ht]
    \centering
    \includegraphics[width=0.8\textwidth]{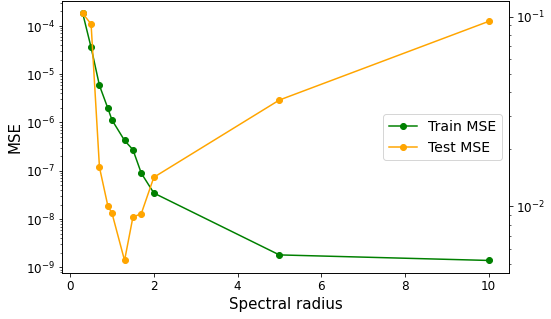}
    \caption[Train and test performance of the reservoir computing algorithm for the different
spectral radius of the reservoir.]{Mean squared error (MSE) of the true and predicted values by the reservoir computing algorithm for the different spectral radius of the reservoir $W$, for both training and test sets.}
    \label{fig:RC_spectral_radius}
\end{figure}

To improve the predicting capacity of the model, we increase the spectral radius until the performance in the test set stops increasing, as shown in Fig.~\ref{fig:RC_spectral_radius}. The optimal spectral radius in this case is $\rho(W) = 1.3$.  The results of this new model are shown in Fig.~\ref{fig:RC_predict_periodic} b). Now the predictions resemble much more the true series.

To further improve such predictions, some of the training modifications presented in the previous section are implemented. In particular, we select a leaking rate of $\alpha = 0.8$, add random noise $\nu(t) \sim U(-0.1,0.1)$ to the output $y_\text{teach}$, and use a ridge regression model with $\gamma = 0.01$ to learn the values of $W^\text{out}$. With these modifications, we obtain the results in Fig.~\ref{fig:RC_predict_periodic} c), which are even much closer to the true output series. 

The choice of the reservoir $W$ highly influences the performance of the ML algorithm. Apart from the spectral radius, it is also important that the graph $W$ is sparse and inhomogeneous. In this case, the trajectories of the internal states $\vec{x}(t)$ are diverse and thus can generalize to unseen data. We check the influence of the homogeneity of the matrix $W$, considering four different scenarios:

\begin{enumerate}
    \item \textbf{Inhomogeneous and sparse reservoir}: This is the reservoir we have used so far. The weights are sampled from the set $\{-a, a\}$, where $a$ is chosen so that $W$ has a certain spectral radius. Only 5\% of the weights have non-zero values.
    
    \item \textbf{Inhomogeneous and dense reservoir}: In this case the reservoir is dense, and all the weights are sampled uniformly from the set $\{-a,a\}$, where $a$ is chosen to achieve a certain spectral radius. 
    
    \item \textbf{Homogeneous and sparse reservoir}: The reservoir is sparse, with only 5\% of non-zero connections. All the non-zero weights have the same constant value.
    
    \item \textbf{Homogeneous and dense reservoir}: In this case the reservoir is dense, and all the weights have the same constant value.
\end{enumerate}

The performance of each of the RC models in the test set is shown in Table~\ref{tab:RC_homogeneous}. The results show that it is very important to use a sparse reservoir in order to make accurate predictions. In the cases where a dense reservoir is used, all the trajectories of the internal states are very similar, and thus the network cannot generalize to unseen predictions. The fact that the weights of the graph $W$ are homogeneous is slightly less important in terms of the accuracy of the predictions, but also relevant. 

\begin{table}[!t]
    \centering
    \begin{tabular}{cccc}
    \hline \hline
      \makecell{Inhomogeneous\\ and sparse} & \makecell{Inhomogeneous\\ and dense} & \makecell{Homogeneous\\ and sparse} & \makecell{Homogeneous\\ and dense} \\
      \hline
       0.0032  & 0.13 & 0.017 &  0.060\\
       \hline 
    \end{tabular}
    \caption{Mean squared error for the prediction phase of the four types of reservoirs, regarding homogeneity and density.}
    \label{tab:RC_homogeneous}
\end{table}

\subsection{Solving a partial differential equation}

The previous example illustrates the important factors and techniques to take into account when training a RC algorithm. We aimed to predict an output series $y_\text{teach}(t)$ which was a function of the input $u(t)$. RC can also be used to solve partial differential equations (\gls{PDE}). We now illustrate the differences in RC when learning to integrate a PDE. To do so, we show an example of how to train a well-known PDE, the heat equation.
\begin{equation}
    \left\{
\begin{array}{lll}
      y_t = & \frac{1}{2} y_{zz} & \quad t>0, 0<z<l\\
      y(z,0) = & \sin(z) & \quad 0<z<l\\
      y(0,t) = & y(l,t) = 0 & \quad t>0\\
\end{array} 
\right.
\end{equation}
The heat equation describes the heat diffusion along a unidimensional bar $x$ of length $l$, which in this case is $l= \pi$. The solution of this equation for $0<t<5$, $0<z<2\pi$ is
\begin{equation}
    \vec{y}(z,t) = \sin(z) e^{-t}.
\end{equation}
In this example, we want the RC algorithm to predict the series $\vec{y}(z,t)$ based on the value of the series at the previous time step $vec{y}(z,t-1)$ for all values of $z$. That is, the RC model should learn to propagate the series in time. Since we want the network to learn from its output, feedback weights $W^\text{back}$ are clearly needed. The space and time domain need to be discretized, so that $\vec{y}_\text{teach}(k,t) = \sin(k) e^{-t}$, where $k$ represents the spatial position and $t$ the time step. For this example, $k \in \{0, \frac{1}{L}, \cdots, \frac{\pi}{L}\}$ is a discretization of the spatial domain ($L=100$) and $t \in \{0, \frac{1}{t_\text{step}}, \cdots, \frac{5}{t_\text{step}}\}$ is a discretization of the time domain ($t_\text{step} = 500$). 

The values of the matrix $W^\text{back}$ are randomly chosen from the set $[-0.5, 0.5]$. A priori we do not need to provide any input series to the reservoir since we only want to learn from past values of $\vec{y}(z,t)$. We thus set $\vec{u}(z,t)=0$, and use no input connections $W^\text{in}$. The reservoir consists of a network with N = 400 nodes, where only 2\% of the nodes are connected. The values of the weights are chosen so that the matrix $W$ has a spectral radius of 0.9. Just as with the previous example we set $f = f^\text{out} = \tanh$. We train the network for $t_\text{step} = 500$ steps and discard the first $t_{\min} = 200$ points. 

\begin{figure}[!ht]
    \centering
    \includegraphics[width=0.85\textwidth]{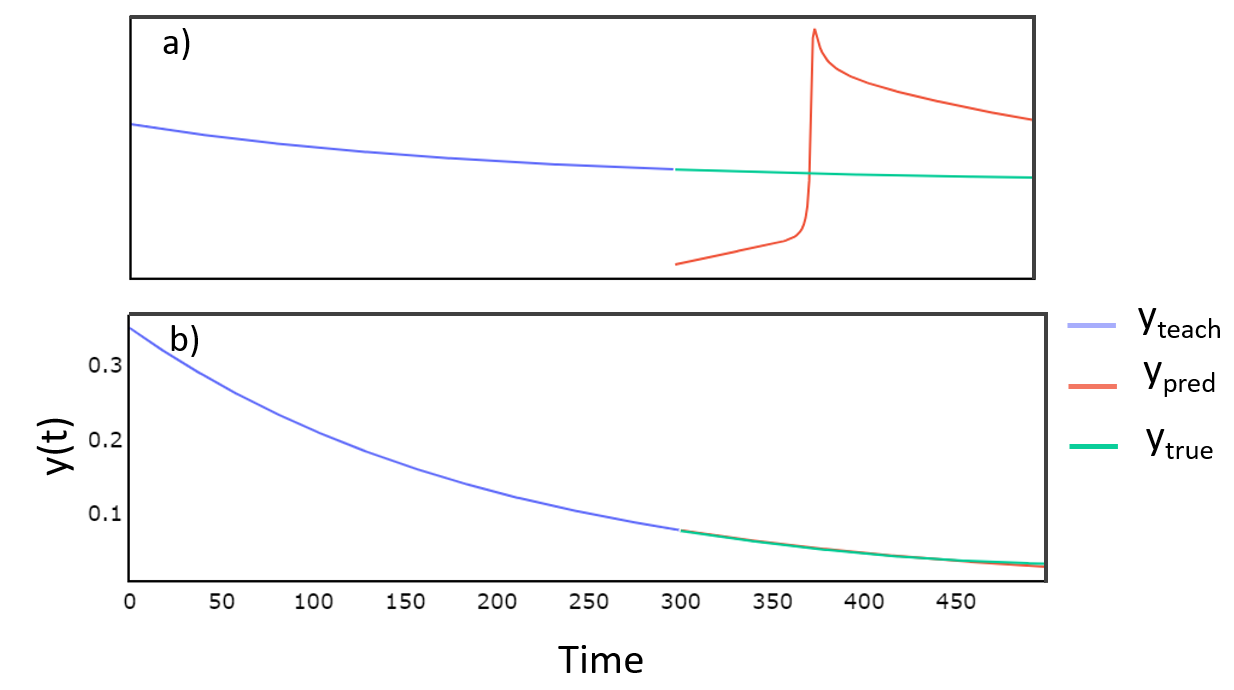}
    \caption[Prediction of the reservoir computing algorithm for the heat equation with a) vanilla training and b) with training improvements.]{Prediction of the reservoir computing algorithm for the heat equation with a) vanilla training and b) training with noisy outputs, ridge regression and adding a constant input. The figures show the output series at $k = \frac{\pi}{10}$.}
    \label{fig:RC_prediction_heat}
\end{figure}

With this training strategy, we obtain the results in Fig.~\ref{fig:RC_prediction_heat} a). The RC model is not able to generate accurate predictions. To improve the performance of the model we have considered the following improvements: adding white noise $\nu(t) \sim U(-0.001, 0.001)$ to the output, training the linear model with regularization $\gamma = 1 \times 10^{-12}$ and using a constant input $vec{u}(k,t) = 1$ with dimension $K=5$. The input weights $W^\text{in}$ are chosen randomly from the set $\{0, -0.14, 0.14\}$ with probabilities $0.5, 0.25, 0.25$ respectively. With these techniques, we obtain the results in Fig.~\ref{fig:RC_prediction_heat} b). As can be seen, now the reservoir can autonomously predict the evolution of the physical system, effectively reducing the MSE in the predictions from $0.13$ to $1 \times 10^{-6}$. The other training modifications did not improve the overall performance of the model for this example.

\subsection{Comparing reservoir computing and neural networks}
In this chapter, we have studied so far two relevant ML algorithms: NNs and RC. NNs can solve many different tasks because of their versatility. The network architecture is chosen to adapt to the problem of interest. For time series forecasting, RNNs are a commonly used deep learning algorithm. RNNs have memory since they use information from past input samples to influence the current input and output.

In contrast to other types of NNs, such as feed-forward or CNNs, RNNs do not assume that the data samples are independent of each other, but that there are correlations between sequential inputs. RC is another algorithm used to predict sequential data, where a network is designed to learn to reproduce the input-output dynamics. In this section, we emphasize the main differences between RNNs and RC.

\begin{enumerate}
    \item \textbf{Model complexity:} One of the main differences between RNNs and RC is the number of training parameters. RNNs use training parameters to learn how to use the information from previous inputs to predict future outputs. This results in a highly complex model, with many parameters to be learned during the training phase. On the other hand, RC uses the echo state properties to ensure that the internal network evolution only depends on the inputs and outputs and thus can learn to reproduce the input-output dynamics. The complexity of RC is reduced to fitting a linear model, which requires learning many fewer parameters. Models with lower complexity require fewer data samples during the training phase. Thus, RC is especially useful when working with small training sets.
    
    \item \textbf{Training times:} RNNs learn the values of their training parameters using the back-propagation algorithm. On the other hand, training a RC model only requires performing a linear regression. For this reason, RNNs require more computational resources and longer training times than RC.
    
    \item \textbf{Network design:} NNs are usually composed of layers of neurons that transform the information sequentially. This layered structure is used to make the training process easier. The network used in RC has a more flexible architecture, where all types of connections are allowed. In particular, the internal network is not usually divided into layers. A random, sparse matrix is used as the internal network, which provides more versatility to the algorithm.
\end{enumerate}
For these reasons, RC has become a popular algorithm when predicting the evolution of time series or dynamical systems. In the following chapters, we will show how to apply RC to integrate PDEs as well as to perform time series forecasting. The RC algorithm, with some modifications, will be adapted to predict the dynamics of a high-dimensional system without producing much overfitting. Examples of these high-dimensional series are molecular quantum systems, where the dimensionality of the data increases dramatically with the number of degrees of freedom. 

Even though RC can in principle work with some of these data sets, the scaling of these systems is a clear limitation for any ML algorithm. This fact motivates us to explore another novel technology, which has the potential to outperform many classical ML algorithms while requiring less computational resources: quantum computing.

 %%%%%%%%%%%%%%%%%%%%%%%%%%%%%%%%%%%%%%%%%%%%%%%%%%%%%%%%%%%%%
 % QUANTUM COMPUTING
 %%%%%%%%%%%%%%%%%%%%%%%%%%%%%%%%%%%%%%%%%%%%%%%%%%%%%%%%%%%%%

\section{Quantum computing}
\label{sect:QC}
 Quantum mechanics is a mathematical framework used for the construction of physical theories. Quantum computing exploits the properties of quantum mechanics to perform information-processing tasks. The laws of quantum mechanics provide great computational power and are yet very different to the classical information rules. In this section, we introduce the terminology and fundamental concepts behind quantum mechanics (for a more detailed explanation see Ref.~\citep{Nielsen}). Then, we introduce how this framework can be used to design quantum algorithms. In particular, we discuss the different trends in QML. Finally, we present the quantum algorithm that will be studied in this thesis, \gls{QRC}.

\subsection{Introduction to quantum mechanics}
\label{sect:postulates}
The theory of quantum mechanics provides a connection between mathematical formalism and physical phenomena. For this reason, it sets the rules for designing any QML algorithm. In this section, we introduce the basic terms and tools used to describe quantum systems.

\begin{figure}
    \centering
    \includegraphics[scale=0.8]{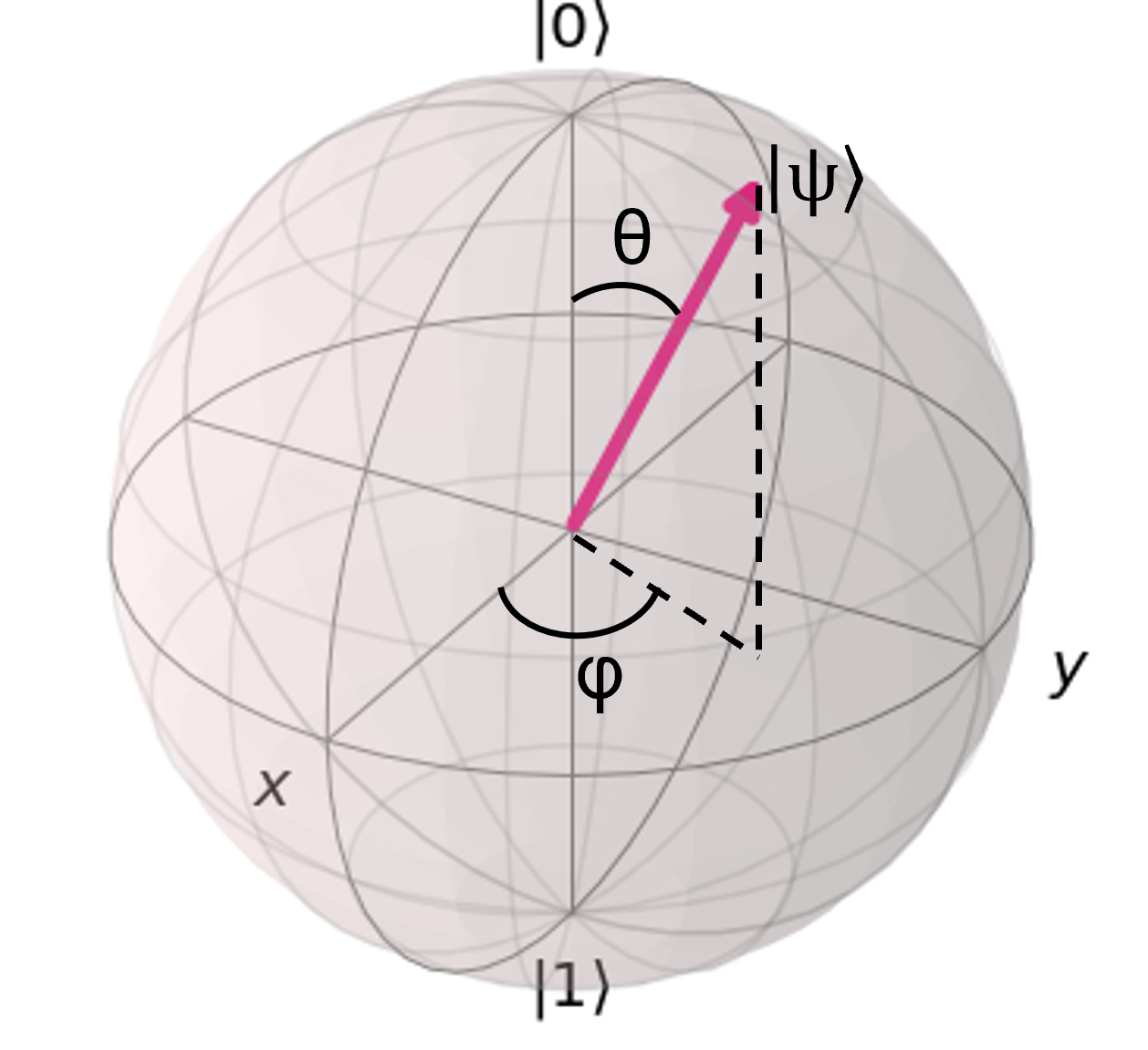}
    \caption{Bloch sphere standard representation.}
    \label{fig:bloch_sphere}
\end{figure}

Any closed quantum system is associated with a Hilbert space $\mathcal{H}$ known as the \emph{state space} of the system. The elements of this Hilbert space are called \emph{state vectors}. The simplest physical system is the \emph{qubit}, whose state space is $\mathcal{H}_2 = \mathbb{C}^2$. The term qubit should not be confused with the term \emph{bit}, which refers to the fundamental unit of classical information. A bit has a state of either 0 or 1, while a qubit can have states $\ket{0}$, $\ket{1}$ or any \emph{superposition} (linear combination) of these two states. That is, the most general state of a qubit is 
\begin{equation}
    \ket{\psi} = a\ket{0}+b\ket{1}, \quad a,b \in \mathbb{C},
\end{equation}
where $\{\ket{0}, \ket{1}\}$ form a complete orthonormal basis of $\mathcal{H}_2$. The state $\ket{\psi}$ must be a unit vector, $\braket{\psi} =1$, which implies that the coefficients must fulfill $|a|^2 + |b|^2 = 1$. A useful geometric representation of single qubits states is the \emph{Bloch Sphere} representation, shown in Fig.~\ref{fig:bloch_sphere}. A qubit can be represented as a point in the Bloch sphere by using two parameters $(\phi, \theta)$ in the following way 
\begin{equation}
    \ket{\psi} = \cos \frac{\theta}{2} \ket{0} + e^{i\phi} \sin \frac{\theta}{2} \ket{1}.
\end{equation}
A set of very useful operators that can transform one qubit state to any other qubit state are the \emph{Pauli matrices}. The matrix form of these operators is 
\begin{equation}
\mathbb{I} = \begin{pmatrix}
1 & 0\\
0 & 1
\end{pmatrix},\quad
    X = \begin{pmatrix}
    0 & 1 \\
    1 & 0
    \end{pmatrix}, \quad
    Y = \begin{pmatrix}
    0 & -i \\
    i & 0
    \end{pmatrix},\quad
    Z = \begin{pmatrix}
    1 & 0 \\
    0 & -1
    \end{pmatrix}.
    \label{eq:Pauli_matrices}
\end{equation}
These matrices make transformations between qubits.  $X$ transforms the state $\ket{\psi} = a\ket{0} + b\ket{1}$ to $a\ket{1} + b\ket{0}$, $Y$ transforms $\ket{\psi}$ to $-ia\ket{1} + ib \ket{0}$, $Z$ transforms $\ket{\psi}$ to $a\ket{0} - b\ket{1}$ and $\mathbb{I}$ leaves $\ket{\psi}$ unchanged. 

Apart from making transformations of quantum states, we can also perform measurements, which allows extracting information from the physical system. Quantum measurements are characterized by a collection $\{M_m\}$ of measurement operators which satisfy
\begin{equation}
    \sum_m M_m^\dag M_m = \mathbb{I}.
    \label{completenessM}
\end{equation}
  The index $m$ indicates the possible measurement outcomes that may occur in the experiment. The probability of result indexed by $m$ occurring when measuring state $\ket{\psi}$ is
 \begin{equation}
     p(m|\psi) = \bra{\psi}M_m^\dag M_m \ket{\psi},
 \end{equation}%
and the state after the measurement is 
\begin{equation}
    \frac{M_m \ket{\psi}}{\sqrt{p(m|\psi)}}.
\end{equation}
Equation~\ref{completenessM} implies that the probabilities add up to one, thus forming a valid probability distribution
\begin{equation}
    \sum_m p(m|\psi) = \bra{\psi} \Big(\sum_m M_m^\dag M_m\Big) \ket{\psi}  = \braket{\psi} = 1 .
\end{equation}
The simplest example of a measurement is the measurement of a qubit on the computational basis. The observable in this case would be the Pauli matrix $Z = \ketbra{0} - \ketbra{1}$. The measurement operators are then 
\begin{equation*}
    P_{+1} = M_{+1}^\dag M_{+1} =  \ketbra{0}, \quad P_{-1} = M_{-1}^\dag M_{-1} =  \ketbra{1}.
\end{equation*}
If the general state $\ket{\psi} = a \ket{0} + b\ket{1}$ is measured, then the probability of having the result $+1$ is 
\begin{equation*}
    p(+1|\psi)= \bra{\psi} P_{+1} \ket{\psi} = \bra{\psi}\ketbra{0}\ket{\psi} = |a|^2,
\end{equation*}
and similarly the probability of getting the output $-1$ is $p(-1|\psi) = |b|^2$. The state after the measurement in each case is 
\begin{equation*}
    \frac{\ketbra{0} \ket{\psi}}{|a|} = \frac{a}{|a|}\ket{0} = \ket{0}, \quad  \frac{\ketbra{1} \ket{\psi}}{|b|} = \ket{1},
\end{equation*}
where we have used that the global phase of a quantum state can be ignored~\citep{Nielsen}.

In order to make useful quantum computations, systems with more than one quantum state (qubit) are needed. The state of a composite physical system is described by means of the tensor product of the single-qubit states. Numbering the systems from 1 to n, and denoting the state in the system $i$ as $\ket{\psi_i}$, the joint state is 
\begin{equation}
    \ket{\Psi} = \ket{\psi_1} \otimes \ket{\psi_2} \otimes \cdots \otimes \ket{\psi_n}.
    \label{post4}
\end{equation}
Note that this definition is consistent with the superposition principle. If a bipartite system can be in state $\ket{\psi_1} \otimes \ket{\psi_2}$ and in state $\ket{\psi_1'}\otimes\ket{\psi_2'}$, then, because of superposition, in principle it should be allowed to be in state $a\ket{\psi_1}\otimes\ket{\psi_2} + b\ket{\psi_1}\otimes\ket{\psi_2}$. This property only makes sense in the tensor product space. Not all states, however, can be written as a tensor product of states (as in Eq.~\ref{post4}).  Consider, for example, the state $\frac{1}{\sqrt{2}} (\ket{01} + \ket{10})$ (where the notation $\ket{01}$ refers to $\ket{0} \otimes \ket{1}$). It is easy to prove that this state cannot be written as a product state $\ket{\psi}\otimes \ket{\phi}$. 

We say that any state that cannot be written as the tensor product of states is an \emph{entangled} state. Interestingly, entangled states play a fundamental role in quantum computing and quantum information, as they are the key to many algorithms and protocols. Entanglement allows multiple qubits to be connected so that the state of one qubit depends on the state of the others. This allows quantum computers to perform certain operations that are not possible on classical computers, allowing for much more efficient computation. 

In quantum computing, entanglement is used to perform quantum algorithms, such as Shor's algorithm for factorization~\citep{Shor}, and Grover's algorithm for search problems~\citep{Grover}, where the use of entangled states allows for significant speed-ups compared to classical algorithms. Entanglement also plays a critical role in QML, where the use of entangled states allows the processing of large amounts of data with fewer quantum computational resources.

\subsection{The Schrödinger equation}
\label{sect:schrodinger}

So far, we have seen how to describe the state of a single and composite quantum system, and how to perform measurements to extract information from them. The last key point in the theory of quantum information is how to perform quantum transformations on the initial states. In this respect, the evolution of a closed quantum system is described by a unitary operator $U$, which transforms a state $\ket{\psi}$ to the state $\ket{\psi'}$ by
\begin{equation}
    \ket{\psi'} = U \ket{\psi}.
    \label{eq:unitary}
\end{equation}
Notice that the laws of quantum mechanics do not specify what unitary operators describe the dynamics of a particular quantum system, and it is the physicist's job to figure out this description. Another way to represent the evolution of a quantum state is through the use of the \emph{Schrödinger equation}
\begin{equation}
    i \ \hbar \frac{\partial \ket{\psi}}{\partial t} = \hat{H} \ket{\psi},
    \label{eq:postulate2p}
\end{equation}
where $\ \hbar$ is the reduced \emph{Planck constant} and $\hat{H}$ is an Hermitian operator known as \emph{Hamiltonian} giving the energy of the physical system. Equations~\ref{eq:unitary} and~\ref{eq:postulate2p} are equivalent because the unitary operator $U$ can always be written as
\begin{equation}
    U = e^{-it\hat{H}/\hbar}
\end{equation}
due to the Hermitian character of the Hamiltonian operator.

 Until now, all the quantum states presented in this section belonged to a discrete Hilbert space. For example, a single qubit belongs to a two-dimensional Hilbert space, where the basis states are $\ket{0}$ and $\ket{1}$. Since any state $\ket{\psi}$ can be written as a linear combination of the basis states, $\ket{\psi}= a\ket{0}+ b \ket{1}$, the state $\ket{\psi}$ is completely determined by a \emph{finite} set of complex numbers $\{a,b\}$. Nonetheless, there are situations where a quantum state is only completely described when using a \emph{continuous} set of coefficients. For example, consider the vibrational Hamiltonian of a diatomic molecule, which depends on the distance between the two nuclei $x$. In this case, the quantum state $\ket{\psi}$ is represented by a function
 \begin{align*}
  \psi \colon \ \   \mathbb{R} &\to \mathbb{C}\\
  x &\mapsto \psi(x)
\end{align*}
that provides the amplitude of the quantum state $\psi(x)$ at every point in space $x$. 
 The normalization condition for continuous states requires that the function $\psi(x)$ is square-integrable 
 \begin{equation}
     \int_\mathbb{R} |\psi(x)|^2 dx = 1.
 \end{equation}
The function $\psi(x)$ is called \emph{wavefunction} since it is the solution to a wave equation, where the term wave simply denotes oscillatory behavior. The square of the wavefunction $|\psi(x_0)|^2$ gives the probability density of finding a particle in the position $x \in [x_0, x_0 + dx]$. This probability density can also be used to calculate the expected value of an operator $A$ 
\begin{equation}
    \expval{A} = \int \psi^*(x) A \psi(x) dx = \int A(x) |\psi(x)|^2 dx.
\end{equation}
In this example, the wavefunction only depends on a one-dimensional variable $x$. In more general cases, the wavefunction can depend on a vector $\vec{x}$, for example, $\vec{x} = (x,y,z)$. Then, $\psi: \mathbb{R}^n \rightarrow \mathbb{C}$, and the previous equations would be integrated over $ \mathbb{R}^n$ instead of $ \mathbb{R}$.

The time evolution of a quantum wavefunction is also determined by the Schrödinger equation. In continuous systems, Eq.~\ref{eq:postulate2p} is transformed to
\begin{equation}
    i \ \hbar \frac{\partial \psi(\vec{x},t)}{\partial t} = \hat{H}(\vec{x}) \psi(\vec{x},t),
    \label{eq:time-dependent}
\end{equation}
which is known as the \emph{time-dependent Schrödinger equation}. The solution of the previous equation (together with an initial condition $\psi(\vec{x},0) = \psi_0$) describes the time evolution of an initial quantum state under the effect of a Hamiltonian $H(\vec{x})$. Consider a quantum system describing a particle of mass $m$. If the Hamiltonian does not depend on time and the kinetic energy is decoupled to the potential energy, the Hamiltonian operator can be written as
\begin{equation}
    \hat{H} = \frac{\hat{p}^2}{2m} + \hat{V},
    \label{eq:Hamiltonian}
\end{equation}
where $\hat{V}$ is the potential energy operator and $\hat{p}$ is the momentum operator, given by
\begin{equation}
    \hat{p} = -i \ \hbar \nabla,
\end{equation}
where $\nabla$ is the Laplacian operator. In Cartesian coordinates,
\begin{equation}
    \nabla^2 = \frac{\partial^2}{\partial x^2} + \frac{\partial^2}{\partial y^2} + \frac{\partial^2}{\partial z^2}.
\end{equation}
Under these circumstances, Eq.~\ref{eq:time-dependent} becomes
\begin{equation}
    i \ \hbar \frac{\partial \psi(\vec{x},t)}{\partial t} =\Big( - \frac{\hbar^2}{2m}\nabla^2 + V(\vec{x}) \Big) \psi(\vec{x},t).
    \label{eq:time-dependent2}
\end{equation}
Since the Hamiltonian does not depend on time, Eq.~\ref{eq:Hamiltonian} gives the total energy of the particle. Thus, Eq.~\ref{eq:time-dependent2} becomes an eigenvalue problem, leading to the \emph{time-independent Schrödinger equation}
\begin{equation}
    \Big( - \frac{\hbar^2}{2m}\nabla^2 + V(\vec{x}) \Big) \phi_n(\vec{x}) = E_n \phi_n(\vec{x}),
    \label{eq:time-independent}
\end{equation}
where $E_n$ are the \emph{eigenenergies} of the particle. The wavefunctions $\phi_n(\vec{x})$ that solve Eq.~\ref{eq:time-independent} are called \emph{eigenfunctions} of the Hamiltonian $\hat{H}$ associated with energy $E_n$. Usually, there is only a discrete number of solutions of Eq.~\ref{eq:time-independent}, so that the set of eigenfunctions can be denoted as $\{\phi_n\}_{n \in \mathbb{N}}$. In this case, the wavefunctions $\{\phi_n\}$ correspond to bounded states, which are square-integrable and localized in a region of space. In other situations, the energies can change continuously $E(k)$, with $k \in \mathbb{R}$. Then, eigenfunctions will be free states, which extend over the whole space and cannot be normalized with the $L^2$ norm. In any case, the time evolution of an eigenfunction $\phi_n(\vec{x},t)$ is given by 
\begin{equation}
    \phi_n(\vec{x},t) = \phi_n(\vec{x},0) e^{-i E_n t/\hbar}.
\end{equation}
The set of eigenfunctions forms a complete basis of the Hilbert space. Therefore, an arbitrary quantum function $\psi$ can be written as a linear combination of eigenfunctions $\{\phi_k\}$. Thus, the time evolution of $\psi$ is
\begin{equation}
    \psi(\vec{x},t) = \sumint c_k \phi_k(\vec{x}) e^{-i E_k t/\hbar} dk,
\end{equation}
where the notation $\sumint$ indicates the sum over all the elements in the basis independently of whether they belong to a discrete set, a continuous set or a mixture of both. The coefficients $c_k$ are given by the overlap
\begin{equation}
    c_k = \braket{\phi_k}{\psi(t=0)} = \int \phi_k^*(\vec{x}) \psi(\vec{x},0) d\vec{x}.
\end{equation}
Notice that the coefficient $|c_k|^2$ gives us the probability of the energy level $k$ being occupied, that is, the probability of getting $E_k$ when measuring the energy of the system. Therefore, understanding the properties of a quantum system is highly related to obtaining the spectrum and eigenstates of the corresponding Hamiltonian $\hat{H}$.

\subsubsection{The Born-Oppenheimer Approximation}
The Born-Oppenheimer approximation~\citep{levine} is one of the basic concepts underlying the description of wavefunctions in molecular physics and quantum chemistry. It involves separating the motion of the electrons and atomic nuclei in a molecule by neglecting the movement of the nuclei when describing the electrons. This is possible because the mass of an atomic nucleus in a molecule is much greater than the mass of an electron, which means the electrons experience much faster acceleration and respond to forces more quickly than the nuclei. As a result, the Born-Oppenheimer approximation allows us to describe the electronic states of a molecule by considering the nuclei to be fixed at stationary positions. However, the nuclei can actually be at different positions and the electronic wavefunction can depend on their positions. The molecular wavefunction can therefore be expressed as a product of the electronic wavefunction and the nuclear wavefunction
\begin{equation}
    \psi_{ne}(\vec{r}, \vec{R} ) = \phi_{ne}(\vec{R}) \varphi_{e}(\vec{r}),
\end{equation}
where $\vec{r}$ are the coordinates of the electrons and $\vec{R}$ are the coordinates of the nuclei. The function $\phi_{ne}(\vec{R})$ is the \emph{vibrational wavefunction}, which is a function of the nuclear coordinates $\vec{R}$ and depends on both the vibrational and electronic quantum numbers $n$ and $e$. The \emph{electronic wavefunction}, $\varphi_{e}(\vec{r})$, is a function of both the nuclear and electronic coordinates but only depends upon the electronic quantum number or electronic state, $e$. These functions can be calculated independently by solving the Schrödinger equation with the electronic and the vibrational Hamiltonian, respectively. 

Chapter~\ref{chapter3} will be focused on solving only the \emph{vibrational} Schrödinger equation for different Hamiltonians, using NN~\citep{Domingo_DL} and classical RC~\citep{Domingo_adaptingRC, Domingo_Morse, Domingo_scars}. On the other hand, Chapter~\ref{chapter5} will be devoted to solving the \emph{electronic} Schrödinger equation using QRC~\citep{Domingo_optimalQRC, Domingo_QRCNoise}.

\subsection{The density matrix}
It is often the case that only partial knowledge of the state of a physical system is available. This can happen for many reasons, such as the imperfections of experimental devices or the coupling of the system of interest with its environment. In these situations, the quantum system can be described through the \emph{density matrix}. 

Suppose that a quantum system in a discrete Hilbert space is in state $\ket{\psi_i}$ with probability $p_i$, for $i=1, \cdots, N$. The set $\{p_i, \ket{\psi_i}\}$ is called an ensemble of pure states. The density matrix is defined as
\begin{equation}
    \rho = \sum_i p_i \ketbra{\psi_i},
\end{equation}
and we say that $\rho$ is a \emph{mixed state}. On the other hand, when the quantum state $\ket{\psi}$ is known exactly (i.e. with $p=1$) we say that $\ket{\psi}$ is a \emph{pure state}. The density matrix of a pure state is simply $\rho = \ketbra{\psi}$. Now an interesting question arises. How do we differentiate between pure and mixed states? A density matrix $\rho$ represents a pure state if and only if $\tr \rho^2 = 1$. Otherwise, $\tr \rho^2 < 1$. For this reason, the quantity $\tr \rho^2$ is called the \emph{purity} of a density matrix. It can be easily shown that any matrix $\rho$ is a density matrix if and only if $\rho$ is positive with trace equal to one. 

The laws of quantum mechanics can also be adapted to the density matrix formalism. In particular, the evolution of a quantum state is described using a unitary operator $U$, such that
\begin{equation}
    \rho' = U \rho U^\dag.
\end{equation}
Moreover, quantum measurements are also characterized by a collection $\{M_m\}$ of measurement operators. The probability of result $m$ occurring when measuring state $\rho$ is
 \begin{equation}
     p(m|\rho) = \tr (M_m^\dag M_m \rho),
 \end{equation}
and the state after the measurement is 
\begin{equation}
    \frac{M_m \rho M_m^\dag}{\sqrt{p(m|\rho)}}.
\end{equation}
Finally, the state of a composite system is also described by the tensor product of individual states
\begin{equation}
    \rho = \rho_1 \otimes \rho_2 \otimes \cdots \otimes \rho_n.
\end{equation}

One of the most useful aspects of the density matrix formalism is the description of a subsystem of a composite quantum system. Consider a bipartite system composed of systems A and B. Suppose we do not have access to subsystem B (maybe because it is too far away). Then, we describe the state of system A using the \emph{reduced} density operator
\begin{equation}
    \rho^A \coloneqq \tr_B(\rho^{AB}),
    \label{partial_trace}
\end{equation}
where $\tr_B(\cdot)$ is the \emph{partial trace} over system B. The partial trace is a linear map defined as
\begin{equation}
    \tr_B(\ketbra{\psi_1}{\psi_2} \otimes \ketbra{\phi_1}{\phi_2} ) = \ketbra{\psi_1}{\psi_2} \tr(\ketbra{\phi_1}{\phi_2} ).
\end{equation}
The partial trace is chosen as the reduced density operator because it provides the correct outputs of the measurements made in the reduced subsystem. That is, given an operator $H \otimes \mathbb{I}$ acting only on system A, the definition in Eq.~\ref{partial_trace} satisfies $\tr(H\rho^A) = \tr((H \otimes \mathbb{I})\rho^{AB})$. Let us consider two examples of composite states:

\begin{enumerate}
    \item Consider the state $\rho^{AB} = \rho^A \otimes \rho^B$, with $\rho^A = \ketbra{\psi}$. In this case, $\tr_B\rho^{AB} = \rho^A\tr \rho^B = \ketbra{\psi} $. In this situation, system B does not affect system A. Even ignoring the knowledge of system B, system A is already a pure state. There is no action that we can do to system B that can affect system A. And even though the original state was a mixed state, when tracing out system B we obtain a pure state in system A.
    
    \item Consider the opposite situation. Let the state of the system AB be
    \begin{equation*}
        \ket{\psi}_{AB} = \frac{1}{\sqrt{2}}(\ket{00} + \ket{11}).
    \end{equation*}
    The density matrix is 
    \begin{equation*}
        \rho = \frac{1}{2}(\ket{00} + \ket{11})(\bra{00} + \bra{11}) = \frac{1}{2} (\ketbra{00} + \ketbra{11}{00} + \ketbra{00}{11} + \ketbra{11}),
    \end{equation*}
    and the reduced state of system A is then
    \begin{equation*}
        \rho^A = \tr_B(\rho^{AB}) = \frac{1}{2}(\ketbra{00} + \ketbra{11}) = \frac{1}{2}\mathbb{I}.
    \end{equation*}
    The state of system A is therefore a \emph{maximally mixed state} (a random mixture of the orthonormal states $\{\ket{00}, \ket{11}\}$ with the same probability). Notice that the joint state $\ket{\psi}_{AB}$ is an entangled state. For a maximally entangled state, the correlations of the two subsystems are so strong that if we do not have access to the joint system, we do not know anything about the individual subsystems.
    
\end{enumerate}

\subsection{Quantum channels}
\label{sect:noise_models}
 As stated before, the density matrix formalism allows studying open systems, that is, physical systems that interact with their environment. Real systems suffer from interactions with the world surrounding them, which produce noisy experimental results. Understanding the noise mechanisms and learning how to control them is key to building useful quantum algorithms. 
 
 We have seen in the previous section how a pure state in a bipartite system AB can turn into a mixed state if we observe only system A. Therefore, it is natural to wonder how the evolution of system A if the bipartite system AB undergoes a unitary evolution can be described. The postulates of quantum mechanics give us the answer to this question: if system A starts being unentangled from B, the resulting state after the evolution will be
 \begin{equation}
     \rho_A = \tr_B\Big(U(\rho_A \otimes \rho_B)U^\dag\Big).
 \end{equation}
 This description allows us to model the noisy system A, but it requires explicitly modelling the evolution of the environment. In reality, most of the time we might not know the exact description of the environment, since many different interactions can give the same result in the subsystem of interest. Moreover, if we are only interested in studying system A, we should find a description which only acts on such system.
 
 The operator-sum representation is an elegant formalism to describe quantum operations that circumvents this problem. Given an initial state $\rho$, the final state is described by a linear map $\epsilon$, called a \emph{quantum channel} such that
 \begin{equation}
     \epsilon(\rho) = \sum_m M_m \rho M_m^\dag,
     \label{eq:quantum_channel}
 \end{equation}
 where the operators $\{M_m\}$ follow the completeness relation $\sum_m M_m^\dag M_m = \mathbb{I}$. These quantum channels are called trace-preserving. There are also non-trace-preserving channels, where $\sum_m M_m^\dag M_m < \mathbb{I}$, which apart from unitary evolution also involve measurements. In this case, Eq.~\ref{eq:quantum_channel} has to be normalized so that $\epsilon(\rho)$ has trace 1. The operators $\{M_m\}$ are called \emph{Kraus operators}. 
 
 An interpretation of the operator-sum representation is the following. Imagine we measure system B in the basis $\{\ket{m}\}$ but we fail to record the outcome of the measurement. Then, we are forced to describe the state after the measurement as an ensemble of post-measurement states, weighted by their probabilities, which leads to Eq.~\ref{eq:quantum_channel}. Notice that we have already seen two examples of quantum channels: unitary evolution $\epsilon(\rho) = U\rho U^\dag$ and measurements $\displaystyle\epsilon(\rho) = \frac{M_m \rho M_m^\dag}{\tr (M_m^\dag M_m \rho)}$. \\

 Quantum channels are relevant since they allow us to describe the process of \emph{decoherence}~\citep{preskill1998quantum, decoherence, Nielsen, Domingo_QRCNoise}, which refers to the transition from pure states to mixed states due to interactions of the system with its environment. As the dimension of a quantum system increases, so does the complexity of its interactions with the environment, making decoherence a major challenge for quantum computing. Decoherence can cause errors in quantum computations, leading to a loss of accuracy and efficiency in the quantum algorithms. To mitigate the effects of decoherence in quantum hardware, various techniques have been developed in the field of error correction~\citep{errorCorrecting, Domingo_RL}. These methods involve encoding quantum information so that it can be protected from errors caused by decoherence. Notice that when Eq.~\ref{eq:quantum_channel} has more than one operator the quantum channel is not unitary, and thus not reversible. Once system A is entangled with system B, we cannot undo the decoherence of system A if we do not have access to system B. In this case, system A leaks information into its environment, and since we cannot control the environment, it is impossible to recover this information. Let us present three relevant quantum channels used to model noisy quantum systems.

 \subsubsection{Amplitude damping}
 
 The amplitude damping channel models the effect of energy dissipation, that is, the loss of energy of a quantum system to its environment. It provides, for example, a model of the decay of an excited two-level atom due to the spontaneous emission of a photon. Let $\ket{0}_A$ be the atomic ground state and $\ket{1}_A$ the atomic excited state. The environment is the electronic magnetic field, which is assumed to be initially in its vacuum state $\ket{0}_E$. There is a probability $p$ of the environment emitting a photon and thus changing from $\ket{0}_E$ to $\ket{1}_E$. The unitary evolution of the joint system can be described as 
 \begin{equation}
\begin{split}
    \ket{0}_A\otimes \ket{0}_E & \rightarrow \ket{0}_A\otimes \ket{0}_E,\\
    \ket{1}_A\otimes \ket{0}_E & \rightarrow \sqrt{1-p} \ket{1}_A\otimes \ket{0}_E + \sqrt{p}\ket{0}_A\otimes \ket{1}_E.
\end{split}
\label{eq:unitary_amplitude_damping}
 \end{equation}
 By tracing out the environment in Eq.~\ref{eq:unitary_amplitude_damping} we obtain the following Kraus operators 
\begin{equation}
    M_0 = 
    \begin{pmatrix}
    1 & 0 \\
    0 & \sqrt{1-p}
    \end{pmatrix}, \quad
    M_1 = 
    \begin{pmatrix}
    0 & \sqrt{p} \\
    0 & 0
    \end{pmatrix}.
\end{equation}
 The operator $M_1$ transforms $\ket{1}$ to $\ket{0}$, which corresponds to the process of losing energy to the environment. The operator $M_0$ leaves $\ket{0}$ unchanged, but reduces the amplitude of $\ket{1}$. The quantum channel is thus
 \begin{equation}
     \epsilon(\rho) = M_0 \rho M_0^\dag + M_1 \rho M_1^\dag = \begin{pmatrix}
     \rho_{00} + p \rho_{11} & \sqrt{1-p}\rho_{01}\\
     \sqrt{1-p}\rho_{10} & (1-p)\rho_{11}
     \end{pmatrix}.
 \end{equation}
 \subsubsection{Phase damping}
 The phase damping channel models the loss of quantum information without loss of energy. Physically, it describes, for example, the evolution of a quantum system for an undetermined period of time. In this case, the relative phase between the energy eigenstates changes, thus losing information without losing energy. The unitary representation of this channel is the following
 \begin{equation}
     \begin{split}
         \ket{0}_A  & \rightarrow \sqrt{1-p} \ket{0}_A \otimes \ket{0}_E  + \sqrt{p} \ket{0}_A \otimes \ket{1}_E, \\
         \ket{1}_A &\rightarrow \sqrt{1-p} \ket{1}_A \otimes \ket{0}_E  + \sqrt{p} \ket{1}_A \otimes \ket{2}_E.
     \end{split}
 \end{equation}
 Notice that the qubit does not make any transition between the $\{\ket{0}, \ket{1}\}$ basis. It is the environment which evolves, with probability $p$, to states $\ket{1}$ or $\ket{2}$. The Kraus operators for this process are
 \begin{equation}
     M_0 = \sqrt{1-p} \mathbb{I}, \quad M_1 =  \begin{pmatrix}
     \sqrt{p} & 0\\
     0 & 0
     \end{pmatrix}, \quad
     M_2 = \begin{pmatrix}
     0 & 0 \\
     0 & \sqrt{p}
     \end{pmatrix}.
 \end{equation}
 The quantum channel is then
 \begin{equation}
     \epsilon(\rho) = M_0 \rho M_0^\dag + M_1 \rho M_1^\dag + M_2 \rho M_2^\dag = (1-\frac{p}{2})\rho + \frac{p}{2} Z\rho Z = \begin{pmatrix}
     \rho_{00} & (1-p)\rho_{01}\\
     (1-p)\rho_{10} & \rho_{11}
     \end{pmatrix}.
 \end{equation}
 Thus, an alternative interpretation of the phase damping channel is that the state $\rho$ is left intact with probability $\displaystyle 1-\frac{p}{2}$, and a $Z$ operator is applied with probability $\displaystyle\frac{p}{2}$.
 
 \subsubsection{Depolarizing}
 
 The depolarizing channel can be described by saying that, with probability $1-p$ the qubit remains untouched, while with probability $p$ an error occurs. The error can be, with equal probability, one of these three:
 
 \begin{enumerate}
     \item Bit-flip error: $\ket{\psi} \rightarrow X \ket{\psi}$,
     \item Phase-flip error: $\ket{\psi} \rightarrow Z \ket{\psi}$,
     \item Both bit-flip and phase-flip errors: $\ket{\psi} \rightarrow Y \ket{\psi}$.
 \end{enumerate}
 Thus, if an error occurs, the initial state $\ket{\psi}$ evolves to an ensemble of the three states: $X\ket{\psi}, Y\ket{\psi}, Z\ket{\psi}$. For this reason, the Kraus operators are 
 \begin{equation}
     M_0 = \sqrt{1-p}\mathbb{I}, \ M_1 = \sqrt{\frac{p}{3}}X, \ M_2 = \sqrt{\frac{p}{3}}Y, \ M_3 = \sqrt{\frac{p}{3}}Z,
 \end{equation}
 and the quantum channel is
 \begin{equation}
     \epsilon(\rho) = (1-p)\rho + \frac{p}{3}(X\rho X + Y \rho Y + Z \rho Z) = (1-p) \rho + \frac{p}{2} \mathbb{I}.
 \end{equation}
 Notice that the depolarizing channel transforms the state $\rho$ to the maximally mixed state with probability $p$. This allows generalizing the depolarizing channel to higher dimensional systems (more than one qubit). In that case, the quantum channel of a $d$-dimensional quantum system would be
  \begin{equation}
     \epsilon(\rho) =  (1-p) \rho + \frac{p}{d} \mathbb{I}. 
 \end{equation}
\subsection{Distance between quantum states}
\label{sect:fidelity}
We have seen that quantum channels can be used to model noisy systems. To measure how noisy a system is, we need a way to measure the distance between the noisy state and its noiseless version. A common distance measure between quantum states is the \emph{fidelity}. Let $\rho$, $\sigma$ be two density matrices. The fidelity of $\rho$ and $\sigma$ is defined as
\begin{equation}
    F(\rho, \sigma) = \sqrt{\rho^{1/2} \sigma \rho^{1/2}}.
\end{equation}
At first glance, fidelity may not seem a useful similarity measure. However, it can be proven that it is symmetric with respect to its inputs, that $0 \leq F(\rho, \sigma) \leq 1$ with $F(\rho,\sigma)=0$ if and only if $\rho$ and $\sigma$ are orthogonal, and with $F(\rho, \sigma)=1$ if and only if $\rho=\sigma$. In the particular case where $\sigma = \ketbra{\psi}$ is a pure state, the fidelity becomes
\begin{equation}
    F(\ket{\psi}, \rho) = \tr \sqrt{\bra{\psi} \rho \ket{\psi} \ketbra{\psi}} = \sqrt{\bra{\psi} \rho \ket{\psi}},
\end{equation}
that is, the fidelity is equal to the square root of the overlap between $\ket{\psi}$ and $\rho$.

\subsection{Quantum circuits and quantum operations}
\label{sect:Quantum_gates}
Quantum computing allows manipulating quantum states to design useful algorithms by means of creating \emph{quantum circuits}. Analogous to classical computing, quantum circuits consist of wires and gates, which transform quantum states and manipulate quantum information. With this framework, we can create algorithms that are fundamentally different to classical ones and hope to find solutions to problems that are intractable for classical computers. A relevant example of a quantum algorithm is Grover's algorithm~\citep{Grover}, which reduces the complexity of doing an unstructured search of a list with $N$ items from $\mathcal{O}(N)$ to $\mathcal{O}(N^{1/2})$. An even more impressive algorithm is Shor's algorithm~\citep{Shor}, which provides a $N$-digit integer factorization method with complexity $\mathcal{O}(N^3)$, instead of the classical complexity $\mathcal{O}(e^{N^{1/3}})$. That being said, not all classical algorithms have a quantum counterpart which improves its complexity. It is very hard to design quantum algorithms that outperform their classical version. There are at least two reasons for this. The first one is that designing algorithms, both quantum and classical, is not an easy task, especially if we aim to outperform an existing algorithm. Moreover, the laws of quantum mechanics are rather counterintuitive, and most quantum algorithms do not follow our native intuitions. In this section, we will describe the basic elements to design quantum circuits, the quantum gates.

\subsubsection{Quantum gates}
Single-qubit gates are the simplest quantum operations. Operations between qubits must preserve the norm and thus are represented by $2 \times 2$ unitary matrices.
We have already discussed the Pauli matrices in Sect.~\ref{sect:postulates}. They represent rotations in the Bloch Sphere of angle $\pi$ around the $x$, $y$ and $z$ axis, respectively. Another fundamental quantum gate is the \emph{Hadamard} gate, which allows to create a superposition between $\ket{0}$ and $\ket{1}$
\begin{equation}
    H = \frac{1}{\sqrt{2}}\begin{pmatrix}
    1 & 1 \\
    1 & -1
    \end{pmatrix}, \quad
    H\ket{0} = \ket{+} = \frac{1}{\sqrt{2}}(\ket{0} + \ket{1}), \quad
    H\ket{1} = \ket{-} = \frac{1}{\sqrt{2}}(\ket{0} - \ket{1}).
\end{equation}
The Hadamard gate can be used to switch between $X$ and $Z$ operators, since $X = HZH$ and $Z=HXH$. 

Pauli matrices give rise to another set of gates, called \emph{rotation gates} around the $x$, $y$ and $z$ axis, respectively.
\begin{equation}
    \begin{split}
        R_x(\theta) =e^{-i(\theta \frac{X}{2})} &= \cos \Big(\frac{\theta}{2}\Big) \mathbb{I} -i \sin\Big(\frac{\theta}{2}\Big) X =  \begin{pmatrix}
        \cos \frac{\theta}{2} & -i\sin \frac{\theta}{2}\\
        -i \sin \frac{\theta}{2} & \cos \frac{\theta}{2}
        \end{pmatrix}, \\
        R_y(\theta) = e^{-i(\theta \frac{Y}{2})} &= \cos\Big(\frac{\theta}{2}\Big) \mathbb{I} -i \sin \Big(\frac{\theta}{2}\Big) Y =  \begin{pmatrix}
        \cos \frac{\theta}{2} & -\sin \frac{\theta}{2}\\
         \sin \frac{\theta}{2} & \cos\frac{\theta}{2}
        \end{pmatrix}, \\
        R_z(\theta) = e^{-i(\theta \frac{Z}{2})} &= \cos \Big(\frac{\theta}{2}\Big) \mathbb{I} -i \sin \Big(\frac{\theta}{2}\Big) Z =  \begin{pmatrix}
        e^{-i \frac{\theta}{2}} & 0\\
        0& e^{i \frac{\theta}{2}}
    \end{pmatrix}. 
\end{split}
\end{equation}
 Another set of fundamental gates is the \emph{phase} gates. These are parametrized gates since one needs to specify the angle $\phi$ to fully define the gate 
\begin{equation}
	P = \begin{pmatrix}
	1 & 0 \\
	1 & e^{i\phi}
	\end{pmatrix}, \ \phi \in \mathbb{R}.
\end{equation}
Notice that the $Z$ gate is a phase gate with $\phi=\pi$, and the identity gate is a phase gate with $\phi=0$. Other relevant phase gates are the $S$ gate and the $T$ gate. The $S$ gate, also called the $\sqrt{Z}$ gate, is a phase gate with $\phi=\pi/2$. It does a quarter turn in the Bloch Sphere around the $z$ axis. Notice that unlike the other gates introduced so far, the $S$ gate is not its own inverse.
\begin{equation}
S = \begin{pmatrix}
1 & 0 \\
0 & e^{i \pi/2}
\end{pmatrix}, \quad S^\dag = \begin{pmatrix}
1 & 0 \\
0 & e^{-i\pi/2}
\end{pmatrix}.
\end{equation}
 In fact, the reason why it is also called $\sqrt{Z}$ gate is because
\begin{equation}
SS \ket{\psi} = Z \ket{\psi}, \quad \forall \ket{\psi}.
\end{equation}
The $T$ gate is a phase gate with $\phi=\pi/4$, also known as the $\sqrt[4]{Z}$ gate
\begin{equation}
T = \begin{pmatrix}
1 & 0 \\
0 & e^{i \pi/4}
\end{pmatrix}, \quad T^\dag = \begin{pmatrix}
1 & 0 \\
0 & e^{-i\pi/4}
\end{pmatrix}.
\end{equation}
The most general single qubit gate is the $U$ gate, which can be decomposed as a series of rotation gates
\begin{equation}
    U(\theta, \phi, \lambda) = R_z(\phi)R_y(\theta)R_z(\lambda) = \begin{pmatrix}
    \cos \frac{\theta}{2} & -e^{i\lambda}\sin \frac{\theta}{2} \\
    e^{i\phi} \sin \frac{\theta}{2} & e^{i(\phi + \lambda)} \cos \frac{\theta}{2}
    \end{pmatrix}.
\end{equation}
Every single qubit gate can be written as a $U$ gate. For example, $U(\pi/2, 0, \pi) = H$ and $U(0,0,\phi) = P(\phi)$. There are also quantum gates that affect more than one qubit at the same time. These gates are very relevant because they allow the creation of entangled states, which are the key to many quantum algorithms. 

\begin{figure}
    \centering
    \includegraphics[width=0.9\textwidth]{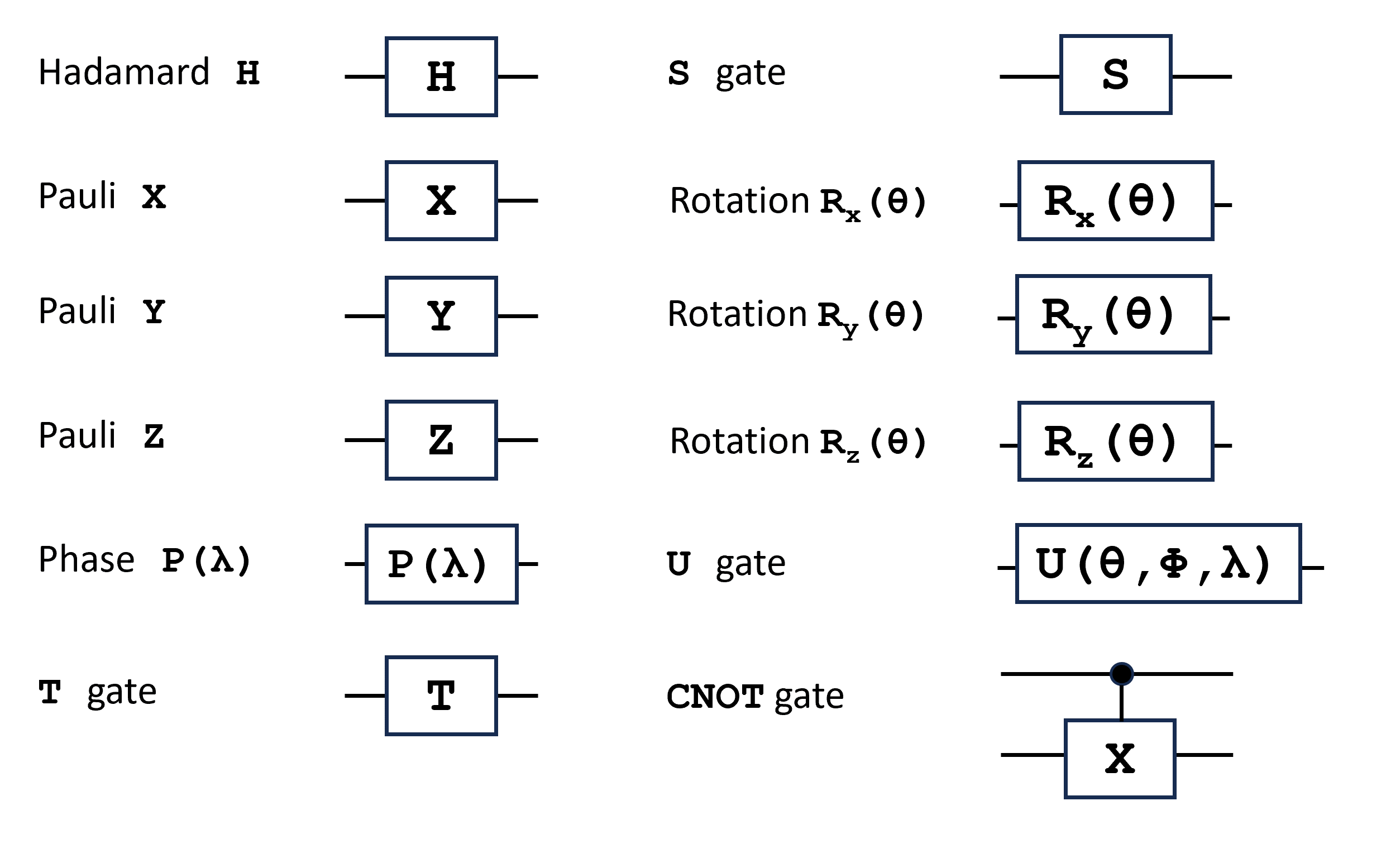}
    \caption{Names and symbols for the most common quantum gates.}
    \label{fig:quantum_gates}
\end{figure}

\emph{Controlled operations} are very useful gates, both in classical and quantum computing. The most relevant controlled operation is the controlled-NOT or $CNOT$ which takes two input qubits, known as the \emph{control qubit} and \emph{target qubit}, respectively. If the control qubit is $\ket{1}$, then the target qubit is flipped, leaving the control qubit intact. 
\begin{equation}
    CNOT = \begin{pmatrix}
    1 & 0 & 0 & 0 \\
    0 & 1 & 0 & 0 \\
    0 & 0 & 0 & 1 \\
    0 & 0 & 1 & 0
    \end{pmatrix}.
\end{equation}
The symbols for the quantum gates introduced in this section are shown in Fig.~\ref{fig:quantum_gates}.
\subsubsection{Universality}
 There are infinitely many quantum gates. However, it is not necessary to implement them all, since there exists a small set of gates that can generate any quantum gate on any number of qubits with arbitrary precision. These sets are called \emph{universal} because they can produce universal quantum computation. That is, they can reproduce anything a quantum computer can do. 
 
 Before presenting some universal sets, it is relevant to first introduce the \emph{Clifford gates}. We have already seen that the Hadamard gate can be used to switch $Z$ and $X$ gates. The $S$ gate has a similar behaviour
 \begin{equation}
     S X S^\dag  = Y, \quad SYS^\dag  = -X, \quad SZS^\dag = Z.
 \end{equation}
 So the $S$ gate swaps $X$ and $Y$ instead of $X$ and $Z$. Thus, combining Hadamard and $S$ gates we can swap $Y$ and $Z$ as well. Clifford gates have the property to transform Pauli gates into Pauli gates via conjugation. That is, if $U$ is a Clifford gate and $P$ is a Pauli matrix then $UPU^\dag$ is a Pauli matrix. For multi-qubit gates, Clifford gates transform tensor product of Pauli matrices to another tensor product of Pauli matrices. For example, the CNOT is a Clifford gate since
 \begin{equation}
     CNOT (X \otimes \mathbb{I}) CNOT = X \otimes X.
 \end{equation}
 Not all gates are Clifford gates. The rotation gates, for example, follow the property
 \begin{equation}
     U R_P(\theta) U^\dag = e^{i \frac{\theta}{2} UPU^\dag}, \quad P  \in \{X,Y,Z\}.
 \end{equation}
 Therefore, by conjugating a rotation with a Clifford gate we can achieve rotations on a different axis. Now, combining $R_x(\theta)$ with $CNOT$ we get
 \begin{equation}
     CNOT(R_x(\theta) \otimes \mathbb{I})CNOT = e^{i\frac{\theta}{2} X \otimes X}.
     \label{eq:CNOT_rot}
 \end{equation}
 Thus, the $CNOT$ gate allows the transformation of a simple one-qubit rotation to a more powerful two-qubit gate. Notice that this is not equivalent to performing the rotation independently in the two qubits, since the gate in Eq.~\ref{eq:CNOT_rot} can produce entanglement. Now, if we combine the output with a $S$ gate, we can obtain rotations in a different axis
 \begin{equation}
     (S \otimes \mathbb{I})e^{i\frac{\theta}{2} X \otimes X} (S^\dag \otimes \mathbb{I}) = e^{i \frac{\theta}{2} Y \otimes X}.
 \end{equation}
 These relationships give us the intuition to introduce a universal set of gates: $CNOT$ with single qubit gates. For a more formal proof refer to Ref.~\citep{Nielsen}. Moreover, since any single qubit gate can be expressed as a combination of rotations, it is sufficient to use $CNOT$ gates with $x$ and $z$ rotations. However, this set still has infinite quantum gates, since $R_x(\theta)$ and $R_z(\theta)$ can take infinitely different values of $\theta$. 
 
 A simpler universal set is composed of the Clifford gates $\{CNOT, H\}$ and the non-Clifford gate $T$ (again, see Ref.~\citep{Nielsen}). There are, of course, infinitely many sets of universal gates. Notice, however, that Clifford gates alone cannot form a universal set because they cannot approximate non-Clifford gates. 
   
\subsection{Quantum machine learning}
\label{sect:QML}
Many ML methods could potentially be improved by combining classical and quantum methods to design hybrid algorithms. In this section, we present a brief overview of relevant QML algorithms. For a more exhaustive explanation of QML check Refs.~\citep{NISQ, QML}. 

\begin{figure}[!ht]
    \centering
    \includegraphics[width=1.0\textwidth]{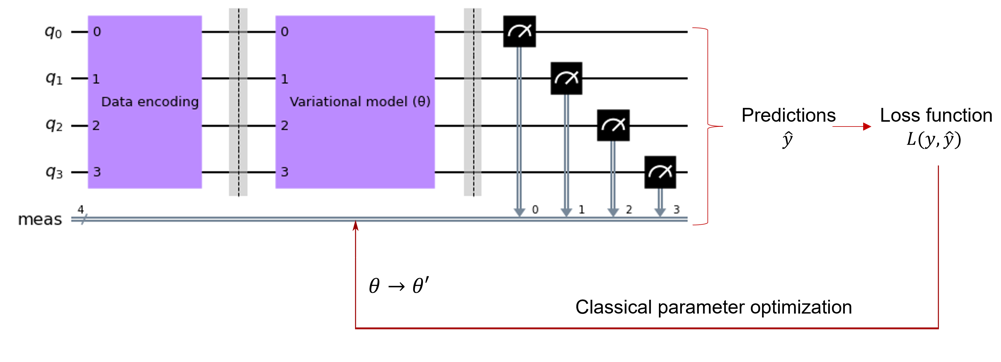}
    \caption{Example of a variational quantum circuit.}
    \label{fig:variational_circuit}
\end{figure}

One of the fundamental NISQ algorithms for QML is parameterized quantum circuits (\gls{PQC}) ~\citep{PQC}. These models consist of circuits containing quantum gates with tunable parameters, which are updated during the training process. PQCs are usually composed of two blocks: data encoding and a quantum variational model. In the first block, the quantum gates map the training data to a quantum circuit. In the second block, parametrized quantum operations followed by measurements are used as a learning algorithm to perform a machine learning task. PQCs are mostly used for supervised QML algorithms. 

Figure~\ref{fig:variational_circuit} shows a schematic representation of the training steps of a PQC in a supervised task. After the (possibly random) initialization of parameters, the model calculates the predictions of the training data. These predictions are used to calculate the loss function, which we aim to minimize. The parameter optimization process can be done with classical optimization methods, such as gradient descent. PQC are the key to multiple QML algorithms, such as the variational quantum eigensolver~\citep{VQE} which uses the variational principle to compute the ground state energy of a Hamiltonian. Also, the quantum approximate optimization algorithm~\citep{QAOA} uses PQC to solve discrete combinatorial optimization. Finally, hybrid quantum-classical NNs use PQC to replace some of the classical layers of the network~\citep{QNNMedical}.

Another set of QML algorithms is based on \emph{kernel methods}. Many machine learning algorithms map the input data to a higher-dimensional space which allows to extract relevant properties from the data. This mapping is usually done via a kernel function such that
\begin{equation}
    k(\vec{x}_i, \vec{x}_j) = \expval{f(\vec{x}_i), f(\vec{x}_j)}.
\end{equation}
with $k$ the kernel function, $\vec{x}_i, \vec{x}_j$ the input samples, $f$ a mapping to a higher dimensional space and $\expval{\cdot}$ the inner product. In QML, the function $f$ can be replaced by a quantum operation, usually a PQC. If the quantum operation is hard to simulate classically, quantum kernels can potentially provide an advantage over classical methods~\citep{SVM}. The quantum kernel can be written as
\begin{equation}
    k(\vec{x}_i, \vec{x}_j) = |\braket{\phi(\vec{x}_j)}{\phi(\vec{x}_i)}|^2 = |\bra{0}^{\otimes n} U^\dag(\vec{x}_j) U(\vec{x}_i) \ket{0}^{\otimes n}|^2,
\end{equation}
where the unitary $U(\cdot)$ represents the quantum circuit. Kernel methods are useful for many ML algorithms, the most popular being support vector machines~\citep{QSVM}. This algorithm constructs a plane in the quantum hyperspace that is used for classification or regression tasks. Other algorithms include Principal Component Analysis~\citep{QPCA} and Gaussian Processes~\citep{QGaussianProcess}.

Quantum algorithms have also been proposed to tackle unsupervised problems. In particular, quantum computing is useful for dealing with generative models, such as Quantum Generative Adversarial Networks~\citep{QGAN}. In this case, instead of predicting the target value for new data, generative models learn the underlying distribution of the training data to generate new samples from the same distribution.

Until now, we have only discussed gate-based quantum computing, that is, quantum computing based on qubits and quantum gates. Another relevant universal quantum computing paradigm is \emph{adiabatic quantum computing}. This formalism relies on the adiabatic theorem, which states that if a quantum system starts at $t=0$ in the ground state of a Hamiltonian $H(t)$ which evolves slowly with time, the system will continue to be at the ground state of the Hamiltonian for $t>0$. 

Adiabatic quantum computation exploits this phenomenon by initially preparing a system in the ground state of a simple, well-known Hamiltonian. Then, the system evolves adiabatically towards a Hamiltonian whose ground state corresponds to a solution to the problem at hand. The challenge with this method is knowing how slowly we need to evolve the state towards the solution of our problem. The value of $\Delta(t)$ depends on the specific Hamiltonian and thus it is generally unknown. For this reason, a relaxation of adiabatic quantum computing is usually used, called \emph{quantum annealing}~\citep{QuantumAnnealing}. This paradigm drops the requirement of the system evolving adiabatically (at a certain speed) and instead repeats the Hamiltonian evolution many times. Then, we select the final state with the lowest energy, which should approximate the actual ground state of the system. It turns out that this paradigm is suitable for solving optimization problems. In particular, for solving quadratic unconstrained binary optimization problems of the form
\begin{equation}
    s^* = \underset{s \in \{+1,-1\}^n}{\text{argmin}} s^T Q s + q^Ts, \quad Q \in \mathbb{R}^{n \otimes n}, \ q \in \mathbb{R}^n,
\end{equation}
where $Q$ models internal interactions and $q$ represents additional constraints.

The properties and dynamics of quantum systems can also be used for ML purposes, which is the case of QRC. In this thesis, we study in depth the properties and designs of the QRC algorithm. The next subsection provides a detailed explanation of the basic architecture of QRC.

\subsection{Quantum reservoir computing}
\label{sect:QRC}

The main advantage of RC, compared to traditional NNs, is the simplicity of the learning model, which usually only requires training a linear regression model. To this end, classical RC uses a network with fixed random weights to learn the input-output dynamics. In contrast, Quantum Reservoirs (\gls{QR}s) are quantum systems with random parameters, which are used to learn from the input data. 

The motivation for using quantum devices lies in the large number of degrees of freedom of quantum systems. A quantum system with $N$ qubits has a Hilbert space of size $2^N$, which means that quantum states have the potential to encode exponentially larger data than classical systems. QRs are particularly useful when dealing with quantum data, such as molecular data. In this case, since the data already model a quantum system, it is natural to use quantum operations to learn its hidden patterns. Moreover, even though quantum computers suffer from noise and decoherence, the easy training strategy of QRC makes it a suitable algorithm for NISQ devices~\citep{Domingo_optimalQRC}. In fact, it has been shown that the amplitude damping noise can be beneficial for the training of the algorithm~\citep{Domingo_QRCNoise}.

QRC has been employed for both temporal~\citep{Domingo_quantum_zucchini} and non-temporal tasks~\citep{Domingo_optimalQRC}. For non-temporal tasks, the term \emph{quantum extreme learning machines} has also been used to denote ML tasks that use QRs~\citep{reviewQRC}.  

\subsubsection{Quantum reservoir computing for non-temporal tasks}
The idea of QRC lies in
using a Hilbert space, where quantum states live, as an
enhanced feature space of the input data. In this way,
the extracted features enhanced by quantum operations are used to feed a classical machine learning model, which predicts the desired target.  Consider the dataset $\{(\vec{x},\vec{y})_i\}$. The first step for designing a QR is selecting a data encoding method, which maps the input data into a quantum circuit. By convention, the initial state of the quantum system is always $\ket{0}^{\otimes n}$. The data encoding process maps
\begin{equation}
    \ket{0}^{\otimes n} \rightarrow V(\vec{x})\ket{0}^{\otimes n},
\end{equation}
where $V(x)$ is a unitary transformation that depends on the input sample $x$. There are multiple ways to design the data encoding mapping. If the input data is a normalized vector $\vec{x}= (x_1, \cdots, x_n), \ x_i \in [0,1]$, then a popular data encoding method generates the state
\begin{equation}
    \ket{x} = (\sqrt{1-x_i}\ket{0} + \sqrt{x_i}\ket{1})^{\otimes 1 \cdots n},
\end{equation}
which uses a $n$-qubit state to encode a $n$-dimensional vector. Another encoding for classical data is the \emph{amplitude encoding}~\citep{amplitudeEncoding}, which only uses $\lceil \log_2(n) \rceil $ qubits to encode a $n$-dimensional vector. In this case, the mapping is
\begin{equation}
    \ket{x} = \frac{1}{|\vec{x}|} \sum_{i=1}^n x_i \ket{i},
\end{equation}
where $\ket{i}$ is the $i$-th computational basis state. This encoding is particularly useful to encode high-dimensional data in a low-dimensional Hilbert state. However, this state preparation requires many more gates than the previous encoding. When the input data is already quantum, the data encoding consists of simply preparing the initial state of the quantum circuit as the quantum state of the data sample. For example, in Chapter~\ref{chapter5} we will see how we encode the ground state of an electronic Hamiltonian as the initial state of the QR. 

After encoding the data in a quantum state, a unitary transformation $U$ is applied to extract useful features from the input data
\begin{equation}
    V(\vec{x})\ket{0}^{\otimes n} \rightarrow UV(\vec{x})\ket{0}^{\otimes n}.
\end{equation}
The operator $U$ is chosen to create enough entanglement to generate useful extracted features from the input data while being experimentally feasible. Then, the expected value of single-qubit observables is measured. These observables are usually the Pauli operators $\{X_0, Z_0, \cdots X_j, Z_j, \cdots X_n, Z_n\}$, where $X_j, Z_j$ represent the Pauli operators $X,Z$ applied to qubit $j$. Notice that, in general, a $n$-qubit unitary $U$ transforms a simple observable $Z$ (or $X$) into a linear combination of Pauli operators 
\begin{equation}
    UZ_1U^\dag = \sum_i \alpha_i P_i, \quad \alpha_i \in \mathbb{C}
\end{equation}
where $\{P_i\}$ are tensor products of local Pauli operators. Therefore, measuring single Pauli operators of a state that has received a unitary transformation could produce complex nonlinear outputs, which could be represented as a linear combination of exponentially many nonlinear functions~\citep{QRC}. 

Finally, the extracted features $\hat{x} = (\expval{X_1}, \expval{Z_1}, \cdots \expval{X_n}, \expval{Z_n})$ are fed to a classical ML algorithm, usually a linear model. Even though more complex models can be used, the QR should be able to extract valuable features so that a simple ML model can predict the target $\vec{y}$. Usually, a simple linear model with regularization, such as ridge regression (see Eq.~\ref{eq:ridge}), is enough to learn the output. 

In conclusion, QRC for non-temporal tasks consists of a linear regression applied to randomly chosen quantum-enhanced features, which come from an exponentially large Hilbert space. Moreover, for some unitaries $U$, the linear model can approximate any continuous function of the input. This property is known as the \emph{universal approximation property}, which implies that QRC can realize machine learning tasks with at least the same power as classical RC~\citep{QRC}.

\subsubsection{Quantum reservoir computing for temporal tasks}

In this setting, instead of having independent data samples, we aim to predict the time evolution of a series $\{\vec{y}(t)\}$. The difference in the QRC algorithm is that for temporal tasks the input has to be fed to the quantum system sequentially. That is, the quantum state is re-initialized at each step of the computation, which requires the quantum state to be a mixed state. For a system of $N>n$ qubits, the first $n$ qubits would be initialized, at each time step $t$ as
\begin{equation}
    \ket{y(t)} = (\sqrt{1-y(t)_i}\ket{0} + \sqrt{y(t)_i}\ket{1})^{\otimes 1 \cdots n},
    \label{eq:input}
\end{equation}
and the full, $N$-qubit state is
\begin{equation}
    \rho(t) = \ketbra{y(t)} \otimes \tr_n\Big(\rho(t-\Delta t)\Big),
\end{equation}
where $y(t)_i$ is the $i$-th component of the time series, and the partial trace $\Tr_n(\cdot)$ is done on the first $n$ qubits. Notice that we require $N>n$ so that the state at time $t$ has information from its past states. The first $n$ qubits are used to encode the data, and the remaining $N - n$ qubits are used to gather information from the previous values of the time series, which is captured by the partial trace in Eq.~\ref{eq:input}.  

After encoding the quantum state at time $t$, the system evolves under a unitary transformation $U$ until time $t + \Delta t$. The unitary transformation can be a quantum circuit, or equivalently, the evolution under a Hamiltonian. In the QRC framework, instead of fine-tuning the dynamics of a 
physical system, the natural, inherent dynamics of the system are used to perform ML tasks. Therefore, the quantum operation consists of a \emph{random} unitary $U$ sampled from a carefully selected family. The state after the unitary evolution is then
\begin{equation}
    \rho(t + \Delta t) = U \rho(t) U^\dag .
\end{equation} 
A commonly used unitary transformation is the evolution under the transverse field Ising model, $\displaystyle U = e^{-i\hat{H}_{\text{Ising}} t}$, whose Hamiltonian is
\begin{equation}
    \hat{H}_{\text{Ising}} = \sum_{i,j=0}^{N-1} J_{ij} Z_iZ_j + \sum_{i}^{N-1} h_{i} X_i, \quad J_{ij}, h_i \in \mathbb{R}.
\end{equation}
At the end of each timestep $t$, the expected value of some local Pauli operators $\{P_i\}$ are calculated, where
\begin{equation}
    \expval{P_i}(t) = \tr \Big( P_i \rho(t+\Delta t)\Big).
\end{equation} 
Finally,  after a sufficiently long training time, the extracted features $\hat{x}(t) = \{\expval{P_i}(t)\} = (\expval{X_1}(t), \expval{Z_1}(t), \cdots \expval{X_n}(t), \expval{Z_n}(t))$ are fed to a classical ML algorithm which predicts the target $y(t + \Delta t)$. The ML algorithm is usually a ridge regression, which again minimizes the expression in Eq.~\ref{eq:ridge}. 

Once the output layer has been trained, the system can evolve autonomously. In this case, the linear model outputs the prediction $\hat{y}(t)$ that will then be fed as the input state at time $t+\Delta t$. Moreover, when the number of qubits of the system $N$ is small, the size of the extracted features containing the expectation values of the local Pauli operators may be quite small. For this reason, one can gather the readout signals from intermediate times between $[t, t+ \Delta t]$. In this case, the input timescale $\Delta t$ is divided into $N_v$ time steps, such that the intermediate steps are given by $t^k = t + k \Delta t/N_v$, $k=0 \cdots N_v$. Then, the feature vector $\hat{x}(t)$ contains the 2$N$ expected values from the $N$ qubits and the intermediate times $t^k$, leading to a total dimension of $2N\times N_v$.

\chapter{Reservoir computing for time series forecasting}
\label{chapter4}

\begin{fquote}[Albert Einstein][1879-1955]
    The important thing is not to stop questioning. Curiosity has its own reason for existing.
\end{fquote}
 This chapter focuses on investigating the adaptation of the RC algorithm to address a highly relevant real-life problem: forecasting the prices of agricultural products. 
 
Understanding the agri-food industry~\citep{FW,scirep1,nfood,nfood2} is becoming increasingly more important every day, especially in connection with making food systems sustainable and more equitable as indicated by the United Nations Development Goals~\citep{UN2015}. Indeed, the agri-food industry is very relevant in the European global market~\citep{dash}, particularly in Spain~\citep{agri-food-Spain}, where it still remains a fundamental pillar of the national economy. 

The agri-commodities market is especially susceptible to food waste due to price crises, where food prices get so low that its marketing is no longer profitable. In fact, one of the biggest issues regarding food loss is that around 14\% of the food produced in the world is lost between harvest and retail~\citep{FoodLoss}. Such price crises are common in Europe, where the commercialization of some products is not profitable, automatically translating into tons of thrown-away food. 

Therefore, accurately predicting changes in the agri-food market is critical to guarantee both the sustainability of the food system and ensure food security. However, this is not an easy task, since the problem implies analyzing small and very volatile time series, which are highly influenced by external factors. In this chapter, we tackle the current food loss problem by exploring the predictability of the agri-food market, focusing on anticipating extreme price changes that can lead to severe price crises.

The organization of this chapter is as follows. In the first part of this chapter (Sect.~\ref{paper:RC_calabacines}) we analyze the use of RC for forecasting the prices of three different vegetables. To do so, the performance of five variants of the RC algorithm are evaluated when predicting future prices of these products. The results show that RC, together with a clever decomposition of the data in components, can outperform traditional time series forecasting approaches. This method is used to predict the time evolution of the prices of three agricultural products (zucchini, tomato and aubergine), in the markets in the southeast of Spain, which is one of the largest supplier regions to the European market. 

Then, in Sect.~\ref{sect:discussion_agro} we discuss two relevant challenges that persist in the realm of forecasting agri-commodities prices, and that will be further developed in future works. The first challenge, discussed in Sect.~\ref{sect:confidence_intervals}, consists in predicting the confidence intervals of future prices, instead of solely relying on deterministic predictions. The second challenge, presented in Sect.~\ref{sect:cross-correlation}, is the design of multivariate RC models, which take into account the prices of multiple agricultural products to make more accurate predictions.

\section{Reservoir computing for agricultural prices prediction }
\label{paper:RC_calabacines}

 In this section, we examine how the RC algorithm can be used to forecast the prices of different agricultural products. Traditionally, research on agricultural prices forecast has heavily relied on Seasonal Autoregressive Integrated Moving Average (\gls{SARIMA})~\citep{sarima,processes} models (see, for example, Refs.~\citep{jadhav2017application,pollos,cerdos}). Other studies have focused on using NNs for price forecasting. For example, in Ref.~\citep{x1} such method was used to simulate  China’s food security early warning system, and in Ref.~\citep{x2} four NNs were developed to predict China's vegetable market prices~\citep{agris}. However, whether the models can anticipate or not the direction of the market is not properly addressed in this type of studies.

One of the main challenges in the prediction of agri-commodity prices is that the highest time resolution available for fruit and vegetable prices is daily, something that severely limits the size of the training samples. Another important challenge is that markets are highly connected networks, 
where prices are influenced by many factors, such as climate, crop production, and exports, among others. Thus, all these factors have to be taken into account to obtain accurate models. Apart from using regressor variables, one also needs to consider the time component of the prices, which provides additional information 
but also implies dealing with new complexities, not present in other ML tasks. For this reason, a ML method that can be trained with few data samples and consider both the regressor variables and the time evolution of the training data is required. RC is a perfect fit to overcome these limitations. RC uses both the past value of the time series and regressor variables to extrapolate the time series. Moreover, the simple training framework of RC reduces the chances of overfitting the training data when working with small datasets. 

In this section, the performance of five variants of the 
RC model applied to the agri-food industry is assessed, 
by comparing their performance with that of the two benchmark algorithms, namely SARIMA and LSTM.  The main conclusion of this comparison is twofold. First, RC clearly outperforms the benchmark models in predicting price time series, especially in anticipating sudden changes in their tendency. Second, the study provides an optimal RC architecture, based on decomposing the time series, 
to anticipate the evolution of food prices under the conditions prevailing in the agri-food market.

Apart from finding an appropriate learning algorithm, it is highly important to correctly evaluate the quality of the predictions. Since prices always tend to be autocorrelated in time, a model can output the price at time $t-1$ as the prediction for the price at time $t$ and still have a low error metric. Such models are obviously useless since they cannot anticipate changes in the 
direction of the market. Although this fact is extremely important when addressing price prediction, it has been often ignored in the scientific literature. 

In the present study, we overcome this problem by evaluating our models not only using the MAE (Mean Absolute Error, introduced in Sect.~\ref{sect:train_NN}), but also using a new metric introduced by us, called the Market Direction Accuracy (\gls{MDA}). This metric checks if a ML model can correctly predict a change of tendency in the price time series (see Sect.~\ref{sect:results_MDA} for details). Therefore, a model rendering minimal MAE but low MDA totally lacks interest, because it would not be able to predict abrupt price changes, which are crucial to prevent food crises. As a striking illustration of this, the results will show that the LSTM, which does not consider external factors, is unable to anticipate the direction of the zucchini market, giving results that are only barely better than pure random guessing. In contrast, the most accurate RC method achieves a $\sim$80\% MDA, which shows that the method is an excellent and profitable ML model to make predictions in this setting.

The organization of the next subsections is as follows. Section~\ref{subsec:H} describes the datasets used for this work. Then, Sect.~\ref{RC_zucchini} presents the standard and variations of the RC models considered in this study. The results in terms of the MAE are presented in Sect.~\ref{sect:results_MAE}, while the results of the MDA are discussed in Sect.~\ref{sect:results_MDA}.

\subsection{Dataset}
  \label{subsec:H}
The present work is based on data concerning the prices at origin in Spain of three different vegetable varieties: zucchini, aubergine and tomato. These prices are published daily after the different auctions have taken place and refer to the price in euros per kilogram. To build up the time series, we have aggregated the prices of each variety weekly, coming up with the average price of the week. This is done because weekly predictions are more representative, and then valuable than the daily ones for the agri-food market. The dataset includes prices from March 2013 to March 2022. Notice that this period includes the COVID-19 pandemic lock-up, which will be reflected in the results. 

These time series are split into training, validation and test sets. The training set contains the prices until December 2019, resulting in 356 data points. The validation set contains the prices of 2020 (53 data points), 
and the test set contains the prices from January 2021 onwards (62 data points). When training the models, the time series is standardised to the $[-1, 1]$ domain using a linear scale.

The validation set is used to select the optimal hyperparameters of each model. The test set is then used to evaluate the performance of the selected models on unseen data. Apart from the time series, the RC models used in this work also use 16 regressor variables to predict the price's time evolution. Such variables gather information about production volumes and international trade that provide complementary information for price prediction.

\subsection{Reservoir computing models}
\label{RC_zucchini}
\begin{figure}[!ht]
\centering
\includegraphics[width=1.0\textwidth]{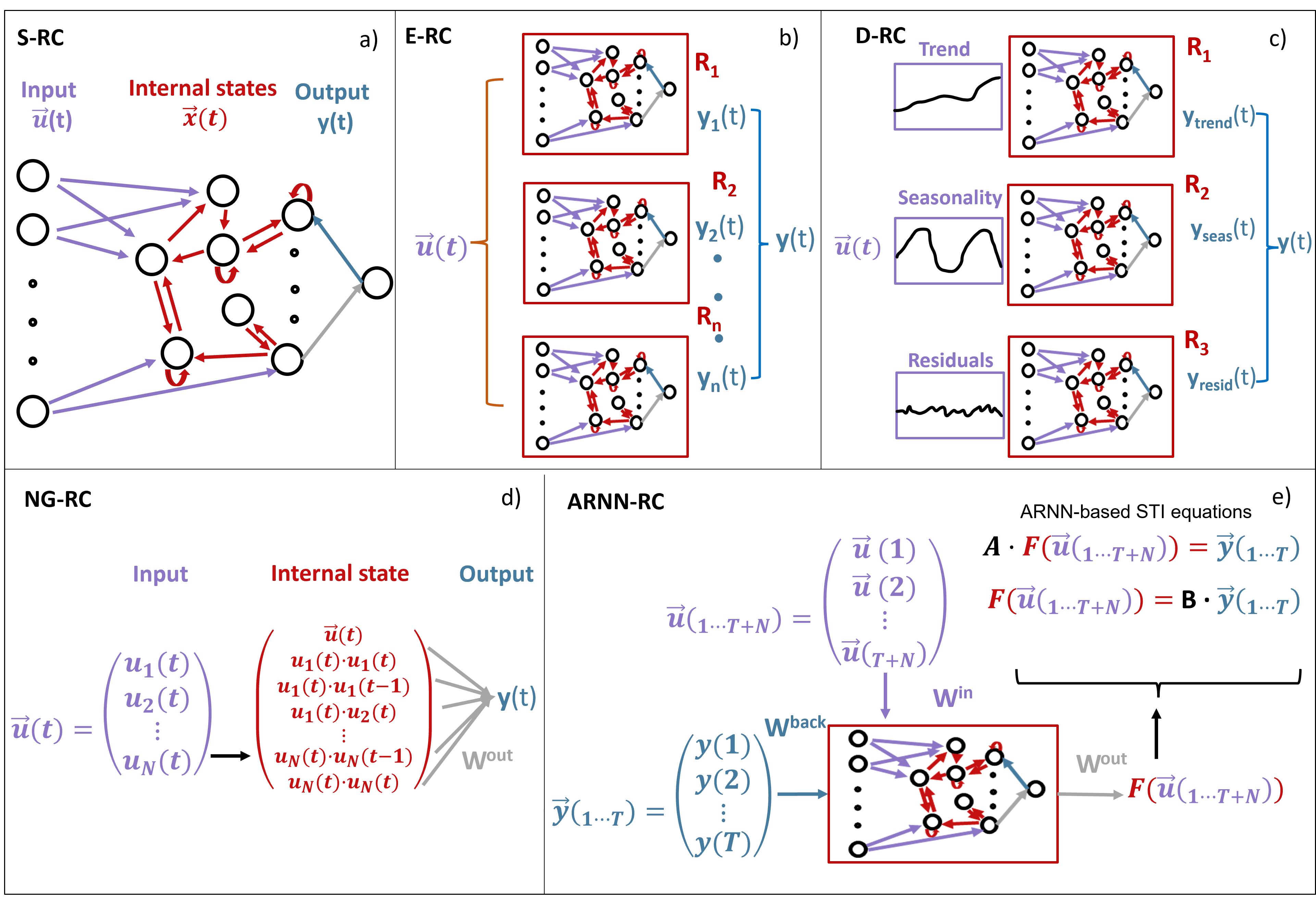}
    \caption[Schematic architecture of the five reservoir computing based models studied in this work: 
    a) S-RC, b) E-RC, c) D-RC, d) NG-RC and e) ARNN-RC.]{Schematic architecture of the five reservoir computing-based models studied in this work: 
    a) represents the original or standard reservoir computing (S-RC) model introduced in Ref.~\citep{ESN}, b) shows an ensemble of reservoir computing (E-RC) models, whose outputs are averaged to construct the final prediction, in c) the time series is decomposed (D-RC) into three parts: trend, seasonality and residuals, and then model E-RC is used to predict each of the series components, d) is the next generation reservoir computing (NG-RC) model presented in Ref.~\citep{next-genRC}, which only uses linear and non-linear functions of the output to predict its dynamics, finally, e) represents the auto recurrent neural network reservoir computing (ARNN-RC) model presented in Ref.~\citep{STI}, which obtains the predictions by mapping the future high-dimensional regressor variables to the associated time series prediction.
    }
\label{fig:RC_models}
\end{figure}

This section describes four variations of the RC algorithms, which will be compared with the two benchmark models (LSTM and SARIMA) and the standard RC architecture, which will from now on, in this section, be referred to as S-RC (see Fig.~\ref{fig:RC_models} and Sect.~\ref{sect:RC}).

\subsubsection{Ensemble Reservoir Computing (\gls{E-RC})}
  \label{subsec:B}
When the size of the training data is small, the performance of the reservoir 
depends strongly on its actual realisation. 
The choice of matrices $W^{\text{in}}, W^{\text{back}}$, and $W$, even when maintaining 
the spectral radius and sparsity, can affect the results of the model. 
Therefore, different runs of the same algorithm may give different results, 
unless the exact same matrices are used. 
In these circumstances, the robustness of the model can be increased by training 
multiple reservoirs with the same hyperparameters (given in  Table~\ref{tab:hyperparams}). 
The different predictions are then averaged to produce the final output. 
This setting is schematically depicted in Fig.~\ref{fig:RC_models} b). 
The final model, in this case, produces more stable results 
with lower variance in the generalization error.

\subsubsection{Trend-seasonality decomposition Reservoir Computing (\gls{D-RC})}
  \label{subsec:C}
When the consecutive values of a time series are too similar, 
many predictive models tend to (wrongly) predict the previous value of the series. 
That is, if $y(t) = y(t-1) + \epsilon$ with $|\epsilon|<<1$, the model might predict 
$\hat{y}(t) = y(t-1)$, and still produce a small error $\epsilon$ in the prediction. 
One way to prevent the model from mimicking the previous value is to decompose the 
series into \textit{trend, seasonality} and \textit{residuals}, 
and then predict each component separately, as shown schematically in Fig.~\ref{fig:RC_models} c). 
Notice that this is very relevant for the price series that are being studied in this work
since they have a strong seasonal periodicity; 
therefore this decomposition is useful to remove temporal patterns. Moreover, the RC models may use different information when forecasting each time series component. For example, the RC predicting the seasonal component may be more influenced by climate or the average price at a particular time. On the other hand, the RC model predicting the residuals may be more influenced by external factors such as the COVID-19 pandemic.

In this case, an additive time series decomposition is suitable to decompose the time series. Indeed, we have performed the Augmented Dickey–Fuller test~\citep{dickey} on the residuals, which is a statistical test used to determine whether a time series has a unit root, indicating non-stationarity. If the $p$-value of this test is high, it means that the residuals are not stationary and thus need to be further decomposed. The $p$-values of this test are $3\times 10^{-11}, 2 \times 10^{-14}$ and $3 \times 10^{-14}$, for the zucchini, aubergine and tomato prices time series, respectively, this meaning that we can reject with high confidence the hypothesis of the residual series needing further differentiating. Once the decomposition is done, an E-RC model is trained with each of the time series components. The final prediction is obtained as the sum of the trend, seasonality and residual predictions.

%Table 
\begin{table}
    \centering
    \begin{tabular}{c|cccccc}
    \hline\hline \\[-0.3cm]
        Model                    & $\alpha$ & $N$ & $\gamma$ & $\rho(W)$ & $D(W)$ & $N_e$ \\  
    \hline \\[-0.2cm]
       S-RC                    & 0.81     & 120 & 2.6      & 0.52      & 0.012 & -- \\
       E-RC                    & 0.98     & 80  & 0.51     & 1.4       & 0.011 & 240 \\
       D-RC (trend)         & 0.58     & 60  & 0.011    & 1.1       & 0.016 & 80 \\
       D-RC (seasonality) & 0.74     & 20  & 0.011    & 0.32      & 0.023 & 100 \\
       D-RC (residuals)    & 0.91     & 60  & 7.24     & 0.85      & 0.022 & 120 \\
       ARNN-RC              & --        & 30  & --        & 0.50       & 0.03  & -- \\ 
       \hline
    \end{tabular}
    \caption[Optimal hyperparameters used to train the different reservoir computing models.]{Optimal hyperparameters used to train the different reservoir computing models: 
    $\alpha$ is the leaking rate, $N$ the number of neurons of the reservoir, 
    $\gamma$ is the regularization parameter, $\rho(W)$ and $D(W)$ the spectral radius 
    and density of $W$, and $N_e$ the number of reservoirs used in the ensemble.}
    \label{tab:hyperparams}
\end{table}

\subsubsection{Next Generation Reservoir Computing (\gls{NG-RC})}
  \label{subsec:D}
In Ref.~\citep{next-genRC}, a novel RC framework was presented, which requires no random matrices and fewer hyperparameters to define the model. 
Its performance was evaluated in that work by studying the evolution of two benchmark chaotic 
dynamical systems: the Lorentz attractor and the double-scroll electronic circuit. 
The operation of the model is schematically depicted in Fig.~\ref{fig:RC_models} d). 
Instead of using a random complex network as the reservoir, 
the $k$ time-delay observations of the dynamical system and nonlinear functions 
of these observations are used to predict the output. 
That is, to predict a dynamical system $(x(t), y(t), z(t))$, one would use linear 
terms of the form $x(t), \cdots, x(t~-~k), y(t), \cdots, y(t-k), z(t), \cdots, z(t-k)$ 
and non-linear terms of the form $x(t)x(t), \cdots, x(t)x(t~-~k), x(t)y(t), 
\cdots x(t)y(t~-~k), \cdots, y(t)z(t~-~k)$ to perform a linear regression to predict 
the future values $(x(t+1), y(t+1), z(t+1))$.
Therefore, the prediction is just a linear function of these time-delay observations, 
and there is no need to define the random matrices  $W^{\text{in}}, W^{\text{back}}$, 
and $ W$. 
In our case, the time series is one-dimensional, so the terms of the $k$-delay vector 
are all of the form $y(t)y(t-k)$. 

\subsubsection{Auto-Reservoir Neural Network (\gls{ARNN-RC})}
  \label{subsec:E}
Another RC framework was developed in Ref.~\citep{ARC}, which allows making multiple-step-ahead predictions of high-dimensional time series with small training sizes. 
This is the ARNN-RC algorithm, which uses the regressor variables as the reservoir 
by obtaining a mapping from the future high-dimensional regressor variables 
to their associated time series prediction. 
This mapping is based on a spatiotemporal information transformation (see Ref.~\citep{STI}). 
The algorithm is schematically illustrated in Fig.~\ref{fig:RC_models} e).

\subsubsection{Hyperparameters}
  \label{subsec:G}
The hyperparameters used for the RC models are reported in Table~\ref{tab:hyperparams}. In all cases, $t_\text{dismiss} = 10$, $f= \tanh$ and $f^\text{out} = \text{id}$ were used. Notice that the NG-RC model does not appear in the table, since random matrices are not used in this method. In this case, optimal parameters were $k=3$ and $\gamma = 0.56$. 
Regarding the benchmark models, LSTM used 128 neurons and 5000 training epochs, and the best SARIMA model was a SARIMA(0,0,1)(1,0,1,45).

\subsection{Qualitative analysis}
\begin{figure}[!ht]
    \centering
    \includegraphics[width=1.0\textwidth]{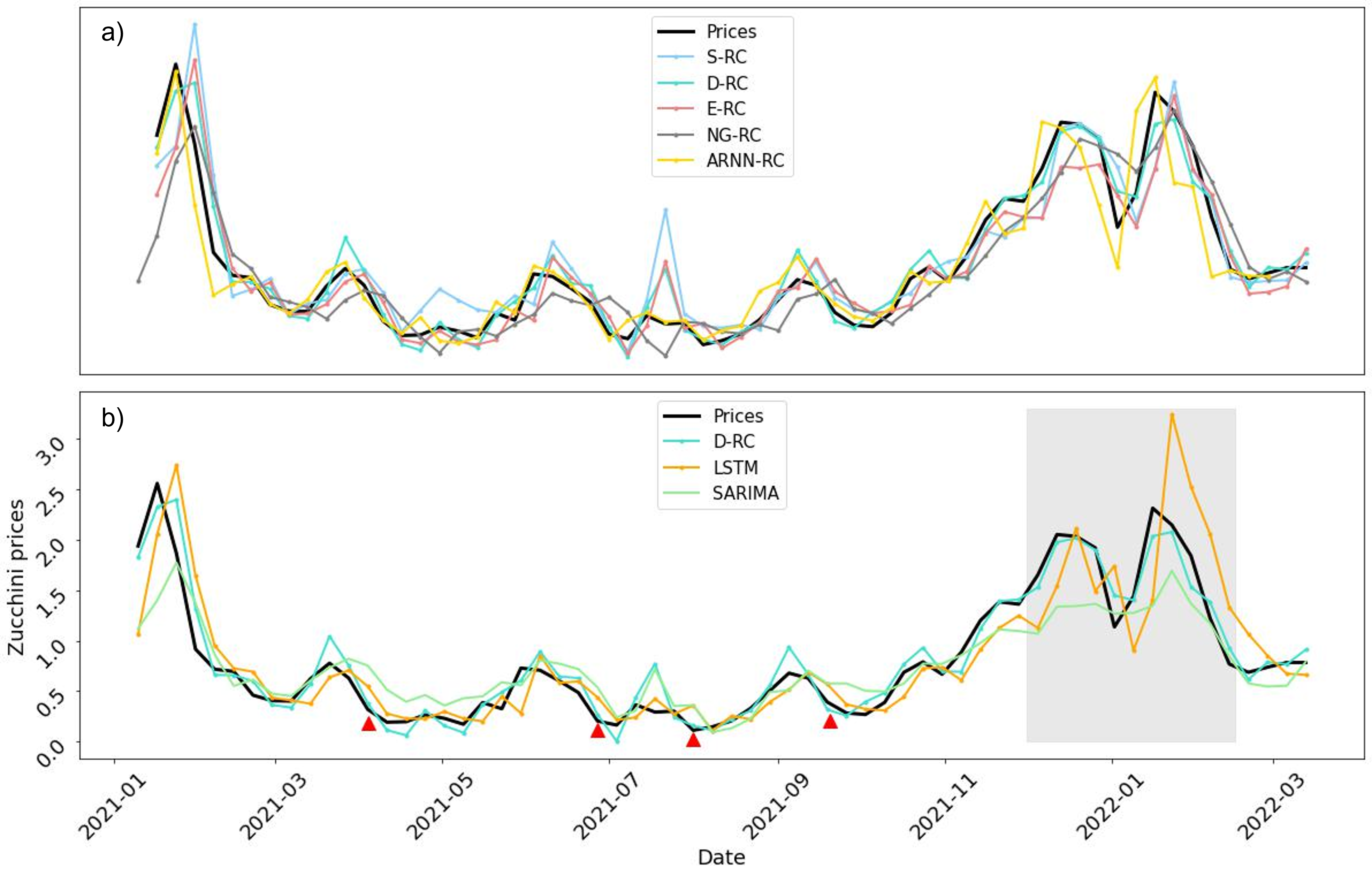}
    \caption[Prediction of the zucchini price time evolution for a) the RC models and b) the SARIMA and LSTM models.]{Prediction of the zucchini price time evolution in the test set 2021-2022 campaign for:
    a) the five reservoir computing-based methods considered in this work, and
    b) the best reservoir computing method compared with SARIMA and LSTM.
    The actual price times series is shown in the black line.}
\label{fig:time_evolution}
\end{figure}

The results obtained with all the methods outlined before are shown in Fig.~\ref{fig:time_evolution}, where the predictions (colored lines) are compared with the actual time series (black line) 
for the zucchini prices in the test set. To aid in the interpretation, the computed results have been separated into two parts. Fig.~\ref{fig:time_evolution} a) shows a comparison among the results produced by the five RC-based algorithms, while Fig.~\ref{fig:time_evolution} b) presents the comparison of the results obtained with the best RC method and those rendered by the benchmark models.

A qualitative eye examination of the results in Fig.~\ref{fig:time_evolution} a) shows that among the RC models, the D-RC (cyan line) is the one having the best performance. The decomposition of the signal into three parts is the key reason for its excellent performance. More interestingly, by comparing the models in Fig.~\ref{fig:time_evolution} b) we see that the D-RC model clearly outperforms the standard LSTM and SARIMA algorithms (yellow and green lines). 
In particular: \\
\begin{enumerate}
    \item The D-RC algorithm is the one predicting more accurately the market falls. To emphasize this result, the red triangles refer to those points in the time series (black line) where the price dropped to minimal values.
    \item More importantly, in the regions near the minima, the LSTM and the SARIMA models predict higher prices than the real ones, while the D-RC can correctly capture the corresponding changes in tendency.
    \item In the region of high price volatility (the shaded region in the plot), where the behavior of the time series significantly differs from the rest, the D-RC model is the only one able to make accurate predictions. On the other hand, the SARIMA forecasts prices with very small variations over time, and the LSTM, despite predicting high volatility, appears largely uncorrelated with the actual prices.
\end{enumerate}
As a result, the D-RC is the only forecasting model that captures the dynamics of the time series during the whole testing period.

\subsection{Results: Mean Absolute Error}
\label{sect:results_MAE}

\begin{table}[!t]
    \centering
    \begin{tabular}{ccccccc}
    \hline \hline \\[-0.3cm]
                      & \multicolumn{6}{c}{ \textbf{Z U C C H I N I} } \\
                       \cmidrule(lr){2-7}
                      & \multicolumn{3}{c}{Mean absolute error} &  \multicolumn{3}{c}{Market direction accuracy}\\
                        \cmidrule(lr){2-4}\cmidrule(lr){5-7}
        %\hline
         Model     & Train & Validation & Test & Train & Validation & Test\\
        \hline\\
        LSTM       & 0.217 & 0.199 & 0.154    & 0.414 & 0.455 & 0.377 \\
        SARIMA   & 0.124 & 0.213 & 0.16      & 0.505 & 0.523 & 0.453 \\
        S-RC        & 0.123 & 0.214 & 0.141    & 0.586 & 0.545 & 0.604 \\
        E-RC        & 0.128 & 0.166 & 0.123    & 0.617 & 0.705 & 0.604 \\
        D-RC        & 0.135 & 0.132 & 0.089   & 0.563 & 0.795 & 0.774 \\
        NG-RC      & 0.167 & 0.182 & 0.15     & 0.476 & 0.605 & 0.463 \\
        ARNN-RC  & --      & 0.197  & 0.122   & --      & 0.659 & 0.529 \\
         \hline
    \end{tabular}
    \caption{Mean absolute error and market direction accuracy of the price predictions for the training, validation and test data sets
    for the seven machine learning models used in predicting the zucchini prices series.}
    \label{tab:calabacines}
\end{table}

%Table II
\begin{table}[!t]
    \centering
    \begin{tabular}{ccccccc}
    \hline \hline \\[-0.3cm]
                      & \multicolumn{6}{c}{ \textbf{A U B E R G I N E}}\\
                       \cmidrule(lr){2-7}
                      & \multicolumn{3}{c}{Mean absolute error} &  \multicolumn{3}{c}{Market direction accuracy}\\
                        \cmidrule(lr){2-4}\cmidrule(lr){5-7}
        %\hline
         Model     & Train & Validation & Test & Train & Validation & Test\\
        \hline \\
        LSTM       & 0.185 & 0.194 & 0.153    & 0.427 & 0.364 & 0.372 \\
        SARIMA   & 0.175 & 0.214 & 0.151    & 0.481 & 0.5     & 0.509 \\
        S-RC       & 0.107  & 0.171 & 0.129    & 0.695 & 0.705 & 0.623 \\
        E-RC       & 0.134  & 0.157 & 0.117    & 0.644 & 0.636 & 0.66 \\
        D-RC       & 0.12   & 0.141 & 0.09      & 0.654 & 0.659 & 0.698 \\
        NG-RC     & 0.147 & 0.18   & 0.182    & 0.529 & 0.535 & 0.481 \\
        ARNN-RC & --       & 0.185 & 0.121    & --       & 0.545 & 0.549 \\
         \hline
    \end{tabular}
    \caption{Same as Table~\ref{tab:calabacines} for aubergine prices.}
    \label{tab:berenjenas}
\end{table}

%Table III
\begin{table}[!t]
    \centering
    \begin{tabular}{ccccccc}
    \hline \hline\\[-0.3cm]
                       & \multicolumn{6}{c}{ \textbf{T O M A T O}}\\
                        \cmidrule(lr){2-4}\cmidrule(lr){5-7}
                       \cmidrule(lr){2-7}
                       & \multicolumn{3}{c}{Mean absolute error} &  \multicolumn{3}{c}{Market direction accuracy}\\
                        \cmidrule(lr){2-4}\cmidrule(lr){5-7}
        %\hline
         Model     & Train & Validation & Test & Train & Validation & Test\\
        \hline \\
        LSTM       & 0.092 & 0.127   & 0.177 & 0.502 & 0.5    & 0.333 \\
        SARIMA   & 0.107 & 0.124 & 0.138   & 0.454 & 0.545 & 0.389 \\
        S-RC        & 0.088 & 0.106 & 0.114   & 0.58   & 0.636 & 0.574 \\
        E-RC        & 0.086 & 0.105 & 0.123   & 0.58   & 0.727 & 0.574 \\
        D-RC        & 0.091 & 0.094 & 0.093  & 0.614 & 0.659  & 0.704 \\
        NG-RC      & 0.097 & 0.128 & 0.162  & 0.529 & 0.488  & 0.364 \\
        ARNN-RC  & --       & 0.126 & 0.131  & --      & 0.545  & 0.481 \\
         \hline
    \end{tabular}
    \caption{Same as Table~\ref{tab:calabacines} for tomato prices.}
    \label{tab:tomates}
\end{table}

\begin{figure}[!ht]
\centering
\includegraphics[width=1.0\textwidth]{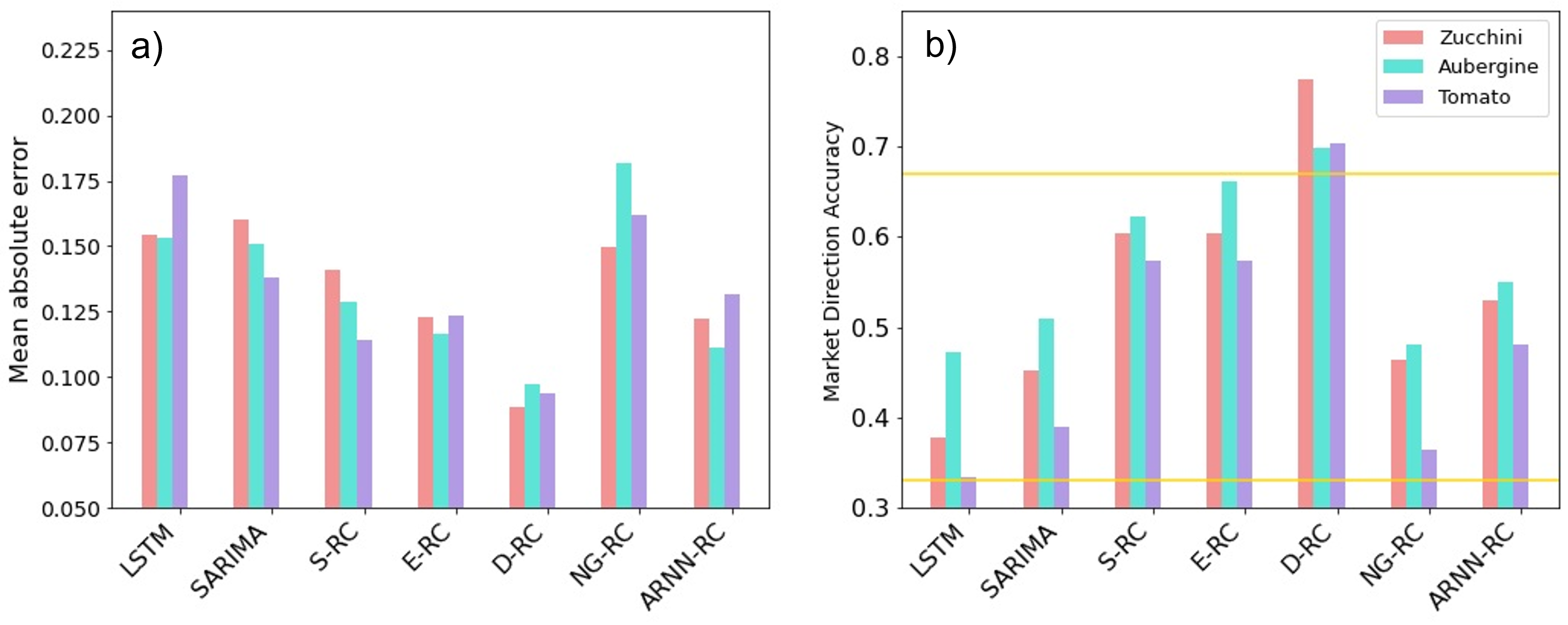}
\caption[Mean absolute error a) and market direction accuracy b) in the test set of zucchini, aubergine, and tomato time series.]{Mean absolute error a) and market direction accuracy b) of the price predictions in the test set of zucchini (in red), aubergine (in green), and tomato (in purple) time series for the seven models studied in this work.}
\label{fig:MAE_MDA}
\end{figure}

Let us discuss now the error metrics for each of the seven models and the three vegetables considered in this work. We begin by calculating the MAE of the predictions in the training, validation and test sets. Since we aim to evaluate the model prediction capacity, we focus on analyzing only the errors observed in the test sample, whose results are given in Fig.~\ref{fig:MAE_MDA}~a). The metrics for training and validation are also provided in Tables~\ref{tab:calabacines},~\ref{tab:berenjenas} and~\ref{tab:tomates}.

%Figure 

The results clearly show that the best-predicting model for this error indicator is the D-RC,
which presents significantly lower errors in all sets. Actually, the MAE for the three vegetable varieties is always under 0.10, 
being the D-RC the only model breaking down this barrier. Next, we find that E-RC and S-RC both show similar performance,
with slightly higher errors, ranging between $0.11-0.13$. 
However, the analysis shows that our E-RC is much more robust to the choice of the random matrices, and therefore its performance is much less prone to oscillations for different realizations of the algorithm. Overall, the third best model is ARNN-RC, whose error in the test set is similar to the MAEs of E-RC and S-RC, even though for validation the errors are slightly higher. Note that by construction, in this method there is no training prediction, and thus the corresponding value does not appear in Tables~\ref{tab:calabacines},~\ref{tab:berenjenas} and~\ref{tab:tomates}. Finally, in contrast to the other RC configurations, the NG-RC exhibits similar performance to that of the benchmark models, LSTM and SARIMA. Taking tomato as an example, NG-RC presents an error of 0.16, which is between those of SARIMA~(0.14) and LSTM (0.18), and it is 78\% higher than the MAE of the best model, the D-RC.

\subsection{Results: Market Direction Analysis}
\label{sect:results_MDA}
Most time series that describe the evolution of real data do not exhibit large variations between consecutive time steps, and thus present a significant degree of autocorrelation. Despite their high volatility, agri-food price time series are no exception. 
As a consequence, a common mistake of many ML models in this scenario, that is not captured by indicators such as MAE, $R^2$, or the relative error, is to use the current value as a reasonable prediction for the next time step, as discussed before. For this reason, we need to further evaluate the RC models using complementary metrics that are robust against this autocorrelation.

In this work, we present the MDA, which assesses the model's ability to accurately predict changes in the trend of the price time series. This metric is defined as follows. First, we consider that in our case prices have increased or decreased within consecutive weeks if the price change is greater than $\pm0.05$\texteuro /kg, respectively. Otherwise, we say that the prices have remained constant. Hence, the target function takes a value of $1$ to indicate an uptrend, $-1$ to indicate a downtrend and $0$ when the price remains constant.

The MDA obtained for the RC models in predicting the new categorical target are shown in Fig.~\ref{fig:MAE_MDA} b) and Tables~\ref{tab:calabacines},~\ref{tab:berenjenas} and~\ref{tab:tomates}. Notice that the target function for calculating the MDA has three labels, thus, assuming the same probability for all classes, a random model would guess correctly (on average) one-third of the times, leading to a 33\% accuracy. This value (as well as 67\% accuracy) has been added, as a yellow line, in Fig.~\ref{fig:MAE_MDA} b) to guide the eye as a further help to interpret the results. As can be seen, all results are above the random accuracy, being the D-RC the one obtaining the best scores, since it renders an accuracy of almost 80\% for zucchini and 70\% for tomato and aubergine. More interestingly, in the case of zucchini, the D-RC method correctly anticipates changing trends in 77.4\% of the cases, as compared to the poor 45.3\% of SARIMA or 37.7\% for LSTM. Accordingly, we can conclude that the performance of this RC-based method is quite impressive, since it gets 2.1 and 1.7 times higher performance than the widespread LSTM and SARIMA, respectively, for the prediction metric considered in this subsection. Moreover, this effect is not particular to zucchini, since similar results are obtained for aubergine and tomato, as can be seen in Tables~\ref{tab:berenjenas} and~\ref{tab:tomates}.

The second-best models are again E-RC and S-RC, which have similar MDA on the test set, even though, as expected, E-RC is slightly better. The MDA for both models is in general over 10 points behind the performance of D-RC. According to this complementary MDA metric, the ARNN-RC and NG-RC perform also much better than LSTM, and slightly better than SARIMA. Also, ARNN-RC has higher accuracy than NG-RC, being the third-best model in terms of MDA as well. At this point, it should be emphasized that in the original paper for the NG-RC~\citep{next-genRC}, the authors evaluated the model for multi-dimensional time series. 
The predictor vectors contained time-delayed products of the time series components. In this work, the time series is just one-dimensional and therefore there are fewer time-delayed products to predict the time evolution, which may be the cause of having higher errors in the test set.

One last interesting point about the results is that all models 
 perform worse, in terms of MAE, in the validation set compared to the performance in the test set. The most likely reason for this behavior
is that the validation set contains data from 2020, which is fairly different from the training data due to the start of the global COVID-19 pandemic. Indeed, prices present an abrupt high increment around March 2020, unrelated to any previous behavior of the series or the regressor variables. This explains why it is more difficult to predict the time series evolution during this period. However, we also noticed that, in terms of MDA, the RC models do not perform significantly worse in the validation set. 
The interpretation for this divergence is that, even though the reservoir cannot predict such extremely high prices (which produce a high MAE) since they never occurred in the past,
it is still capable of following the trend of the signal, 
leading to high accuracy when predicting price rises and falls.

In this study, we have compared different RC variants and showed the benefit of decomposing the time series and modeling each component separately. The studied time series present high volatility, are influenced by external factors and are small in size. For this reason, the conclusions of this work are not limited to the agri-food market but can be extended to many other domains where the same characteristics are also present. The results have shown that RC represents an excellent solution to model and predict small complex data, that would be difficult to train with traditional RNN without producing too much overfitting.

\section{Discussion and future work}
\label{sect:discussion_agro}
In the previous section, the RC algorithm was successfully adapted to tackle the complex task of forecasting agricultural product prices. It was shown that RC methods clearly outperform traditional statistical methods, such as SARIMA, as well as state-of-the-art RNNs like the LSTM. Notably, among the five RC algorithm variants explored, the D-RC method emerged as the undisputed champion, exhibiting unparalleled performance. 

However, two persisting challenges still lie ahead in achieving precise agri-commodity price forecasting: the accurate prediction of confidence intervals and multivariate price forecasting. In this section, we will engage in a comprehensive discussion of these challenges, seeking innovative solutions to further enhance the accuracy and reliability of agri-commodity price predictions.

\subsection{Confidence intervals prediction}
\label{sect:confidence_intervals}
The task presented in the previous section consisted in performing deterministic forecasting of the prices of multiple products, since only a single prediction value was given by the RC model. Deterministic forecasting provides a single-point estimate of future prices, lacking information about the uncertainty associated with the forecasts. While deterministic forecasting can be useful for providing a simple estimate of future values, it does not capture the inherent variability and uncertainty present in real-world situations. It is limited in its ability to account for unforeseen events, fluctuations in market conditions, or other factors that may impact the accuracy of the forecast.

In contrast, probabilistic forecasting approaches, such as confidence interval prediction~\citep{interval_prediction}, explicitly consider uncertainty by providing a range within which a future price value is likely to fall, based on a given level of confidence. Such method gives the predicted lower and upper limits at a particular nominal confidence level. High-quality prediction intervals are defined as those that cover a specified proportion of the observations while still being as narrow as possible. Indeed, under the same level of confidence, narrower prediction intervals can provide more accurate information with fewer uncertainties. 

Confidence intervals provide a measure of uncertainty and variability in the predictions, allowing for a more comprehensive understanding of the volatility of the market and the performance of the forecasting model. This information is valuable for making informed decisions, assessing risks, and planning appropriate strategies. Simply relying on single-point predictions is not enough to make accurate business decisions based on forecasted prices; indeed, it is essential to know the range within which future prices are likely to fall. As an example, imagine that the current zucchini price is of $1.5$€/kg and our goal is to predict next week's price. If the RC model only outputs a single value, $y=1.6$€/kg, we may think that the current trend is for the price to increase. However, if the RC model generates a confidence interval prediction of $y \in [1.45,1.65]$€/kg, we would be aware that the price could increase, remain stable, or even decrease slightly. This additional information empowers us to design appropriate strategies. Confidence intervals not only provide information about the accuracy and confidence of our RC model but also about the volatility of the market. For this reason, it is of vital importance to study forecasting methods that provide confidence interval prediction instead of only offering single-point predictions.

\begin{figure}[!ht]
\centering
\includegraphics[width=1.0\textwidth]{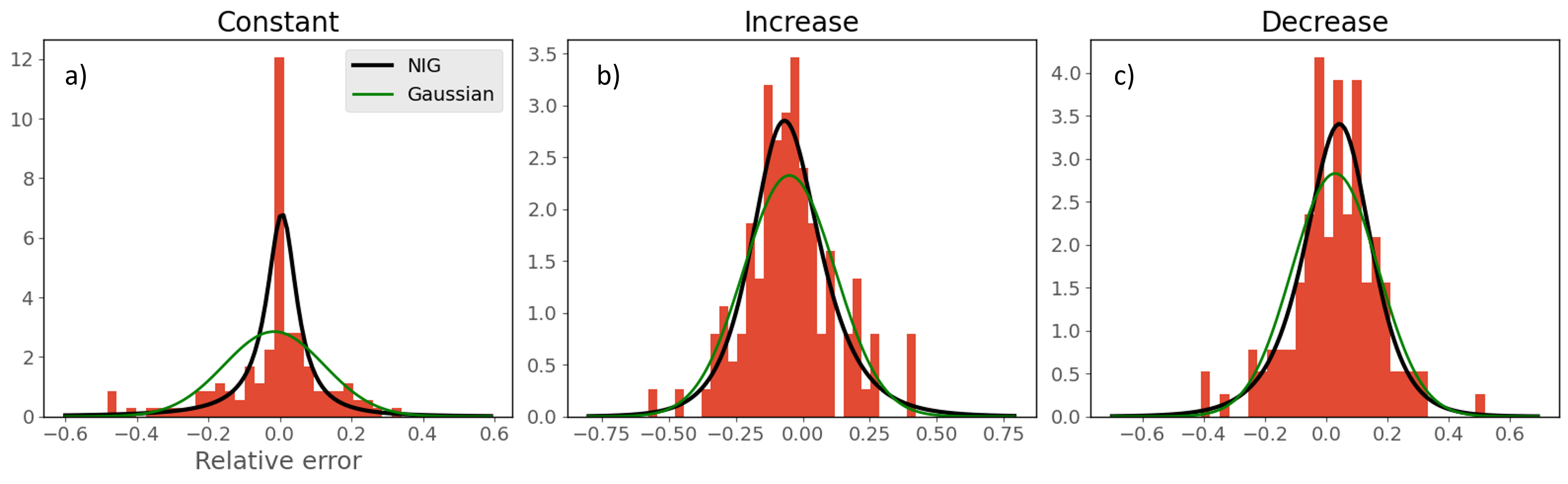}
\caption[Distribution of the relative error in the training set for the a) \emph{Constant} set, the b) \emph{Increase} set and the c) \emph{Decrease} set.]{Distribution of the relative error in the training set for the a) \emph{Constant} set, the b) \emph{Increase} set and the c) \emph{Decrease} set. See Eq.~\ref{eq:interval_definition} for the definition of the three sets. }
\label{fig:dist_zucchini}
\end{figure}
There are many ways to design a confidence interval prediction method. One of the simplest approaches involves assessing the uncertainties related to deterministic forecasts. For instance, by modeling the statistical distribution of errors associated with the D-RC algorithm, it becomes possible to estimate the confidence intervals associated with the D-RC predictions. Let us illustrate how this method would work, using the D-RC prediction of the zucchini time series presented in the previous section. Figure~\ref{fig:dist_zucchini} shows the error $\epsilon(t)$ of the D-RC predictions in the training set, defined as $\epsilon(t) = y_\text{teach}(t) - y(t)$, where $y_\text{teach}(t)$ is the actual price at time $t$ and $y(t)$ is the prediction of the D-RC model. 

Notice that the error distribution has been studied for three different sets in Fig.~\ref{fig:dist_zucchini}. The zucchini prices are very volatile, presenting abrupt changes of tendency. As a result, we expect the error distribution to exhibit different behavior depending on whether or not there is a significant change of tendency in the predictions. For this reason, the training data has been separated into three sets according to the relative increase in the prices. The first set contains the data with a relative increase higher than +10\% (Increase). The second class contains the data with a relative increase lower than -10\% (Decrease). Finally, the third set contains the data with a relative increase between -10\% and +10\% (Constant):
\begin{eqnarray}
\label{eq:interval_definition}
 &\text{Increase} \rightarrow &\displaystyle \frac{y_\text{teach}(t-1) - y(t)}{y(t)} \geq 0.1, \\ \nonumber
   &\text{Decrease} \rightarrow &\displaystyle \frac{y_\text{teach}(t-1) - y(t)}{y(t)} \leq -0.1, \\ \nonumber
   &\text{Constant} \rightarrow & -0.1 < \displaystyle \frac{y_\text{teach}(t-1) - y(t)}{y(t)} < 0.1.
\end{eqnarray}
Notice that the classification in Eq.~\ref{eq:interval_definition} only depends on the \emph{predicted} prices at time $t$, $y(t)$, and not on the true prices at time $t$, $y_\text{teach}(t)$). In other words, the \emph{Increase} set contains data points where the RC model predicted a price increase compared to the previous true price $y_\text{teach}(t-1)$. Similarly, the \emph{Decrease} set comprises points where the RC model predicted a price decrease relative to the previous known price.

As can be seen in Fig.~\ref{fig:dist_zucchini}, the relative error is significantly smaller for the set containing the \emph{Constant} distribution. Moreover, the \emph{Increase} distribution has positive skewness, while the \emph{Decrease} distribution has negative skewness. This skewness agrees with the definition of the three sets and means that the predictions tend to underestimate big changes of tendency in the data.

After plotting the error histogram, it is necessary to find a suitable distribution that accurately describes the data. Notice that Fig.~\ref{fig:dist_zucchini} contains two curves, representing the best fit of a Gaussian distribution, in green, and the best fit of the Normal Inverse Distribution (\gls{NIG}), in black. The Gaussian distribution is the most standard one, however, it does not always represent the correct behavior of the data. For example, in the \emph{Constant} set, the Gaussian distribution does not capture the sharp peak around 0. On the other hand, the best fit for the error distribution is provided by the NIG distribution, which is a normal mixture of the inverse Gaussian distribution~\citep{norminvgaus}. It is particularly useful for modeling price data, where heavy tails and skewness are commonly observed. The probability density function of the NIG distribution is given by
\begin{equation}
    f(x,a,b) = \frac{a K_1(a\sqrt{1+y^2})}{\pi \sqrt{1 + y^2}}e^{\sqrt{a^2 - b^2} + by}, \quad y= \frac{x - x_0}{\sigma},
    \label{eq:NIG}
\end{equation}
where $x$ is a real number, the parameter $a$ is the tail heaviness and $b$ is the asymmetry parameter satisfying $a>0$ and $|b| \leq a$, $x_0$ and $\sigma$ are the location and scale factors and $K_1$ is the modified Bessel function of second kind~\citep{bessel}. The best fit of the parameters are $a=0.12$, $b=-0.048$, $x_0=0.0069$, $\sigma=0.051$ for the \emph{Constant} set, $a=1.36$, $b=0.22$, $x_0=-0.083$, $\sigma=0.20$ for the \emph{Increasing} set and $a=1.66$, $b=-0.25$, $x_0=0.056$, $\sigma=0.18$ for the \emph{Decreasing} set.

After modeling the error distribution, the confidence intervals are calculated, for a given confidence level $\alpha$ as:
\begin{eqnarray}
    U_t = y(t) + \text{NIG}_{x_0, \sigma}^{a,b}(1-\alpha/2), \quad L_t = y(t) + \text{NIG}_{x_0, \sigma}^{a,b}(\alpha/2).
\end{eqnarray}
Notice that since each of the three training sets (\emph{Constant, Increase, Decrease}) is modeled with a different distribution, the width of the interval varies depending on the specific set to which a given prediction belongs.
\begin{figure}[!ht]
\centering
\includegraphics[width=1.0\textwidth]{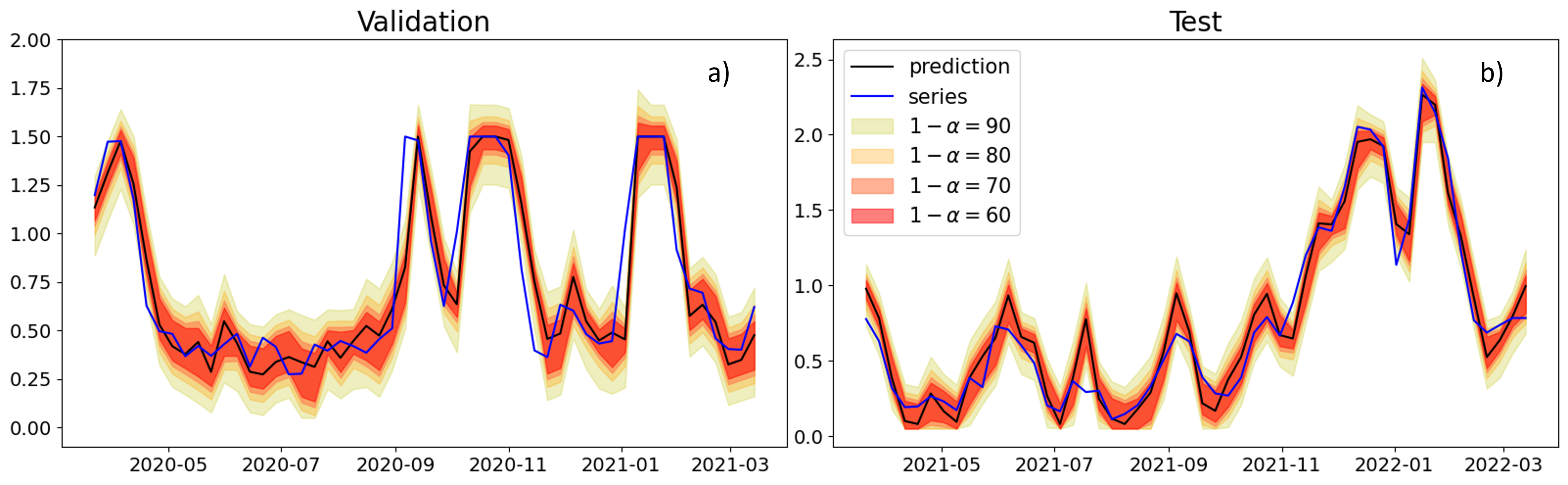}
\caption{Predicted confidence intervals computed with the normal inverse Gaussian distribution (see Eq.~\ref{eq:NIG}) for the zucchini time series for a) the validation and b) the test set.}
\label{fig:intervals_zucchini}
\end{figure}
The predicted confidence intervals, for the validation and test set, and four different values of $(1-\alpha)$ = 90\%, 80\%, 70\% and 60\% are displayed in Fig.~\ref{fig:intervals_zucchini}. Notice that most of the target values are enclosed by these constructed prediction intervals. This observation highlights the accuracy of the confidence intervals generated by the proposed method in estimating the uncertainty associated with the D-RC predictions. 

Moreover, the intervals predicted in Fig.~\ref{fig:intervals_zucchini} are significantly narrow while covering the specified proportion of samples, meaning that they do not overestimate the error of the RC model. The width of the interval depends on the level of confidence; for a confidence level of 90\%, the average width is 0.46~€/kg, while for a confidence level of 60\%, the average width decreases to 0.20~€/kg. Therefore, the NIG distribution is suitable for providing narrow intervals with the appropriate coverage. In contrast, the confidence intervals resulting from modeling the error with a Gaussian distribution (see distribution in Fig.~\ref{fig:dist_zucchini}) are wider. Indeed, for a confidence level of 60\%, the width of the intervals increases to 0.25~€/kg, meaning that using the Gaussian distribution provides less informative confidence intervals than those obtained with the NIG distribution.

While the method described above is useful to quantify the uncertainty in the predictions of the RC algorithms, it is not the most accurate method to predict confidence intervals in general. One limitation of this method is that the confidence intervals solely depend on the predicted value $y(t)$ and do not take into account the values of external regressor variables, which are essential for ensuring the optimal performance of the RC models. Therefore, a more sophisticated model that considers the intricate patterns of the time series and incorporates the relevant additional variables is required to achieve a precise estimation of the confidence intervals.

Recently, a novel algorithm was proposed~\citep{RC_CI} to train a RC model to directly predict confidence intervals. This method involves a modified RC model where the readout layer is implemented using a three-layer NN instead of the traditional linear model. The internal states of the reservoir are calculated with the same procedure as the standard RC (see Eq.~\ref{eq:update_x}). Then, the NN, which acts as the readout layer $W^\text{out}$, is trained to predict the upper and lower bounds of the confidence intervals. 

The training process of the NN incorporates a quality-driven loss function~\citep{RC_CI}, specifically designed to ensure that the predicted intervals have minimum width while maintaining the desired coverage probability. For instance, if the confidence interval is intended to encompass a probability of $1-\alpha$, the predicted interval should at least cover this probability while being as narrow as possible. Consequently, this methodology harnesses the power of the RC framework to accurately determine the upper and lower bounds of the confidence intervals.

It is important to note that this RC-based approach inherits the advantages of the RC framework. One of the key advantages is its ability to capture nonlinear relationships between time series data and external variables. Additionally, the simple training strategy of the method enables the effective handling of small and volatile datasets. As a result, this approach offers a significantly improved and more precise prediction of confidence intervals, enhancing the overall reliability of the forecast.

This method was originally developed to predict the confidence intervals of wind power, which is essential for improving the availability of wind power, developing a power generation plan, and ensuring the safe operation of a power system~\citep{RC_CI}. In future studies, we aim to investigate the adaptation of this methodology to predict confidence intervals for agri-food prices. Moreover, we will examine the impact and correlation between the regressor variables and the width of the predicted intervals, to identify the most influential factors contributing to market variability and the accuracy of forecasting models. Additionally, we intend to expand the presented method to predict confidence intervals at multiple confidence levels, which will provide a more comprehensive understanding of how confidence intervals expand with increasing coverage. Finally, we will extend this framework to multivariate price forecasting (discussed in the next subsection), by considering the influence of multiple products in the forecast of individual prices. Such extensions hold great potential for delivering more precise and insightful forecasts, thereby enabling improved decision-making processes in the agricultural and food sectors.

\subsection{Multivariate price forecasting}
\label{sect:cross-correlation}
Let us now discuss the challenge of designing multivariate RC models. In Sect.~\ref{paper:RC_calabacines}, we used the RC algorithm to predict the prices of three different agricultural products (zucchini, aubergine and tomato) independently. This means that a single RC model was used for each of the three products so that one RC model did not have any information about the prices of the other products. In that setting, this approach made sense because the three studied products were significantly different so the price of one product did not have a large influence on the prices of the other ones.

\begin{figure}[!ht]
    \centering
    \includegraphics[width=0.5\textwidth]{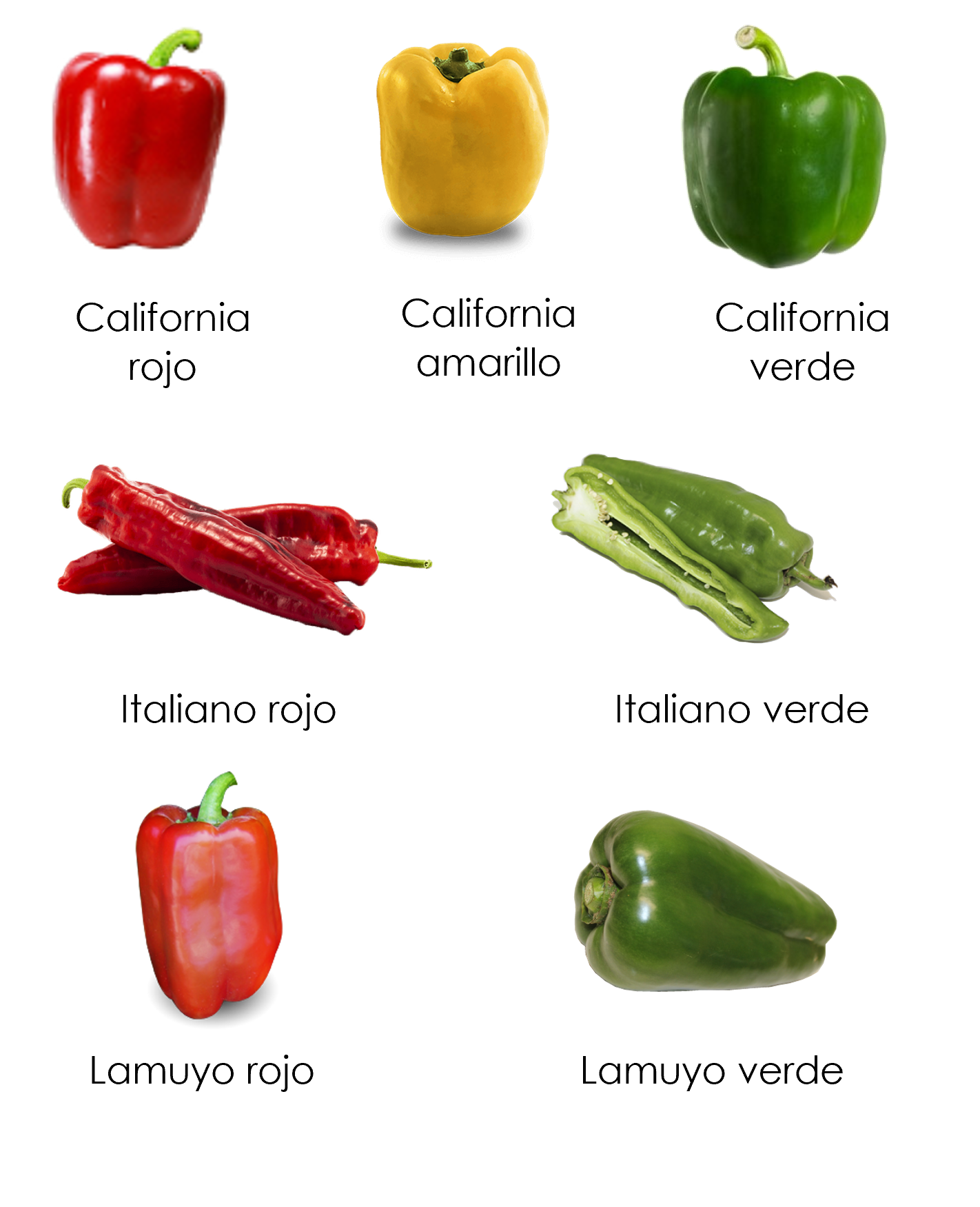}
    \caption{Names and images of the seven pepper varieties used for the cross-correlation forecasting analysis.}
    \label{fig:all_peppers}
\end{figure}
However, when studying the price time evolution of similar products, the series may present a positive or negative correlation with each other. In these situations, incorporating information about the past prices of one product might be useful to improve the forecast of another product. For example, different types of peppers such as green, yellow and red peppers may have positively correlated prices as they are often bought and offered together by grocery stores. On the other hand, similar products, such as different varieties of green peppers, may have negatively correlated prices as consumers tend to purchase only one type of them. Therefore, incorporating information on the past price of the green pepper type, for example, could be used to improve the prediction of the price of the yellow pepper.

For this reason, multivariate forecasting models, that take into account the relationships between the different products, are likely to outperform individual predictions. To illustrate this fact, in this section, we briefly compare the performance of multivariate D-RC models, which use information from different time series to obtain the predictions, with the performance of individual D-RC models. To this end, we analyze the weekly prices, again in euros per kilogram, of various types of peppers from the southeast region of Spain. The seven varieties of peppers, which are shown in Fig.~\ref{fig:all_peppers}, are denoted by their Spanish name: \emph{California Rojo} (\gls{CR}), \emph{California Amarillo} (\gls{CA}), \emph{California Verde} (\gls{CV}), \emph{Italiano Verde} (\gls{IV}), \emph{Italiano Rojo} (\gls{IR}), \emph{Lamuyo Rojo} (\gls{LR}) and \emph{Lamuyo Verde} (\gls{LV}). The dataset includes prices from 2013 to 2021, where the training set contains the samples from 2013 to 2019, the validation set contains the data from 2020 and the test set contains the data from 2021.

The performance of two sets of D-RC models in forecasting the pepper's time series will now be compared. The first one is the D-RC method introduced in Sect.~\ref{paper:RC_calabacines}, which predicts the prices of the seven peppers individually, that is, one D-RC model is trained for each pepper time series. Therefore, we train seven different D-RCs in total.

\begin{figure}[!ht]
    \centering
    \includegraphics[width=0.6\textwidth]{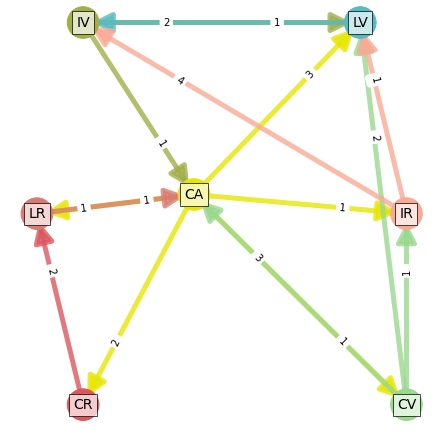}
    \caption[Performance improvement network.]{Performance improvement network. A link from one node to another means that the time-delayed series of the origin node improves the reservoir computing prediction of the destiny node. Each node refers to a pepper variety denoted by its Spanish name (see main text).}
    \label{fig:improvement_network}
\end{figure}
The second set of models uses a multivariate RC approach. In this scenario, each D-RC model is employed to predict a specific pepper series, resulting in a total of seven distinct D-RC models. However, these models incorporate past prices of other pepper varieties to enhance the reservoir's predictive capabilities. To identify the relevant past values for forecasting each series, we analyzed the linear correlations among time-delayed series. Specifically, we calculated the correlation between past prices of one pepper variety and the current price of another. When a significant correlation was observed, it indicated that the past prices of the first pepper could be valuable for predicting the prices of the second pepper. Subsequently, the D-RC models were trained by including the prices of other pepper types as additional variables. Figure~\ref{fig:improvement_network} presents a network diagram illustrating the correlations between the pepper time series that contributed to enhancing the reservoir's performance. Nodes in the network are connected only if the inclusion of time-delayed prices from the first node results in improved performance of the multivariate D-RC algorithm in the second node.

\begin{figure}[!ht]
    \centering
    \includegraphics[width=1.0\textwidth]{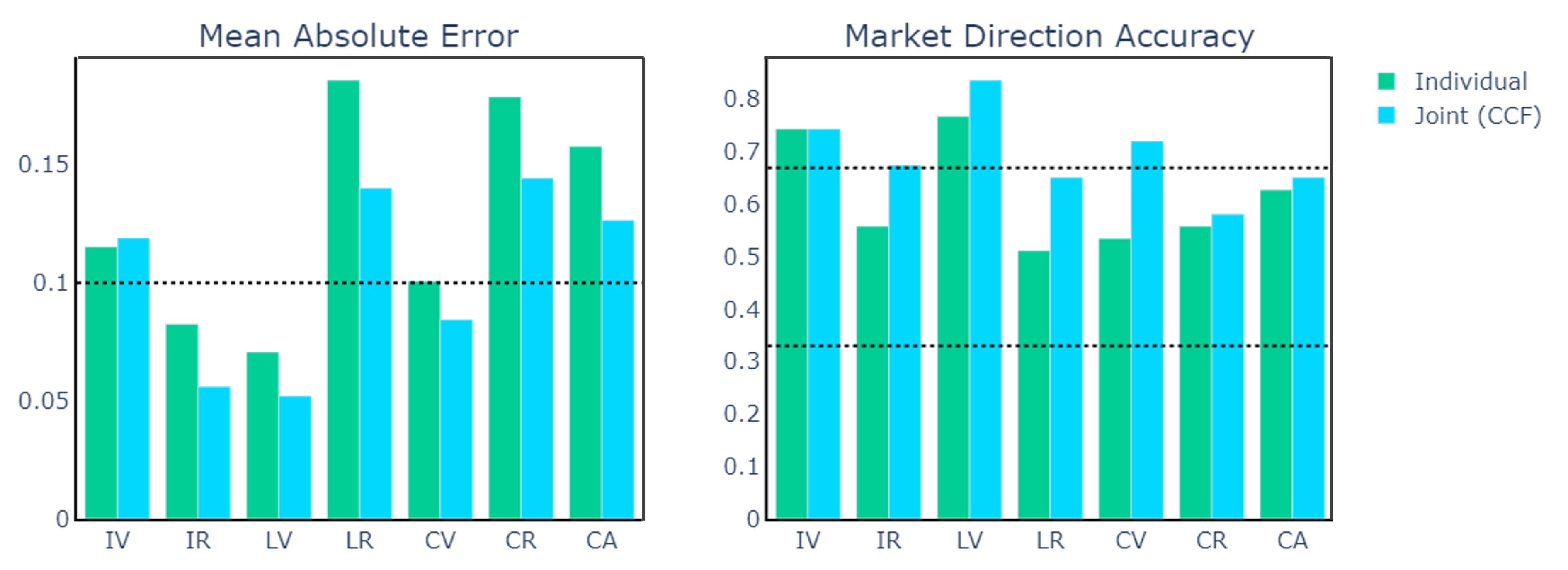}
    \caption[Mean absolute error and market direction accuracy for the multivariate reservoir computing models.]{Mean absolute error and market direction accuracy for the multivariate reservoir computing models compared to the individual reservoir computing models, applied to the seven pepper prices, in the test set.}
    \label{fig:separate_CCF}
\end{figure}
After identifying the optimal set of additional regressor variables, the next step is to evaluate the performance of the multivariate D-RC models on the test set and compare it with the performance of the individual D-RC models. The corresponding results are presented in Fig.~\ref{fig:separate_CCF}. As can be seen, the MAE of all pepper varieties is lower for the multivariate D-RC model compared to the individual D-RC models. Moreover, the MDA is higher for all series when using the multivariate D-RC models. These findings provide compelling evidence for the effectiveness of employing multivariate models in predicting agricultural product prices. It is worth noting that this methodology is not limited to the agri-commodities market but can be applied to any dataset that comprises cross-correlated time series.

%!TEX root = ../thesis.tex
%*******************************************************************************
%****************************** Third Chapter **********************************
%*******************************************************************************
\chapter{Reservoir computing for quantum systems}
\label{chapter3}
\begin{fquote}[Erwin R.J.A.  Schrödinger][1887-1961]
    The task is not to see what has never been seen before, but to think what has never been thought before about what you see every day.
    %The present is the only thing that has no end.
\end{fquote}

In the previous chapter, we discussed the details involved in the adaptation of RC to forecast the prices of agricultural products. In this chapter, we show how the same RC algorithm can be tailored to address a completely distinct challenge: the propagation of quantum wavefunctions over time. It is worth noting that the task of predicting agricultural commodity prices differs significantly from the task of propagating quantum states in time. While the latter requires dealing with complex-valued, high-dimensional data, the former studies low-dimensional, real-valued, volatile time series, which are heavily impacted by external factors. Consequently, the adaptability of the RC algorithm to address both problems shows its remarkable power and versatility.

In the field of quantum chemistry, ML (Machine Learning, see Sect.~\ref{sect:MLNN}) methods are recently proving to have advantages over the standard computational chemistry approaches~\citep{SpecialIssue}, especially in the case
of high-dimensional systems. In a quantum-mechanical scenario, the physical system is described by means of the \emph{Schrödinger equation} (see Eq.~\ref{eq:time-dependent})~\citep{Schrodinger}, which determines the properties and dynamics of such systems. Therefore, being able to efficiently solve this equation is crucial to determining molecular structural properties and molecular dynamics ~\citep{Lanyon}.

Traditionally, numerical integration methods~\citep{CompChem} have been used to solve the Schrödinger equation. However, when the size and/or complexity of the system increases, finding such solutions becomes challenging, requiring expensive computational resources and suffering from error propagation.  For this reason, data-based approaches, such as ML, are recently becoming more popular to tackle quantum problems. These methods do not rely on information about the underlying physical model of the system, but only use data obtained from observations or related calculations, making them more versatile to study multiple systems in quantum chemistry. 

In this chapter, we show how to use ML methods to solve the Schrödinger equation for different quantum systems. The first part of this chapter is devoted to finding solutions to the \emph{time-independent} Schrödinger equation (see Eq.~\ref{eq:time-independent}), whose solutions are the eigenstates associated with certain eigenenergies. While the main focus of this thesis is studying the RC algorithm, in this first project of this chapter we use a NN approach instead, which will motivate the use of RC to study more complex quantum systems. In the second part of this chapter, we extensively compute the time evolution of quantum states by means of solving the \emph{time-dependent} Schrödinger equation  (see Eq.~\ref{eq:time-dependent}). For this purpose, we propose a novel RC-based methodology that allows an accurate and efficient propagation of quantum wavefunctions in time. 

Thus, the organization of this chapter is as follows. Section~\ref{sect:paper1} introduces a NN method to integrate the time-independent Schrödinger equation, which allows to efficiently calculate high-energy states for multiple Hamiltonians. In Sect.~\ref{sect:paper2}, the RC algorithm is adapted to propagate an initial wavefunction through time, thus solving the time-dependent Schrödinger equation. This method is illustrated by integrating three simple 1D systems and one 2D system. The Morse oscillator, which has been widely used to model the vibrational dynamics of the H$_2$O molecule, is studied in Sect.\ref{sect:paper3}. Finally, in Sect.~\ref{sect:paper4} we apply this method to study a more complex and realistic quantum system, the coupled quartic oscillator, which has been of great interest in the field of quantum chaos~\citep{chaotic, Fabio2, Fabio1}.

\section{Neural networks for the computation of vibrational wavefunctions}
\label{sect:paper1}
Computing the eigenstates and eigenenergies of a quantum system accurately is a significant challenge in molecular physics and computational chemistry. 
Traditional numerical methods~\citep{CompChem} rely on the variational principle, which allows numerically calculating the first $N$ eigenstates of a quantum system. This in turn means that to approximate the $N$-th eigenstate, the $N-1$ lower-lying ones must also be calculated. This makes the task particularly difficult when seeking excited states. 

Recently, in computational chemistry, ML has been widely used~\citep{SpecialIssue, Manzhos_2020, hermann2020solving} to solve the electronic Schrödinger equation~\citep{Noe, manyElectron} and, to a lesser extent, the \emph{vibrational} Schrödinger equation~\citep{RigoMLTST,Schnet, Schnet2}. Deep learning methods have been successful in predicting the ground energy of multiple Hamiltonians for the vibrational Schrödinger equation~\citep{DLSchrodinger, MLSchrodinger, CompetingLosses}. The calculation of excited states is more challenging due to anharmonicities and mode couplings present in the eigenfunctions~\citep{Domingo_DL}, which may result in the so-called "scarred" functions~\citep{TesisFabio}, studied in Sect.~\ref{sect:paper4}. 

The first work of this chapter aims to train NNs to generate the ground and excited eigenfunctions of different molecular vibrational Hamiltonians. Instead of only predicting the mean energy of an eigenstate, we also obtain the associated wavefunction, which provides complete information about the system's state. In addition, we focus on obtaining high-energy states, which correspond to more intricate wavefunction topologies, rather than just the ground state of the Hamiltonian. 

Two different scenarios will be considered. In the first one, random polynomial Hamiltonians are used to train the network, where the energy function is a polynomial function of the space coordinates ($x$ and $y$). Such Hamiltonians, together with their associated eigenfunctions, are used to train a NN. Once the training is completed, the network is asked to generalize to non-polynomial potentials, in particular \emph{(decoupled) Morse Hamiltonians}.

In the second scenario, more complex and realistic Hamiltonians are studied. The NN is trained using molecular potentials (decoupled Morse Hamiltonians) where the eigenstates are an analytical solution of the Schrödinger equation. Then, the performance of the NN is tested using complex perturbed potentials, in particular \emph{coupled Morse Hamiltonians}, which have no analytical solution. 

In the next subsections, we present the studied quantum systems, the NN architecture and data generation strategies together with the results obtained for both scenarios.

\subsection{First scenario: Random polynomial Hamiltonians}
\label{sect:Morse_Hamiltonian_paper1}
\subsubsection{Training data: Random polynomial Hamiltonians}
In the first scenario, the training data is a set of Hamiltonians with random polynomial potentials of (up to) degree four. Both one-dimensional (1D) and two-dimensional (2D) potentials are considered. For the 1D case, the Hamiltonians are given by
\begin{equation}
H(x) = \frac{p^2}{2m} + V(x), \qquad \mbox{with} \quad 
          V(x) = \sum_{i\leq4} \alpha_i \ x^i,
\label{eq:poly_1D}
\end{equation}
and for 2D case,
\begin{equation}
H(x,y) = \frac{p_x^2 + p_y^2}{2m} + V(x,y), \qquad \mbox{with} \quad V(x,y) = \sum_{i+j\leq4} \alpha_{ij} \ x^i y^j .
\label{eq:poly_2D}
\end{equation}
Each Hamiltonian has an associated set of eigenfunctions, which are the solution of the corresponding time-independent Schrödinger equation defined in Eq.~\ref{eq:time-independent}. For the training step, the eigenfunctions are obtained with a standard numerical method based on the variational principle~\citep{variational_original}, which will be outlined in the next section. The wavefunctions obtained with the variational method are considered to be the \textit{exact} ones, which are then compared with the \textit{predicted} wavefunctions obtained with the NN.

Since the kinetic energy operator is the same for all Hamiltonians, the NN can be trained using only the potential function.
This allows passing an easy representation of the Hamiltonian to the network. 
With this procedure, the training data consists of a set of pairs $\{V_i, \psi_i\}_i$, where $V_i$ is the $i$-th training potential and $\psi_i$ 
the associated exact wavefunction. The 1D (2D) representation of both potentials and wavefunctions consists of a grid on a linear (rectangular) domain
so that $V_i$ is a matrix containing the values of $V_i(\vec{x})$ with $\vec{x}$ belonging to a closed interval (or rectangular lattice for the 2D systems). Similarly, $\psi_i$ is a matrix containing the values of $\psi(\vec{x})$ in such domain. For each particular energy vibrational state (either the ground or excited state), the NN is trained to predict the eigenfunction associated with a polynomial Hamiltonian. Then, we ask the network to reproduce the eigenfunctions for some other, more general, non-polynomial Hamiltonians. 

\subsubsection{Test data: Morse Hamiltonians}
In order to test the performance of the NN, we will consider three types of Hamiltonians: random polynomial potentials (generated in the same way as the training data), the standard harmonic oscillator and Morse potentials. The first two types of Hamiltonians are included in the training set so that it should be easy for the NN to accurately predict the corresponding eigenfunctions. However, the third class of potentials, the Morse potentials, are sufficiently different from the training potentials since they are non-polynomial in their coordinates. Such Hamiltonians also adequately represent the vibrational potential interaction of a diatomic molecule. We write the 1D Morse potential as
\begin{equation}
V(x) = D_e \left[e^{-2a(x-x_e)} - 2e^{-a(x-x_e)}\right],
\end{equation}   
where $x$ is the distance between nuclei, $x_{e}$ is the corresponding equilibrium bond distance,  $D_{e}$ is the well depth (defined relative to the dissociated atoms), and $a$ is a parameter controlling the width of the potential 
(the smaller $a$ is, the deeper the well). This potential approaches zero at $x \rightarrow \infty$ and equals $-D_{e}$ at its minimum at $x=x_e$. 
The Morse potential is the combination of a short-range repulsion term (the former) and a long-range attractive term (the latter).

\begin{figure}[!ht]
\centering
  \includegraphics[scale=0.65]{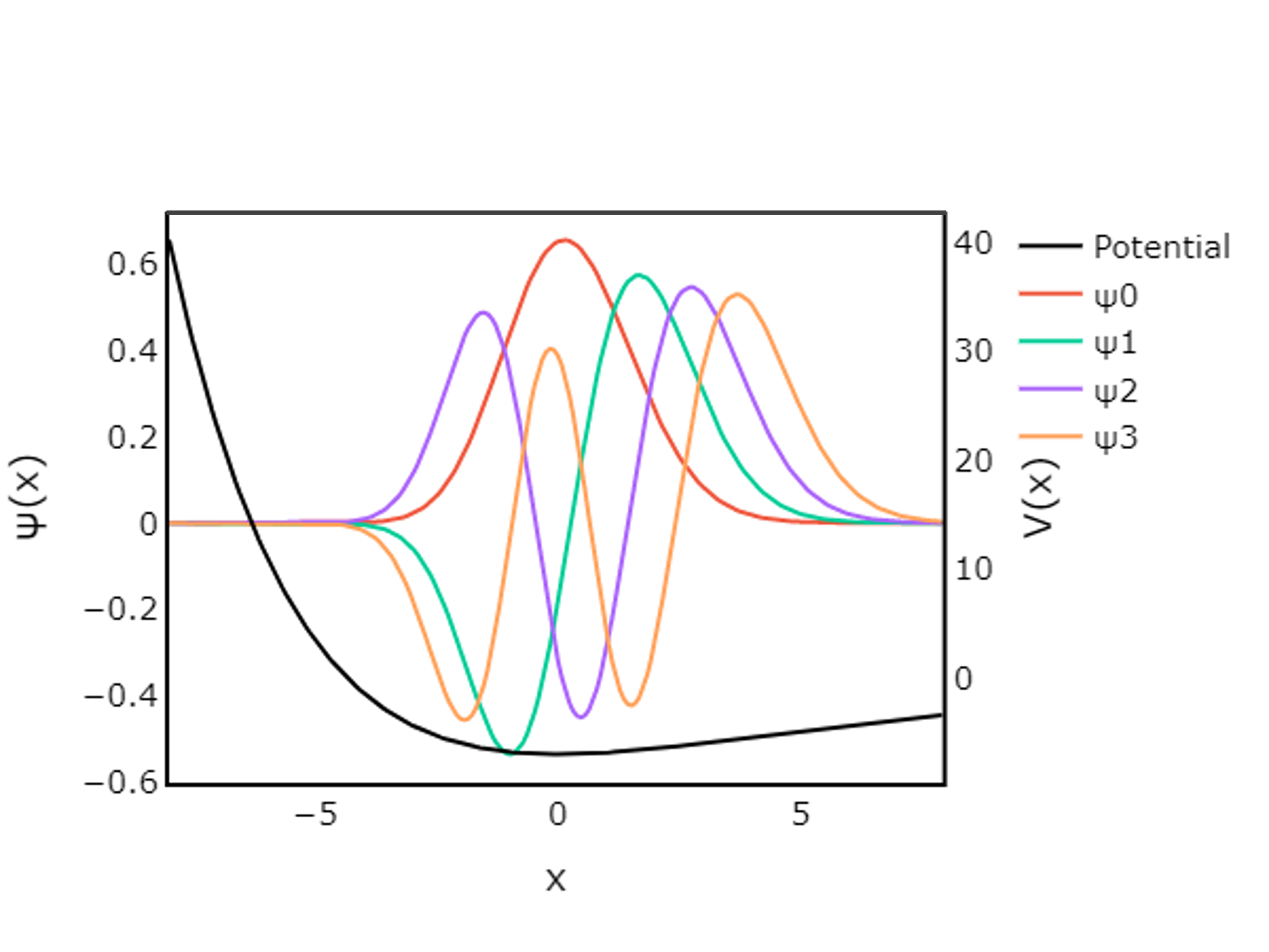}
 \caption{Example of a Morse potential with $D_e=7$, $a=0.16$ and $x_e = 0$, and the corresponding eigenfunctions 
 for the energy levels $n=0,1,2,3$. }
 \label{fig:example_morse}
\end{figure}

The Hamiltonian associated with the Morse potential has an analytical solution for eigenenergies $\{E_n\}$ and eigenfunctions $\{\phi_n(x)\}$,
$n$ being the corresponding quantum number, which is given by
\begin{equation}
E_n = -\frac{a^2 \hbar^2}{2m} \left(\lambda - n - \frac{1}{2}\right)^2, \quad n=0,1,2, \cdots, \lfloor \lambda - \frac{1}{2}\rfloor,
\label{eq:energy_morse}
\end{equation}
and
\begin{equation}
\phi^M_n(z) = N_n z^{\lambda - n - 1/2} e^{-\frac{1}{2}z} L_n^{(2\lambda - 2n -1)}(z) ,
\label{eq:wavefunction_morse}
\end{equation}
respectively, where
\begin{equation}
\lambda = \displaystyle\frac{\sqrt{2mD_e}}{a \hbar}, \qquad %\vspace{0.07in}\\
z = 2\lambda e^{-a(x-x_e)}, \qquad  \mbox{and} \quad%\vspace{0.07in}\\
N_n = \Big(\displaystyle\frac{n!(2\lambda -2n -1)}{\Gamma(2\lambda -n)} \Big)^{1/2},\vspace{0.07in}\\
\end{equation}
and $L_n^{(\alpha)}$ is a generalized Laguerre polynomial. 
Figure~\ref{fig:example_morse} shows an example of a Morse potential, together with the eigenfunctions for the first four lowest energy levels.

For 2D systems, we consider decoupled Morse Hamiltonians, whose potential energy is given by
\begin{equation}
V(x,y) = U_M^{D_1,a_1}(x) + U_M^{D_2,a_2}(y),
\label{eq:decoupled_Morse}
\end{equation}
where $U_M^{D,a}$ represents a Morse potential with well depth $D$ and width $a$. Since the potential is separable in the $(x,y)$ coordinates, the resulting eigenfunctions are analytical and are a product of the 1D Morse eigenfunctions in Eq.~\ref{eq:wavefunction_morse}. 

The performance of the network is also tested with the standard harmonic oscillator Hamiltonian
\begin{equation}
    H_{\text{HO}}(x) = \frac{p^2}{2m} + \frac{1}{2}m\omega^2 x^2,
\end{equation}
where $\omega$ is the characteristic frequency of the system. This Hamiltonian also has analytical eigenenergies and eigenfunctions, given by
\begin{equation}
    E_n = \hbar \omega (n +1/2), \quad \chi_n(x) = \displaystyle \frac{1}{\sqrt{2^n n!}}\Big(\frac{m\omega}{\pi \hbar}\Big)^{1/4} e^{-\frac{m \omega x^2}{2 \hbar}} H_n(\sqrt{\frac{m \omega}{\hbar}} x),
\end{equation}
where $H_n(x)$ are the Hermite polynomials. In this thesis, we always set $m=1, \ \hbar=1$, unless stated otherwise.

\subsubsection{Data generation}
\label{sect:data_poly}

\begin{table}[!t]
    \begin{tabular}{ccc}
        \hline \hline
        $\alpha_i$ & min & max\\
        \hline
        $\alpha_0$ & -4.5 & 1.5\\
        $\alpha_1$ & -0.65 & 0.65\\
        $\alpha_2$ & 0.2 & 1.0\\
        $\alpha_3$ & -0.01 & 0.01\\
        $\alpha_4$ & 0 & 0.1\\
        \hline
    \end{tabular}
    \qquad
    \begin{tabular}{ccc|ccc|ccc}
        \hline \hline
        $\alpha_{ij}$ & min & max    & \quad$\alpha_{ij}$   & min    & max & \quad$\alpha_{ij}$  & min   & max\\
        \hline 
        $\alpha_{00}$ & -3 & 0.1      & \quad$\alpha_{02}$ & 0.2    & 1.0  & \quad$\alpha_{04}$ & 0      & 0.2 \\
        $\alpha_{10}$ & -0.2 & 0.1    & \quad$\alpha_{21}$ & -0.02 & 0.02 & \quad$\alpha_{13}$ & -0.01 & 0.01 \\
        $\alpha_{01}$ & -0.2 & 0.1    & \quad$\alpha_{12}$ & -0.01 & 0.01 & \quad$\alpha_{22}$ & 0      & 0.04 \\
        $\alpha_{11}$ & -0.02 & 0.02 & \quad$\alpha_{03}$ & -0.01 & 0.01 & \quad$\alpha_{31}$ & -0.01 & 0.01 \\
        $\alpha_{20}$ & -0.05 & 0.05 & \quad$\alpha_{30}$ & -0.01 & 0.01 & \quad$\alpha_{40}$ & 0      & 0.02 \\
        \hline
    \end{tabular}
\caption{Lower and upper bounds for the coefficients of the polynomial potentials in one dimension 
of Eq.~(\ref{eq:poly_1D}) (left) and in two dimensions of Eq.~(\ref{eq:poly_2D}) (right). }
\label{table:coefs_poly}
\end{table}

\begin{figure}
\centering
  \includegraphics[width=1.0\textwidth]{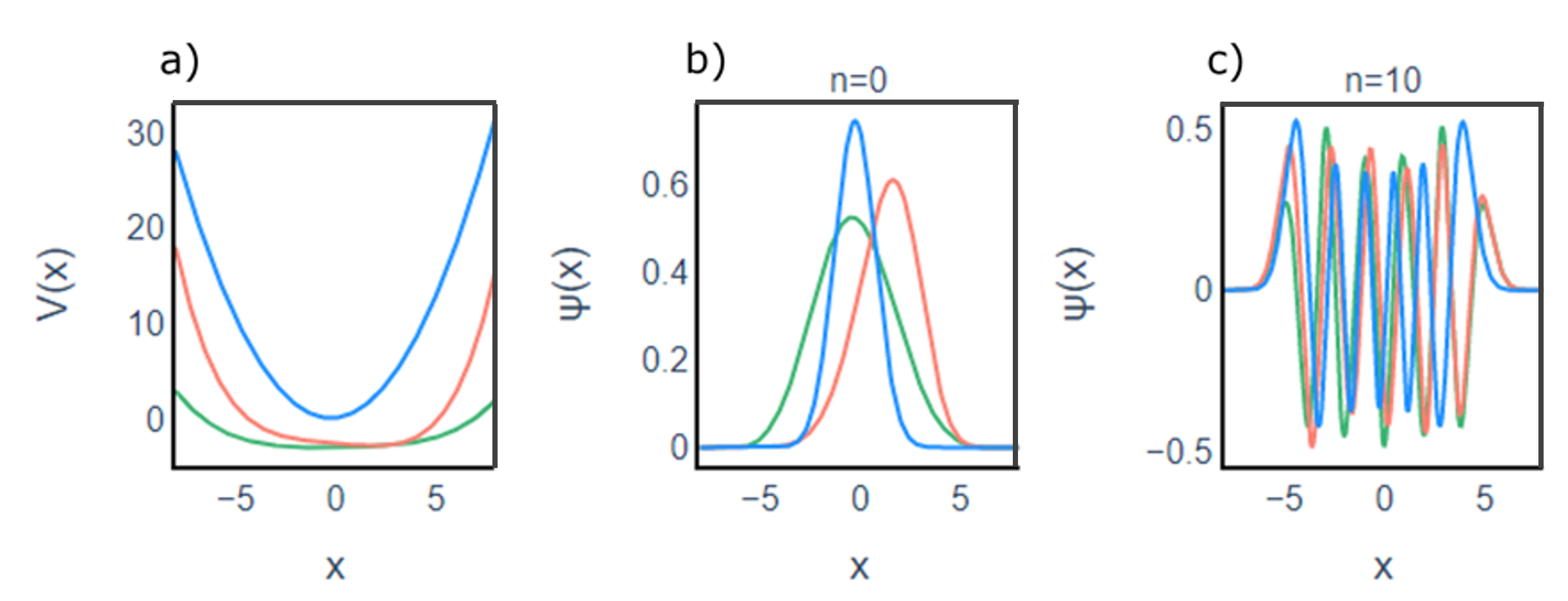}
 \caption{a) Three examples of random polynomial potentials, b) their associated eigenfunctions for the ground
 state and c) the 10-th excited state obtained with the variational method described in the text.}
 \label{fig:example_poly}
\end{figure}

This section describes the details of the training set generation. In this first scenario, the studied potentials are randomly generated with polynomials up to degree four. However, to ensure that the eigenstates have discrete energies, and thus are physically bound states (as opposed to free states), some restrictions on the polynomial coefficients need to be imposed. The first condition ensures that the even terms ($x^2$ and $x^4$) dominate over the odd terms ($x$ and $x^3$). Also, the potential is allowed to be negative and non-centred by including negative values for $\alpha_0$ and $\alpha_1$ [see Eq.~(\ref{eq:poly_1D})]. Finally, we use small values of the coefficients so that the potential does not achieve very high values, which can lead to numerical instability. The values of $\{\alpha_i\}$ (for 1D potentials) and $\{\alpha_{ij}\}$ (for 2D potentials), chosen according to the previous conditions, are shown in Table~\ref{table:coefs_poly}. Figure~\ref{fig:example_poly} shows some examples of random polynomial potentials and their associated wavefunctions for two values of the vibrational number $n=0$ and $n=10$.

Apart from the polynomial potentials, the training set also contains the eigenfunctions associated with such Hamiltonians. 
These polynomial Hamiltonians do not usually have an analytical solution, hence a numerical solver needs to be used. 
In this work, we use the variational method using harmonic oscillator eigenfunctions $\chi_n(x)$ as a basis set
to generate the eigenfunctions of an arbitrary Hamiltonian $\hat{H}$. 
That is, since $\{\chi_n(x)\}$ form a complete basis set for the Hilbert space $\mathcal{H}$  we can write any 
wavefunction $\psi(x)\in\mathcal{H}$ as a linear combination of the harmonic oscillator eigenfunctions
\begin{equation}
\psi(x) = \sum_{n=0}^\infty a_n \chi_n(x), \quad a_n \in \mathbb{C} \ \forall n.
\end{equation}
Therefore, the problem reduces to finding the values of $\{a_n\}$ which minimize the mean energy 
\begin{equation}
\expval{\hat{H}} = \expval{\hat{H}}{\psi} = \int_{-\infty}^\infty \left[\sum_{n=0}^\infty a_n \chi_n(x)\right] \hat{H} 
     \left[\sum_{m=0}^\infty a_m \chi_m(x)\right] \, dx,
\end{equation}
where it is assumed that the eigenfunction $\psi$ is normalized in the standard way, i.e.,~$\langle\psi|\psi\rangle=1$. Therefore, the task of finding the eigenenergies and eigenfunctions is reduced to an eigenvalue problem. For full details of the variational method, and its applications to 2D potentials, see Appendix~\ref{appendix1}.

\subsubsection{Neural network}
 \label{sec:NN_poly}
%----------------------------------------------------
\begin{figure}[!ht]
\centering
  \includegraphics[width=\textwidth]{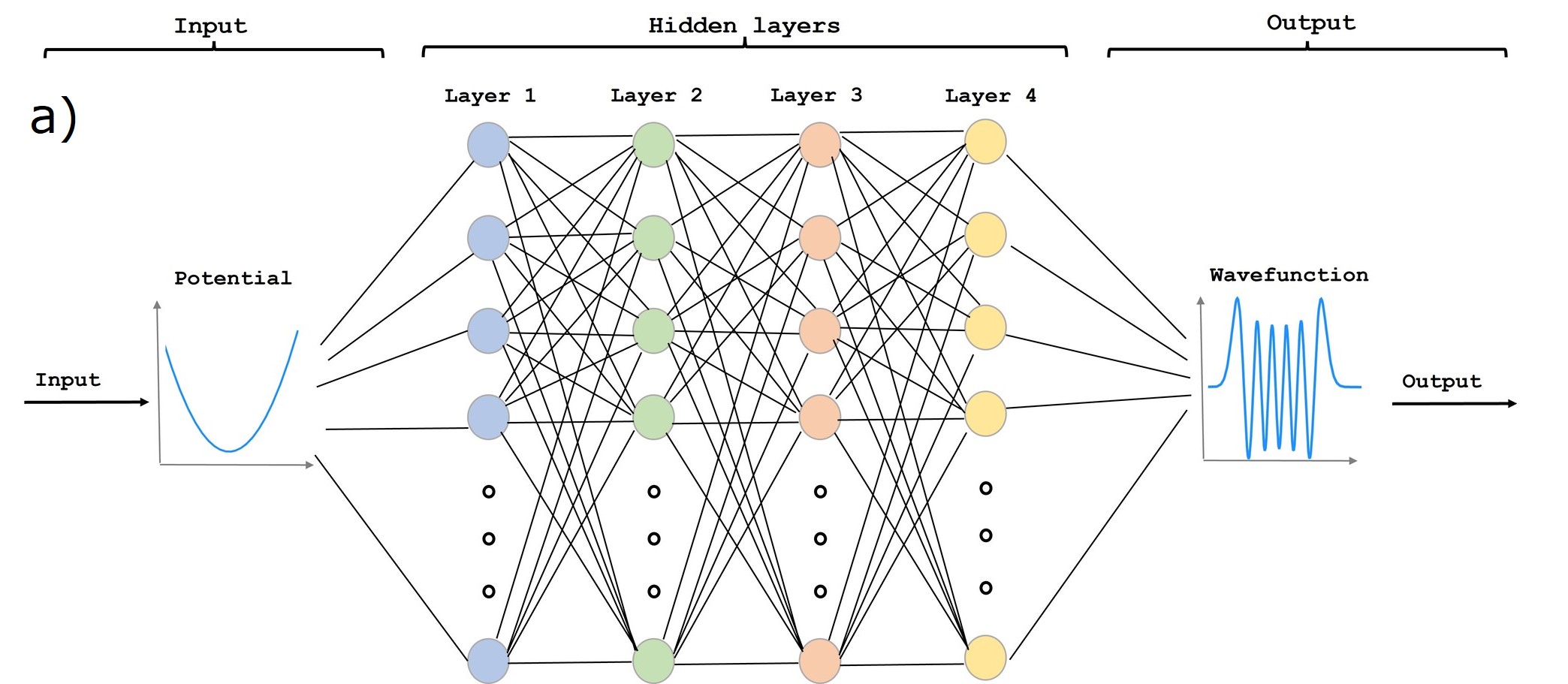}\\ \vspace*{0.5cm}
  \includegraphics[width=\textwidth]{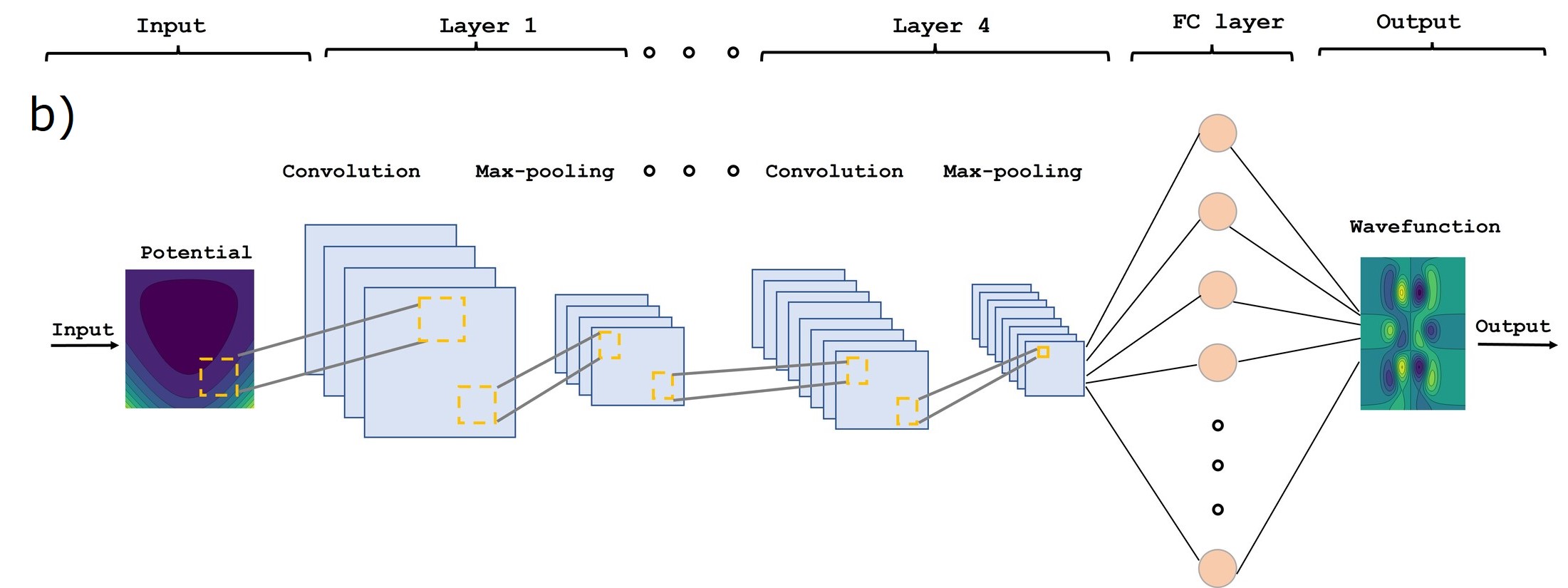}
 \caption[Schematic plots of the two types of neural networks (fully connected and convolutional neural network) considered in this work.]{Schematic plots of the two types of neural networks considered in this work:
   a) Fully connected neural network used  for the 1D potentials,
   and b) convolutional neural network used for 2D potentials. 
   The fully-connected network is composed of an input layer, four fully-connected layers and an output layer. 
   The input is a 1D array containing the potential $V(x)$, and the output is its associated wavefunction $\psi_{n}(x)$ for a particular quantum number $n$. 
   The convolutional network is composed of the input layer, four convolutional and max-pooling layers, a fully-connected layer and an output layer. 
   The input is a 2D array containing the potential $V(x,y)$ and the output is its associated wavefunction $\psi_{n_xn_y}(x,y)$ for a pair of quantum 
   numbers ($n_x,n_y$).  }
 \label{fig:NN_potentials}
\end{figure}

After obtaining the training data with the variational method, the architecture and learning process of the NNs need to be defined. For 1D potentials, we use a fully connected network. The input is an array of 200 points containing the values of the potential in the spatial domain $x\in[-8,8]$ for the 
fundamental state and $x\in[-20,20]$ for the excited states. 
Our network consists of four fully connected layers with 256, 256, 128 and 128 neurons, respectively, and RELU activation functions. Every fully-connected layer is followed by a dropout layer with parameter 0.2. These dropout layers help prevent overfitting and thus help the network to generalize to unseen potentials. The output layer is a linear layer with 200 neurons, which predicts the wavefunction for the given potential. A schematic plot of our NN is displayed in Fig.~\ref{fig:NN_potentials} a).
This network is trained using an Adam optimizer~\citep{Adam} with a learning rate of 0.0005 for 1000 iterations. The network is trained with 5000 samples with a batch size of 64. 

For 2D potentials, a CNN is used instead. This model is used to extract a lower-dimensional embedding of the original data which is then used to make the predictions. CNNs are known to be very useful to extract local patterns from the input data which results in valuable features for the learning task. 
The input is a 2D array of size 100$\times$100 in the spatial domain $(x,y)\in[-10,10] \times [-10,10]$. 
The network consists of four convolutional layers with 64, 64, 32 and 32 filters respectively. The kernel size is 3 for all four layers, and the stride is (2,2). All activation functions are also RELU here. After each convolutional layer, a max-pooling layer is added to reduce the dimensionality of the embedding. A pooling size of (2,2) and a stride of (1,1) is used. Moreover, after each max-pooling layer, we add a dropout layer with parameter 0.2 to avoid overfitting the network. Then, two fully connected layers with 128 neurons each are used to produce the final prediction. 
The output layer is a linear layer of the same size as the input. The training is performed in the same way as in the 1D case. A schematic representation of the CNN is displayed in Fig.~\ref{fig:NN_potentials} b).

\subsection{Results}
After the training process, we test the ability of the NNs to generalize to unseen Hamiltonians. In particular, the network is evaluated with three types of Hamiltonians: random polynomial potentials, harmonic oscillators and Morse oscillators. 

The parameters of the different potentials are chosen so that the associated eigenfunctions fit in the spatial domain of interest. The random polynomial potentials are generated according to the parameters in Table~\ref{table:coefs_poly}. For the ground state, we choose $\omega \in [0.2, 1]$ for the harmonic oscillator and  $D_e \in [3,7]$, $a \in [0.05, 0.1]$, $x_e \in [-0.5,0.5]$ for the Morse Hamiltonian. For the excited states, we choose $\omega \in [0.65, 0.9]$ for the harmonic oscillator and $D_e \in [3,7]$, $a \in [0.05, 0.09]$, $x_e \in [-5,-4]$ for the Morse Hamiltonian. Notice that both the polynomial potentials and harmonic oscillator belong to the same distribution as the training set. On the other hand, the Morse potentials are not a polynomial function of the spatial coordinates, and thus it should be more challenging for the NN to appropriately predict their corresponding eigenfunctions. 

\begin{figure}[!ht]
\centering
  \includegraphics[width=1.1\columnwidth]{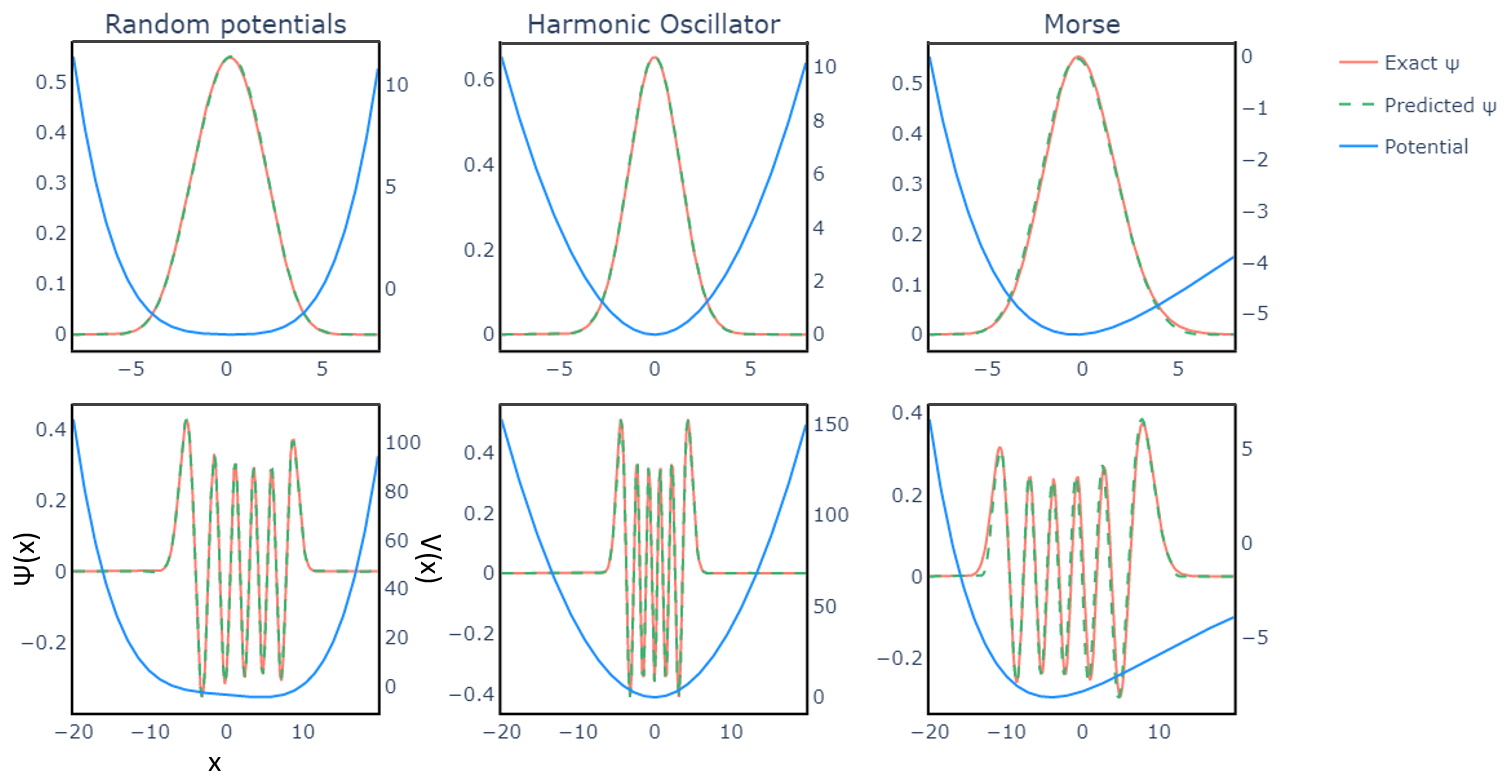}
 \caption[Example of the neural network prediction for three different 1D potentials: random polynomial potential (left), 
 harmonic oscillator (middle), and Morse potential (right).]{Example of the neural network prediction for three different 1D potentials: random polynomial potential (left), 
 harmonic oscillator (middle), and Morse potential (right). 
 Each plot displays the potential (blue), the prediction of the eigenfunction (green) and the exact eigenfunction (orange). The first row corresponds to the fundamental state and the second row corresponds to the 10th excited state.} 
 \label{fig:pred_1D}
\end{figure}

Two networks are used for the testing process, the first one to reproduce the ground states, and the second to reproduce excited eigenfunctions corresponding to $n=10$. Table~\ref{table:MSE_1D} shows the MSE (Mean Squared Error) obtained for the predicted eigenfunctions, defined as
\begin{equation}
  MSE[\psi_n(\vec{x})] = \sum_{\vec{x}} |\psi_n(\vec{x})-\tilde{\psi}_n(\vec{x})|^2,
  \label{eq:MSEpsi}
\end{equation}
where $\tilde{\psi}(\vec{x})$ is the predicted eigenfunction for the ground ($n=0$) or excited state ($n=10$), and $\psi_n(\vec{x})$ is the associated exact eigenfunction obtained with the variational method. Table~\ref{table:MSE_1D} also shows the MSE for the mean energies of such vibrational states. These results are also shown graphically in Fig.~\ref{fig:pred_1D}. Moreover, Table~\ref{table:MSE_1D} also shows the overlap between the ground and excited wavefunctions, $|\bra{\psi_0}\ket{\psi_{10}}|$, which should be close to zero since the wavefunctions should be orthogonal to each other. 

\begin{table}
\begin{tabular}{l|ll}
\hline \hline \\[-0.3cm]
Potential Type       & MSE($\psi$)         & MSE(E)\\
\hline\\[-0.2cm]
Polynomial  & $6\times 10^{-6}$ & $6\times 10^{-8}$\\
H.O   & $3\times 10^{-5}$ & $3 \times 10^ {-6}$\\
Morse      & $1\times 10^{-5}$ & $2 \times 10^{-6}$\\
\hline
\end{tabular}
\quad
\begin{tabular}{l|lll}
\hline \hline \\[-0.3cm]
Potential Type         & MSE($\psi$)        & MSE(E)  & $| \bra{\psi_0}\ket{\psi_{10}}|$\\
\hline \\[-0.2cm]
Polynomial  & $3\times 10^{-5}$ & $1 \times 10^{-6}$ & $2 \times 10^{-4}$ \\
H.O   & $2\times 10^{-6}$ & $2 \times 10^{-7}$ & $5 \times 10^{-4}$\\
Morse       & $6\times 10^{-3}$ & $9 \times 10^{-6}$ & $7 \times 10^{-4}$\\
\hline
\end{tabular}
\caption[Mean squared error for the eigenfunctions and energies for the 1D systems for the fundamental state (left) and the 10-th excited state (right).] {Mean squared error for the eigenfunctions and energies for the 1D systems -polynomial potentials, harmonic oscillator and Morse potentials- for the fundamental state (left) and the 10-th excited state (right). The second table also contains the overlap of the fundamental and 10-th excited states.}
\label{table:MSE_1D}
\end{table}

In the case of the Morse Hamiltonian ground state, the MSE for both energy and wavefunction is similar to those 
obtained for the polynomial potentials. 
This means that the NN can effectively generalize to non-polynomial potentials when trained with polynomial potentials. 

The MSE of the harmonic oscillator potentials is also similar to the MSE of the random polynomial potentials, 
which is not an unexpected result since the harmonic oscillator is a polynomial potential as well. Finally, it should be remarked that, since the MSE of the wavefunction is similar for all potentials, the NN is not producing much overfitting. 

Regarding the excited eigenfunctions, the MSE for both the energy and wavefunction is small for the three types of potentials. 
However, the predicted wavefunction for the Morse potentials presents a higher MSE than the MSE of the polynomial potentials. 
In particular, Fig.~\ref{fig:pred_1D} shows that the tails of the wavefunction are not correctly reproduced. 
This is a consequence of training the network only with polynomial potentials, whose wavefunctions have a significantly different decay. 
However, the error in the tails of the wavefunction does not affect much the value of the mean energy, 
since the MSE for the energy is similar to the MSE of the polynomial potentials. 

Finally, we calculate the overlap between the ground and excited wavefunctions, $|\bra{\psi_0}\ket{\psi_{10}}|$, for all the Hamiltonians. Since the exact wavefunctions are orthogonal to each other, the overlap of the predicted wavefunctions should be close to zero. We observe that for all the Hamiltonians, the overlap is of the order of $10^{-4}$, which is quite close to zero. Notice that this quantity can be used as a measure of the performance of the method. This result is important since the NNs are not explicitly told to reproduce the orthogonality of the wavefunctions. In fact, different NNs are used to predict each eigenstate, and thus the network has no information about the other eigenstates. Therefore, the NN learns to reproduce a quantum property without being explicitly programmed to do so. 

\begin{figure}[h!]
\centering
  \includegraphics[width=1.0\columnwidth]{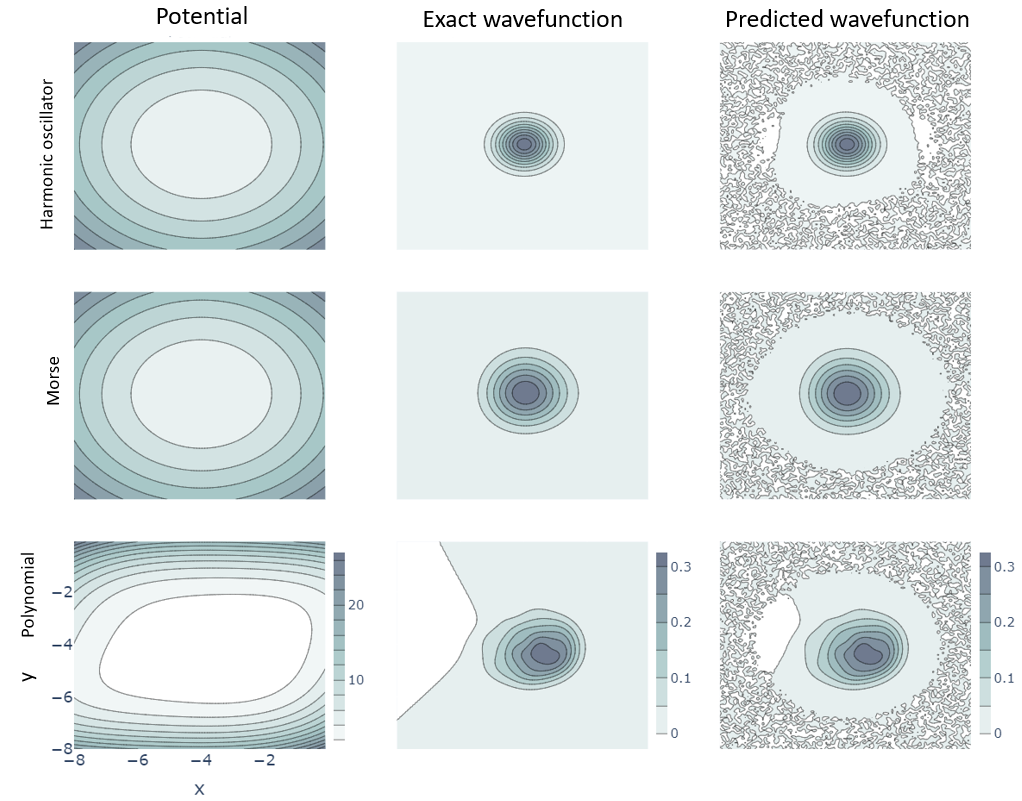}
 \caption{Example of potential (left), exact eigenfunction (middle) and predicted eigenfunction (right) for the ground state of a harmonic oscillator (top), the Morse (middle) and random polynomial (bottom) Hamiltonian.} 
 \label{fig:example_2D_poly}
\end{figure}

A different NN is trained to reproduce the fundamental wavefunction of several 2D (harmonic, Morse and polynomial) potentials. Again, the training data consists of random polynomial potentials, while the test data contains also decoupled Morse potentials. The MSE results for the three cases are summarized in Table~\ref{table:MSE_2D}. 
Figure~\ref{fig:example_2D_poly} shows examples of the potential, the exact ground state and the predicted ground state 
for the three types of potentials. We observe that the network is also able to reproduce the fundamental wavefunctions for 2D Morse oscillators, as the MSE for both the energy and the wavefunctions are similar to the MSE of the random polynomial Hamiltonians.

These results show the advantages of using ML instead of a model-based approach. Even though the data generation process is very different for the polynomial potentials (training) and Morse oscillators (prediction), the ML model demonstrates its superiority by accurately reproducing the Morse eigenfunctions.

\begin{table}
\begin{center}
    \begin{tabular}{l|ll}
        \hline \hline \\[-0.3cm]
        Potential Type         & MSE($\psi$)        & MSE(E) \\
        \hline \\[-0.2cm]
        Polynomial potentials & $2\times 10^{-7}$ & $5\times 10^{-5}$  \\
        Harmonic oscillator   & $9\times 10^{-6}$ & $2 \times 10^ {-4}$ \\
        Morse potentials     & $9\times 10^{-6}$ & $3 \times 10^{-4}$   \\
        \hline
    \end{tabular}
    \caption{Mean squared error for the eigenfunctions and eigenenergies for three types of 2D potentials for the ground state.}
    \label{table:MSE_2D}
\end{center}
\end{table}
 
\subsection{Second scenario: Coupled Morse Hamiltonians}
\label{sect:Coupled_Morse_paper1}
In the previous section, we validated the use of NNs to predict the eigenfunctions of new Hamiltonians for a certain quantum number $n$. The goal of the second scenario is to study the performance of a NN, trained with the solutions of an analytically solvable Hamiltonian $\hat{H}_0$, in reproducing the wavefunctions of a perturbed,
non-separable Hamiltonian $\hat{H}$. Mathematically, let $\hat{H}$ be the Hamiltonian whose eigenfunctions we want to find, and suppose that it can be written \textit{\`a la Kolmogorov-Arnold-Moser}~\citep{Llave} as
\begin{equation}
\hat{H} = \hat{H}_0 + \hat{H}_1,
\label{eq:sum_Hamiltonians}
\end{equation}
where $\hat{H}_0$ is a Hamiltonian whose eigenstates are analytically known. 
If the perturbation $\hat{H}_1$ is small, then $\hat{H}_0$ is a good approximation of $\hat{H}$, and it can be expected that a NN can generalize the wavefunctions for $\hat{H}_0$ to those for $\hat{H}$. However, this is not obvious \textit{a priori} since resonance between modes in the excited states can change significantly the
topology of these wavefunctions. Because of the complexity of such associated wavefunctions, this problem is more challenging for the NN. For this reason, some training strategies will be implemented to ensure the convergence of the algorithm. In this subsection we introduce the Hamiltonian at study, the training data and the details of the training process.

\subsubsection{Test data: Coupled Morse potentials}
We consider the following kinetically coupled 2D Morse oscillator, which has been extensively studied in the past
as a model for the stretching vibrations of the H$_2$O molecule~\citep{Jaffe,H2O2,H2O,H20_NN}:
\begin{equation}
H(x_1, x_2, p_1, p_2) = \frac{1}{2}(G_{11}p_1^2 + G_{11}p_2^2) + G_{12}p_1p_2 + U_M(x_1) + U_M(x_2).
\label{eq:coupled_morse}
\end{equation}
with $G_{12} \ll G_{11}$. Functions $U_M$ are 1D Morse potentials characterized by parameters $a$ and $D_e$, and $p_1, p_2$ are the corresponding conjugate momenta. Since $G_{12} \ll G_{11}$, the
Hamiltonian in Eq.~\ref{eq:coupled_morse} can be written in the form of Eq.~\ref{eq:sum_Hamiltonians} with
\begin{equation}
  H_0(x_1, x_2, p_1, p_2) = \frac{1}{2}(G_{11}p_1^2 + G_{22}p_2^2) + U_M(x_1) + U_M(x_2), 
    \quad H_1(x_1, x_2, p_1, p_2) = G_{12}p_1p_2.
\end{equation}
The goal of this scenario is to train a NN that is able to predict the eigenfunctions for multiple coupled Morse oscillators with different values of parameters $a$, $D_e$ and $G_{12}$. 

In the previous scenario, we provided only the potential function to the NN, instead of the whole Hamiltonian. This allows an easy representation of the Hamiltonian as a grid containing the values of the potential energy on a rectangular spatial domain. This representation was possible in the cases presented before because all the Hamiltonians had the same kinetic energy. 
Nonetheless, the coupled Morse Hamiltonian contains a coupling term in the momentum coordinates instead of in the spatial coordinates. 
Therefore, to be able to provide only the potential energy to the NN, we have to make a change of coordinates in the Hamiltonian in such a way that the coupling appears only in the spatial term, i.e.,~ the potential.
To this end, we can rewrite the Hamiltonian in generalized coordinates so that the coupling appears in the spatial coordinates instead of in the momentum coordinates. The details of this transformation can be found in Appendix~\ref{appendix2}. The resulting Hamiltonian is the following
\begin{multline}
K(x,y,p_x,p_y) = \frac{1}{2}(p_x^2 + p_y^2) + 
   U_M\left(\frac{1}{\sqrt{2}}\left[\sqrt{G_{11} + G_{12}}y- \sqrt{G_{11} - G_{12}}x\right]\right) + \\
   U_M\left(\frac{1}{\sqrt{2}}\left[\sqrt{G_{11} + G_{12}}y+ \sqrt{G_{11} - G_{12}}x\right]\right).
 \label{eq:canonical_transform}
\end{multline}
Let us remark that this equivalent Hamiltonian has no kinetic coupling in the momenta $p_x$ and $p_y$, 
but it has been moved to the potential term, between spatial coordinates $x$ and $y$,
which is more adequate for our computational purposes when using NNs, as indicated before. 

\subsubsection{Training data: Decoupled Morse Hamiltonians}
Now that we have defined the coupled Morse potential in generalized coordinates, we have to define the training data of the NN. In this case, we use \emph{decoupled} Morse potentials to train the network, which acts as a first-order approximation of the \emph{coupled} Morse potential. 
The training potentials will then be of the form of Eq.~\ref{eq:decoupled_Morse} in the new coordinates. Notice that the Morse parameters of this Hamiltonian are different in each spatial coordinate.

Recall that the Hamiltonian associated with this potential is separable, and the Schrödinger equation has an analytical solution in terms of 
Eqs.~\ref{eq:energy_morse} and~\ref{eq:wavefunction_morse}. 
Thus, there is no need to use a numerical solver to train the NN, which makes the method more convenient. 
In this case, the training data are pairs $\{V_{ij},\psi_{ij}\}$, where $V_{ij}$ is a matrix containing the values of the potential in a rectangular domain, 
and $\psi_{ij}$ is the associated wavefunction corresponding to the vibrational quantum numbers $n_x,n_y$. 

We train different models for different quantum numbers to reproduce multiple excited eigenstates of the coupled Morse Hamiltonian. 
Notice that here the excited states are identified by their quantum numbers ($n_x$, $n_y$) instead of their energies $E(n_x,n_y)$. 
This choice is made because when the difference between eigenenergies is small, the $n$-th energy level of two similar potentials 
can correspond to very different quantum numbers ($n_x$, $n_y$), and consequently completely different wavefunctions. 
This fact would confuse the network since two similar inputs would have very different outputs. In the next subsection, we give further details about how the training data are generated.

\subsubsection{Data generation}

NNs are useful methods for extracting features of complex data. 
However, if the training and test sets are too different, the network may not be able to produce accurate predictions. For this reason, the training data must resemble as much as possible the test data. Since the NN is trained using separable Morse Hamiltonians and their eigenfunctions, we need to carefully select the parameters of such Hamiltonians to optimize the training process. To this end, two strategies are used to generate useful training data:

\begin{itemize}
\item \textbf{Selecting the decoupled Morse parameters $(D_1,a_1)$, $(D_2,a_2)$:} 
The parameters of the decoupled potential used in the training phase are chosen in two different ways:

\begin{itemize}
\item By performing curve fitting and finding the parameters $(D_1,a_1)$, $(D_2,a_2)$ which best approximate the coupled Morse 
potential in Eq.\ref{eq:canonical_transform}.
\item By choosing the parameters $(D_1,a_1)$, $(D_2,a_2)$ which have the same quadratic order Taylor expansion coefficients 
as the coupled Morse potential. 
Considering Eq.\ref{eq:Tayor_morse} the choice should be
\begin{equation}
    D_1 = D_2 =D_e,  \qquad \mbox{and} \quad 
    a_1 = \sqrt{G_{11} - G_{12}}a, \quad 
    a_2 = \sqrt{G_{11} + G_{12}}a.
\end{equation}
\end{itemize}
To obtain the training data, we first generate samples of the parameters of the coupled Morse potential 
$a\in [0.09, 0.12]$, $D_e \in [1,10]$, and then find the decoupled Morse parameters $(D_1,a_1)$, $(D_2,a_2)$, according to the previous strategies. For all Hamiltonians, we set $G_{11}=1.05$, and we test two different values of the coupling parameter, $G_{12} = 0.015$ and $G_{12} = 0.35$.  

\item \textbf{Selecting the $x$ and $y$ ranges:} Once the Morse parameters have been chosen, we try to improve the resemblance 
with the coupled Morse potential by changing the values of the spatial domains. 
The input of the NN is a grid containing the values of the potential in a certain spatial domain, but such domain is not specifically given. 
Therefore, if we change the range of this domain the network will not notice the difference, as long as the number of points remains constant. 
This fact allows us to stretch the spatial domain so that the decoupled potential is more similar to the coupled potential. 
We perform a grid search to find the domain ranges that best approximate the coupled potential energy. 
The only constraint is that in a given spatial domain, the associated wavefunction fits into that domain. 
Otherwise, the sample will not be useful for training. 
Recall that this technique could only be used because the learning algorithm (the NN) is a data-based approach
instead of a model-based approach, which means that uses no information about the underlying physical model of the system. 
\end{itemize}

\subsubsection{Neural network}
The architecture of the NN is the same as the one used with the 2D random polynomial potentials (see Fig.~\ref{fig:NN_potentials} b)).
The only difference in the training process is the choice of the loss function.
When the energy of the system increases, the eigenfunctions of the coupled Morse Hamiltonian show significant differences to any of the eigenfunctions of the decoupled Morse Hamiltonian, due to the effect of the different nonlinear resonances existing in the system~\citep{Jaffe,H2O2}. 
For example, the number of nodes of the wavefunction may not be well-defined. 
In this case, training the NN with only the decoupled wavefunction does not give optimal results. For this reason, we add a custom loss function to help train the network, defined in the following way 
\begin{equation}
  S_{loss} = \norm{\hat{H}\widetilde{\psi}^c - \widetilde{E}\widetilde{\psi}^c}^2 + \lambda_\text{norm}|\norm{\widetilde{\psi}^c}^2-1|,
  \label{eq:custom_loss}
\end{equation}
where $\widetilde{E}$ is the predicted mean energy for the quantum numbers $(n_x, n_y)$, which is calculated using the input potential and the predicted 
wavefunction for the \emph{coupled} Morse oscillator $\widetilde{\psi}^c$. The left term of the loss function ensures that the predicted wavefunctions follow the Schrödinger equation, and the right term enforces the normalization condition of the wavefunctions. Therefore, the total loss function is
\begin{equation}
\mathcal{L} =  MSE_d + \lambda \bar{S}_\text{loss,c} =  \frac{1}{N_d} \sum_{i=0}^{N_d} \norm{\widetilde{\psi}_i^d - \psi_i^d}^2 
   + \lambda \sum_{i=0}^{N_c} \left(\norm{\hat{H}\widetilde{\psi}_i^c - \widetilde{E}\widetilde{\psi}_i^c}^2 + \lambda_\text{norm}|\norm{\widetilde{\psi}_i^c}^2 -1|\right),
\label{custom_loss}
\end{equation}
where $\{\psi_i^d\}_i$ are the wavefunctions of the decoupled Hamiltonian and $\{\psi_i^c\}_i$ are the wavefunctions of the 
coupled Hamiltonian. 
Parameter $\lambda_{norm}$ is chosen so that the two terms of the custom loss have the same order of magnitude. 
In this case, we choose $\lambda_{norm}=10^4$ for all the training process. 
We train the network for 300 iterations. 
During the first 100 iterations, we set $\lambda=0$ so that the model learns to reproduce the wavefunctions of multiple 
decoupled Hamiltonians. 
Then, we choose $\lambda$ so that the two loss functions have the same order of magnitude. 
In this case, we set $\lambda=10^5$. 
Notice that since the decoupled Morse potential already gives a fair approximation of the exact wavefunction, 
we do not need to put any constraints on the energy of the system. 
The NN converges to the exact solution of the Schrödinger equation.

\subsection{Results}

\begin{figure}[!ht]
\centering
  \includegraphics[width=1.0 \columnwidth]{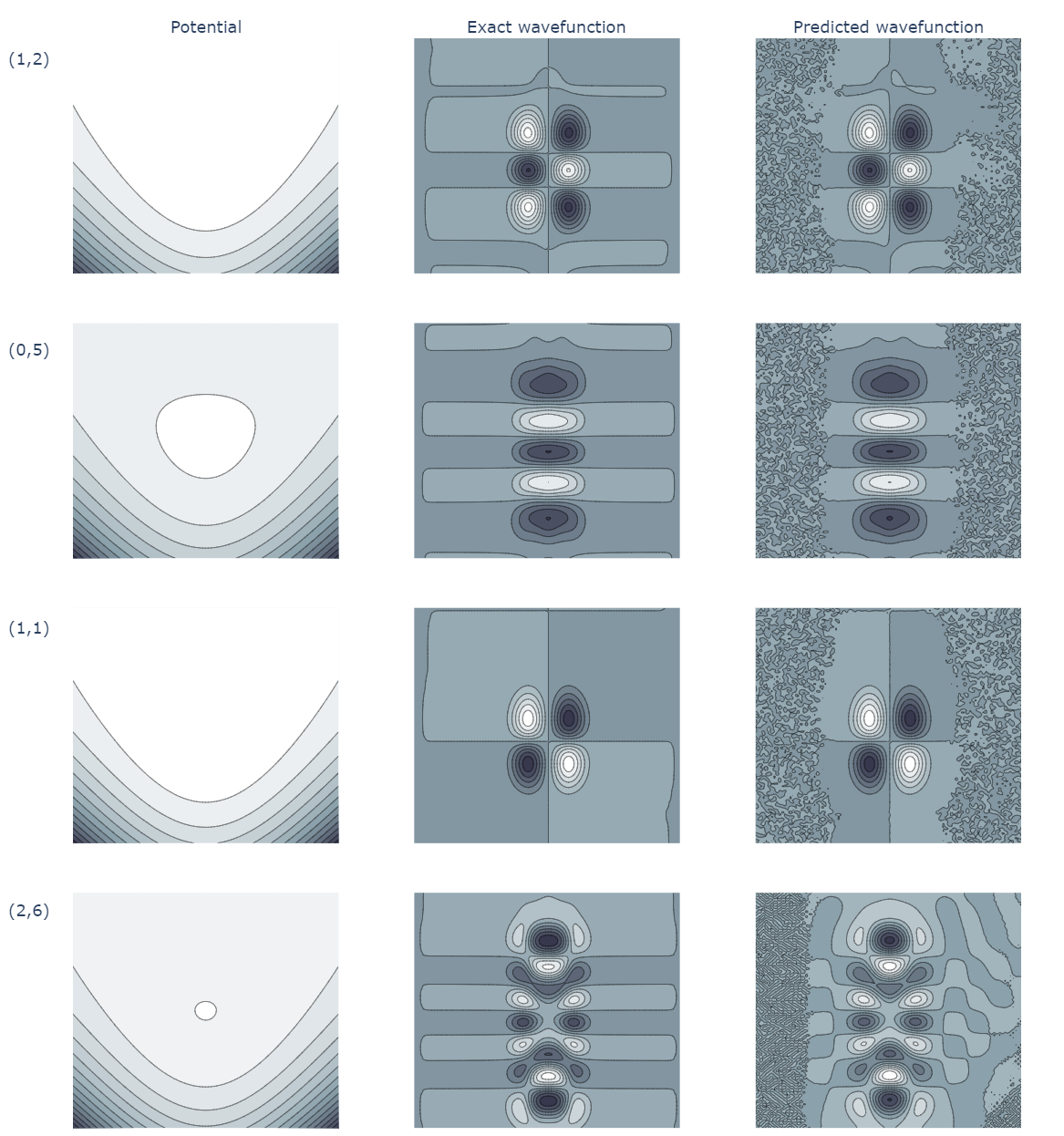}
 \caption[Coupled Morse potential from Eq.~\ref{eq:canonical_transform} (left), exact eigenfunction (middle) and predicted eigenfunction (right) for multiple excited eigenfunctions.]{Coupled Morse potential from Eq.~\ref{eq:canonical_transform} (left), exact eigenfunction (middle) and predicted eigenfunction (right) for multiple excited eigenfunctions. The quantum numbers $(n_x, n_y)$ and Morse parameters $(D_e, a)$ are: $(n_x, n_y) = (1,2), \ (D_e,a) = (2.5, 0.095)$ for the first row, $(n_x, n_y) = (0,5), \ (D_e,a) = (2.1,0.095)$ for the second row, $(n_x, n_y) = (1,1), \ (D_e,a) = (2.4, 0.097)$ for the third row and $(n_x, n_y) = (1,2), \ (D_e,a) = (1.6, 0.098)$ for the fourth row, respectively. In all cases, $G_{11} = 1.05, G_{12} = 0.015$.} 
 \label{fig:coupled_eigenfunctions}
\end{figure}

\begin{figure}[!ht]
\ContinuedFloat
\centering
  \includegraphics[width=1.0\columnwidth]{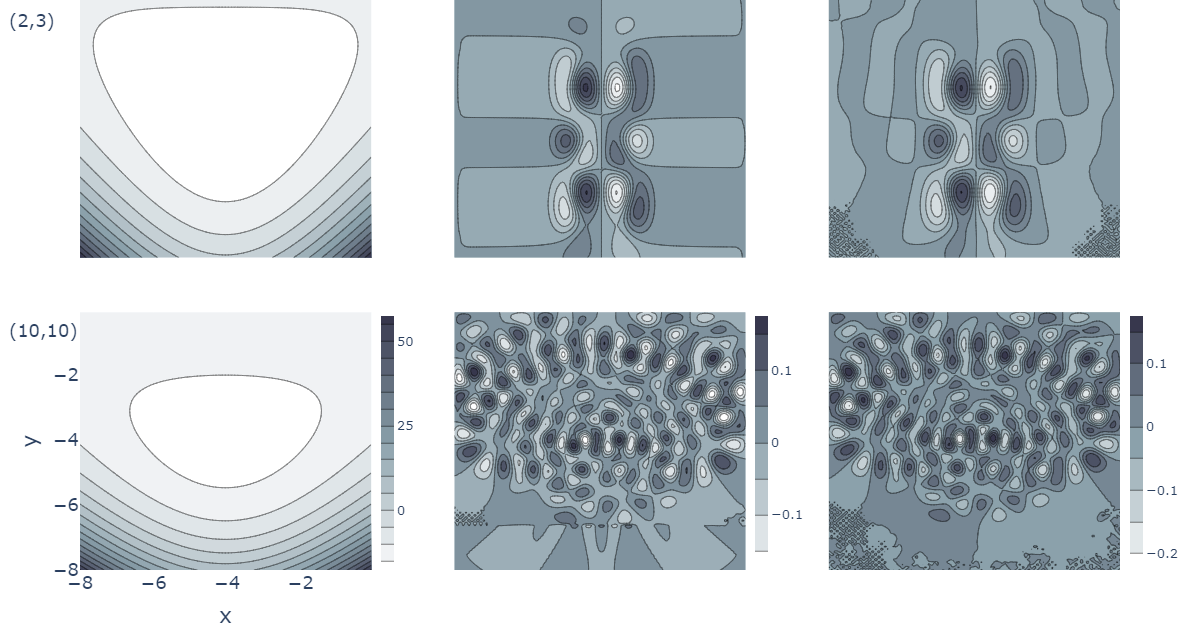}
 \caption[Continued  Fig.~\ref{fig:coupled_eigenfunctions}]{\textit{(continued)} Coupled Morse potential from Eq.~\ref{eq:canonical_transform}  (left), exact eigenfunction (middle) and predicted eigenfunction (right) for multiple excited eigenfunctions. The quantum numbers $(n_x, n_y)$ and Morse parameters $(D_e, a, G_{12})$ are  $(n_x, n_y) = (2,3), \ (D_e,a, G_{12}) = (1.1, 0.093, 0.015)$ for the first row and $(n_x, n_y) = (10,10), \ (D_e,a, G_{12}) = (7.8,0.098, 0.35)$ for the second row, respectively.}  
\end{figure}

\begin{figure}[!ht]
\centering
  \includegraphics[width=1.0\columnwidth]{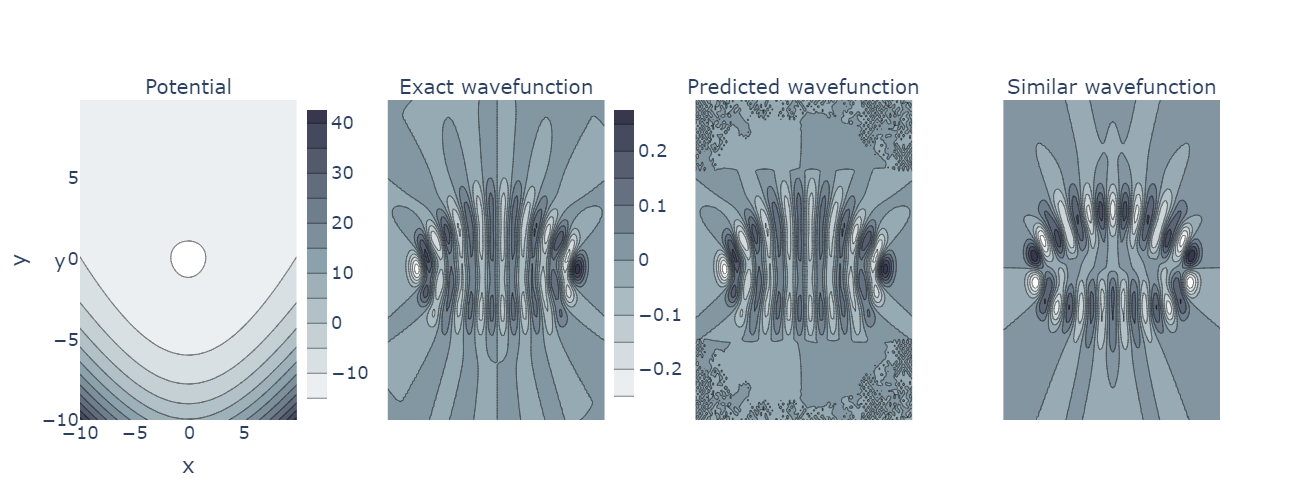}
 \caption[Coupled Morse potential from Eq.~\ref{eq:canonical_transform}  (first), exact eigenfunction (second) and predicted eigenfunction (third) for the excited 
 eigenfunction.]{Coupled Morse potential from Eq.~\ref{eq:canonical_transform}  (first), exact eigenfunction (second) and predicted eigenfunction (third) for the excited 
 eigenfunction with quantum numbers $(n_x, n_y) =(17, 1)$. The fourth plot displays a similiar eigenfunction to the predicted one. 
 The Morse parameters are $(D_e, a, G_{12} )= (7.1, 0.096 , 0.35)$. }
 \label{fig:coupled_eigenfuctions2}
\end{figure}

The NN is trained to reproduce the eigenfunctions of coupled Morse Hamiltonians for different quantum numbers ($n_x$, $n_y$). At first, the value of the coupling parameter is set to $G_{12} =0.015$. With this configuration, some examples of exact and predicted eigenfunctions are shown in Figs.~\ref{fig:coupled_eigenfunctions} and~\ref{fig:coupled_eigenfuctions2}.

 As can be seen, for low energies, the eigenfunctions of the coupled Morse potential exhibit a well-defined nodal pattern which leads
to an unambiguous quantum numbers assignment; see, for example, the rows 1-3 in Fig.~\ref{fig:coupled_eigenfunctions}
which correspond to states $(n_x,n_y)=(1,2), \ (0,4)$, and $(1,1)$, respectively. These wavefunctions are very similar in topology to the eigenfunctions of the decoupled Morse potential, which is used to train the NN. Therefore, it is not unexpected that the network can easily generate the eigenfunctions for the case of the coupled Morse potential. For these low values of energy, the strategies used to generate the training data are enough to make the NN able to generate the eigenfunctions for coupled Morse potentials with high accuracy. 

However, for higher energy levels; see rows 4-5 in Fig.~\ref{fig:coupled_eigenfunctions}, the eigenfunctions present sizeable distortions with respect to the uncoupled ones, and the quantum numbers cannot be defined so easily. In these cases, it is necessary to add the loss function in Eq.~\ref{eq:custom_loss} to make the network able to predict the corresponding wavefunctions.

To increase the distortion of the eigenfunctions for the coupled Morse potential and make them more challenging for being found by the NN, we increase the coupling factor to $G_{12}=0.35$. With this new value of the coupling parameter, high energy states with a high distortion were also reproduced: two representative examples are shown in the last row of Fig.~\ref{fig:coupled_eigenfunctions} and Fig.~\ref{fig:coupled_eigenfuctions2}. 
The latter contains an additional plot, showing a wavefunction with similar energy and similar quantum numbers. 

Indeed, when the energy levels increase, there may be more than one wavefunction with very close energy and similar quantum numbers. 
In these cases, using the eigenfunctions of decoupled Morse potentials to train the network only allows the reproduction of one of the wavefunctions, 
i.e.,~that which is more similar to the decoupled wavefunction. 
If we wanted to distinguish between these eigenstates, we would have to provide a finer approximation to the target wavefunction. 
One option to perform this task is to gradually increase the coupling factor $G_{12}$ to approach the final desired Hamiltonian, using at each step in the approximation the coupled Morse potential with for the previous (smaller) coupling factor to approximate the wavefunctions corresponding to higher-coupled potentials.

\begin{table}[!t]
\begin{center}
    \begin{tabular}{l|lllll}
        \hline \hline
         & $\psi_{11}$ & $\psi_{05}$ & $\psi_{12}$ & $\psi_{23}$ & $\psi_{26}$ \\
        \hline \\[-0.2cm]
        $\psi_{11}$  & 1.0 & $1\times 10^{-4}$ & $3\times 10^{-4}$ & $9 \times 10^{-4}$ & $3 \times 10^{-7}$  \\
        $\psi_{05}$    &  & 1.0 & $2 \times 10^{-3}$ & $4\times 10^{-4}$ & $4\times 10^{-4}$ \\
        $\psi_{12}$ & & & 1.0 & $1\times 10^{-3}$ & $1\times 10^{-4}$\\
        $\psi_{23}$ & & & & 1.0 & $1\times 10^{-4}$ \\
        $\psi_{26}$ & & & & & 1.0 \\
        \hline  
    \end{tabular}
\end{center}
\caption{Overlap $|\bra{\psi_k}\ket{\psi_l}|$ of the excited eigenfunctions of the coupled Morse Hamiltonian with parameters $(D_e,a) = (2.1,0.095)$ and $G_{12} = 0.015$.}
\label{table:4}
\end{table}

Finally, we have also calculated the overlap between the excited wavefunctions of the same Hamiltonian. The results are shown in Table~\ref{table:4}. We see that the results are close to zero even though the wavefunctions were generated with different NNs so they had no information about other eigenstates. 

All in all, these results show that NNs trained with the appropriate learning algorithm can reproduce high-energy eigenstates of complex Hamiltonians. This can be considered a good first step towards the computation of vibrational wavefunctions in high-dimensional systems, where the ML methods bear a clear advantage over the standard computational chemistry approaches.

%%%%%%%%%%%%%%%%%%%%%%%%%%%%%%%%%%%%%%%%%%%%%%%%%%%%%%%%%%%
% RC TO SOLVE THE SCHRODINGER EQUATION
%%%%%%%%%%%%%%%%%%%%%%%%%%%%%%%%%%%%%%%%%%%%%%%%%%%%%%%%%%%
\section{Adapting Reservoir computing to solve the Schrödinger equation }
\label{sect:paper2}
The previous study shows that NNs trained with the appropriate strategy can be used to solve the time-independent Schrödinger equation to calculate the ground and high-energy states of multiple molecular Hamiltonians. The limitation of the previous approach is that a NN needs to be trained for every quantum number associated with an energy level. As a result, if one aims to obtain all the excited states within a certain energy range, a separate NN needs to be trained for each eigenstate within that range. This process is time-consuming and computationally demanding, and thus it is the bottleneck of the previous method.

In the second part of this chapter, we introduce a novel approach, based on RC, that allows us to obtain all the eigenenergies and eigenstates of a Hamiltonian within an energy range~\citep{Domingo_adaptingRC}. The first step of this new method consists of integrating the time-dependent Schrödingr equation (see Eq.~\ref{eq:time-dependent}). Given an initial wavefunction  $\psi_0(\vec{x})$, the solution of the Schrödinger equation provides the time-evolution of the associated quantum state. The propagation of a quantum state with certain (mean) energy allows the calculation, by Fourier transformation, of the eigenenergies and eigenfunctions of the quantum system around the same energy, using the following methodology:

\begin{enumerate}
    \item Prepare an initial wavefunction $\psi_0(\vec{x})$ as a minimum uncertainty Gaussian wave packet or coherent state with (mean) energy $E_0$. For 1D systems, the initial wavepacket is of the form
    \begin{equation}
        \psi_0(x)  =  \left(\frac{1}{\pi}\right)^{1/4} 
        e^{-\frac{(x-x_{0})^2}{4(\Delta x)^2}}  \times \, e^{-i\,p_{0}(x-x_{0})},
        \label{eq:initial_wavepacket}
    \end{equation}
    where $x_0$ is the central position, $\Delta x$ is the spread or width, and $p_0$ is the momentum (or phase) of the wavepacket. The energy of the wavepacket in Eq.\ref{eq:initial_wavepacket} corresponds to the value of the Hamiltonian function at position $x_0$ and momentum $p_0$. That is, if the Hamiltonian is of the form $H(p,x) = \frac{p^2}{2m} + V(x)$, the energy of the wavepacket will be $E_0 = H(p_0, x_0) = \frac{p_0^2}{2m} + V(x_0)$. For higher dimensional systems, the initial wavepacket will be a product of coherent states from Eq.~\ref{eq:initial_wavepacket} in the different coordinates. For example, for a 2D system, the initial wavepacket will be of the form $\psi_0(x,y) = \psi^x_0(x)\psi^y_0(y)$, with $\psi^x_0(x)$ associated with parameters ($x_0, \Delta x, p_{x_0})$ and $\psi^y_0(y)$ associated with $(y_0, \Delta y, p_{0_y}$).
    \item Calculate the time evolution $\psi(\vec{x}, t)$ of $\psi_0(\vec{x})$ under the Hamiltonian $H(\vec{p},\vec{x})$ by solving the time-dependent Schrödinger equation until time $T$.
    \item Calculate the energy spectrum $I(E)$. 
    First, we calculate the time-correlation function 
    \begin{equation}
        A(t) = \braket{\psi(t)}{\psi(0)} = \int_{\vec{x}} \; \psi(\vec{x},t)^* \psi_0(\vec{x}) \; d\vec{x}.
        \label{eq:time-corr}
    \end{equation}
Then, we obtain the energy spectrum $I(E)$ as the Fourier transform of the time-correlation function
    \begin{equation}
        I(E) = \int_0^T \; A(t) \; e^{i E t/ \hbar} \; dt .
    \end{equation}
    The eigenenergies $E_n$ correspond to the peaks of the modulus of the energy spectrum function $|I(E)|^2$. 
    
    \item Calculate the eigenfunctions  $\phi_n(\vec{x})$ by performing the Fourier transform of
    $\psi(\vec{x},t)$ at the eigenenergies $E_n$ obtained in step 3.
    \begin{equation}
        \phi_n(\vec{x}) = \int_0^T \; \psi(\vec{x},t) \; e^{-i E_n t/\hbar} \; dt.
        \label{eq:spectrum}
    \end{equation}
    The integral in Eq.~\ref{eq:spectrum} should be carried out, in principle, for infinite time, but for practical purposes, it is done for a large interval of time $T$.
\end{enumerate}

Therefore, the problem is reduced to propagating initial wavepackets with time by solving the time-dependent Schrödinger equation. Notice that the symmetry of the initial wavepacket will determine the set of eigenfunctions that will be obtained with this method. That is, if the initial wavepacket in Eq.~\ref{eq:initial_wavepacket} has a certain symmetry (for example around the $x$ axis), the resulting wavefunctions will also have this same symmetry. If one wants to obtain a different set of eigenfunctions, the previous steps need to be repeated with a different $\psi_0(\vec{x})$. 

In this thesis, we propose to use the RC algorithm to propagate the initial wavefunction in time under the effect of a certain Hamiltonian. Thus, we aim to develop a RC model that can be trained with the time evolution of the wavepacket until time $t_\text{train}$ and then used to further integrate the Schrödinger equation until time $T = t_\text{train} + t_\text{test}$.

To do so, the RC method needs to be adapted to propagate wavefunctions, which are complex-valued high-dimensional matrices. That is, the wavefunction $\psi(\vec{x},t)$ is a complex-valued function of the spatial coordinates $\vec{x}$, which are usually multi-dimensional. 
In the RC framework, $\psi(\vec{x},t)$ is represented as a set of matrices $\{\psi(\vec{x},t)\}_t$, where each matrix contains the values of $\psi(\vec{x})$ at time $t$ in a grid of points spanning
$\vec{x}$. Because of the computationally cheap training process of RC, we expect this algorithm to be significantly more efficient than traditional numerical integration methods and NNs such as LSTMs. The use of  $\psi(\vec{x},t)$ as the target of the reservoir presents two challenges in the RC original formulation, which are described in the following subsection.
 
\subsection{Reservoir computing challenges }
\subsubsection{Complex numbers}
The RC framework is usually done with real numbers. 
That is, all the inputs $\vec{u}(t)$, outputs $\vec{y}(t)$, internal states $\vec{x}(t)$, and weight matrices $W,
W^\text{in}, W^\text{back}$ and $W^\text{out}$ are real-valued. 
However, the target time series in our quantum problem is a wavefunction $\psi(\vec{x},t)$ which, in general, takes complex values. 
Therefore, we need to generalize the update of the internal states and the learning algorithm to complex-valued data.

\subsubsection{High dimensional data}
In quantum scenarios, the target data is represented as a matrix, where each entry is the value of the wavefunction in a discretized spatial grid. The size of this matrix increases exponentially 
with the dimension of the physical system, that is, the dimension of $\vec{x}$. In the usual RC setting, we use a large reservoir $W$, compared to the dimension of the input. However, in quantum problems, the target can have a very high dimension, and thus it is not feasible to design a reservoir much larger than the target size. 
Moreover, the bigger the reservoir, the more data are needed to train the linear model. If the complexity of the method is too large, the linear model may overfit the training data. 
Therefore, we need to adapt the algorithm to avoid this overfitting when dealing with high-dimensional data.

\subsection{Reservoir computing advances for quantum data}
This subsection presents the modifications done in the standard RC framework in order to adapt it to the quantum setting.

\subsubsection{Complex-valued ridge regression}
\label{sec:complex}
We begin by presenting the extension to complex-valued arrays. 
The update of the internal state consists of matrix multiplications and an application of a non-linear function. Recalling Eq.~\ref{eq:update_x_leaking_train}:
\begin{eqnarray*}
     \vec{\tilde{x}}(t) & = &f[(W^{\text{in}} \vec{u}(t)+W\vec{x}(t-1)+W^{\text{back}}\vec{y}_{\text{teach}}(t-1)],\nonumber\\
     \vec{x}(t) & = & (1-\alpha)\;\vec{x}(t-1) + \alpha \vec{\tilde{x}}(t).
     \label{eq:internal_states}
\end{eqnarray*}
As long as the function $f$ is defined for complex numbers, 
the previous equation holds for complex-valued arrays. 
In this work, we propose the following activation function:
\begin{equation}
    f(x) = \tanh(\Re(x)) + i \; \tanh(\Im(x)).
\end{equation}
Also, we set $f^\text{out} = id$, where $id$ is the identity function. 
Once the internal states have been calculated, the matrix $W^{\text{out}}$ is calculated 
by minimizing the MSE with $L^2$ regularization. 
This corresponds to performing a complex-valued ridge regression
\begin{eqnarray}
    MSE_r(\vec{y}, \vec{y}_{\text{teach}}) = &\frac{1}{T - t_{\text{min}}} \displaystyle\sum_{t=t_{\min}}^T \Big|f_{\text{out}}^{-1}(\vec{y}_{\text{teach}}(t))   - W^{\text{out}} \vec{x}(t)\Big|^2 + \gamma |W^{\text{out}}|^2,
\end{eqnarray}
where $|\cdot|$ denotes the $L^2$ norm in complex values. 
In matrix form this equation becomes
\begin{eqnarray}
    MSE_r(\vec{y}, \vec{y}_\text{teach}) = &(f_\text{out}^{-1}(\vec{y}_\text{teach}) - W^\text{out} X)^* \times (f_\text{out}^{-1}(\vec{y}_\text{teach}) - W^\text{out} X) + \gamma |W^\text{out}|^2,
\end{eqnarray}
where $^*$ denotes the conjugate transpose, and $X$ is the matrix containing the internal states. 
In the case of real values, the conjugate transpose is just the transpose. 
This linear model has a closed solution
\begin{equation}
    W^\text{out} = \Big(X^* X + \gamma \mathbb{I}\Big)^{-1} \times \Big(X^*f_\text{out}^{-1}(\vec{y}_\text{teach})\Big),
\end{equation}
and therefore, an analytical solution for the linear model with complex-valued data can be computed. 

\subsubsection{Multiple-step training}
\label{sec:Multi-step}

\begin{figure}[!ht]
\centering
\includegraphics[width=0.8\textwidth]{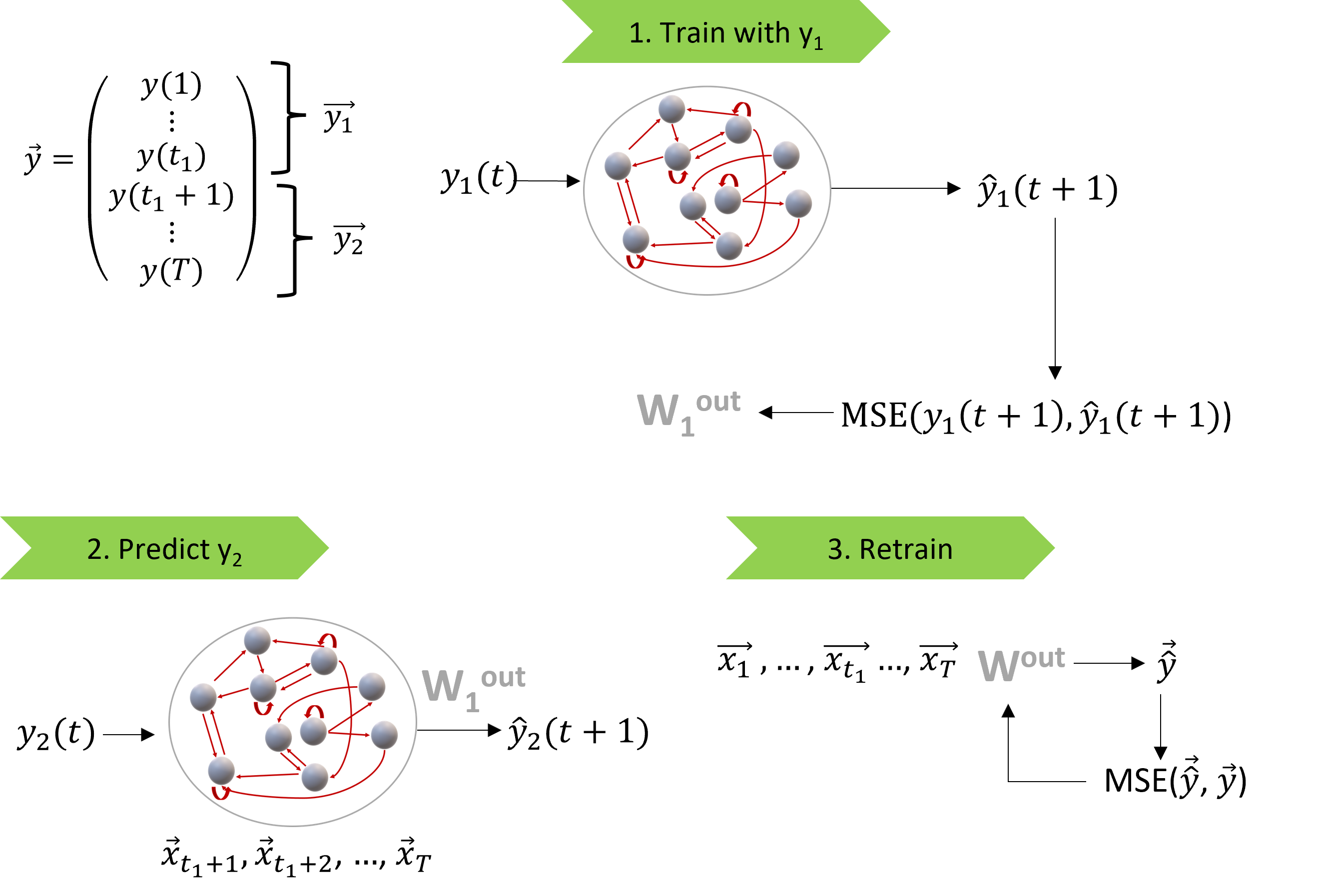}
\caption{\label{fig:MultiRC}Architecture of multi-step training of the reservoir computing model.}
\end{figure}

The second and main challenge in dealing with our quantum wavefunctions is the high dimensionality of the target matrices. 
To capture all the underlying information of the dynamical system, the size of the reservoir $W$ should be at least of the order of magnitude of the target data. According to Hoeffding's theorem~\citep{MLBook}, models with high complexity require a large amount 
of data to reduce the variance of the model predictions, following the (Hoeffding's) inequality 
\begin{equation}
    E_{out}\leq E_{in} + \mathcal{O}\left(\sqrt{\frac{C}{N}}\right),
    \label{Hoeffding}
\end{equation}
 where $E_{out}$ is the error in the test set, $E_{in}$ is the error in the training set, $C$ is an indicator of complexity of the ML model and $N$ is the number of data samples. In our case, the size of the reservoir is similar to the number of data samples, and thus the method is likely to produce overfitting. To reduce this effect, we propose the use of a multi-step learning training. We call this method \emph{multi-setp RC} as opposed to the original RC formulation, which will be referred to here as \emph{standard RC}. The training steps of the multi-step RC are depicted schematically in Fig.~\ref{fig:MultiRC}, and described as: 
 
 \begin{enumerate}
     \item Split the training input and output into two parts. 
     \begin{eqnarray}
         \vec{y} = & 
         \begin{pmatrix}
         \vec{y}(1) \\
         \vdots\\
          \vec{y}(t_1)\\
          \vec{y}(t_1+1)\\
         \vdots \\
          \vec{y}(t_\text{train})
         \end{pmatrix} = 
         \begin{pmatrix}
          \vec{y}_1(1)\\
         \vdots\\
          \vec{y}_1(t_1)\\
          \vec{y}_2(t_1 +1)\\
         \vdots\\
          \vec{y}_2(t_\text{train})\\
         \end{pmatrix} = 
         \begin{pmatrix}
         \vec{y_1}\\
         \vec{y_2}\\
         \end{pmatrix}, \nonumber \\
         \vec{u} = &
         \begin{pmatrix}
          \vec{u}(1) \\
         \vdots\\
          \vec{u}(t_1)\\
          \vec{u}(t_1+1)\\
         \vdots \\
          \vec{u}(t_\text{train})
         \end{pmatrix} = 
         \begin{pmatrix}
          \vec{u}_1(1)\\
         \vdots\\
          \vec{u}_1(t_1)\\
          \vec{u}_2(t_1 +1)\\
         \vdots\\
          \vec{u}_2(t_\text{train})\\
         \end{pmatrix} = 
         \begin{pmatrix}
         \vec{u_1}\\
         \vec{u_2}\\
         \end{pmatrix}.
     \end{eqnarray}
     \item Find the internal states of the reservoir $\vec{x}(1), \cdots ,\vec{x}(t_1)$ with $\vec{y_1}$ and 
     $\vec{u_1}$, and perform a ridge regression to find $W^\text{out}$.
     \item Evolve the internal states of the reservoir $\vec{x}(t_1 + 1), \ldots, \vec{x}(t_\text{train})$ by letting the 
     reservoir predict the values of $\vec{y_2}$.
     \item Using all the internal states $\vec{x}(1), \cdots, \vec{x}(t_1),\vec{x}(t_1+1), \ldots \vec{x}(t_\text{train})$ retrain the linear model to correct $W^\text{out}$. 
     That is, perform again a ridge regression to find the \emph{new} $W^\text{out}$ using the \emph{old} values of $W^\text{out}$ as the initial guess for the coefficients of the linear model.
 \end{enumerate}
When the linear model overfits the training data, small changes in the internal state evolution due to prediction errors produce high errors in the future prediction of the wavefunctions. Therefore, the errors tend to propagate fast until the reservoir is not able to correctly propagate the quantum state. 
In our proposed method, we allow the reservoir to evolve after training by predicting a part of the training data. This simulates the real situation found during testing. During this predicting phase, the internal states are modified using the \textit{predicted} output instead of the \textit{exact} ones (see Eq.~\ref{eq:update_x_leaking}). The predicted outputs have some errors, which are transmitted to the internal states, which are then used to retrain the linear model. This learning strategy shows the reservoir how small predicting errors modify the internal states during the test phase. In fact, in Sect.~\ref{sect_training_improvements}, we showed that adding white noise to the target value of the reservoir $\vec{y}_\text{teach}$ could help improve the performance of RC. In the multi-step RC, instead of adding white noise, the internal states in step 2 are modified using the real noise that will be present in the prediction step. This effect reduces overfitting, decreasing the test error significantly. In the following sections, we will compare the performance of the multi-step RC with the performance of the standard RC model. 

In addition, a standard RNN, i.e. LSTM (see Sect.~\ref{sect:RNN}) is also used as a benchmark model. In particular, a LSTM with 1024 neurons has been trained for 1000 epochs, with the same training and test data as the other RC models. The training data is generated by numerically integrating the Schrödinger equation. For this purpose, the widespread Fast Fourier Transform (\gls{FFT}) method proposed by  Kosloff and Kosloff~\citep{kosloff} is used. The solutions of the Schrödinger equation generated by the FFT method will be considered to be the \emph{exact} ones, and will be compared with the solutions generated by the RC methods and the LSTM.

\subsection{Studied quantum systems}
Now we present the different quantum systems used for this study, which include three 1D quantum systems and one 2D system. We use 5000 time steps both for training and for testing. 
The spatial domains are chosen so that they contain the wavefunctions at all times $t$. The systems under study are the following:

\begin{enumerate}
    \item \textbf{Harmonic oscillator}. In this case, the Hamiltonian $\hat{H}$ is
    \begin{equation}
        H(x) = \frac{1}{2}\frac{\partial^2}{\partial x^2} +  \frac{1}{2} x^2.
        \label{eq:HOsystem}
    \end{equation}
    The (computationally effective) spatial domain is $x \in [-10,10]$ covered with a grid of 200 equidistant points. 
    That is, $\psi(x,t)$, for a fixed $t$, is a vector of size 200. 
    The time interval between time steps is $\Delta t=0.002$.
    The initial wavefunction $\psi_0(x)$ is the minimum uncertainty Gaussian wave packet (see Eq.~\ref{eq:initial_wavepacket}), centered at $x_0=0$, with a width of $\displaystyle\Delta x = \frac{1}{2\sqrt{2}}$, 
    a phase/momentum $p_0 = 3.35$, which gives a (mean) energy $E=6$.

    \item \textbf{Morse potential}. The Morse Hamiltonian, presented in Sect.~\ref{sect:Morse_Hamiltonian_paper1}, adequately represents the vibrational interaction of a diatomic molecule. Recall that we write the corresponding Hamiltonian $\hat{H}$ as
    \begin{equation}
        H(x) = \frac{1}{2}\frac{\partial^2}{\partial x^2} +  D_e\left(e^{-2a(x-x_0)} - 2 e^{-a(x-x_0)}\right).
        \label{eq:Morse}
    \end{equation}
    We choose $D_e=7$, $a=0.09$ and $x_0=0$. 
    The spatial domain is $x \in [-10,25]$ with 140 points. 
    The time interval between time steps is $\Delta t=0.007$. 
    The initial wavefunction $\psi_0(x)$ is again the minimum uncertainty packet 
    (Eq.~\ref{eq:initial_wavepacket}) centered at $x_0=0$, with a phase characterize by $p_0 = 2$, giving an energy $E=-4.8$.

    \item \textbf{Polynomial potential} The last quantum system considered corresponds to a 
    quartic polynomial potential
    \begin{equation}
        H(x) = \frac{1}{2}\frac{\partial^2}{\partial x^2} +  \alpha_0 + \alpha_1x + \alpha_2x^2 + \alpha_3x^3 + \alpha_4x^4,
        \label{eq:polynomial}
    \end{equation}
    where $\alpha_0 = -0.5, \alpha_1 = 0.14, \alpha_2 = 0.09, \alpha_3 = -0.01,$ and $\alpha_4 = 0.001$. 
    The spatial domain here is $x \in [-15,25]$ covered with 150 points. 
    The time interval between time steps is $\Delta t=0.007$. 
    The initial wavefunction $\psi(x,0)$ is again the minimum uncertainty packet 
    (Eq.~\ref{eq:initial_wavepacket}) centered at $x_0=0$, with a phase parameter $p_0 = 2$, giving an energy $E=1.8$

\end{enumerate}
Apart from the 1D systems we also study one 2D system, consisting of a 2D harmonic oscillator potential, whose Hamiltonian is given by
\begin{equation}
    H(x,y) = \frac{1}{2}\left(\frac{\partial^2}{\partial x^2}+\frac{\partial^2}{\partial y^2}\right)+  
    \frac{\omega_x^2}{2} (x - x_0)^2 + \frac{\omega_y^2}{2} (y - y_0)^2. 
\end{equation}
We choose $\omega_x^2 = 1,$ $\omega_y^2=2$ and $x_0=y_0=0$. 
Notice that this choice of frequencies prevents the appearance of resonances in the dynamics, and then degeneracies in the eigenenergies.
The initial wavefunction $\psi_0(x,y)$ is the minimum uncertainty Gaussian wave packet with energy 
$E=3.2$,
\begin{equation}
    \psi_0(x,y) = \left(\frac{1}{\pi}\right)^{1/8} e^{-\frac{(x-x_0)^2}{(4 \Delta x)^2} -\frac{(y-y_0)^2}{(4 \Delta y)^2}}\;
    e^{i\,p_{0}(x - y)},
    \label{eq:2DHO}
\end{equation}
which is centered at $x_0=y_0=0$, with widths of $\Delta x = \Delta y=0.9$, and a phase $p_0 = \pm 1.75$.
The spatial domain is $(x,y) \in [-5,5] \times [-5,5]$ with 50 points in each dimension. 
Therefore, $\psi(x,y,t)$ for each $t$ is a matrix of size $50 \times 50$. 
Notice that the dimension of the data is of the order of the training data. 
Therefore, the RC model is very likely to produce overfitting, 
which motivates the multi-step learning we propose.

Table~\ref{tab:params} shows the training parameters, as defined in Sect.~\ref{sect:RC} and Sect.~\ref{sect_training_improvements}, used for each of the quantum systems considered in this work. The same parameters are used both for the standard RC and multi-step RC models. We follow the criteria given in Ref.~\citep{TrainingParams} to choose the appropriate hyperparameters. For example, the leaking rate $\alpha$, which determines the velocity of the dynamics of the internal states, is chosen so that the internal states evolve with the same velocity as the input-output dynamics. For the multi-step RC, we used 85\% of the training data for the first training step and 15\% for the second step. 
\begin{table}[!t]
    \centering
    \begin{tabular}{l|ccccccc}
    \hline \hline \\[-0.3cm]
        Quantum system & $\rho(W)$ & $\alpha$ & $t_{\text{min}}$ & $\gamma$ & $N$ & $W$ density   \\
         \hline  \\[-0.2cm]
         Harmonic oscillator 1D & 1.10 & 0.015 & 300 & 0.1 & 2500 & 0.015 \\ 
         Morse potential & 0.75 & 0.015 & 500 & 0.5 & 1500 & 0.015 \\
         Polynomial potential & 0.75 & 0.017 & 50 & 0.05 & 1500 & 0.008 \\
         Harmonic oscillator 2D & 1.10 & 0.06 & 300 & 0.5 & 2000 & 0.015\\
         \hline 
    \end{tabular}
    \caption{Reservoir computing training parameters, as defined in Sect.~\ref{sect:RC}, for the four 
    studied quantum systems studied in this work.}
    \label{tab:params}
\end{table}

\subsection{Results}
\begin{table}
    \centering
    \begin{tabular}{l|lcc}
    \hline \hline
         Quantum system & Model & MSE wavefunctions & MSE energies  \\
         \hline \\[-0.2cm]
         Harmonic oscillator 1D & Standard RC   &  $4 \times 10^{-5}$ & $ 3 \times 10^{-4}$\\
                                & Multi-step RC &  $3 \times 10^{-5}$ &  $ 1 \times 10^{-5}$\\
                                & LSTM &  $4 \times 10^{-4}$ &  $2 \times 10^{-2}$\\
         Morse                  & Standard RC   &  $7 \times 10^{-5}$ & $ 4 \times 10^{-5}$\\
                                & Multi-step RC &  $2 \times 10^{-5}$ & $2 \times 10^{-5}$\\
                                & LSTM &  $4 \times 10^{-2}$ & $3 \times 10^{-3}$ \\
         Polynomial             & Standard RC   &  $2 \times 10^{-4}$ & $ 3 \times 10^{-5}$\\
                                & Multi-step RC &  $1 \times 10^{-4}$ & $ 8 \times 10^{-6}$\\
                                & LSTM &  $2 \times 10^{-2}$ &  $5 \times 10^{-3}$\\
         Harmonic oscillator 2D & Standard RC   &  $1 \times 10^{-4}$ & $ 4 \times 10^{-3}$\\
                                & Multi-step RC &  $2 \times 10^{-5}$ & $8 \times 10^{-5}$\\
                                & LSTM &  $2 \times 10^{-3}$ &  $2 \times 10^{-2}$\\
         \hline 
    \end{tabular}
    \caption[Mean squared error for the wavefunctions $\psi(\vec{x},t)$ and eigenenergies for four quantum systems.]{(Averaged) mean squared error for the wavefunctions $\psi(\vec{x},t)$ (see Eq.~\ref{eq:MSEpsi_t}) and eigenenergies for the four quantum systems considered in this work using the standard reservoir computing model and the multi-step RC model. The table also contains the mean squared error for the long-short-term memory network, our benchmark model.}
    \label{tab:mse}
\end{table}

\begin{table}[!ht]
    \centering
    \begin{tabular}{l|lc}
    \hline \hline
        Quantum system & Model & Execution time (min) \\
        \hline\\[-0.2cm]
         Harmonic oscillator 1D & Standard RC   &  7\\
                                & Multi-step RC &  10 \\
                                & LSTM &  51 \\
                                & FFT &  44 \\
         Morse  & Standard RC   &  6\\
                                & Multi-step RC &  7 \\
                                & LSTM &  40 \\
                                & FFT &  51 \\
         Polynomial  & Standard RC   &  4\\
                                & Multi-step RC &  6 \\
                                & LSTM &  36 \\
                                & FFT  &  48 \\
         Harmonic Oscillator 2D  & Standard RC   &  10\\
                                & Multi-step RC &  15 \\
                                & LSTM &  198 \\
                                & FFT  &  66 \\
    \end{tabular}
    \caption[Training times for the four quantum systems considered in this work. ]{Training times for the four quantum systems considered in this work. Standard reservoir computing and multi-step reservoir computing are the two models studied in this work. The long-short-term memory network (LSTM) and Fast Fourier Transform (FFT)~\citep{kosloff} are the two benchmark models: a standard deep learning model and a standard numerical solver respectively.}
    \label{tab:time_RC}
\end{table}

Table~\ref{tab:mse} shows the MSE obtained for the wavefunctions $\psi(\vec{x},t)$, averaged over the time period $[0,T]$, given by
\begin{equation}
  MSE[\psi(\vec{x},t)] = \frac{1}{T |\vec{x}|}\sum_t\sum_{\vec{x}} |\psi(\vec{x}, t)-\tilde{\psi}(\vec{x},t)|^2,
  \label{eq:MSEpsi_t}
\end{equation}
where $\tilde{\psi}(\vec{x}, t)$ is the predicted wavefunction obtained with the RC algorithm, $\psi(\vec{x},t)$ is the exact wavefunction obtained with the FFT method, and $|\vec{x}|$ is the dimension of the spatial domain. 

Table~\ref{tab:mse} also shows the MSE of the corresponding (mean) energies, of the different systems chosen to study. 
These results serve as a rigorous assessment of the performance of the method, since the quantum properties are evaluated in all space using $\psi^*\psi$ as the probability density. Therefore, only the regions where the modulus of the wavefunction is high are important for most purposes, i.e.~ for the computation of observable mean values.

We see that, for all cases, the MSE of both the wavefunctions and energies is smaller when using the multi-step RC model. 
The largest difference in performance appears in the 2D harmonic oscillator, where the MSE is one order of magnitude smaller 
for the wavefunctions and two orders of magnitude smaller for the energies. 
Therefore, only the multi-step learning RC algorithm can correctly recover the eigenfunctions and eigenenergies of such quantum system.

Notice that the training data for the 2D harmonic oscillator consists of matrices of size $50 \times 50$ (2500 entries), while the 1D data are vectors of size 100 to 200. Therefore, it is harder to train the RC model on 2D data, since the reservoir size is of the same order of magnitude as the amount of training data. 
Thus, the standard RC method overfits the training data and does not generalize well to the test data. On the contrary, the multi-step learning strategy refits the linear readout $W^\text{out}$ by showing the reservoir how to adapt to unseen test data. This prevents fast error propagation during the test phase. 

Table~\ref{tab:mse} also shows the MSE of the standard LSTM for both wavefunctions and energies. We see that the MSE of the LSTM is much larger than the MSE of the RC-based models (one to three orders of magnitude). An explanation for this fact is that the RC models only need to train the last layer of the network, and thus have many less trainable parameters than the LSTM. For example, for the 1D harmonic oscillator, the LSTM has more than 12 million trainable parameters, while the RC has 500,000 trainable parameters. For this reason, the LSTM takes longer to train and tends to overfit the training data, thus leading to large generalisation errors.

\begin{figure*}
\includegraphics[width=1.05\textwidth]{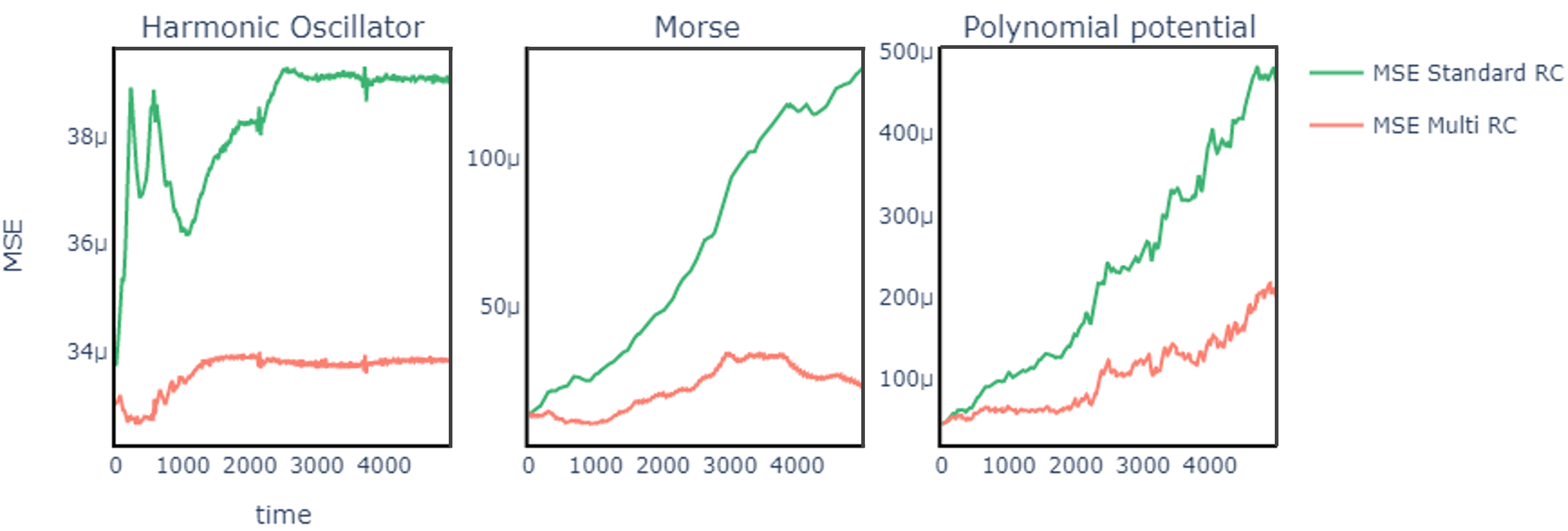}
\caption{Time evolution of the mean squared error for the harmonic, Morse, and polynomial 
oscillator wavefunctions computed with the standard reservoir computing and multi-step reservoir computing methods.}
\label{fig:mse1D}
\end{figure*}

\begin{figure}[b!]
\centering
\includegraphics[width=0.7\columnwidth]{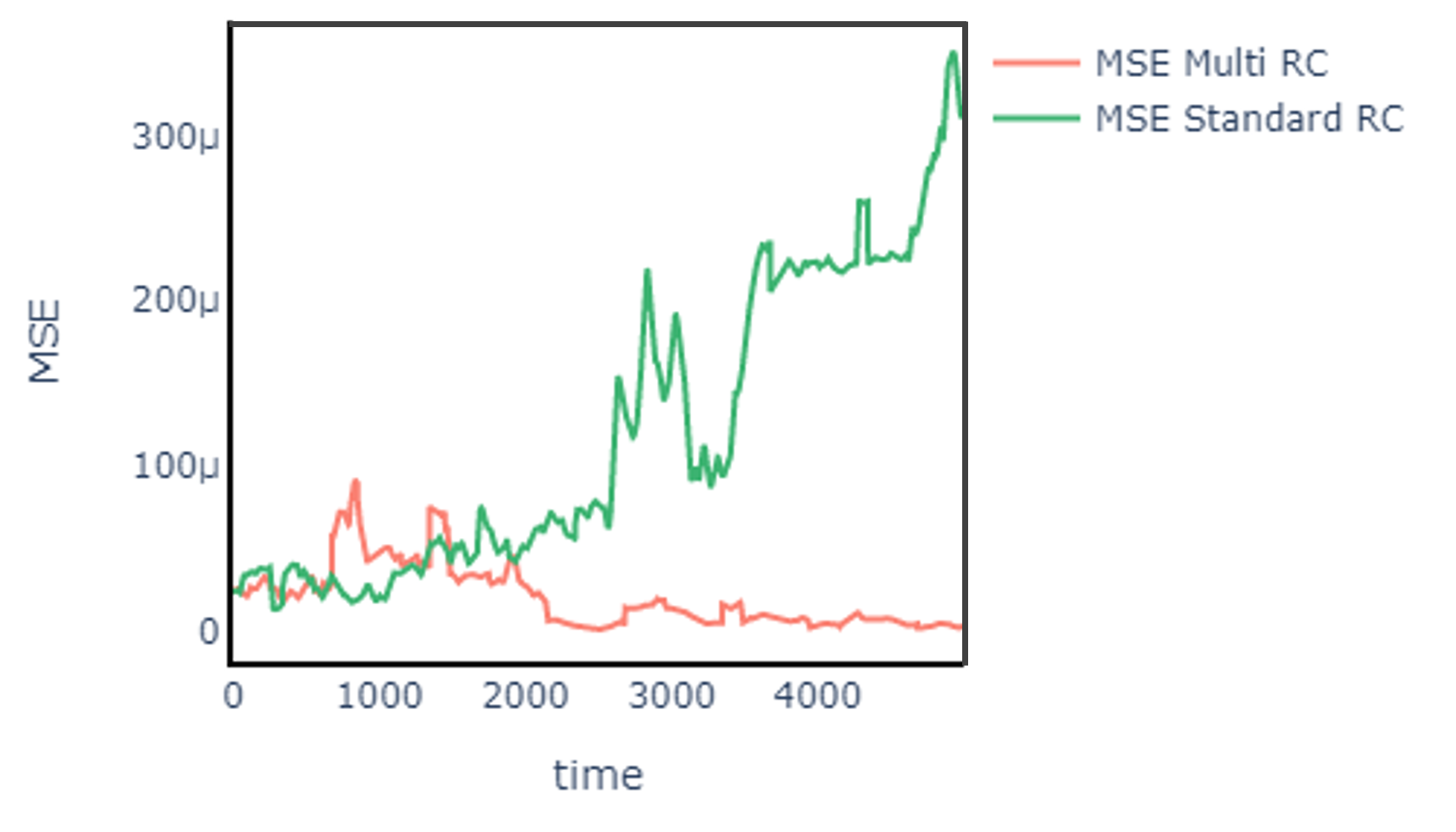}
\caption{Time evolution of the mean squared error of the 2D harmonic oscillator wavefunctions 
for the standard reservoir computing and the multi-step reservoir computing.}
\label{fig:mse2D}
\end{figure}

\begin{table}[t!]
    \centering
    \begin{tabular}{l|l}
    \hline \hline \\[-0.3cm]
        Quantum system & Predicted eigenenergies \\         \hline \\[-0.2cm]
        Harmonic oscillator 1D & 2.5, 3.5, 4.5, 5.5, 6.5, 7.5, 8.5, 9.5 \\[0.2cm]
        Morse & -6.8326, -6.5040, -6.1834, -5.8710, -5.5666, \\
              & -5.2704, -4.9822, -4.7022, -4.4302, -4.1664, \\
              & -3.9106, -3.6630, -3.4234, -3.1920  \\[0.2cm]
        Polynomial potential & 0.1556, 0.6187, 1.0800, 1.5426, 2.0087, \\
                             & 2.4801, 2.9582, 3.4453 \\[0.2cm]
        Harmonic oscillator 2D & 3.0946, 3.2838, 3.4730  \\
        \hline
    \end{tabular}
    \caption{Eigenenergies obtained from the spectra calculated from the predicted wavefunctions $\psi(\vec{x},t)$ 
    computed using the multi-step reservoir computing, for all four studied quantum systems.}
    \label{tab:energies}
\end{table}

\begin{figure*}
\includegraphics[width=1.0\textwidth]{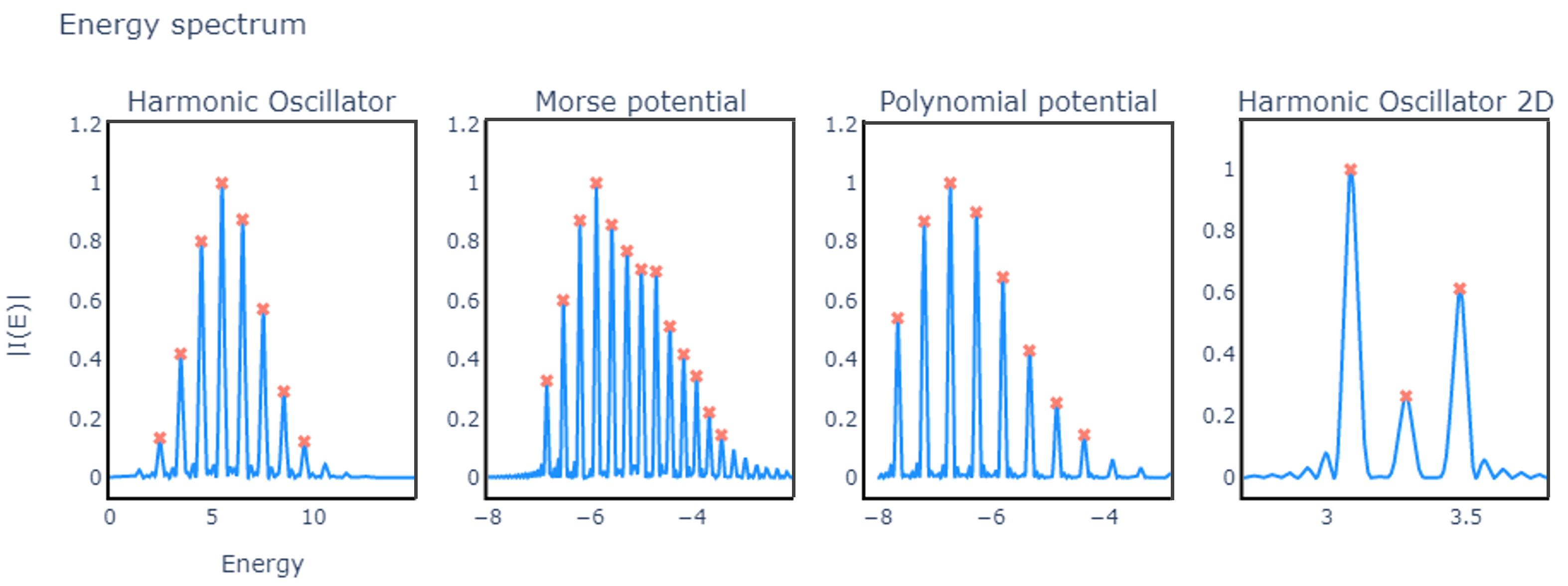}
\caption[Energy spectra obtained using the predicted wavefunctions from the multi-step RC, for all four studied quantum systems.]{Energy spectra obtained using the predicted wavefunctions from the multi-step RC, for all four studied quantum systems. 
The orange peaks indicate the eigenenergies of each quantum system. 
Moreover, the numerical values of the eigenenergies are given in Table~\ref{tab:energies}.}
\label{fig:spectrum}
\end{figure*}

In Figs.~\ref{fig:mse1D} and~\ref{fig:mse2D}, we present the results of Table~\ref{tab:mse} for the RC-based models in a graphic way, 
by plotting the MSE of the predicted wavefunctions $\psi(\vec{x},t)$ as a function of time. We see that due to error propagation, the MSE tends to increase with time, aside from some random fluctuations. For all the studied systems, the MSE increases slower with time when using the multi-step RC.  Moreover, we see that for the 1D and 2D harmonic oscillators and the Morse Hamiltonian, the MSE from the multi-step RC seems to stabilize, while the MSE of the standard RC keeps increasing. Therefore, for a fixed error tolerance, the multi-step RC method allows us to predict a longer time evolution than the standard RC. 

Table~\ref{tab:time_RC} shows the execution times for the three ML models studied in this work: the standard RC, the multi-step RC and the LSTM. The table also shows the execution time of a standard PDE solver, the FFT method. We see that the multi-step RC is slightly slower than the standard RC, but still much faster than both the LSTM and the FFT.

\begin{figure}[!ht]
    \includegraphics[width=1.0\textwidth]{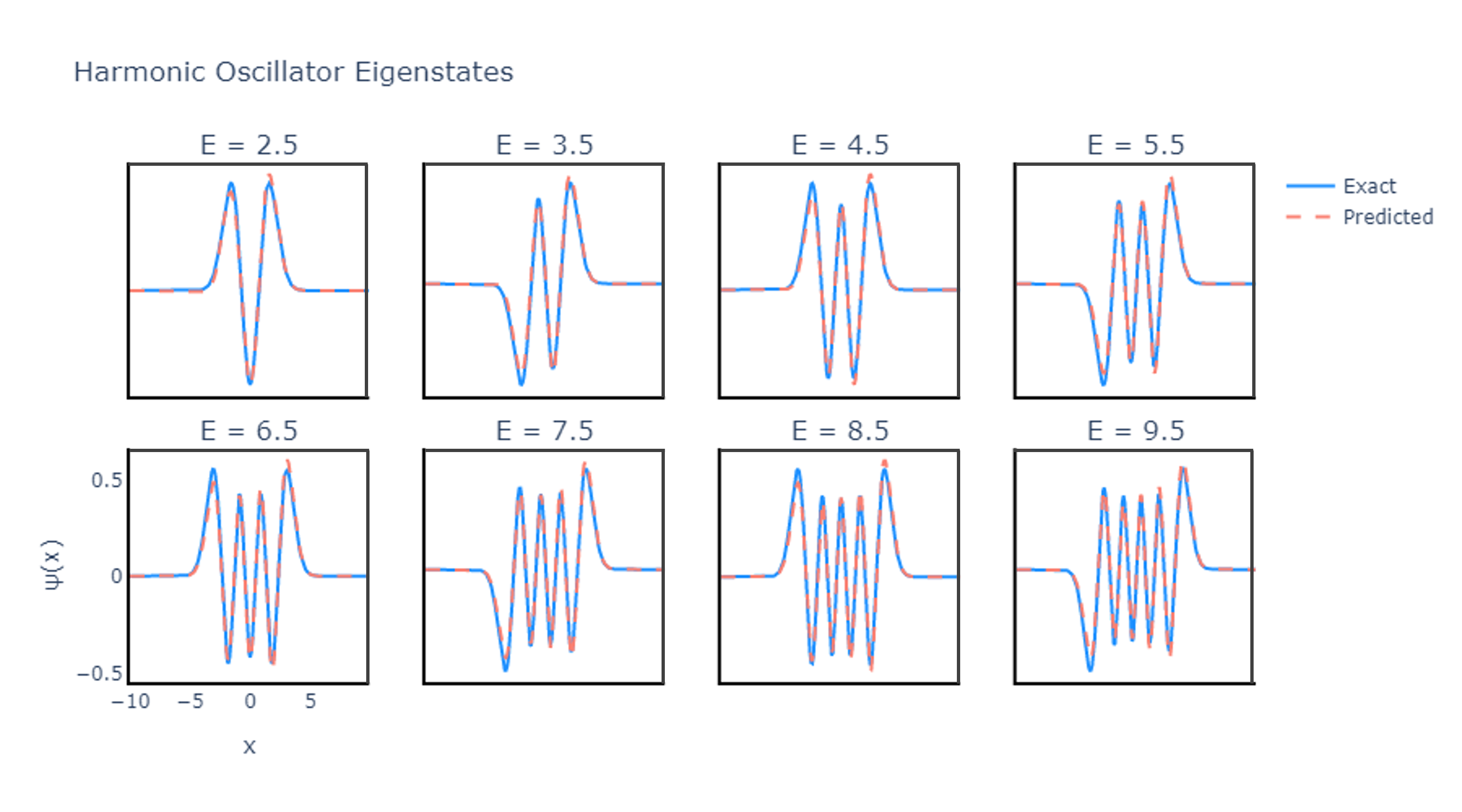}
    \caption{Exact and predicted eigenfunctions for the harmonic oscillator system (Eq.~\ref{eq:HOsystem}).}
    \label{fig:wavesHO}
\end{figure}

\begin{figure}[!ht]
    \includegraphics[width=1.05\textwidth]{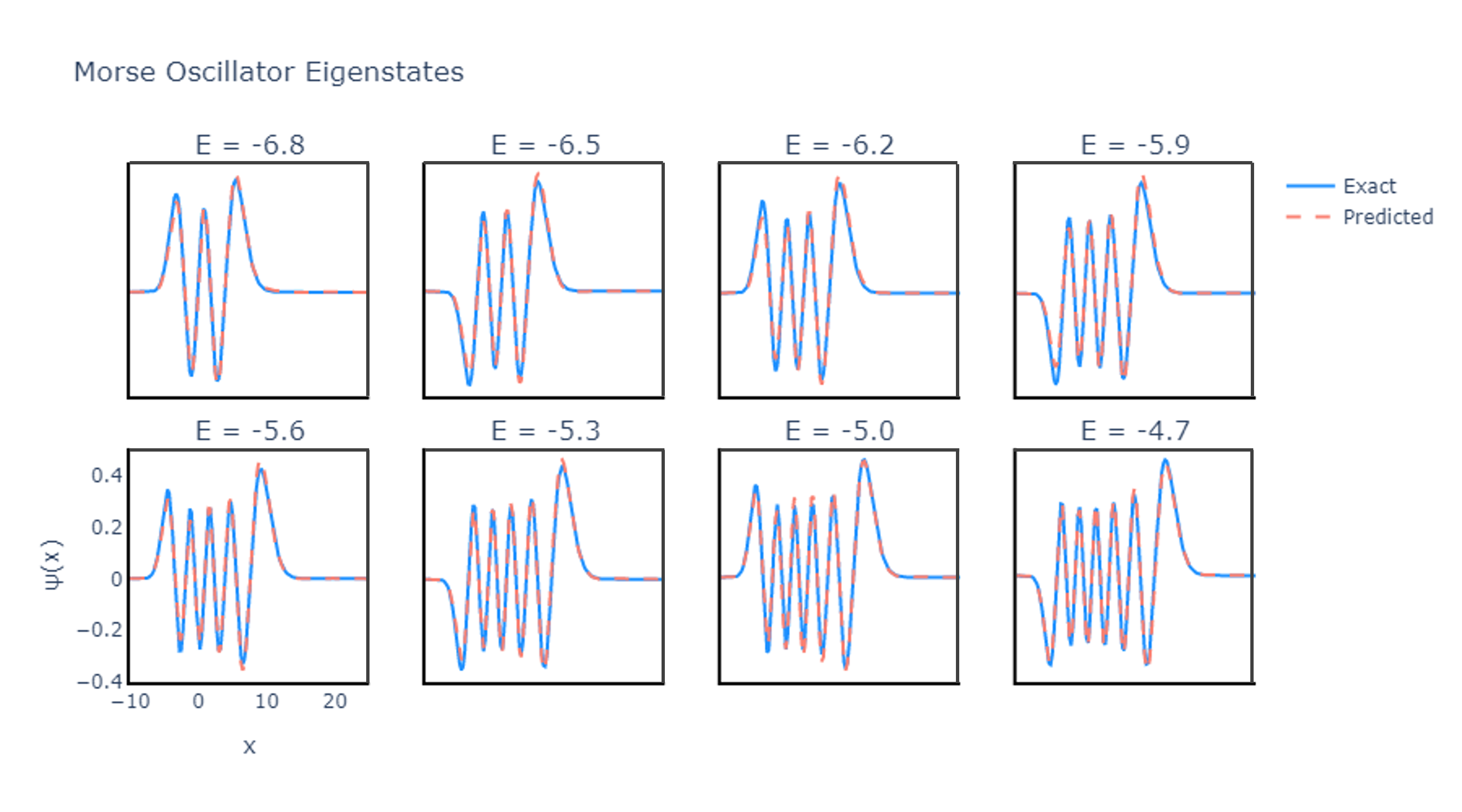}
    \caption{Same as Fig.~\ref{fig:wavesHO} for the Morse Hamiltonian system (Eq.~\ref{eq:Morse}).}
    \label{fig:wavesMorse}
\end{figure}

\begin{figure}[!ht]
    \includegraphics[width=1.0\textwidth]{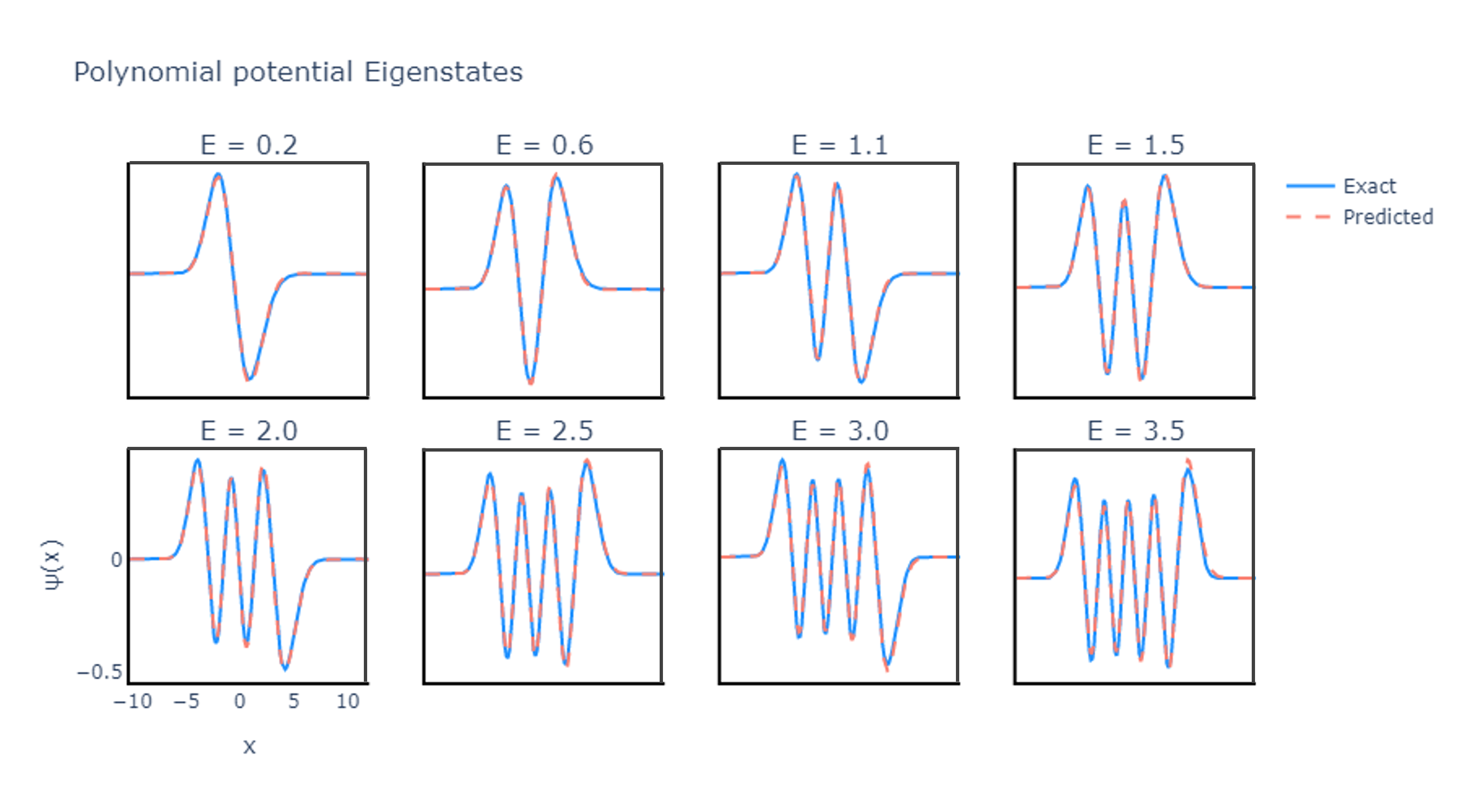}
    \caption{Same as Fig.~\ref{fig:wavesHO} for the polynomial Hamiltonian system (Eq.~\ref{eq:polynomial}).}
    \label{fig:wavesPoly}
\end{figure}

\begin{figure}[!ht]
\includegraphics[width=1.0\textwidth]{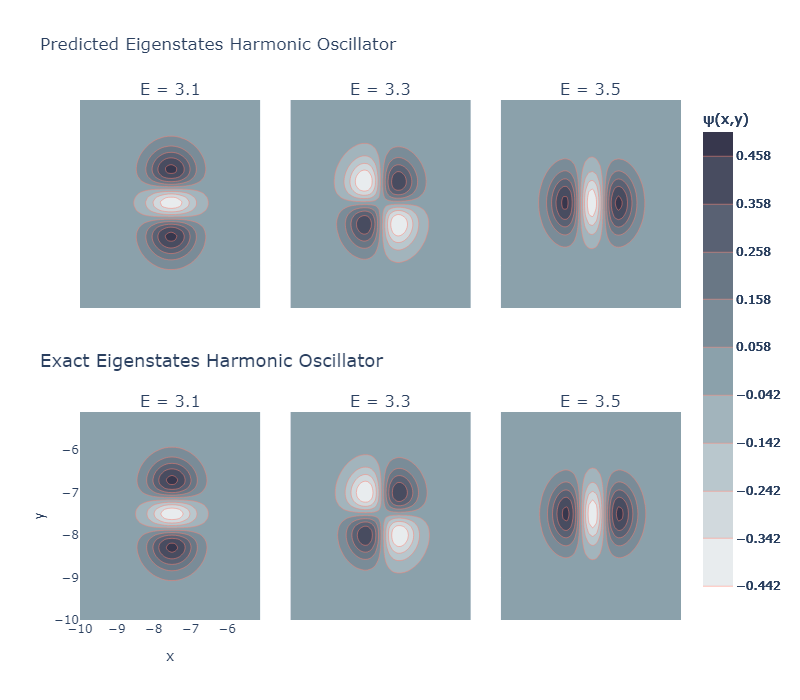}
\caption{Predicted (top row) and exact (bottom row) wavefunctions for the 2D harmonic oscillator system of
   Eq.~\ref{eq:2DHO}.}
\label{fig:wavesHO2D_pred}   
\end{figure}
After studying the performance of the RC models when propagating the wavefunctions in time, we can recover the eigenenergies and eigenfunctions around a certain energy. Figure~\ref{fig:spectrum} shows the spectra for the different studied systems obtained using the predicted $\psi(\vec{x},t)$ with the multi-step RC model. 
The peaks of the spectrum appear at the eigenenergies around the mean energy of the initial wavepacket $\psi_0(\vec{x})$. For example, the initial state of the 1D harmonic oscillator is a minimum uncertainty Gaussian wavepacket 
(see Eq.~(\ref{eq:initial_wavepacket})) with mean energy $E=6$.  
Accordingly, the obtained eigenenergies go from $E=2.5$ to $E=9.5$, which are centered around the initial energy $E=6$.  The values of all the eigenenergies are given in Table~\ref{tab:energies}.

The results show that integrating the time-dependent Schrödinger equation allows recovering the eigenstates around certain energy, 
without needing to compute all the eigenstates with lower energy, as happens in the usual variational method. 
Using the eigenenergies, we recover the associated eigenfunctions by computing the Fourier transform of the wavefunction
$\psi(\vec{x},t)$. 
Figures~\ref{fig:wavesHO},~\ref{fig:wavesMorse}, and~\ref{fig:wavesPoly} show the predicted compared to the exact eigenfunctions 
for the 1D harmonic oscillator, the Morse Hamiltonian and the polynomial potential systems, and
Fig.~\ref{fig:wavesHO2D_pred} 
shows the predicted and exact eigenfunctions for the 2D harmonic oscillator.

The exact eigenstates are calculated analytically for the harmonic oscillator (both 1D and 2D) and the Morse potential, 
and numerically (using the variational method in appendix~\ref{appendix1}) for the polynomial potential. 
We see that in all cases the predicted eigenfunctions are in very good agreement with the exact ones.  
This fact confirms that the multi-step RC method can correctly propagate a quantum wavepacket with time.

It is worth mentioning that the RC models do not impose the energy conservation of the solution of the Schrödinger equation, which is a fundamental property of any quantum system. However, during the training phase, the reservoir learns to reproduce the input-output dynamics, which follows this property. Since the predictions of the eigenenergies and eigenfunctions are accurate, the reservoir correctly imposes the energy conservation property during prediction, without being explicitly programmed to do so. This is a further confirmation of the accuracy of the method.

All in all, the multi-step RC method effectively solves the time-dependent Schrödinger equation, yielding accurate eigenenergies and eigenfunctions. 
This is achieved by preventing overfitting when working with high-dimensional 
data thus allowing the use of the adapted RC method for time propagation in quantum systems.

%%%%%%%%%%%%%%%%%%%%%%%%%%%%%%%%%%%%%%%%%%%%%%%%%%%%%%%%%%
% RC MORSE HAMILTONIAN
%%%%%%%%%%%%%%%%%%%%%%%%%%%%%%%%%%%%%%%%%%%%%%%%%%%%%%%%%%
\section{Reservoir computing for the Coupled Morse Hamiltonian}
\label{sect:paper3}
The previous section introduced an adaptation of the RC algorithm to integrate the time-dependent Schrödinger equation. This approach enables the efficient calculation of eigenenergies and eigenstates within a specific energy range, without the need to calculate lower-lying eigenstates. As previously stated, compared to traditional methods based on the variational principle, this method is significantly more efficient when interested in excited states. The algorithm was tested with some simple quantum systems, namely the 1D and 2D harmonic oscillator, the (decoupled) Morse oscillator and a polynomial potential. 

After testing the effectiveness of the method, we aim to apply the algorithm to a more complex and realistic quantum system, the kinetically coupled 2D Morse oscillator Hamiltonian, introduced in Sect.~\ref{sect:Coupled_Morse_paper1}. This Hamiltonian has been extensively studied in the past as a model for the stretching vibrations
of the H$_2$O molecule~\citep{Jaffe,H2O2, H2O, H20_NN}. Recall that the Hamiltonian is given by Eq.~\ref{eq:coupled_morse},
\begin{eqnarray*}
H(x_1, x_2, p_1, p_2) & = &\frac{1}{2}(G_{11}p_1^2 + G_{22}p_2^2) + G_{12}p_1p_2 + U_M(x_1) + U_M(x_2), \nonumber \\
U_M(x)                          & = & D_e (e^{-2a(x-x_e)} - 2e^{-a(x-x_e)}),
\end{eqnarray*}
where $x_1$ and $x_2$ are the stretching coordinates and $p_1, p_2$ the corresponding conjugate momenta. In this study, we will consider the parameters of the Hamiltonian that model the stretching of the H$_2$O molecule, which was first introduced in Ref.~\citep{H2O}. In this case, the $G$-matrix elements are equal to
\begin{equation}
G_{11} = G_{22} = \frac{m_H + m_O}{m_H m_O}, \quad G_{12} = \frac{\cos\alpha}{m_O}<0,
\end{equation} 
being $m_H = 1837.3$ a.u.\  and $m_O = 29166.3$ a.u.\ the H and O atomic masses \footnote{Notice that the atomic units used here should not be confused by the atomic units of mass, where $m_H = 1.00784 $ a.u.m and $m_O = 15.999$ a.u.m.}, respectively, and $\alpha$ the bending angle, kept frozen here at its equilibrium value of 104.5$^\circ$. Functions $U_M$ are 1D Morse potentials in the O--H stretching coordinates, characterized by parameters $a=1.1484$ a.u.\ controlling the width of the well, the well depth  $D_e = 0.20276$ a.u.\ and the equilibrium bond distance $x_e=0$. Recall that in Sect.~\ref{sect:Coupled_Morse_paper1}, we trained a NN to predict the eigenfunctions of coupled Morse Hamiltonians. In that case, a change of coordinates was used so that the coupling part appeared in the spatial coordinates instead of the momentum coordinates. This representation was more appropriate for training the NN. In this scenario, however, we do not need to change the coordinates of the original system, since the RC algorithm can learn the input-output dynamics regardless of the Hamiltonian. For this reason, in this study we will consider the original formulation of the Hamiltonian in Eq.~\ref{eq:coupled_morse}.
%Table I
\begin{table}
    \centering
    \begin{tabular}{ccccccc}
        \hline \hline \\[-0.3cm]
        $E_0$ & $x_{1,0}$ & $x_{2,0}$ & $\Delta x_1$ & $\Delta x_2$ & $p_{1,0}$ & $p_{2,0}$ \\
        \hline \\
        -0.384328 & -0.205528 & -0.505528 & 0.1 & 0.1 & 3 & 3 \\
        -0.271096 & -0.005528 & -0.005528 & 0.1 & 0.1 & 10 & 10 \\
        -0.201995 & 0.394472 & 0.394472 & 0.1 & 0.1 & 10 & 10 \\
        -0.108866 & 0.494472 & 0.494472 & 0.1 & 0.1 & 15 & 15 \\
        \hline
        \end{tabular}
    \caption{Parameters of the initial wavefunction of the coupled Morse oscillator, as described in Eq.\ref{eq:initial_wavepacket}, at four different mean energies $E_0$.}
    \label{tab:initial_waves}
    \end{table}

In the following subsections, the eigenfunctions and eigenenergies will be computed for four different initial conditions with increasing energy. The parameters of the corresponding initial wavepackets  $\psi_0(x_1,x_2)$ are reported in Table~\ref{tab:initial_waves}. The resulting eigenfunctions and eigenenergies obtained with the RC method will be compared with those obtained with the standard numerical integration method based on the variational principle (see Appendix~\ref{appendix1}). The eigenergies $\{\tilde{E}_n\}$ and eigenfunctions $\{\tilde{\phi}_n\}$ computed with the variational method will be considered to be the exact ones.

\subsection{Dimensionality reduction}
\label{sect:dim_red}
In the previous work, in Sect.~\ref{sect:paper2}, we showed how to adapt the RC algorithm to integrate the time-dependent Schrödinger equation. One of the biggest challenges when using RC to propagate quantum states is the size of the input matrices given to the reservoir, which represent the quantum wavefunction $\psi(x_1,x_2,t)$ at a given time $t$. The size of these matrices increases substantially with the energy of the physical system since higher precision levels are required to appropriately describe the wavefunctions. Then, large reservoirs are needed to train the RC algorithm, which is likely to overfit the training data. For this reason, we proposed a new RC training strategy, the multi-step RC, that helped reduce overfitting.

In this section, instead of reducing the overfitting problem by using the multi-step learning, we propose to represent the coupled Morse oscillator wavefunction $\psi(x_1,x_2,t)$ with the coefficients 
of its decomposition in the decoupled Morse Hamiltonian eigenbasis.  
That is, given a wavefunction $\psi(x_1,x_2,t)$ and the decoupled Morse basis $\{\phi^D_n(x_1,x_2)\}_n$, where
\begin{equation}
    \phi^D_{n_1,n_2}(x_1,x_2)  = \phi^M_{n_1}(x_1)\phi^M_{n_2}(x_2), 
\end{equation}
with $\phi^M_{n_1}(x_1)$ being the 1D Morse eigenfunctions in Eq.\ref{eq:wavefunction_morse}. Then, the coupled Morse wavefunctions can be written as
\begin{eqnarray}
    \psi(x_1,x_2,t) & = & \sum_n a_n(t) \ e^{iE^D_n \frac{t}{\hbar}}  \ \phi^D_n(x_1,x_2),  \text{  with}\nonumber \\
    a_n(t) & = &\langle \psi(t) | \phi^D_n\rangle \, e^{-iE^D_n \frac{t}{\hbar}},  
\end{eqnarray}
where $E^D_n$ are the decoupled eigenenergies of the decoupled Hamiltonian (Eq.~\ref{eq:energy_morse}), and $a_n(t)$ are some coefficients which change with time. Notice that if the $\{\phi^D_n(x_1,x_2)\}$ basis approximates the eigenfunctions of the coupled Morse Hamiltonian, 
the coefficients $a_n(t)$ will change slowly with time. 
Moreover, for a wavepacket with a certain initial mean energy $E_0$, the number of coefficients needed to obtain 
a good approximation of the wavefunction will be significantly smaller than the dimension of the matrix $\psi(x_1,x_2,t)$
represented in a spatial grid. For example, the wavefunctions for the initial wavepacket with energy $E=-0.1089$ are represented by a matrix 
containing 4000 entries, while only 400 coefficients are needed to represent the wavefunction in the decoupled Morse basis set. 

The RC algorithm will be trained to propagate the coefficients $a_n(t)$ in time. This strategy allows us to reduce significantly the dimension of the system so that the standard RC training strategy is enough to learn to propagate the wavefunction without overfitting the training data. Notice that this strategy could only be used in this case because the Hamiltonian of interest could be approximated by its decoupled version, which has an analytical solution.

The training parameters used to train the reservoir computing algorithm are the following: 
In all cases,  $t_\text{train}=15000$, $t_\text{test}=10000$, $t_\text{min} = 100$, $N=500$, $\gamma=5 \times 10^{-5}$ and $\alpha=0.9$. 
The spectral radius of the internal states $W$ is set $\rho(W) = 0.04$, and the density of $W$ is set to 0.025.

\subsection{Results}
%Figure 2
\begin{figure*}[t]
\centering
\includegraphics[width=1.0\textwidth]{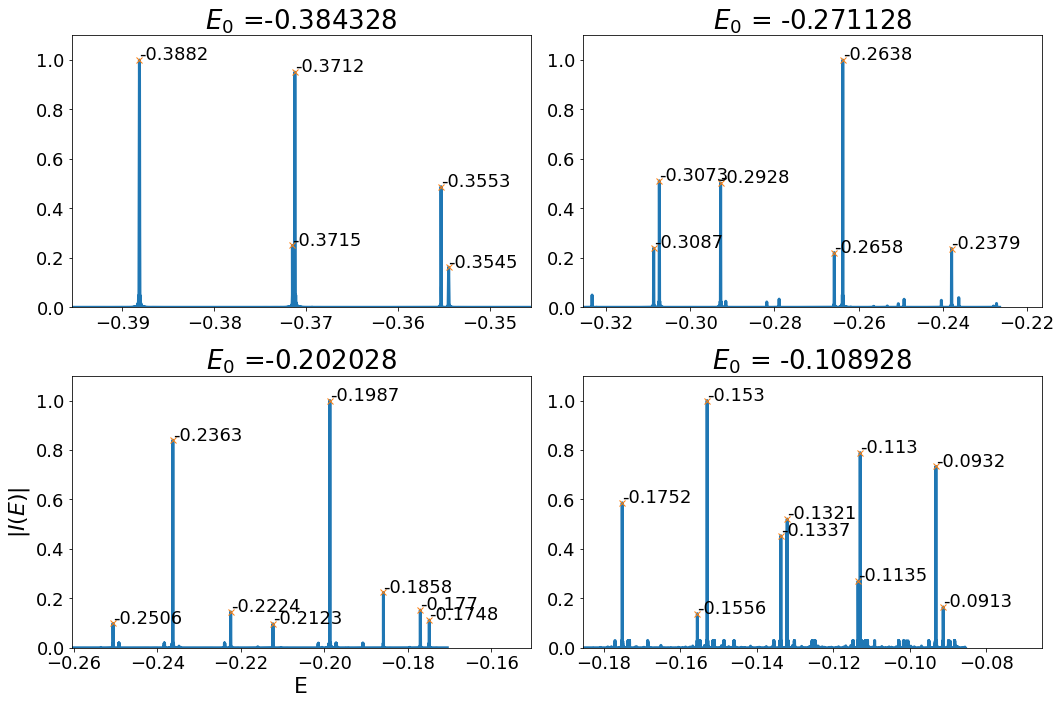}
\caption[Modulus of the energy spectrum $|I(E)|$ for the coupled Morse Hamiltonian with four different initial conditions with associated mean energy $E_0$.]{ Modulus of the energy spectrum $|I(E)|$ for the coupled Morse Hamiltonian with four different initial conditions with associated mean energy ${E_0=-0.384328, -0.271128, -0.202028, -0.108928}$. The peaks of the spectra correspond to the eigenenergies spanned by the initial wavefunction. The associated eigenfunctions are shown in Figs~\ref{fig:Morse_eigenfunctions0},~\ref{fig:Morse_eigenfunctions1},~\ref{fig:Morse_eigenfunctions2}, and~\ref{fig:Morse_eigenfunctions3}.}
\label{fig:Morse_spectra}
\end{figure*}
In this subsection, we present the results obtained with the RC approach and compare their accuracy with those from the usual variational method. A comparison of the execution time needed for the RC approach and the standard FFT method will also be given.

Figure~\ref{fig:Morse_spectra} shows the modulus of the spectra $|I(E)|$ obtained with the procedure described in Sects.~\ref{sect:paper2} and~\ref{sect:dim_red} for the four initial wavepackets described by the parameters in Table~\ref{tab:initial_waves}. The corresponding mean energies are specified at the top of each panel. As can be seen, each one of them consists of a number of very clearly defined sharp peaks centered at the eigenenergies of the eigenstates spanned by the initial wavefunctions (with numerical values also indicated inside the panels).

%Figure 
\begin{figure*}[!ht]
\centering
    \includegraphics[width=0.95\textwidth]{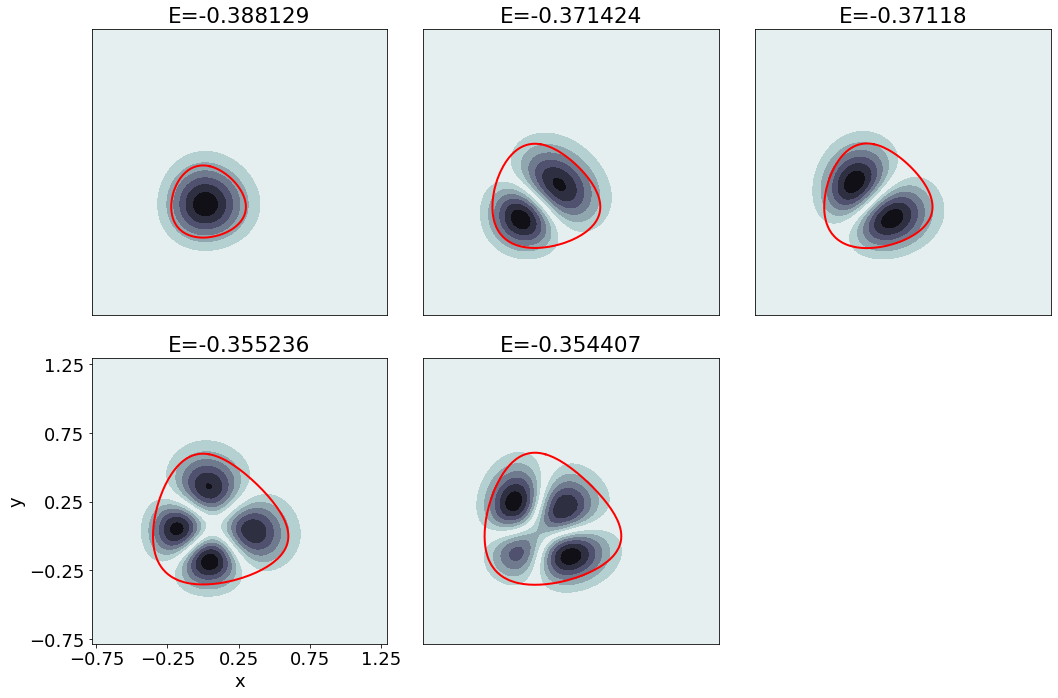}
    \caption[Coupled Morse eigenfunctions obtained with reservoir computing method described in Sects.~\ref{sect:paper2} 
    and~\ref{sect:dim_red} for the wavepacket with mean energy $E_0 = -0.384328$.]{Coupled Morse eigenfunctions obtained with reservoir computing method described in Sects.~\ref{sect:paper2} 
    and~\ref{sect:dim_red} for the wavepacket with mean energy $E_0 = -0.384328$ and parameters given in Table~\ref{tab:initial_waves}.
    For easier visualization, the plots show the norm of the eigenstates $|\phi_k(x_1,x_2)|$. 
    The red curve shows the equipotential line at each eigenenergy, which is indicated at the top of each panel. }
    \label{fig:Morse_eigenfunctions0}
\end{figure*}
%
%Figure 
\begin{figure*}[!ht]
\centering
\includegraphics[width=0.98\textwidth]{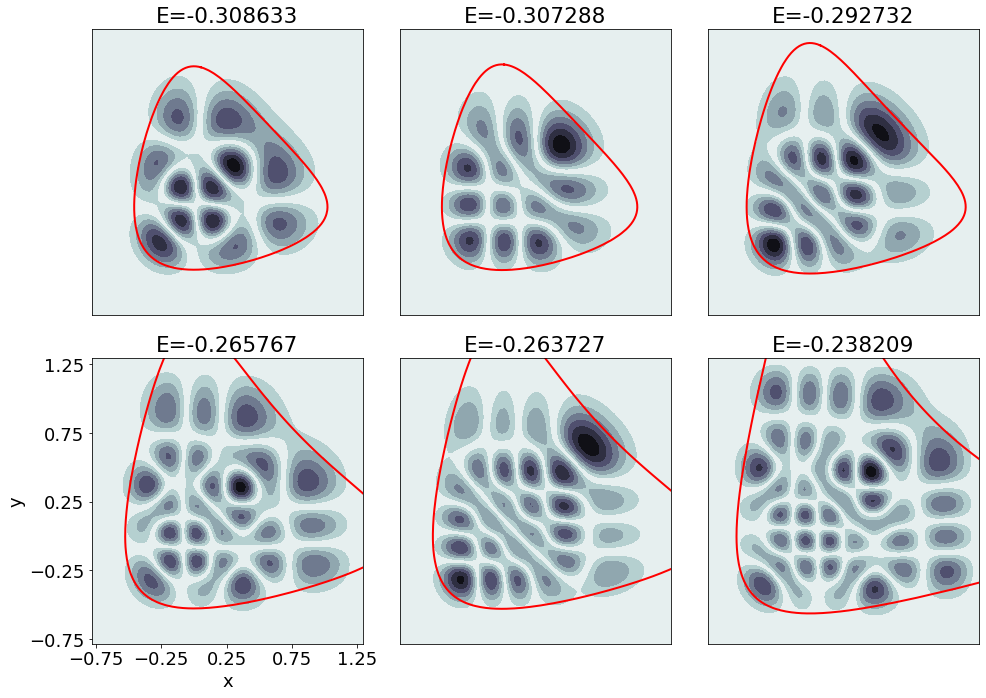}
\caption{Same as Fig.~\ref{fig:Morse_eigenfunctions0} for $E_0 = -0.271096$. }
\label{fig:Morse_eigenfunctions1}
\end{figure*}
%
%Figure 
\begin{figure*}[!ht]
\centering
\includegraphics[width=0.98\textwidth]{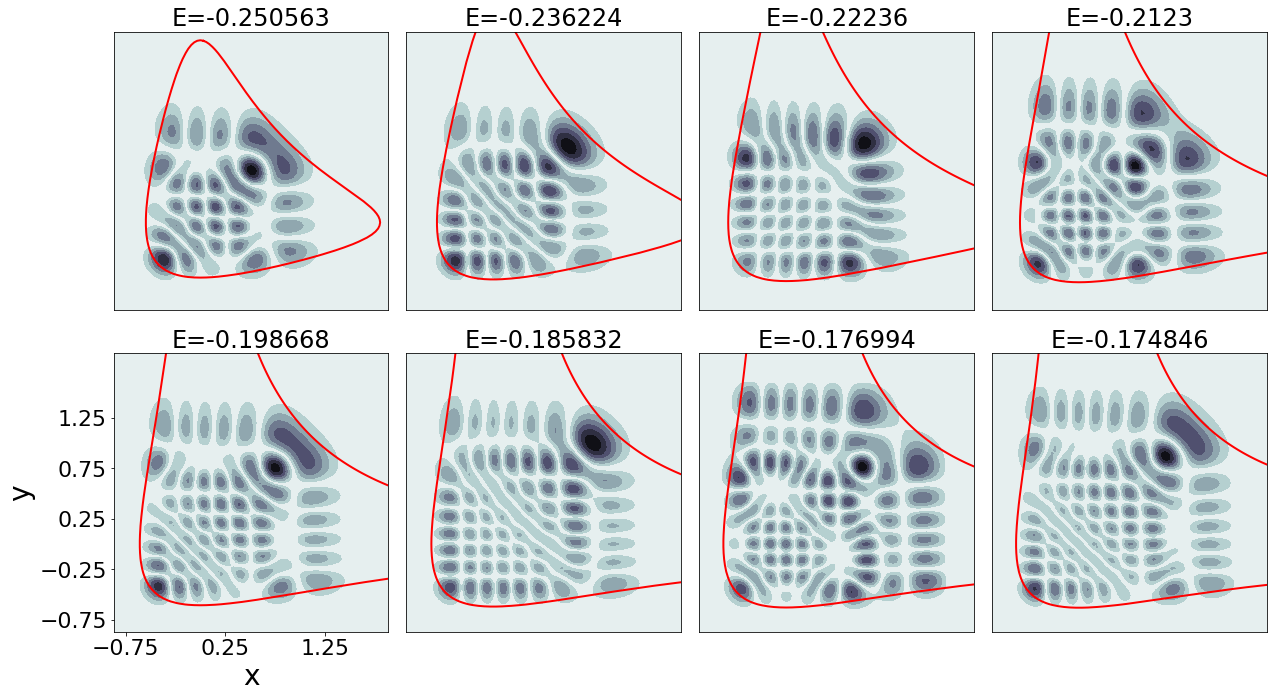}
\caption{Same as Fig.~\ref{fig:Morse_eigenfunctions0} for $E_0 = -0.201995$. }
\label{fig:Morse_eigenfunctions2}
\end{figure*}
%
%Figure 
\begin{figure*}[!ht]
\centering
\includegraphics[width=0.98\textwidth]{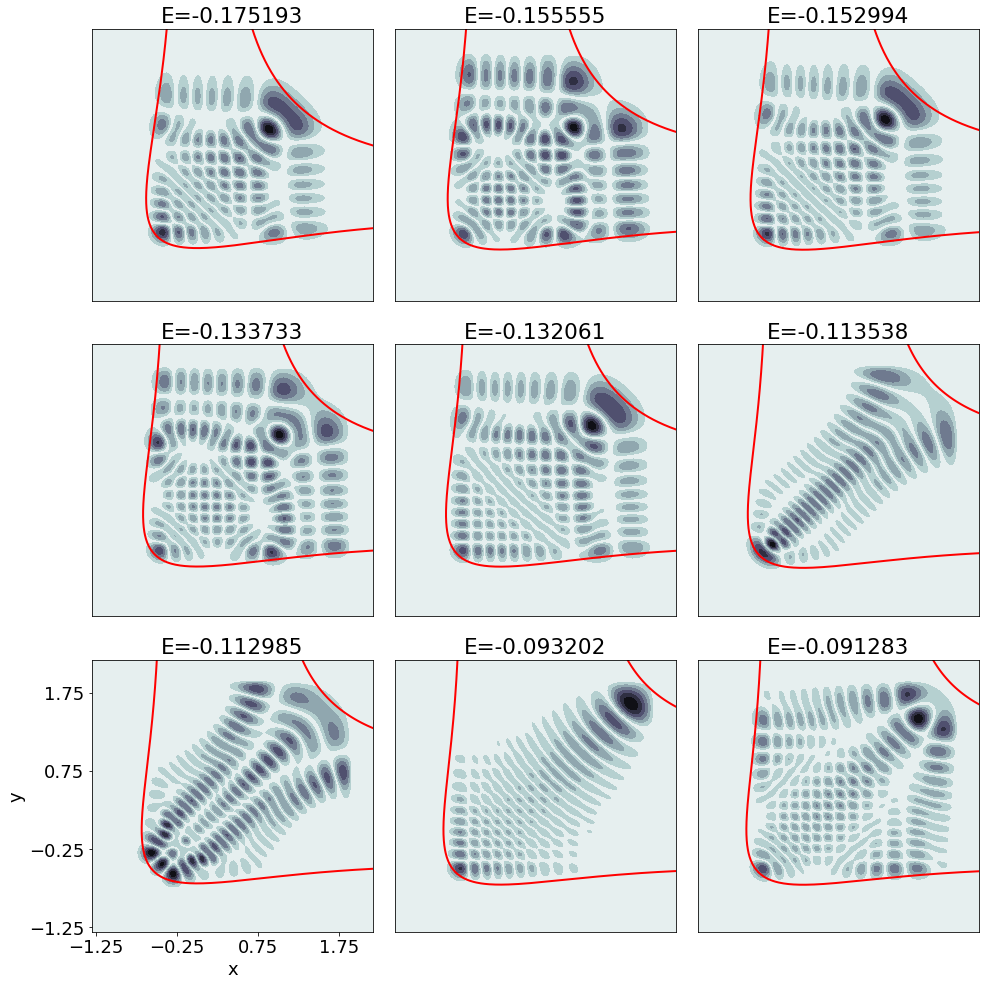}
\caption{Same as Fig.~\ref{fig:Morse_eigenfunctions0} for $E_0 = -0.108866$. }
\label{fig:Morse_eigenfunctions3}
\end{figure*}
\begin{table}[!t]
    \centering
    \begin{tabular}{ccccc}
    \hline \hline \\[-0.3cm]       
         $E_0$ & $E_k$ &  Absolute error in $E_k$ &  Squared error in $E_k$ & MSE $\phi_k$  \\ \\[-0.3cm]
         \hline  \\
    -0.384328 & -0.388129 & 3.1 $\times 10^{-7}$ & 9.8 $\times 10^{-14}$& 8.7 $\times 10^{-6}$ \\
        & -0.371424 & 3.1 $\times 10^{-7}$ & 9.4 $\times 10^{-14}$&4.5 $\times 10^{-5}$ \\
        & -0.371180 & $3.9\times 10^{-7}$ & 1.5 $\times 10^{-13}$ & 1.0 $\times 10^{-5}$\\
        & -0.355256 & $3.6\times 10^{-6}$ & 1.3 $\times 10^{-11}$ & 1.0 $\times 10^{-4}$\\
        & -0.354427 & $1.2\times 10^{-6}$ & 1.5 $\times 10^{-12}$ & 6.6 $\times 10^{-6}$\\
        \hline \\
    -0.270928 & -0.338633 & 9.1 $\times 10^{-7}$ & 8.3 $\times 10^{-13}$ & 6.6 $\times 10^{-8}$ \\
        & -0.307768 & $3.8\times 10^{-6}$ & 1.5 $\times 10^{-11}$ & 2.3 $\times 10^{-9}$ \\
        & -0.292204 & $2.5\times 10^{-7}$ & 6.1 $\times 10^{-14}$ & 8.0 $\times 10^{-8}$ \\
        & -0.265239 & $5.9\times 10^{-7}$ & 3.5 $\times 10^{-13}$ & 6.0 $\times 10^{-6}$ \\
        & -0.263199 & $4.3\times 10^{-4}$ & 1.8 $\times 10^{-7}$ & 5.6 $\times 10^{-7}$ \\
        & -0.238671 & $2.6\times 10^{-5}$ & 6.8 $\times 10^{-10}$ & 9.8 $\times 10^{-5}$ \\
        \hline \\
    -0.202028 & -0.250563 & 3.2 $\times 10^{-7}$ & 1.0 $\times 10^{-13}$ & 1.6 $\times 10^{-8}$ \\
        & -0.236224 & 5.8 $\times 10^{-6}$ & 3.4 $\times 10^{-11}$ & 1.6 $\times 10^{-8}$ \\
        & -0.222360 & 5.2 $\times 10^{-7}$ & 2.7 $\times 10^{-13}$ & 4.9 $\times 10^{-8}$ \\
        & -0.212300 & 7.6 $\times 10^{-5}$ & 5.8 $\times 10^{-9}$ & 3.7 $\times 10^{-7}$ \\
        & -0.198668 & 7.7 $\times 10^{-5}$ & 5.9 $\times 10^{-9}$ & 2.8 $\times 10^{-7}$ \\
        & -0.185832 & 7.7 $\times 10^{-5}$ & 5.9 $\times 10^{-9}$ & 2.7 $\times 10^{-7}$ \\
        & -0.176994 & 7.5 $\times 10^{-4}$ & 5.7 $\times 10^{-7}$ & 1.6 $\times 10^{-5}$\\
        & -0.174846 & 4.0 $\times 10^{-5}$ & 1.6 $\times 10^{-9}$ & 6.3 $\times 10^{-7}$\\
        \hline \\
    0.2966 & -0.175193 & 3.6 $\times 10^{-4}$ & 9.4 $\times 10^{-8}$ & 3.6 $\times 10^{-8}$ \\
        & -0.155555 & 6.1 $\times 10^{-6}$ & 3.8 $\times 10^{-11}$ & 3.8 $\times 10^{-5}$ \\
        & -0.152994 & 2.9 $\times 10^{-4}$ & 8.6 $\times 10^{-8}$ & 2.0 $\times 10^{-6}$ \\
        & -0.133733 & 2.3 $\times 10^{-4}$ & 5.2 $\times 10^{-8}$ & 5.3 $\times 10^{-6}$ \\
        & -0.132001 & 4.0 $\times 10^{-5}$ & 1.6 $\times 10^{-9}$ & 9.6 $\times 10^{-7}$ \\
        & -0.113478 & 6.5 $\times 10^{-4}$ & 4.3 $\times 10^{-7}$ & 3.8 $\times 10^{-3}$ \\
        & -0.112925 & 1.1 $\times 10^{-4}$ & 1.2 $\times 10^{-8}$ & 7.9 $\times 10^{-3}$ \\
        & -0.092202 & 2.3 $\times 10^{-4}$ & 5.3 $\times 10^{-8}$ & 2.1 $\times 10^{-5}$ \\
        & -0.090283 & 3.0 $\times 10^{-3}$ & 9.2 $\times 10^{-6}$ & 1.9 $\times 10^{-4}$ \\
        \hline
    \end{tabular}
    \caption[Eigenenergies $E_k$, their absolute and square errors and the mean squared error of the eigenfunctions
     obtained with reservoir computing for the coupled Morse oscillator.]{Eigenenergies $E_k$, their absolute and square errors and the mean squared error of the corresponding eigenfunctions
     obtained with the reservoir computing algorithm for the coupled Morse oscillator. The table contains the four initial wavepackets with mean energies $E_0$ and parameters
     in Table~\ref{tab:initial_waves}.}
    \label{tab:mse_Morse}
\end{table}
Moreover, Figs.~\ref{fig:Morse_eigenfunctions0},~\ref{fig:Morse_eigenfunctions1},~\ref{fig:Morse_eigenfunctions2}, and~\ref{fig:Morse_eigenfunctions3} 
show the corresponding eigenfunctions computed according to item 4 in Sect.~\ref{sect:paper2}. The red curves in each panel indicate the equipotential contour at the associated eigenenergies values $E_k$. 
Notice that the probability density is mainly concentrated within the region delimited by the equipotential curve. Also, notice that we only obtain eigenfunctions with the same symmetry as that of the initial wavepacket. In this case, all obtained eigenfunctions are symmetric with respect to the main diagonal (see values of the momenta in Table~\ref{tab:initial_waves}). To obtain eigenstates (and eigenenergies) with other symmetries, one only needs to change the symmetry of the initial wavefunction.

The quality of these results, which have been obtained with the RC method and the dimensionality reduction technique described in Sect.~\ref{sect:dim_red}, 
are compared with those obtained with the traditional variational method 
in Table~\ref{tab:mse_Morse}. In it, we report the eigenenergies obtained with each initial wavepacket in Table~\ref{tab:initial_waves},
its absolute and squared error, and the MSE of the associated eigenfunction $\phi_k(x_1,x_2)$ (see Eq.~\ref{eq:MSEpsi}).

As can be seen in Table~\ref{tab:mse_Morse}, the absolute error of the eigenenergies increases with the energy of the initial packet.
This is because the corresponding wavefunctions become more complex and thus it is harder for the RC algorithm to predict the time evolution of the associated dynamics. However, we see that the absolute error of all the eigenenergies is always smaller than $\sim 10^{-3}$, and usually significantly smaller, which means that the 
RC model can correctly propagate the quantum states in time even for high-energy states. It is also worth noticing that the precision of the variational method also decreases as the energy of the system increases. 
For the eigenenergies obtained with the fourth (highest) initial condition, the absolute error of the variational method is of the 
order of $\sim  10^{-3}$ (see Table~\ref{tab:convergence_variational} in Appendix~\ref{appendix1}), meaning that the errors in Table~\ref{tab:mse_Morse} are lower bounded by the errors in Table~\ref{tab:convergence_variational}.  

Regarding the MSE in the eigenfunctions, they are in general higher than the squared errors in $E_k$. This result is not unexpected since the eigenfunctions contain much more information about the quantum state than the corresponding eigenenergies. Anyhow, this is not a terrible result, since only the regions with high probability density contribute to the quantum properties of the system. A proof of this fact is that, despite the errors in the eigenfunctions, the mean energy of the system is correctly reproduced. 

\begin{table}[!t]
    \centering
    \begin{tabular}{clc}
        \hline \hline \\[-0.3cm]
        $E_0$ & Method & Execution time (min) \\ \\[-0.3cm]
        \hline \\
        -0.384328 & Reservoir Computing & 8\\
        & FFT & 75 \\
        -0.271128 & Reservoir Computing & 10\\
        & FFT & 80 \\
        -0.202028 & Reservoir Computing & 13\\
        & FFT & 87 \\
        -0.109928 & Reservoir Computing & 15\\
        & FFT & 102 \\
        \hline
        \end{tabular}
        \caption{Execution training times for the reservoir computing method and the standard Fast Fourier Transform for the four initial wavepackets for the coupled Morse oscillator.}
\label{tab:time_Morse}
\end{table}
Let us conclude this subsection by discussing the execution time efficiency of the RC method compared with the FFT propagation on which it is based, and that could have been used for the whole period of time
needed in the integration to accurately compute the eigenstates of our system. Table~\ref{tab:time_Morse} compares the execution times for the RC approach and the standard FFT propagator. 

As can be seen, the results show that RC is much faster than the classical integrator. Furthermore, a limitation of the latter method is the small time step needed to avoid large numerical errors in the calculation. In contrast, in the RC approach, the velocity of the input-output dynamics is controlled by the $\alpha$ and $\rho(W)$ parameters. The flexibility of the machine learning algorithm allows for the use of time steps up to 300 times larger than the one needed for the FFT method. As a result, the FFT method also requires more memory to store the wavefunctions for the intermediate time steps.

Moreover, the RC algorithm is data-based, so it is agnostic to the underlying physical model of the system. For this reason, it is easily adaptable to any other quantum systems, as we showed for example in Sect.~\ref{sect:paper2}. On the contrary, the traditional integration methods need to be adapted to the Hamiltonian at study. For example, Appendix~\ref{appendix1} shows how to apply the variational method to integrate the Schrödinger equation for the coupled Morse Hamiltonian. This method involves calculating some integral functions that depend on the potential and kinetic energy of the Hamiltonian, and thus a different algorithm needs to be used for each quantum system. 

To summarize, the results show that RC is a suitable algorithm for predicting the time evolution of the coupled Morse Hamiltonian for multiple initial conditions, as it accurately predicts the eigenenergies and eigenfunctions with high precision. This ML algorithm is faster and more adaptable than traditional solvers based on either FFT or the variational method.

%%%%%%%%%%%%%%%%%%%%%%%%%%%%%%%%%%%%%%%%%%%%%%%%%%%%
% RC SCAR FUNCTIONS
%%%%%%%%%%%%%%%%%%%%%%%%%%%%%%%%%%%%%%%%%%%%%%%%%%%%

\section{Constructing eigenfunctions and scar functions by reservoir computing}
\label{sect:paper4}

In the last section of this chapter, we apply the multi-step RC algorithm to a complex and realistic quantum system, the coupled quartic oscillator, which has been of great interest in the field of quantum chaos~\citep{chaotic, Fabio2,Fabio1}. Apart from obtaining the eigenfunctions and eigenenergies by following the steps described in Sect.~\ref{sect:paper2}, we also aim to calculate the so-called \emph{scar functions}, that will be introduced below. 

The organization of the next subsections is the following. First, in Sect.~\ref{sect:quantum_chaos} we introduce the field of quantum chaos and its relevance for understanding the relationship between classical and quantum mechanics. Then, in Sect.~\ref{sect:quartic}, we provide details about the system of interest. In Sect.~\ref{sect:scars_method} we present the method to calculate the scar functions, and finally in Sect.~\ref{sect:scars_Results} we discuss the performance of the multi-step RC model in the prediction of both the eigenfunctions and scar functions. 

\subsection{Introduction to quantum chaos}
\label{sect:quantum_chaos}
Chaotic motion is a well-understood phenomenon in classical systems. Chaotic systems can be optimally characterized by an extreme sensitivity to initial conditions. This means that small differences in the initial conditions of two trajectories can lead to exponential divergence as time advances. This phenomenon is widely known as the \emph{butterfly effect}, which was named after the famous talk given by Lorenz in 1995~\citep{lorenz1995essence}. He suggested that the flapping of a butterfly's wings in one place could potentially influence the formation of a tornado in another location. This metaphor symbolizes that even a small perturbation, such as the subtle motion of a butterfly, can cause a chain of events that amplifies over time and leads to large-scale effects.

There are many quantitative indicators of classical chaos, the most important one being the Lyapunov exponent $\lambda$ ~\citep{Lyapunov_original}. This indicator characterizes the exponential rate of divergence of nearby trajectories in the phase space of a dynamical system. That is, given two nearby initial conditions in phase space, the Lyapunov exponent is defined as the average rate of exponential divergence of these trajectories. %A general trajectory of a particle is described by the phase space vector $\vec{z}(t)$, consisting of both the spatial coordinates $\vec{x}(t)$ and the momentum $\vec{p}(t)$ of such particle

% \begin{equation}
%     \vec{z}(t) = \begin{pmatrix}
%         \vec{p}(t) \\
%         \vec{x}(t)         
%     \end{pmatrix}.
% \end{equation}
% Consider a trajectory $\vec{z}_1(t)$ associated to the initial conditions $\vec{z}_1(0)$. Now, consider a second trajectory $\vec{z}_2(t) = \vec{z}_1(t) + \delta \vec{z}(t)$, associated to an initial condition $\vec{z}_2(0) = \vec{z}_1(0) + \delta \vec{z}(0)$. Notice that this second initial condition is a first-order perturbation of the first initial condition. As the dynamical system evolves in time, the perturbation $\delta \vec{z}(t)$ also evolves. The Lyapunov exponent, denoted as $\lambda$, is defined as the exponential growth rate of this perturbation over time

% \begin{equation}
%     \lambda = \lim_{t \rightarrow \infty} \frac{1}{t} \log \frac{||\delta \vec{z}(t)||}{||\delta \vec{z}(0)||}.
%     \label{eq:Lyapunov}
% \end{equation}

Usually, the Lyapunov exponent of the dynamical system refers to the largest Lyapunov exponent which takes place along one direction, over all initial conditions. This exponent can be either negative, zero or positive. If $\lambda > 0$, two near trajectories diverge exponentially, which indicates that the system exhibits chaotic behavior. Otherwise, the system is a regular non-chaotic system (for $\lambda=0$), or a converging system (for $\lambda <0$). 

While classical chaos is widely understood and precisely defined, there is no direct equivalent in quantum mechanics. The dynamics of quantum systems are governed by the time-dependent Schrödinger equation. That is, an initial wavefunction $\psi_0$ evolves under the effect of a Hamiltonian $\hat{H}$ according to  Eq.~\ref{eq:time-dependent}. Notice that this equation is linear with the wavefunction $\psi(t)$ (i.e. a linear combination of two solutions of Eq.~\ref{eq:time-dependent} is also a solution of this equation). This simple fact makes it impossible to define quantum chaos in a similar way as in classical mechanics. Indeed, the linearity of the Schrödinger equation prohibits the existence of extreme dependence on the initial conditions of the wavefunctions. 

Even though there is not a direct equivalent to classical chaos in quantum mechanics, quantum systems with classically chaotic Hamiltonians do exhibit special properties that arise from the classically chaotic behavior of the system. In this way, nowadays, the field of \emph{quantum chaos}~\citep{Berr89,Berr90,BerryKeating92,Keating92} is devoted to the study of quantum manifestations of classical chaotic systems. The link between the quantum and classical worlds exists through the \emph{quantum-classical correspondence}~\citep{Haake2010}, which states that quantum systems can be associated with their classical analogue, corresponding to the classical limit of $\hbar \longrightarrow 1$. 

The field of quantum chaos has been studied mainly through two different theories. The first one is known as \emph{Random Matrix Theory}. It was first introduced by Wigner~\citep{wigner} to understand the spectra of complex nuclei from a statistical perspective. The idea of this theory is that the chaotic properties of quantum systems can be found within the statistical distribution of the eigenenergies. Following this idea, Bohigas et al.~\citep{bohigas} formulated a conjecture stating that quantum systems with a classical chaotic analogue are described by Random Matrix Theory. That is, if the energy level spacing of arbitrary quantum systems follows the Wigner distribution, the system is a quantum chaotic system. On the other extreme, when the classical Hamiltonian is integrable, the energy level spacing follows a Poisson distribution. In this way, the energy levels of a quantum system are influenced by the chaotic or regular dynamics of the classical analogue. Even though this theory is widely accepted, it only allows studying universal properties of quantum chaotic systems, so that individual quantum systems cannot be described within this theory.

The second theory used to study quantum chaos is based on semiclassical theory and \emph{scar functions}. The scarring phenomenon plays a central role in the study of the correspondence between classical and quantum mechanics in the presence of chaos. The term \emph{scar} was introduced by Heller~\citep{Heller} to describe the high accumulation of the probability density of some eigenfunctions along the periodic orbits of quantum chaotic systems. This result was highly surprising since, even though the chaotic classical dynamics cause the classical trajectories to deviate from periodic orbits, some of its eigenfunctions tend to concentrate near those periodic orbits~\citep{Pol94}. In 1988, Bogomolny~\citep{Bogomolny1988} showed that scarring is not a property of a single eigenstate, but that the localization around periodic orbits can occur by averaging a group of eigenstates in energy and position. In 1989, Berry~\citep{Berr89} extended the scarring analysis to the phase space, by studying the contribution of periodic orbits to the density of states in phase space by averaging the Wigner distribution. 

In Ref.~\citep{Pol94}, Polavieja et al. proposed a semiclassical construction of a wavefunction localized along a periodic orbit as a combination of a few eigenfunctions of a classically chaotic system. That is, by launching a minimum uncertainty coherent wave packet (see Eq.~\ref{eq:initial_wavepacket}) along a periodic orbit and following it for a \emph{short} period of time, the propagated wavefunction was able to collect all the available quantum information on the dynamics around the neighbourhood of the periodic orbit. Then, using the Fourier Transform, one could obtain a wavefunction, which was highly localized around the periodic orbit at study. Such localized wavefunctions could be constructed using only a few eigenfunctions of the quantum system. Later, this construction of localized wavefunctions was further refined, resulting in the so-called \emph{scar functions}~\citep{Ver01}, which are not only localized along the periodic orbit in configuration space but also along its invariant manifolds in phase space. 

 More recently, Revuelta~\citep{TesisFabio} proposed a sophisticated and efficient method, which will be the one used in this thesis, to calculate scar functions. The scar functions obtained with this method have proven to be an efficient basis set for the calculation of the eigenstates of such systems~\citep{Fabio1}, and for the calculation of excited chaotic eigenfunctions in arbitrary energy windows~\citep{Fabio3}. Moreover, in Ref.~\citep{Tino21}, by using a correlation diagram of eigenenergies versus the Planck constant, the authors showed the existence of quantum scars, which represent the quantum manifestations of the classical resonances in a molecular system. Currently, the concept of scars has been extended to many-body problems~\citep{manybodyscars}.
 
 The first step of the method used in this thesis to calculate scar functions consists of preparing an initial wavefunction localized around a periodic orbit, called \emph{tube function}
 \begin{equation}
        \psi_0^\text{tube}(x,y) = \int_0^T dt e^{i \epsilon_n t/\hbar} \psi_0(x,y,t).
        \label{eq:psi_tube}
\end{equation}
where $T$ is the period of the periodic orbit, $\epsilon_n$ are the Bohr-Sommerfeld energies (described below), and the function $\psi_0(x,y,t)$ is a frozen Gaussian centred on the trajectory of the periodic orbit so that its probability density is forced to stay around such periodic orbit, with coordinates $(x_t, y_t)$ and momentum $(p_{xt}, p_{yt})$. 
    \begin{equation}
        \psi_0(x,y,t) = e^{-i\alpha_x(x-x_t)^2 - i\alpha_y (y-y_t)^2} \times  e^{\frac{i}{\hbar}(p_{xt}(x-x_t) + p_{yt}(y-y_t) + \theta(t))},
        \label{eq:psi0_tube}
    \end{equation}
where $\alpha_x$ and $\alpha_y$ are the widths along the two spatial dimensions, which in this thesis are set to $\alpha_x = \alpha_y = 1$ and 
\begin{equation}
    \theta(t) = \frac{S(t)}{\hbar} - \frac{\pi}{2} \mu,
\end{equation}
  where  $\mu$ is the Maslov index~\citep{maslov}, which is used to quantify the phase changes that occur when semiclassical states cross classical boundaries, and $S$ is the classical action 
  \begin{equation}
    S(t) = \int_0^t (p_x \dot{x} + p_y \dot{y})dt.
    \label{eq:action}
\end{equation}
In semiclassical theory, the only allowed wavefunctions around a periodic orbit are those whose cumulative phase around one period is a multiple of $2\pi$. This condition is known as the Bohr-Sommerfeld quantization, which states that the phase $\theta(T)$ must follow the condition
\begin{equation}
        \theta(T) =  \frac{S(T)}{\hbar} - \mu \frac{\pi}{2} = 2\pi n,
        \label{eq:gamma_t}
    \end{equation}
 The condition in Eq.~\ref{eq:gamma_t} ensures a non-destructive interference of the wavefunction as the wavepacket circulates the periodic orbit. The number $n$ refers to the number of excitations along the periodic orbit, and it allows us to label different wavefunctions with different energies. Given the Bohr-Sommerfeld quantization relation, one can obtain the associated Bohr-Sommerfeld energies $\epsilon_n$ (present in Eq.~\ref{eq:psi0_tube}) for a certain Hamiltonian, which are the only allowed semiclassical energies. 

Once the tube function in Eq.~\ref{eq:psi_tube} has been calculated, the wavefunction is propagated under the quantum chaotic Hamiltonian for a \emph{short} period of time, which is known as the Ehrenfest time $t_E$. The Ehrenfest time is the time it takes a localized wavepacket to explore all the available phase space at a certain energy. It is also the time in which the semiclassical approximation is valid since it marks the point at which the expectation values of the position and momentum of the wavepacket start to behave classically. The mathematical expression of $t_E$ is the following
    \begin{equation}
        t_E = \frac{1}{2\lambda} \ln\Big(\frac{\mathcal{A}}{\hbar}\Big),
        \label{eq:Ehrenfest}
    \end{equation}
where $\lambda$ is the Lyapunov exponent of the system and $\mathcal{A}$ is a characteristic area of the classically available phase space~\citep{Fabio3}. The Ehrenfest time can be estimated numerically based on the characteristic scales of each quantum system.

Once the short-time propagation of the tube function is calculated, the low-resolution energy spectrum $I(E)$ is computed. The peaks of this low-resolution spectrum $E_n$ are the energies of the scar functions. Then, the scar functions are calculated by computing the Fourier transform at energies $E_n$. The method to calculate scar functions adapted to the coupled quartic oscillator is described in Sect.~\ref{sect:scars_method}. For more details of semiclassical theory and quantum chaos, see Ref.~\citep{TesisFabio}.

\subsection{The coupled quartic oscillator}
\label{sect:quartic}
The quantum system of interest in this work is the dynamics of a unit mass particle moving in a 2D coupled quartic potential:

\begin{equation}
    H(x,y) = \frac{1}{2}(p_x^2 + p_y^2) + \frac{1}{2}x^2y^2 + \frac{\epsilon}{4}(x^4 + y^4),
    \label{eq:quartic}
\end{equation}
with $\epsilon=0.01$. The classical dynamics of this system are very chaotic~\citep{chaotic}, and for this reason this system has been extensively studied in the field of quantum chaos~\citep{quartic1,quartic2,Fabio2,Fabio1,Fabio3}. The potential in Eq.~\ref{eq:quartic} belongs to the $C_{4v}$ symmetry group. In this work, we will only focus on studying the eigenfunctions belonging to the $A_1$ irreducible representation, which is symmetric around axes $x$ and $y$ and also around the two diagonals $y = \pm x$. In this setting, the initial wavepackets used to calculate the eigenfunctions will be symmetric with respect to the axis and diagonals. Eigenfunctions belonging to other irreducible representations could also be obtained by changing the symmetry of the initial wavepacket, as will be indicated later. 

%Notice that the calculated scar functions will have the symmetry of the periodic orbit associated with them, which do not necessarily belong to an irreducible representation.

Since this potential is homogeneous, the classical trajectories are mechanically similar. That is, any trajectory  $(x(t), y(t), p_x(t), p_y(t))$ at a certain energy $E$ can be scaled to another energy $E'$ with coordinates $(x'(t), y'(t), p'_x(t), p'_y(t))$, by using the following scaling relations:

\begin{equation}
        \left\{\begin{array}{ll}
        & x':= \eta x, \quad p_x':= \eta^2 p_x, \\
        & y' := \eta y, \quad  p_y' := \eta^2 p_y,  \\
        & t' := t/\eta, \quad S' = \eta^3 S, 
    \end{array}\right.
    \label{eq:scaling}
\end{equation}
where $\eta =  \Big(\frac{E'}{E}\Big)^{1/4}$, and $S$ is the classical action defined in Eq.~\ref{eq:action}. 

To calculate the eigenenergies and eigenfunctions, the multi-step RC algorithm will be used to propagate an initial wavefunction $\psi_0(\vec{x})$ under the evolution of the quantum Hamiltonian associated with Eq.~\ref{eq:quartic}. The initial wavefunction is a minimum uncertainty Gaussian wave packet as in Eq.~\ref{eq:2DHO}. Then, the eigenfunctions and eigenenergies will be calculated, employing the Fourier transform, for three different initial conditions at energies $E=1,10$ and $100$ respectively. For simplicity of visualization, all the systems are scaled to energy $E=1$ using the scaling relations in Eq.~\ref{eq:scaling}. The parameters of the initial conditions are shown in Table~\ref{tab:initial_waves_quartic}.

\begin{table}
    \centering
    \begin{tabular}{l|cccccccc}
    \hline \hline \\[-0.3cm]
      Energy & $x_0$ & $y_0$ & $\Delta x$ & $\Delta y$  & $p_0$ & $t_\text{train}$ & $t_\text{test}$ & dt\\
      \hline \\[-0.2cm]
       1  &  0 & 0 & 0.5 & 0.5 & 1 & 1500 & 4650 & 0.0014\\
       10 & 1.86 & 1.86 & $\frac{1}{\sqrt{2}}$ & $\frac{1}{\sqrt{2}}$ & -2 & 5000 & 10000 & 0.0003\\
       100 & 3.665 & 3.665 & 1 & 1 & -3 & 3000 & 8000 & 0.00003\\
       \hline
    \end{tabular}
    \caption[Parameters of the initial wavefunction for the coupled quartic oscillator, as described in Eq.~\ref{eq:2DHO}, at three different energies.]{Parameters of the initial wavefunction for the coupled quartic oscillator, as described in Eq.~\ref{eq:2DHO}, at three different energies. Training and test steps for the multi-step reservoir computing algorithm, and time interval between steps.}
    \label{tab:initial_waves_quartic}
\end{table}

\subsection{Scar function calculation}
\label{sect:scars_method}

\begin{figure}[!ht]
 \includegraphics[width=1.0\columnwidth]{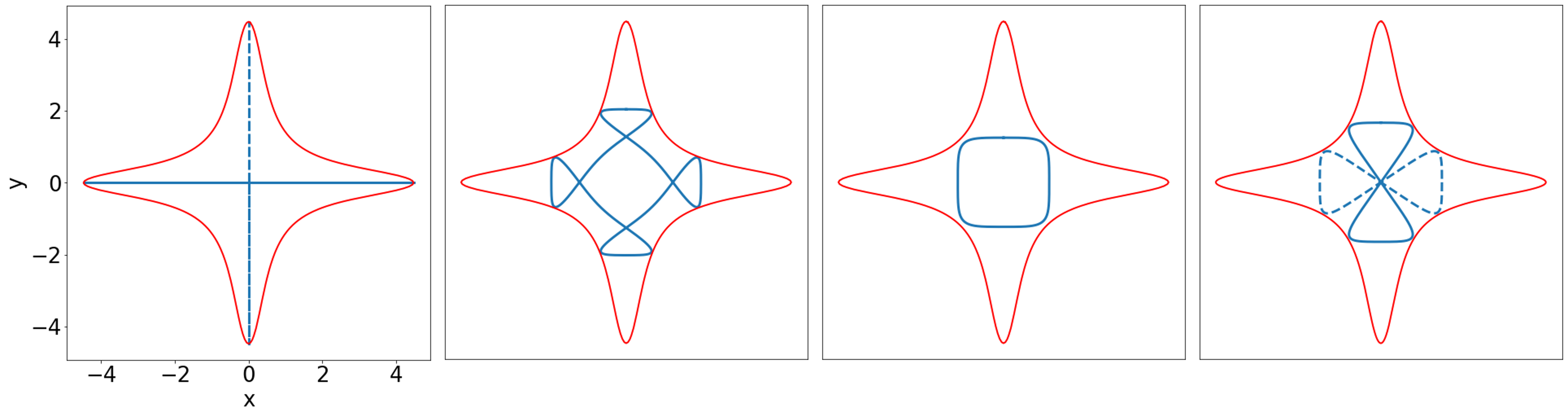}
  \caption{Representation of the four periodic orbits studied in this work at energy $E=1$, together with the equipotential curve at $V(x,y) =1$.  }
 \label{fig:scars_po}
\end{figure}

Apart from the eigenfunctions, we also use the RC algorithm to compute the scar functions along different periodic orbits, at different energies. The description of the four periodic orbits is given in Table~\ref{tab:scars_PO}, together with a graphical representation in Fig.~\ref{fig:scars_po}. For simplicity, we have named the four periodic orbits as \textit{horizontal}, \textit{quadruple-loop}, \textit{square} and \textit{triangle}, respectively. Notice that the \textit{triangle} and \textit{horizontal} orbits are in fact made of two periodic orbits (see the solid and dashed lines in Fig.~\ref{fig:scars_po}). This choice is made so that the resulting scar functions belong to the $A_1$ irreducible representation, that is, that the scar functions are symmetric with respect to the $x$, $y$ axis and the diagonals $y = \pm x$. 
%
%---------------------------------------------------------------------
%Figure: PO
%---------------------------------------------------------------------
%----------------------------------------------------------------------
%----------------------------------------------------------------------------

Given a periodic orbit of the classical Hamiltonian in Eq.~\ref{eq:quartic}, the steps to compute its associated scar functions are the following:

\begin{enumerate}
    \item Calculate the trajectory periodic orbit by solving Hamilton's equations for the Halitonian in Eq.~\ref{eq:quartic}:
    \begin{equation}
        \left\{\begin{array}{ll}
        &\dot{p}_x = -(xy^2 + \epsilon x^3), \\
        &\dot{p}_y = -(x^2y + \epsilon y^3), \\
        &\dot{x} = p_x,\\
        &\dot{y} = p_y, \\
        &\dot{S}_x = p_x\dot{x} = p^2_x,  \\
        &\dot{S}_y = p_y\dot{y} = p^2_y,
    \end{array}\right.
    \end{equation}
     
    with the appropriate initial conditions for the periodic orbit. Notice that the last two equations are added to calculate the classical action, which is used to define the initial wavepacket (see step 2). 
    \item Prepare an initial wavefunction $\psi_0^\text{tube}(x,y)$ defined in Eq.~\ref{eq:psi_tube}. In this case, the Bohr-Sommerfeld energies $\epsilon_n$ are computed using Eqs.~\ref{eq:gamma_t} and~\ref{eq:scaling}:
    \begin{equation}
        \epsilon_n = \Big(\frac{2\pi \hbar}{S(T)}\left(n + \frac{\mu}{4}\right)\Big)^{4/3},
        \label{eq:Bohr-sommerfeld-energies}
    \end{equation}
    where $S(T)$ is the classical action accumulated around a full period of the periodic orbit (as in Eq.~\ref{eq:action}) with energy $E=1$ (calculated in step 1), and $n$ is the excitation number. The values of the coefficients associated with each periodic orbit are shown in Table~\ref{tab:scars_PO}.

    \begin{table}
        \centering
        \begin{tabular}{l|ccccc}
            \hline \hline
            Scar & T &  n & $\mu$ & $(x_0,y_0,p_{x_0}, p_{y_0})$ \\
            \hline    
            \multirow{2}{*}{Horizontal} & \multirow{2}{*}{33.17}  & 12,14,16,18,20,22, & \multirow{2}{*}{16} & \multirow{2}{*}{(0,0, $\sqrt{2}$,$\sqrt{2}$)} \\
             & & 42,47,53, 58,62 &  &\\
            Quadruple-loop  & 18.75  & 4,5,6,7,8,9,10 & 12 & (0, 2.028, 1.384, 0) \\
            Square  & 7.84  & 3,4,5,6,7,8 & 4 & (0, 1.239, 1.410, 0) \\
            Triangle  & 9.54  & 20,22,24 & 4 & (0, 1.655, 1.401, 0) \\
            \hline
        \end{tabular}
        \caption[Period (T), excitation numbers (n), Maslov index ($\mu$) and initial conditions for the four periodic orbits studied in this work, at energy $E=1$.]{Period (T), excitation number (n), Maslov index ($\mu$) and initial conditions for the four periodic orbits studied in this work, at energy $E=1$. These values are taken from Ref.~\citep{TesisFabio}.}
        \label{tab:scars_PO}
    \end{table}
    \item  Compute the time evolution $\psi(x,y,t)$ under the Hamiltonian $H(x,y)$. Again, we use the FFT method to train the multi-step RC model and use the predictions of the multi-step RC to further propagate the wavefunction.
    \item Calculate the low-resolution energy spectrum $I(E)$. 
    First, we calculate the time-correlation function in Eq.~\ref{eq:time-corr}. Then, we obtain the energy spectrum $I(E)$ as the Fourier transform of the time-correlation function for a short period of time
  
    \begin{equation}
        I(E) = \int_{-t_E}^{t_E} \; A(t) \; e^{i E t/ \hbar} \; dt ,
        \label{eq:low-spectrum}
    \end{equation}
      
    where $t_E$ is the Ehrenfest time given by Eq.~\ref{eq:Ehrenfest} with $\lambda \approx  0.385 E^{1/4}$ and $\mathcal{A} = 11.1 E^{3/4}$ for the $A_1$ representation, which were calculated numerically in Ref.~\citep{Fabio3}. The peaks $E_n$ of the low-resolution spectrum $|I(E)|$ are used to compute the scar functions.
    
    \item Calculate the scar functions $\psi^\text{scar}(x,y)$ by performing the Fourier transform of $\psi(x,y,t)$ for a short period of time at the energies $E_n$

    \begin{equation}
        \psi^\text{scar}_n(x,y) = \int_{-t_E}^{t_E} \; \psi(x,y,t) \; e^{-i E_n t/\hbar} \; dt.
        \label{eq:inital}
    \end{equation}
\end{enumerate}

\begin{table}%[H]
    \centering
    \begin{tabular}{l|cccc}
    \hline \hline \\[-0.3cm]
        System &  $\alpha$ & $\gamma$ & $N$   \\
         \hline \\[-0.2cm]
         Eigenfunctions $E=1$ & 0.2 &  0.001 & 2000 \\
         Eigenfunctions $E=10$ & 0.2 &  0.001 & 2000 \\ 
         Eigenfunctions $E=100$ & 0.017 &  0.1 & 10000 \\
         Horizontal scar & 0.1 &  1 & 1500 \\
         Quadruple-loop scar & 0.3 &  0.1 & 2000 \\
         Square scar & 0.3 &  0.1 & 3000 \\
         Triangle scar & 0.3 &  0.1 & 3000 \\
         \hline 
    \end{tabular}
    \caption{Reservoir computing training parameters, as defined in Sect.~\ref{sect:RC}, used for the coupled quartic oscillator. }
    \label{tab:params_scars}
\end{table}    
Table~\ref{tab:params_scars} shows the training parameters used for the different initial conditions of the coupled quartic oscillator. For the multi-step process, we used 80\% of the training data for the first training step and 20\% for the second step. The spectral radius of the internal states $W$ is set $\rho(W) = 0.5$, the density of $W$ is set to 0.005, and $t_\text{min} = 500$.  

%--------------------------------------------------------------------------
% RESULTS
%----------------------------------------------------------------------

\subsection{Results} 
\label{sect:scars_Results}
%----------------------------------------------------------------------
%----------------------------------------------------------------------------
% Table 1: Energies error
%----------------------------------------------------------------------
%----------------------------------------------------------------------------
\begin{table}[!t]
    \centering
    \begin{tabular}{lccc}
    \hline
    \hline \\[-0.3cm]
        Energy & Energy MSE & Eigenfunction MSE  \\
        \hline \\[-0.2cm]
        0.56323  & $4 \times 10^{-10}$ & $5 \times 10^{-7}$\\
        1.8848 & $< 10^{-10}$ & $4 \times 10^{-7}$\\
        2.8638 & $4 \times 10^{-10}$ & $1 \times 10^{-5}$\\
        4.8286  & $1\times 10^{-8}$ & $6 \times 10^{-6}$\\
        5.2584  & $1 \times10^{-8}$ & $7 \times 10^{-6}$\\
        9.3067  & $ 1 \times 10^{-8}$ & $9 \times 10^{-6}$\\
        14.9547 & $ 1 \times 10^{-8}$ & $1 \times 10^{-5}$\\
        28.6843 & $ 4 \times 10^{-8}$ & $8 \times 10^{-6}$\\
        46.8794 & $ 1 \times 10^{-6}$ & $2 \times 10^{-5}$\\
        48.5846 & $ 6 \times 10^{-6}$ & $3 \times 10^{-6}$\\
        50.2605 & $ 6 \times 10^{-7}$ & $8 \times 10^{-7}$\\
        52.5642 & $ 3 \times 10^{-5}$ & $4 \times 10^{-6}$\\
        56.9221 & $ 3 \times 10^{-6}$ & $5 \times 10^{-7}$\\
        83.1969 & $ 1 \times 10^{-5}$ & $6 \times 10^{-7}$\\
        87.9381 & $ 9 \times 10^{-8}$ & $ 6 \times 10^{-7}$\\
        89.3304 & $ 2 \times 10^{-5}$ & $ 9 \times 10^{-6}$\\
        \hline
    \end{tabular}
    \caption[Eigenfunctions for the coupled quartic oscillator.]{Eigenenergies for the coupled quartic oscillator, together with the mean squared error of the eigenenergies and the eigenfunctions calculated with the variational method (see Appendix~\ref{appendix1}).}
    \label{tab:scars_mse}
\end{table}

%
%---------------------------------------------------------------------
%Figure: Wavefunctions E=1
%---------------------------------------------------------------------
\begin{figure}[!ht]
 \includegraphics[width=1.0\columnwidth]{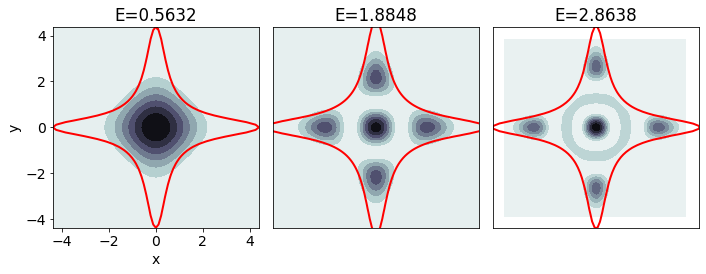}
  \caption{Probability density of the eigenfunctions of the coupled quartic oscillator obtained from propagating an initial wavepacket with energy $E=1$.}
 \label{fig:scars_1}
\end{figure}
%----------------------------------------------------------------------
%---------------------------------------------------------------------------

%---------------------------------------------------------------------
%Figure: Wavefunctions E=10
%---------------------------------------------------------------------
\begin{figure}[!ht]
 \includegraphics[width=1.0\columnwidth]{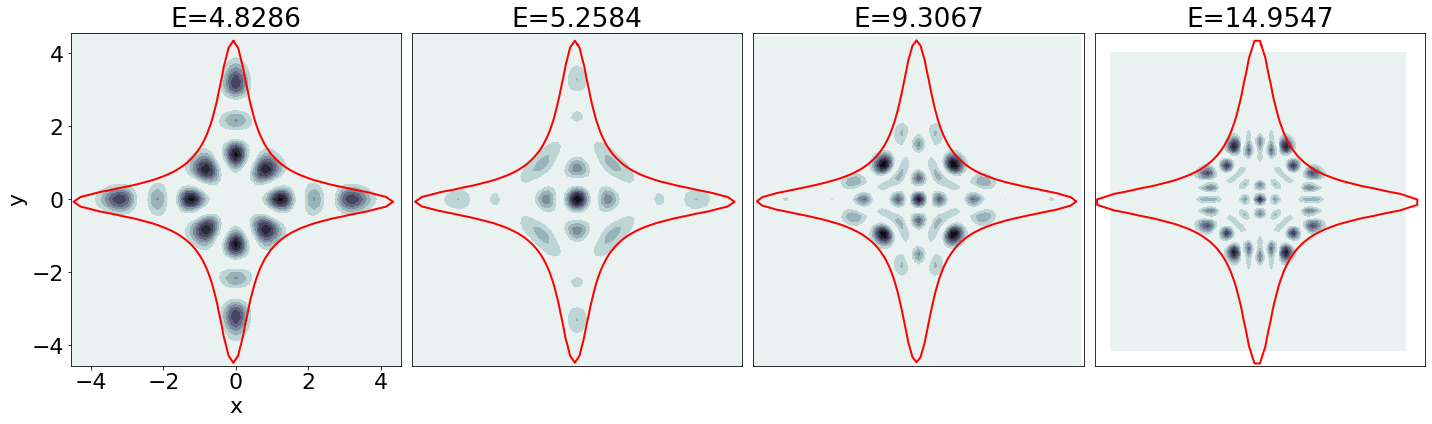}
  \caption[Probability density of the eigenfunctions of the coupled quartic oscillator with mean initial energy E=10.]{Probability density of the eigenfunctions of the coupled quartic oscillator obtained from propagating an initial wavepacket with energy $E=10$, scaled to the domain of $E=1$. The red lines show the equipotential curves at $V(x,y)=1$.}
 \label{fig:scars_2}
\end{figure}
%----------------------------------------------------------------------
%---------------------------------------------------------------------------

%---------------------------------------------------------------------
%Figure: Wavefunctions E=100
%---------------------------------------------------------------------
\begin{figure*}[!ht]
 \includegraphics[width=1.0\linewidth]{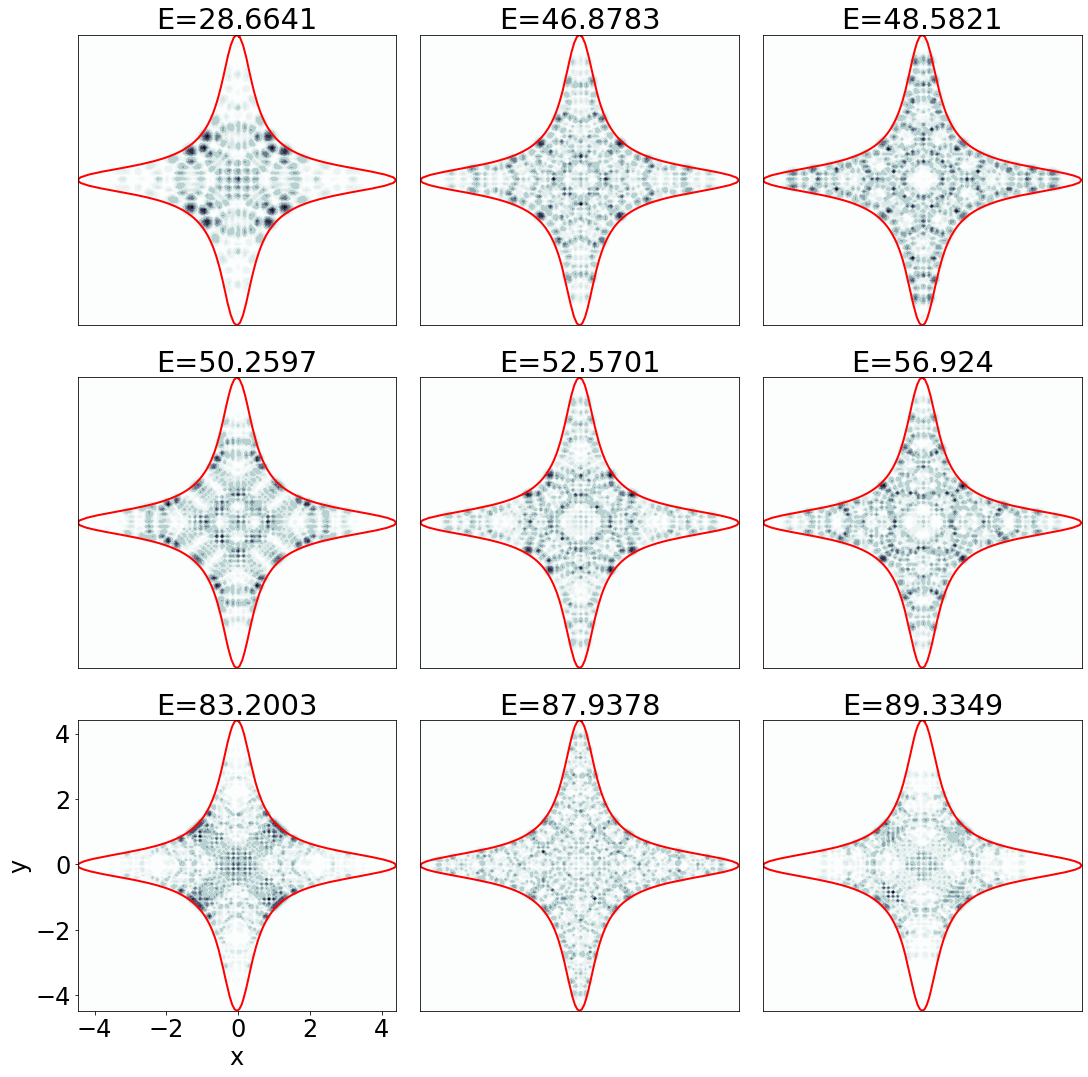}
  \caption{Same as Fig.~\ref{fig:scars_2} for mean initial energy $E=100$.}
 \label{fig:scars_3}
\end{figure*}
%----------------------------------------------------------------------
%----------------------------------------------------------------------------

In this subsection, we examine the performance of the multi-step RC method to obtain both the eigenfunctions and scar functions of the coupled quartic oscillator. 

\subsubsection{Eigenfunctions}
Figures~\ref{fig:scars_1},~\ref{fig:scars_2} and~\ref{fig:scars_3} show the obtained eigenfunctions and eigenenergies for the three initial wavepackets centred at energies $E=1, 10$ and $100$, respectively. Because of the symmetry imposed on the initial wavefunction, we only obtain the totally symmetric (with respect to the axis and the diagonals) eigenfunctions, belonging to the $A_1$ representation of the $C_{4v}$ symmetry group. By changing the symmetry of the initial wavefunctions, eigenfunctions belonging to another symmetry class could be obtained. 

For comparison, we have also calculated the eigenenergies and eigenfunctions using the variational method (see Appendix~\ref{appendix1}). The MSE of the energies and eigenfunctions computed with the multi-step RC method are provided in Table~\ref{tab:scars_mse}. The MSE of the eigenenergies increases with the energy of the system since the eigenfunctions become more complex and thus harder for the multi-step RC algorithm to predict its time evolution. However, it is seen that the MSE of all the energies and eigenfunctions is of the order or smaller than $1\times10^{-5}$, which means that the ML model can correctly propagate the quantum states in time. Therefore, the multi-step RC, being agnostic to the underlying physical model of the system, can also reproduce the dynamics of complex quantum chaotic systems, proving the versatility of the method. 

The advantage of using a ML algorithm instead of a classical numerical integrator is the execution time needed to propagate the initial state in time. Table~\ref{tab:scars_time} shows the execution time for the FFT integrator and the multi-step RC models. These results show that multi-step RC is much faster than the classical integrator, which makes it suitable for predicting the long-time evolution of quantum states. 

\begin{table}[!ht]
    \centering
    \begin{tabular}{l|ccc}
    \hline
    \hline
         System & Method &  Execution time \\
    \hline\\[-0.3cm]
        Eigenfunctions E=1  & FFT & 3h 35min \\
        Eigenfunctions E=1  & multi-step RC & 10 min\\[0.3cm]
  
        Eigenfunctions E=10  & FFT & 4h 50min \\
        Eigenfunctions E=10  & multi-step RC & 36 min\\[0.3cm]
        Eigenfunctions E=100  & FFT & 24h 47min \\
        Eigenfunctions E=100  & multi-step RC & 1h 34min\\[0.3cm]
        Horizontal scar  & FFT & 2h 58min \\
        Horizontal scar  & multi-step RC & 12min\\[0.3cm]
        Quadruple-loop scar  & FFT & 1h 32min \\
        Quadruple-loop scar  & multi-step RC & 7min\\[0.3cm]
        Square scar  & FFT & 3h 44min \\
        Square scar  & multi-step RC & 25min\\[0.3cm]
        Triangle scar  & FFT & 3h 27min \\
        Triangle scar  & multi-step RC & 24min\\      
    \hline
    \end{tabular}
    \caption{Execution times for the Fast Fourier Transform (FFT) integrator and the multi-step RC model, for the eigenfunctions and scar functions for the coupled quartic oscillator.}
    \label{tab:scars_time}
\end{table}

%
%---------------------------------------------------------------------
%Figure: Example spectrum
%---------------------------------------------------------------------
\begin{figure*}[!ht]
\centering
 \includegraphics[width=0.85\textwidth]{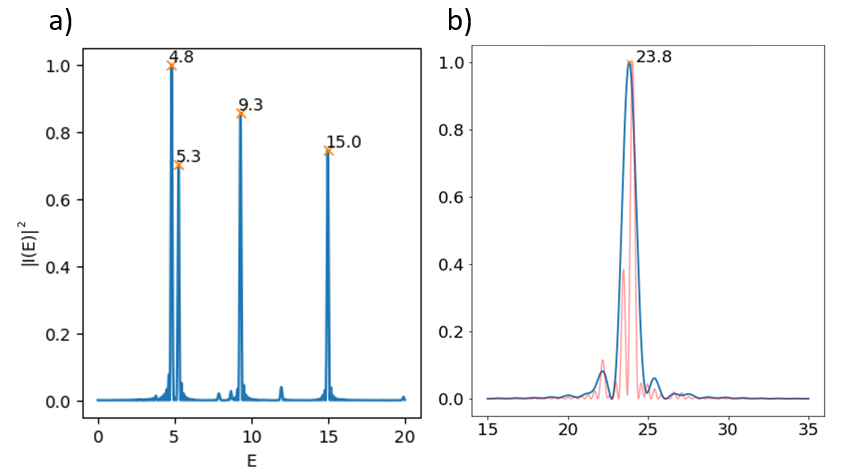}
  \caption[a) Energy spectrum $|I(E)|^2$ for the coupled quartic oscillator at $E=10$. b) Low-resolution energy spectrum.]{a) Example of the modulus of the energy spectrum $|I(E)|^2$ for the coupled quartic oscillator at $E=10$. b) Long-time energy spectrum $|I(E)|^2$ (in red) and its low-resolution version (in blue), for the quadruple-loop scar function at energy $E = 22.824$.}
 \label{fig:scars_spectrum}
\end{figure*}
%----------------------------------------------------------------------

%
%---------------------------------------------------------------------
%Figure 4: Scars horizontal
%---------------------------------------------------------------------
\begin{figure*}[!ht]
 \includegraphics[width=1.0\linewidth]{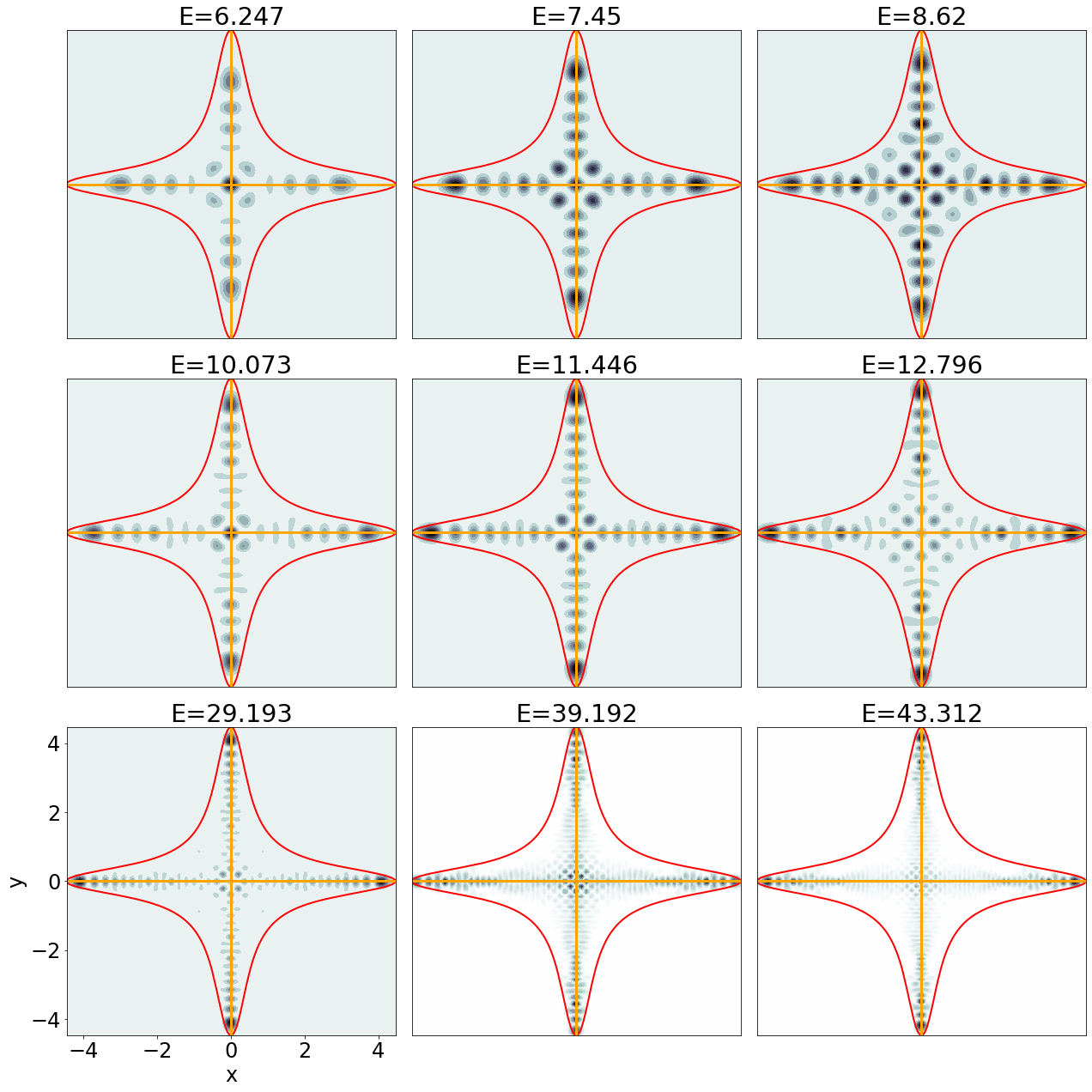}
  \caption{Probability density $|\psi(x,y)|^2$ of the scar function for the horizontal periodic orbit scaled at $E=1$, at different energies (see Table~\ref{tab:scars_PO}), together with the equipotential curve $V(x,y)=1$.}
 \label{fig:scars_4}
\end{figure*}
%----------------------------------------------------------------------
%----------------------------------------------------------------------------

%
%---------------------------------------------------------------------
%Figure 5: Scars Quadruple-loop
%---------------------------------------------------------------------
\begin{figure*}[!ht]
 \includegraphics[width=1.0\linewidth]{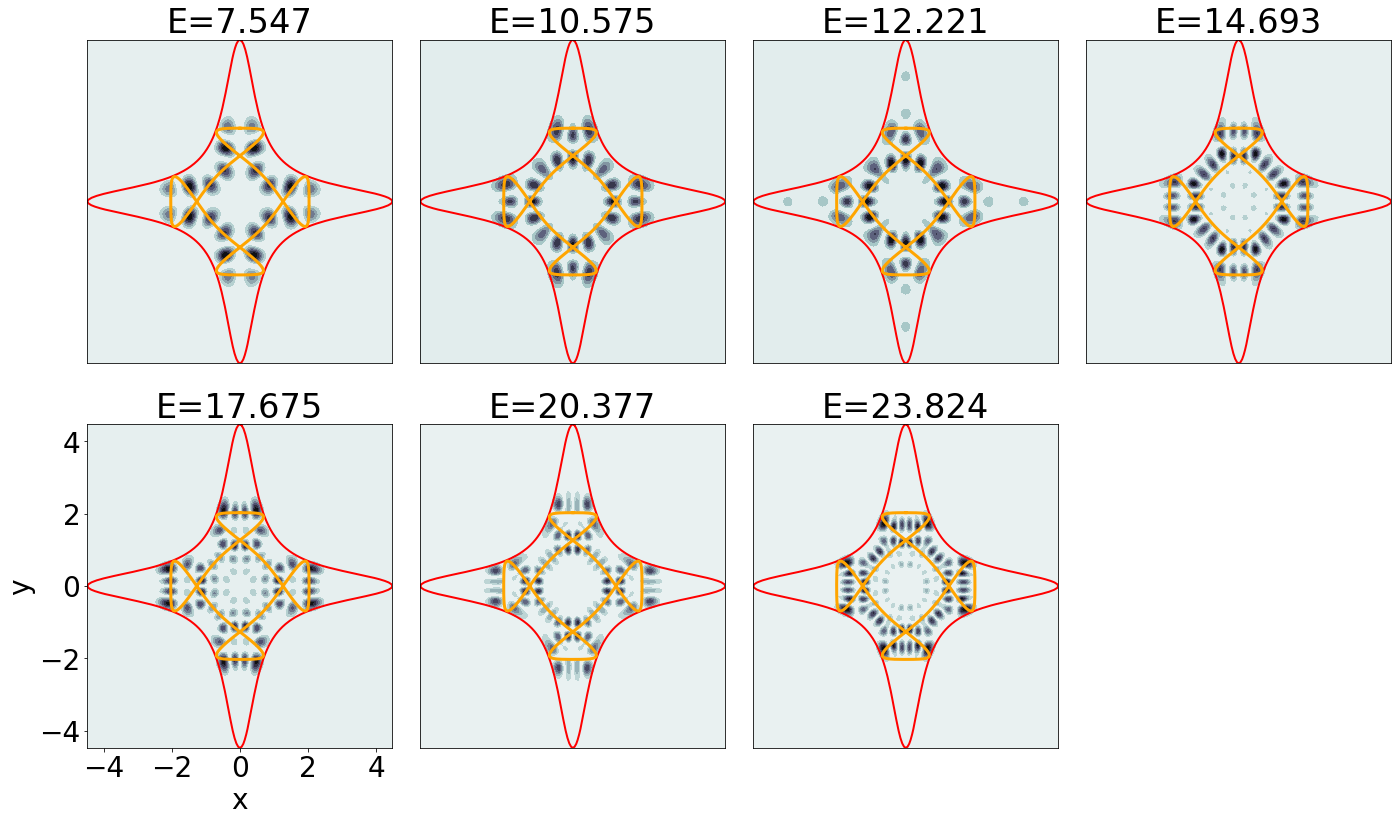}
  \caption{Same as Fig.~\ref{fig:scars_4} for the quadruple-loop periodic orbit.  }
 \label{fig:scars_5}
\end{figure*}
%----------------------------------------------------------------------
%----------------------------------------------------------------------------

%
%---------------------------------------------------------------------
%Figure 6: Scars square
%---------------------------------------------------------------------
\begin{figure*}[!ht]
 \includegraphics[width=1.0\linewidth]{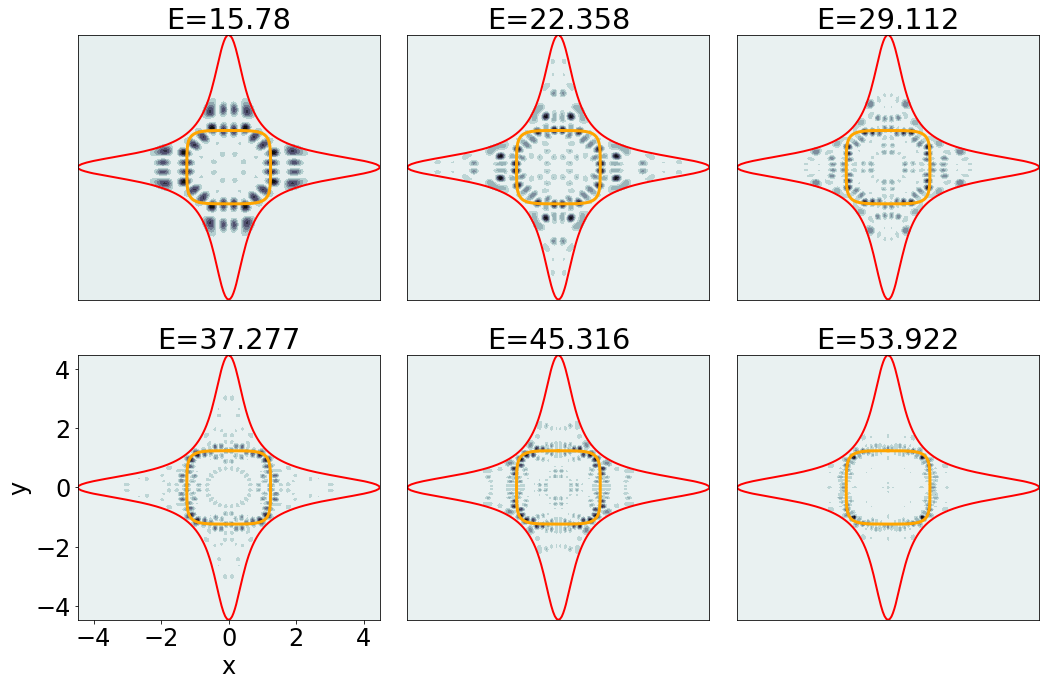}
  \caption{Same as Fig.~\ref{fig:scars_4} for the square periodic orbit. }
 \label{fig:scars_6}
\end{figure*}
%----------------------------------------------------------------------
%----------------------------------------------------------------------------

%
%---------------------------------------------------------------------
%Figure 7: Scars triangle
%---------------------------------------------------------------------
\begin{figure*}[!ht]
 \includegraphics[width=1.0\linewidth]{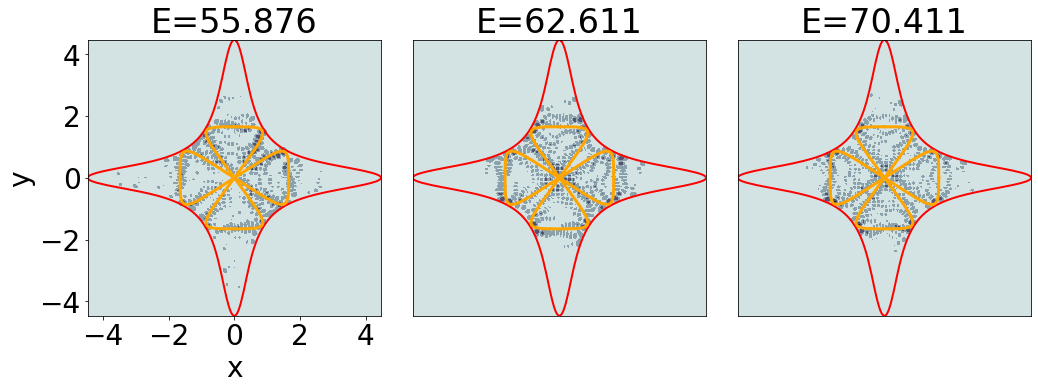}
  \caption{Same as Fig.~\ref{fig:scars_4} for the triangle periodic orbit.  }
 \label{fig:scars_7}
\end{figure*}
%----------------------------------------------------------------------
%----------------------------------------------------------------------------

%
%---------------------------------------------------------------------
%Figure 8: Semiclassical energies
%---------------------------------------------------------------------
\begin{figure*}[!ht]
 \includegraphics[width=1.0\textwidth]{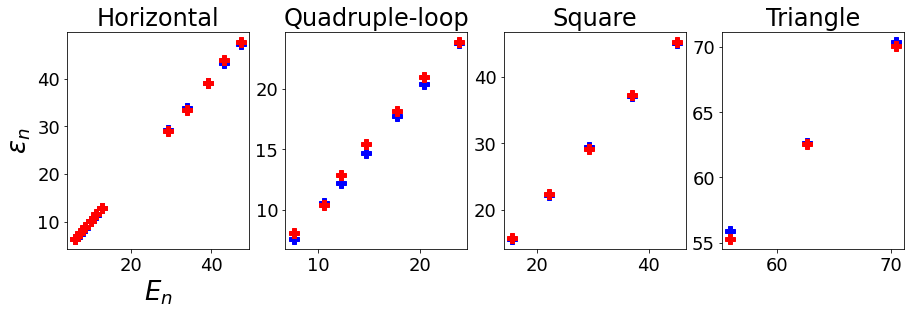}
  \caption[Bohr-Sommerfeld energies and reservoir computing energies for the periodic orbits of the coupled quartic oscillator.]{Bohr-Sommerfeld energies $\epsilon_n$ (see Eq.~\ref{eq:Bohr-sommerfeld-energies}) for the four periodic orbits (see Fig.~\ref{fig:scars_po}) studied in this work for the coupled quartic oscillator, compared to the energy $E_n$ obtained with the multi-step reservoir computing method.}
 \label{fig:scars_8}
\end{figure*}
%----------------------------------------------------------------------
%----------------------------------------------------------------------------

\subsubsection{Scar functions}

Apart from the eigenfunctions, we have also used the multi-step RC to calculate the scar functions at different energies, corresponding to four periodic orbits shown in Fig.~\ref{fig:scars_po}. As explained in Sect.~\ref{sect:quartic}, the energies of the scar functions are obtained by finding the peaks of the low-resolution energy spectrum. An example of the high-resolution and low-resolution spectrum is given in Fig.~\ref{fig:scars_spectrum}. On the left panel, we have an example of a high-resolution energy spectrum, used to calculate the eigenfunctions around energy $E=10$. Figure~\ref{fig:scars_spectrum} b) shows the low-resolution (resolved up to the Ehrenfest time) energy spectrum for the quadruple-loop scar function at energy $E = 22.824$ (in blue), together with a longer-time spectrum (in red). We see that for the longer-time spectrum, the energy peaks are more pronounced, while the low-resolution spectrum has wider peaks, which contain several high-resolution peaks. 

Once the energies of the scar functions have been calculated, we obtain the scar functions as explained in Sect.~\ref{sect:quartic}. Figures~\ref{fig:scars_4},~\ref{fig:scars_5},~\ref{fig:scars_6} and~\ref{fig:scars_7} show the calculated scar functions for the four periodic orbits at different energies, scaled to energy $E=1$ for easier visualization. We see that all the scar functions are localized around their associated periodic orbit. However, we see that there are sometimes interactions between periodic orbits. For example, the scar function obtained in Fig.~\ref{fig:scars_6} for energy $E=15.78$ is localized around the square periodic orbit, but it also has visible and easily countable nodes around the quadruple-loop periodic orbit. This happens because the time-propagation of the initial wavefunction gathers information about the phase space around both periodic orbits, which is then shown in the resulting scar function. Moreover, notice that in Fig.~\ref{fig:scars_5}, there is a quadruple-loop scar function at a similar energy ($E=14.693$) to the square scar function. On the other hand, there are other scar functions totally localized around one single periodic orbit (see, for example, the horizontal scar functions in Fig.~\ref{fig:scars_4}, or the square scar function at energy $E=53.922$ in Fig.~\ref{fig:scars_6}).

We have compared the energies obtained with our method with the Bohr-Sommerfeld quantized energies. The results are shown in Fig.~\ref{fig:scars_8}. As can be seen, the energies obtained with both methods are very similar. We do expect, however, small differences between the Bohr-Sommerfeld energies and the quantum energies obtained with the RC method. First, the Bohr-Sommerfeld energies are a semiclassical approximation of the quantum energies. Moreover, the Bohr-Sommerfeld energies only take into account the influence of a single periodic orbit. We have already seen that some of the resulting scar functions obtained with the RC method contain small interactions between different periodic orbits, which also causes small differences in the obtained energies.

In conclusion, the results of this section show that the multi-step RC method is able to predict the time evolution of the coupled quartic oscillator for different initial conditions, and obtain, with high accuracy, the associated eigenenegies, eigenfunctions and scar functions. This ML algorithm is faster than the traditional FFT solver and is easily adaptable to different systems.

\chapter{Quantum Reservoir Computing}
\label{chapter5}

\begin{fquote}[Seth Lloyd][1960-]
    The universe is a quantum computer, and the substrate of all computation is the universe itself.
\end{fquote}

In Chapter \ref{chapter3}, we studied how ML (Machine Learning, see Sect. \ref{chapter1}) methods and, in particular, RC (Reservoir Computing, see sect. \ref{sect:RC}) can be used to solve the Schrödinger equation for vibrational Hamiltonians. In particular, we developed a RC-based method that allows us to integrate the time-dependent Schrödinger equation. One of the main challenges when dealing with molecular quantum states is the representation of wavefunctions in high-dimensional arrays. That is, in Chapter \ref{chapter3}, each quantum wavefunction was represented with a matrix, whose dimension scales exponentially with the size of the system (the dimension of the vector $\vec{x}$ representing geometrical configurations). For example, the simple 1D Morse oscillator could be described using only a vector of size 140, while a matrix with 4000 entries was required to represent the more complex coupled Morse oscillator (see Sect. \ref{sect:paper3}).

Even though the multi-step RC method has proven to be successful in predicting the eigenstates and even scar functions of complex quantum systems, the high-dimensional scaling of the input data is a clear bottleneck for the RC algorithm and any traditional integration method. As the dimension of the input matrices grows, larger reservoirs are needed to capture the system's dynamics, consequently requiring more extensive training datasets. Consequently, training the RC algorithm effectively would entail solving the Schrödinger equation with a numerical integration method over a longer time period, which would significantly increase the total computational time, thus reducing the advantages of using RC.

The same limitation applies to the NN (Neural Network, see Sect.~\ref{sect:NNs})  approach to calculate the eigenstates of molecular Hamiltonians. Since the input and output data samples are high-dimensional matrices, complex NNs are required to make accurate predictions. This results in having a large number of training parameters, which require long training times and expensive computational resources. 

For this reason, in this thesis, we propose to study a novel technology that solves this limitation and has the potential to dramatically improve the performance of some ML algorithms: quantum computing. Because of the exponential scaling of the Hilbert space, quantum computing can deal with high-dimensional data without requiring exponentially many computational resources. QML (Quantum Machine Learning, see Sect.~\ref{sect:QML}) is a relatively new field compared to ML and deep learning. Current quantum computers have limitations in terms of the number of qubits and experiments available for users and produce noisy outputs when running complex quantum circuits. Therefore, rather than directly applying QML algorithms to solve the Schrödinger equation, we focus on designing QML methods that take into account the limitations of current hardware, that is, that are suitable for the NISQ era (Noisy Intermediate Scale Quantum era, introduced in Chapter~\ref{chapter1}).

The main algorithm studied in this chapter is QRC (Quantum Reservoir Computing, see Sect.~\ref{sect:QRC}). QRC is an adaptation of the classical RC that we have studied in the previous chapters. Because of the versatility of the RC framework, the method can be easily adapted to quantum computing. The resulting QRC algorithm, as we will see in this chapter, is useful for many different applications, such as time series forecasting~\citep{Domingo_quantum_zucchini}, quantum chemistry calculations~\citep{Domingo_optimalQRC,Domingo_QRCNoise} and hybrid quantum-classical NN design~\citep{Domingo_hybrid}. Moreover, just like its classical counterpart, it has a simple learning strategy, which only requires training a linear regression model. Moreover, it requires using only shallow circuits with simple quantum gates, which makes QRC  especially suitable for NISQ devices. For these reasons, QRC is currently a highly relevant quantum algorithm in QML.

The idea behind QRC consists of using the Hilbert space as an enhanced feature space of the input data. In this way, quantum entangling operations are used to extract valuable features from the data, which are then fed to a classical ML model, which predicts the desired target. In gate-based quantum computation, QRC consists of a random quantum circuit, the QR (Quantum Reservoir, see Sect.~\ref{sect:QRC}), applied to an initial state, which encodes the input data. As we just said, the goal of the QR is to extract valuable information from the input data so that the measurements of simple local operators are useful features to predict the output. These features are then fed to a classical ML algorithm, typically a linear model. To ensure that the extracted features contain enough information for learning the input-output relationships, the QR must be a complex quantum circuit. Therefore, careful design of the QR is essential for achieving optimal performance of the QML model. 

For this reason, the first part of this chapter (Sect.~\ref{sect:optimalQRC}) is devoted to studying the optimal design of QRs for QML. In this case, the QML task at study consists of finding excited energies of molecular Hamiltonians in quantum chemistry. Instead of solving the vibrational Schrödinger equation as in Chapter~\ref{chapter3}, the QRs will be used to integrate the \emph{electronic} Schrödinger equation. The reason for this is that there is a direct and standard mapping from the electronic Hamiltonian to the qubit space, which makes this task suitable for quantum computing algorithms.

The presence of noise in quantum devices is one of the biggest challenges in quantum computing, particularly QML. As a result, many efforts are being put into either correcting~\citep{errorCorrecting, Domingo_RL, ErrorCorrecting2} or mitigating~\citep{errorMitigation, errorMitigation2} those errors. In this thesis, however, we study how we can, instead of suppressing the noise in quantum devices, use such noise to improve the performance of QRC. For this reason, after obtaining the optimal QR design, in Sect.~\ref{sect:noise_QRC} we study how the presence of noise affects the performance of QRC. Surprisingly, the results demonstrate that under some circumstances, quantum noise can be used to improve the performance of QRC. That is, certain noise types can be beneficial to QRC, while others should be prioritized for correction~\citep{Domingo_QRCNoise}. The analysis of the optimal reservoir design is extended to temporal tasks in Sect.~\ref{sect:QRC_forecasting}, by studying the performance of different QRs in forecasting the zucchini time series presented in Chapter~\ref{chapter4}. 

Finally, Sect.~\ref{sect:hybrid_CNN} presents an application of QRC that allows acceleration of the drug design process. In particular, a hybrid quantum-classical NN is designed to predict the binding affinity of a drug candidate and its target protein. The results show that the use of hybrid NNs with QRs highly accelerates the training process and reduces the complexity of the network.

\section{Optimal design of quantum reservoirs}
\label{sect:optimalQRC}
QRs are random quantum circuits used to extract valuable features from the input data. These circuits are constructed by randomly selecting quantum gates from a specific family or distribution. The goal is to create a complex quantum circuit that can effectively extract useful features for predicting the output, based on the input data. To illustrate the importance of this fact, let us consider a simple circuit composed only of one-qubit gates. Such circuit lacks the ability to generate entanglement between different inputs, thus failing to capture intricate relationships among the inputs. On the other hand, a circuit that incorporates quantum entangling operations can effectively compute complex relationships among the inputs, discovering hidden patterns that can be used to accurately predict the output.

Therefore, to design optimal QRs in terms of performance in QML, the key consideration is to create circuits that exhibit significant \emph{complexity}. This naturally raises the question: How do we measure the complexity of a quantum circuit? Various approaches have been developed to estimate the complexity of different families of quantum circuits. Notably, the concept of the majorization principle, initially introduced in Ref.~\citep{majorization_original}, has emerged as a compelling indicator of complexity for random quantum circuits. In Ref.~\citep{majorization}, it was demonstrated that this principle accurately captures the complexity of quantum circuits. Moreover, the majorization indicator offers the advantage of requiring few circuit executions to evaluate, making it very suitable for NISQ devices.

 In this thesis, we show that, apart from being an indicator of complexity, the majorization criterion also serves as an indicator of \emph{performance} of the QRs~\citep{Domingo_optimalQRC}. In this way, the majorization criterion will be used to select the optimal quantum circuits that will serve as QRs. The resulting quantum circuits will be easily realized in NISQ computers, representing a significant advantage over other commonly used QRs. 

The performance of QRC will be assessed using different families of quantum circuits, which have different complexity according to the majorization principle. Also, the number of gates needed for each family to obtain optimal performance will be studied. In NISQ devices, this number should be as small as possible to minimize error propagation due to large error rates and short coherence times. The results will show that the optimal quantum circuits provided by the majorization criterion require significantly fewer quantum gates than the transverse-field Ising model, which has been widely used as a QR~\citep{QRC2, quantumchemQRC, OptQRC, QRC}. 

The organization of the next subsections is as follows. In Sect.~\ref{sect:majorization}, the majorization principle is introduced. Then, Sect.~\ref{sect:families_gates} presents the families of quantum circuits used for this study and the implementation details. The QML task is described in Sect.~\ref{sect:chemistry_setting}, which consists of predicting the first and second excited energies of the electronic Hamiltonian for the LiH and H$_2$O molecules. Finally, Sects.~\ref{sect:performance_QRC},~\ref{sect:distribution_QRC} and~\ref{sect:Ising} show the results of this study.
 
\subsection{Majorization criterion}
\label{sect:majorization}
\begin{figure}[!ht]
    \centering
    \includegraphics[width=1.0 \textwidth]{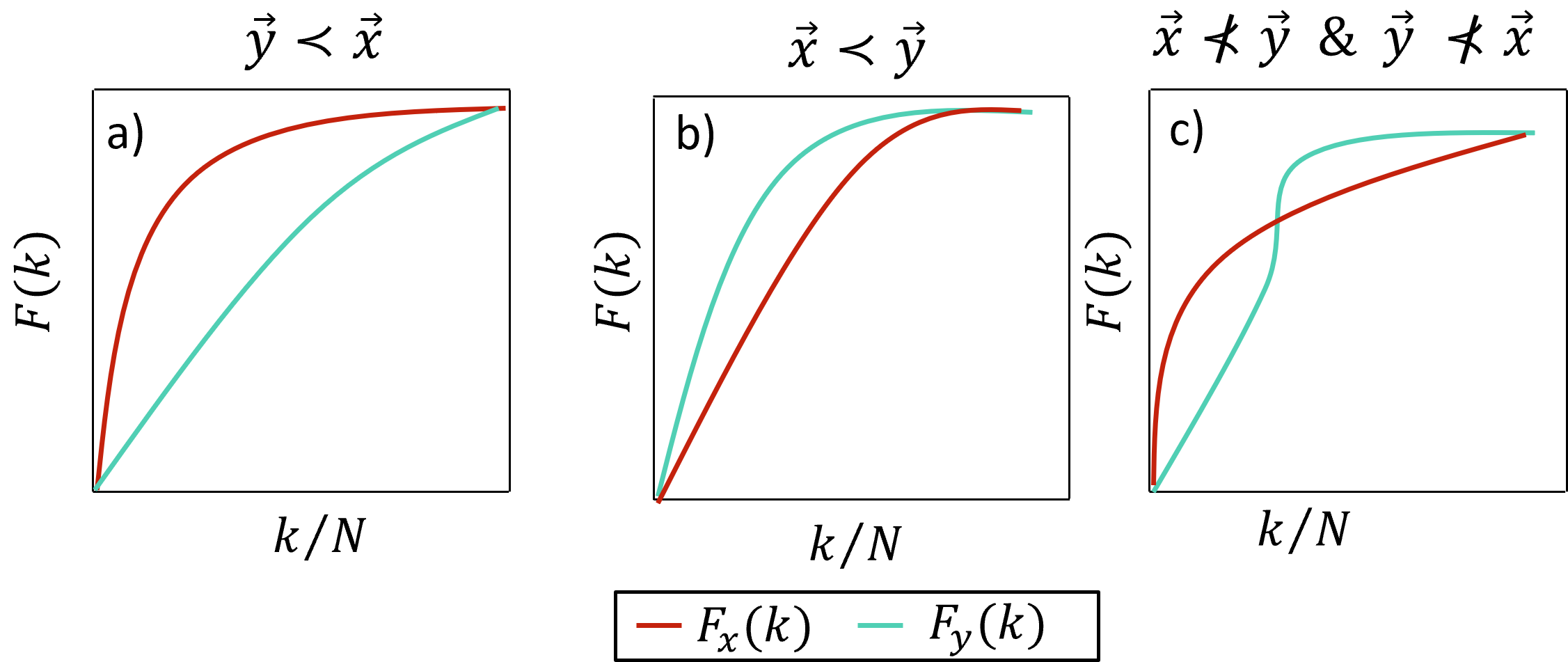}
    \caption[Illustration of the majorization criterion.]{Illustration of the majorization criterion. Consider two probability vectors $\vec{x},\vec{y}$ and their cumulant functions $F_x(k), F_y(k)$. In a), $\vec{y}$ majorizes $\vec{x}$ ($\vec{x} \prec \vec{y}$). In b), $\vec{x}$ majorizes $\vec{y}$ ($\vec{y} \prec \vec{x}$). In c), however, $\vec{x}$ does not majorize $\vec{y}$ and $\vec{y}$ does not majorize $\vec{x}$.}
    \label{fig:majorization}
\end{figure}

Majorization is a mathematical concept that allows one to
decide whether a probability distribution is more disordered than another. Let $\vec{x}, \vec{y} \in \mathbb{R}^n$ be probability vectors, i.e., real vectors of non-negative components and normalized to unity: $0 \leq x_i, y_i \leq 1 \ \forall i \in \{1, \dots N\}$, $\sum_i x_i = \sum_i y_i = 1$. We say that $\vec{y}$ majorizes $\vec{x}$ ($\vec{x} \prec \vec{y}$) if for all $k<N$

\begin{equation}
    \sum_{i=1}^k x_i^\downarrow < \sum_{i=1}^k y_i^\downarrow, \qquad 
    \sum_{i=1}^N x_i^\downarrow = \sum_{i=1}^N y_i^\downarrow,
\end{equation}
where $x^\downarrow$ indicates that the vector $\vec{x}$ has been arranged in non-increasing order. The partial sums in the equation above are called cumulants, where the $k$-th cumulant is denoted as $F_x(k)$. A useful way of visualizing the majorization (partial) ordering is by plotting the Lorenz curves $F_x(k)$ and $F_y(k)$ as a function of $k/N$~\citep{cumulants}, as can be seen in the example of Fig.~\ref{fig:majorization}. Then, $\vec{y} \prec \vec{x}$ if the Lorenz curve for $\vec{x}$ is above the curve for $y$ for all values of $k/N$ (see Fig.~\ref{fig:majorization} a)). On the other hand, if the Lorenz curve for $\vec{y}$ is always above the curve for $\vec{x}$, then $\vec{x} \prec \vec{y}$ (see Fig.~\ref{fig:majorization} b)). Notice that there exist vectors $\vec{x}$ and $\vec{y}$ for which neither $\vec{x}$ majorizes $\vec{y}$ nor $\vec{y}$ majorizes $\vec{x}$ (see Fig.~\ref{fig:majorization} c)). 

Recently, in Ref.~\citep{majorization} the authors used the majorization concept to measure the complexity of random quantum circuits. In their work, they built random circuits and evaluated the final state on the computational basis, obtaining the associated probabilities $\vec{x}$. Then, the authors showed that the fluctuations (standard deviation) of the Lorenz curve $F_x(k)$ for a given \emph{family} of random circuit qualify as an indicator of the complexity of such family. 

In this way, the indicator could discriminate between universal and
non-universal classes of random quantum circuits. Recall, from Sect.~\ref{sect:Quantum_gates}, that a universal class of quantum circuits can be used to approximate any quantum computation with arbitrary precision. On the other hand, a non-universal set of quantum circuits cannot approximate any desired quantum computation and thus they do not have the full computational power of a universal set. As a result, universal families are more complex than non-universal ones. Moreover, the majorization criterion could also detect the complexity of some non-universal but not classically
efficiently simulatable quantum random circuits, making it a very precise complexity indicator.

Compared to other complexity criteria, such as the entanglement spectrum~\citep{entanglement_spectrum}, evaluating the majorization in a quantum circuit requires significantly fewer operations. This makes the majorization principle a criterion especially suitable for the NISQ era, where quantum computation has to be performed with limited quantum resources. 

In this work, we consider seven families of quantum circuits that have different complexity according to the majorization principle. Multiple QRs will be sampled from these seven families, and their performance in QRC will be assessed. The results will show that the QRs with higher complexity according to the majorization criterion provide better results in the QML task.

\subsection{Families of quantum circuits}
\label{sect:families_gates}
Seven families of quantum circuits, with different complexity according to the majorization criterion, are studied. For a given family, the quantum circuit composing the QR is built by adding a fixed number of random quantum gates uniformly sampled from such family. Let us describe the seven families of quantum circuits.

\begin{enumerate}
    \item \textbf{G1:} The quantum circuit is constructed from the generator G1 = \{CNOT, H, X\}, where CNOT is the controlled-NOT gate, H stands for Hadamard, and X is the NOT-gate, as explained in Sect.~\ref{sect:Quantum_gates}. Recall that the Hadamard gate allows to create a superposition between the basis states $\ket{0}, \ket{1}$, and the CNOT gate generates entanglement between two qubits. The set G1 generates a subgroup of Clifford~\citep{G1} group, and thus is non-universal and efficiently classically simulatable. Recall that the Clifford group transforms Pauli matrices to other Pauli matrices, and therefore it cannot generate complex (non-Clifford) gates.
    \item \textbf{G2:} The quantum circuit is constructed from the generator G2 = \{CNOT, H, S\}, with S being the $\pi/4$ phase gate (also known as $\sqrt{Z}$ gate, see again Sect.~\ref{sect:Quantum_gates}). The circuits constructed from G2 generate the whole Clifford group~\citep{G2}, so they are non-universal and classically simulatable, but more complex than G1 circuits.
    \item \textbf{G3:} The quantum circuit is constructed from the generator G3=\{CNOT, H, T\}, where T is the $\pi/8$ phase gate (or $\sqrt[4]{Z}$). The G3 family is universal (as discussed in Sect.~\ref{sect:Quantum_gates}) and thus approximates the full unitary group $U(N)$ to arbitrary precision.
    \item \textbf{Matchgates (\gls{MG}):} MG are two-qubit gates formed 
    from two one-qubit gates, A and B, with the same determinant. A acts on the subspace spanned by $\ket{00}$ and $\ket{11}$, while B acts on the subspace  spanned by $\ket{01}$ and $\ket{10}$. A and B are randomly sampled from the unitary group U(2):

    \begin{equation}
        G(A,B) = 
        \begin{pmatrix} 
            a_1 & 0 & 0 & a_2 \\
            0 & b_1 & b_2 & 0 \\
            0 & b_3 & b_4 & 0 \\
            a_3 & 0 & 0 & a_4
        \end{pmatrix}, \quad |A| = |B|.
    \end{equation}
    Matchgates circuits are also universal (except when acting on nearest neighbor lines only)~\citep{matchgates1, matchgates2}. 
    \item \textbf{Diagonal-gate circuits (\gls{D2}, \gls{D3}, \gls{Dn}):} The last families of gates are diagonal in the computational basis. The diagonal gates are separated into 3 families: $D_2$, $D_3$ and $D_n$ (where $n$ is the number of qubits). Here, $D_2$ gates are applied to pairs of qubits, $D_3$ gates are applied to 3 qubits and $D_n$ gates are applied to all the qubits.
    \begin{equation}
        D_k(\phi_1, \cdots, \phi_{2^k}) = 
        \begin{pmatrix} 
            e^{i\phi_1} & 0 & \cdots & 0  \\
            0 & e^{i\phi_2} & \cdots & 0 \\
            \vdots & \vdots  & \ddots & \vdots \\
             0 & 0 & \cdots & e^{i\phi_{2^k}}
        \end{pmatrix}, \quad k \in \{2,3,n\},
    \end{equation}
    with $\phi_i$ chosen uniformly from $[0, 2\pi) \ \forall i$. The gates are applied on all combinations of $k$ (out of $n$) qubits, the ordering being random. At the beginning and the end of the circuit (after the initialization of the state), Hadamard gates are applied to all qubits. As diagonal gates commute, they can be applied simultaneously. Diagonal circuits cannot perform universal computation but they are not always efficiently classically simulatable~\citep{diagonals}. As opposed to the other families of circuits, which can be of arbitrary depth, the diagonal $D_2$, $D_3$ and $D_n$ families contain a fixed number of gates, being $\binom{n}{2}$, $\binom{n}{3}$ and 1 gates, respectively. 
\end{enumerate}

These seven families (G1, G2, G3, MG, $D_2$, $D_3$ and $D_n$) can be ordered in terms of complexity according to the majorization principle, as it was demonstrated in Ref.~\citep{majorization}. The G3 family has the highest complexity, followed closely by the MG family. This agrees with the fact that both families are universal, and thus approximate the whole space of operators. The diagonal circuits $D_3$ and $D_n$ are next in terms of complexity. The $D_2$ family has a slightly smaller complexity. Finally, G2 and G1 are the less complex families of circuits, the complexity of G2 being slightly higher than the complexity of G1. Again, this agrees with the two families not being universal and the G1 being a subgroup of the Clifford group. 

 Regarding our simulations, we will perform 400 simulations for each of the seven families of quantum circuits, where we train and evaluate a QRC model sampled from a given family, to perform a QML task (described in Sect.~\ref{sect:chemistry_setting}). For this purpose, we construct circuits of 20, 50, 100, 150 and 200 gates for the G1, G2, G3 and MG families, and assess the influence of the number of gates in the final performance of the QRC model. For the G3 family, we also construct circuits with up to 1000 gates to study the QRC performance when using deep circuits (which have a lot of quantum gates). On the other hand, with the MG we construct circuits with 5, 10 and 15 gates. Recall that the diagonal circuits have a fixed number of gates so that the number of gates is kept constant for these circuits. 

Additionally, we compare the studied families with the Ising model, which has been widely used as a QR in the literature~\citep{QRC2, quantumchemQRC, OptQRC,QRC}. The transverse-field Ising Hamiltonian is a model of a many-body system that has an exact analytical solution~\citep{Ising}.  It describes a spin-1/2 chain with only nearest-neighbour interaction with an external magnetic field.  In gate-based quantum computing, the quantum circuit performs the time evolution of a quantum state under the random transverse-field Ising Hamiltonian

\begin{equation}
    \hat{H}_{\text{Ising}} = \sum_{i,j=0}^{N-1} J_{ij} Z_iZ_j + \sum_{i}^{N-1} h_{i} X_i,
    \label{eq:Ising}
\end{equation}
where $X_i$ and $Z_j$ are Pauli operators acting on the site $i, j$-th qubit, and the coefficients $h_i$ and $J_{ij}$ are chosen according to Ref.~\citep{DynamicalIsing}, which provide a state of the art method to select optimal parameters of the Ising model for QRC. In this case, $J_{ij}$ are sampled from the uniform distribution $U(-J_s/2, J_s/2)$ and $h_i = h$ are constant. The optimal parameters in Ref.~\citep{DynamicalIsing} fulfill $h/J_s = 0.1$. In our case, all time evolutions will be performed for a lapse of time $T =10$. With these parameters, in Sect.~\ref{sect:performance_QRC} we compare the performance in QRC of the seven families described above together with the performance of the Ising circuits.

In current NISQ devices, executing simpler quantum circuits is crucial to minimize the effect of noise and thus the errors present in our experiments. For this reason, is extremely valuable to design QRs that can be implemented with the lowest number of (simple) quantum gates. To this end, in Sect.~\ref{sect:Ising} we calculate the number of gates needed to implement the Ising model in a quantum circuit, compared to those needed for the other families of circuits. With these calculations, we will be able to determine which QRs are most suitable for the NISQ era, both in terms of ML performance and efficiency of implementation in current devices.

Before presenting the results of our analysis, let us describe, in the next subsection, the QML task: finding the electronic excited energies for the LiH and H$_2$O molecule.
 
\subsection{Quantum chemistry setting}
\label{sect:chemistry_setting}
To test the performance of each family of quantum circuits we need to define a ML task. In this case, QRC is applied to solve a quantum chemistry problem because of the advantages of using QML to study quantum data~\citep{2ndquant2}. Since the input data is already in a quantum state, quantum algorithms are suitable to infer the properties of the system, 
since they naturally perform quantum operations of the data.

Moreover, the exponential size of the Hilbert space with the number of qubits allows us to study systems that would not be tractable classically. For example, to represent a system with $d$ degrees of freedom, one would need to use a classical vector of dimension $2^d$, while only $d$ qubits would be needed in a quantum computer. NISQ devices can currently work with around hundreds of qubits~\citep{IBMRoadmap}, which classically would require using vectors a size of $2^{100}$, which are currently totally intractable. 

In our case, the quantum chemistry task consists of predicting the excited energies of the electronic Hamiltonian of two molecules, LiH and H$_2$O, given their ground state. This task is relevant since computing the excited states of a Hamiltonian is a harder task than computing its ground state. In fact, the ground state can be obtained as an output of another NISQ quantum algorithm, such as the variational quantum eigensolver~\citep{VQE} (see Sect.~\ref{sect:QML}).

In order to make quantum calculations with the electronic Hamiltonian of a molecule, such Hamiltonian has to be mapped to a qubit space, so that the ground state of the molecule could be the input of a quantum circuit. This mapping is performed in two steps~\citep{2ndquant2, 2ndquant1}. First, the corresponding second-quantized Hamiltonian is obtained, which essentially describes this Hamiltonian as a linear combination of products of fermionic creation and annihilation operators. Then, these fermionic operators are mapped into qubit Pauli operators via Jordan-Wigner transformations~\citep{JW}. 

Once the qubit Hamiltonian has been calculated, the corresponding ground state is calculated by (numerical) exact diagonalization. Such state is given as an input to the QRC algorithm. The algorithm is tested for two molecules, LiH and H$_2$O, in the configuration ranges: $R_{\text{LiH}} \in [0.5, 3.5]$ a.u., $R_{\text{OH}} \in [0.5,1.5]$ a.u., with fixed angle $\phi_{\text{HOH}}=104.45^\circ$. Let us provide details about the mapping from the electronic to the qubit space, as well as the training algorithm.

\subsubsection{Second-quantized Hamiltonian}

The electronic Hamiltonian of a molecule with $K$ nuclei and $N$ electrons, in atomic units, is given by
\begin{equation}
H_e(\vec{r}, \vec{R}) = -\sum_i \frac{\nabla^2}{2} - \sum_{i,I} \frac{Z_I}{|\vec{r}-\vec{R}_I|}  
   +\frac{1}{2} \sum_{i\neq j} \frac{1}{|\vec{r}_i-\vec{r}_j|},
   \label{eq:electronic_He}
\end{equation}
where $Z_I$, $\vec{R}_I$ are the atomic number and position of the $I$-th nucleus and $\vec{r}_i$ is the position of the $i$-th electron. The second quantization method projects the electronic Hamiltonian onto a basis set of $M$ wavefunctions $\{\phi_p(\vec{x}_i)\}_p$, that approximate the electron spin-orbitals,
where $\vec{x}_i=(\vec{r}_i, \sigma_i)$ represents the spatial and spin coordinates of the $i$-th electron. In this work, the electron wavefunctions $\phi_p(\vec{x}_i)$ are described using the standard STO-3G basis, where each orbital is approximated by a linear combination of three Gaussian distributions~\citep{STO-3G}. Then, the many-electron wavefunctions associated with Hamiltonian in Eq.~\ref{eq:electronic_He} are written as a Slater determinant of single electron wavefunctions, which automatically ensures antisymmetrization, as required by Pauli's exclusion principle 
\begin{equation}
    \psi(x_0,\cdots, x_{N-1}) = \displaystyle\frac{1}{\sqrt{N!}} 
   \begin{vmatrix} \phi_0(x_0) & \phi_1(x_0) &\cdots & \phi_{M-1}(x_0)\\
   \phi_0(x_1) & \phi_1(x_1) &\cdots & \phi_{M-1}(x_1)\\
   \vdots & \vdots & \ddots & \vdots\\ 
   \phi_0(x_{N-1}) & \phi_1(x_{N-1}) &\cdots & \phi_{M-1}(x_{N-1})\\
   \end{vmatrix} .
\end{equation}
This representation, where only one determinant is used, is called the Hartree-Fock approximation. Other post Hartree-Fock methods such as the coupled cluster and full configuration interaction use more complex trial wavefunctions, typically consisting of a linear combination of these determinants~\citep{2ndquant2}. The number of spin-orbitals $M$ is, in general, larger than the number of electrons of the system $N$. Then, a convenient way to make that the wavefunction contains only $N$ electrons is the use of an occupation number vector $\ket{f}$, whose elements are $f_p=0,1$ depending on whether the corresponding spin-orbital is empty or full, respectively (absent or present). Accordingly, in the case of a single Slater determinant, we have
\begin{equation}
    \psi(x_0, \cdots, x_{N-1}) = \ket{f_{M-1}, \cdots, f_0} = \ket{f}. 
\end{equation}
Electrons can be excited into single-electron spin-orbitals by the fermionic creation operators $a^\dagger_p$, and  de-excited by the corresponding annihilation operators $a_p$, according to the following expressions:
\begin{eqnarray}
    & & a_p \ket{f_{M-1}, \cdots, f_p, \cdots, f_0} = \delta_{f_p,1} \left[ (-1)^{\sum_{i=0}^{p-1}f_i}\ket{f_{M-1}, \cdots, f_p +1, \cdots, f_0} \right],\nonumber\\
    & & a_p^\dagger \ket{f_{M-1}, \cdots, f_p, \cdots, f_0} =
    \delta_{f_p,0} \left[ (-1)^{\sum_{i=0}^{p-1}f_i}\ket{f_{M-1}, \cdots, f_p +1, \cdots, f_0}  \right],
\end{eqnarray}
where the sum is addition modulo 2. With these definitions, the second quantization form of the electronic Hamiltonian becomes~\citep{2ndquant1}
\begin{equation}
    \hat{H} = \sum_{p,q} h_{pq} a^\dagger_p a_q + 
      \frac{1}{2}\sum_{p,q,r,s} h_{pqrs} a_p^\dagger a_q ^\dagger a_r a_s,
\end{equation}
with
\begin{eqnarray}
    h_{pq} = \int dx \; \phi_p^*(x) \Big(- \frac{\nabla^2}{2} 
    - \sum_{I} \frac{Z_I}{|r-R_I|} \Big)\phi_q(x), \nonumber\\
    h_{pqrs} = \int dx_1 dx_2 \; \frac{\phi_p^*(x_1)\phi_q^*(x_2) \phi_r(x_2) \phi_s(x_1)}{|\vec{r}_1-\vec{r}_2|}.
\end{eqnarray}
The first integral represents the electronic kinetic energy plus the Coulomb interaction with the nuclei, while the second one accounts for the electron-electron Coulomb repulsion.

\subsubsection{Reduction of spin orbitals basis set}

Before mapping the second-quantized Hamiltonian to the qubit space, we study whether it is possible to reduce the size of the spin-orbitals basis, since $N < M$ in general. Indeed, it is often the case that certain orbitals are very likely to be either fully occupied or fully empty (usually referred to as \textit{virtuals}) in all Slater determinants. As calculating the ground state energy is essentially a question of distributing electrons among orbitals, we can simplify the calculation by using this information. Specifically, we can remove spin orbitals from the calculation if their expected occupation number is close to 0 or 2~\citep{2ndquant1}, and reduce the calculation to only the ambiguously occupied ones. This is known as performing the calculation in a \textit{reduced active space}. The occupation of a given spin-orbital can be evaluated, within the natural molecular orbitals philosophy, by calculating the single-particle reduced density matrix $\rho^1$, defined as
\begin{equation}
    \rho^1_{ij} = \braket{\psi| a_i^\dagger a_j}{\psi},
\end{equation}
where $\ket{\psi} = \sum_f \alpha_f \ket{f}$ is any suitable molecular wavefunction. When diagonalized, the eigenvalues of $\rho^1$  are called natural orbital occupation numbers (\gls{NOON}s). Spin orbitals with a NOON close to 0 or 2 can be assumed completely empty or occupied, respectively, and they can be removed from the basis set. For example, in the LiH molecule considered in our calculations, orbital 0 has a NOON of 1.99992, which is almost 2. This means that these orbitals are always doubly occupied, 
and then any term including $a_0^\dagger, a_0, a_1^\dagger, a_1$
can be eliminated. Also, orbital 3 has a very small NOON. Therefore, we assume this orbital is never occupied, and thus the two corresponding fermion operators $a_6^\dagger, a_6, a_7^\dagger, a_7$ can be removed from the Hamiltonian. Thus, the resulting Hamiltonian has 8 spin orbitals, which will be mapped to 8 qubits.  For the $\mathrm{H_2O}$ molecule, the NOONs of the second and fourth orbitals are very large compared to the other orbitals. Thus, we remove the fermion operators $a_2^\dagger, a_2, a_3^\dagger, a_3, a_6^\dagger, a_6,a_7^\dagger$ and $a_7$  from the Hamiltonian, leading to a Hamiltonian with 10 spin orbitals. 

\subsubsection{Jordan-Wigner transformation}
The last step of our procedure is to transform the second-quantized Hamiltonian to the qubit space, using a Jordan-Wigner transformation~\citep{JW}, which maps the occupied state of a spin-orbital to the $\ket{0}$ or $\ket{1}$ state of a qubit. In formal terms, this can be expressed as
\begin{equation}
  \ket{f_{M-1}, \ldots, f_p, \ldots, f_0} \rightarrow \ket{q_{M-1}, \ldots, q_p, \ldots, q_0}, \quad q_p = f_p \in \{0,1\}.
\end{equation}
The fermionic creation and annihilation operators increase or decrease the occupation number of a spin-orbital by 1, and they also introduce a multiplicative phase factor. The qubit mappings of the operators preserve these features and are given by
\begin{eqnarray}
    a_p = Q_p \otimes Z_{p-1} \otimes \cdots \otimes Z_0, \nonumber\\
    a_p^\dagger = Q_p^\dagger \otimes Z_{p-1} \otimes \cdots \otimes Z_0,
\end{eqnarray}
where $Q_p = 1/2(X+iY)$, $Q_p^\dagger = (1/2)(X-iY)$. The $Q$ or $Q^\dagger$ operator change the occupation number of the target spin-orbital, while the string of $Z$ operators recovers the exchange phase factor $(-1)^{\sum_{i=0}^{p-1}f_i}$. Using the Jordan-Wigner transformation, the second-quantized fermionic Hamiltonian is mapped into a linear combination of products of single-qubit Pauli operators, as
\begin{equation}
    \hat{H} = \sum_j h_j \prod_i P_i^j,
    \label{eq:Hamiltonian_qubit}
\end{equation}
where $h_j$ is a real scalar coefficient, $P_i^j$ represents one of the Pauli operators $I, X, Y$ or $Z$, index $i$ denotes which qubit the operator acts on, and index $j$ denotes the term in the Hamiltonian. Finally, the Hamiltonian in Eq.\ref{eq:Hamiltonian_qubit} is diagonalized to obtain the ground state $\ket{\psi_0}_{\vec{R}}$ for a given molecule configuration $\vec{R}$. The mapping to the qubit space is done by using the OpenFermion software~\citep{OpenFermion}.

\subsubsection{Training details}

\begin{figure*}
\centering
\includegraphics[width=1.0\textwidth]{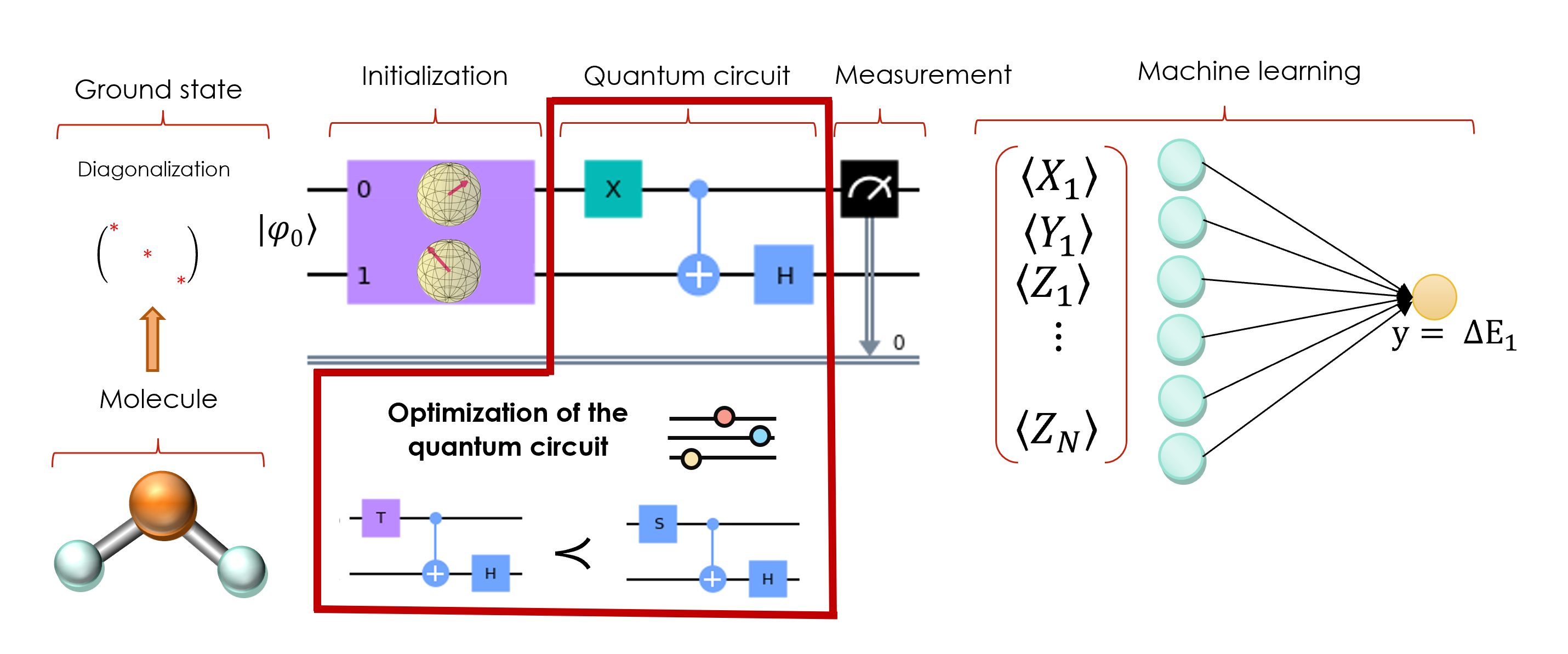}
\caption[Pipeline used to train the quantum reservoir computing model for the prediction of the electronic energy states.]{Pipeline used to train the quantum reservoir computing model. 
    First, the electronic Hamiltonian of the molecule is mapped to the qubit space, 
    and its ground state is calculated by direct diagonalization. 
    Such ground state is fed to the quantum reservoir, 
    which is a random quantum system sampled from one of the seven families studied in this work.
    Local Pauli operators are then measured and fed to the classical machine learning algorithm, 
    which predicts the excited energy of the molecule. 
    The choice of the quantum reservoir is optimized according to the majorization principle
    introduced in Ref.~\citep{majorization}. }
\label{fig:setting_QRC}
\end{figure*}

Given a ground state $\ket{\psi_0}_{\vec{R}}$ of a molecule in a certain configuration $\Vec{R}$, our goal is to predict the excited energy of a molecule, relative to its ground energy, using the seven families of QRs described in the previous section. The relative energy is given by
\begin{equation}
  \Delta E_1(\vec{R}) = E_1(\vec{R}) - E_0(\vec{R}),
  \label{eq:DeltaE}
\end{equation}
where $E_0(\vec{R})$ and $E_1(\vec{R})$ are the ground state and first excited energies, respectively. 

The QRC algorithm is trained using different molecular configurations (i.e. different bond lengths $\vec{R}$) so that the QRs aim to predict $\Delta E_1(\vec{R})$ for a given configuration $\vec{R}$. Then, the QRC algorithm is asked to predict $\Delta E_1(\vec{R})$ for \emph{new} values of $\vec{R}$. 
For the numerical simulation we split the data set $\{\ket{\psi_0}_{\vec{R}}, \Delta E_1(\vec{R})\}_R$ in training and test sets. The latter contains 30\% of the data, $R_{\text{LiH}} \in [1.1, 2.0]$ a.u.\ and $R_{\text{OH}} \in [1.05, 1.35]$ a.u., and it is chosen so that the reservoir has to extrapolate to unseen data. 

Once we have generated the training and test data, we design the pipeline of the QRC algorithm, which is schematically shown in Fig.~\ref{fig:setting_QRC}. 
The input of the QR is $\ket{\psi_0}_{\vec{R}}$. After applying the random circuit to the initial state, we measure the expectation values 
of the local Pauli operators $\{X_0, Z_0, \cdots, X_n, Z_n\}$, 
where $X_i, Z_i$ are the Pauli operators $X, Z$ applied to the $i$-th qubit. Notice that since the Pauli operators $X_1, \cdots, X_n$ 
(similarly for $Z_1, \cdots, Z_n$) commute with one another, 
the associated observables can be simultaneously measured. Therefore, the number of experiments needed to obtain all the observed values does not scale with the number of qubits. The measurements provide a classical vector $X(\vec{R})$ containing the extracted information 
from the ground state: 
\begin{equation}
    X(\vec{R}) = \left( \expval{X_0}, \expval{Z_0}, \cdots,  
        \expval{X_n}, \expval{Z_n} \right)^T ,
\end{equation}
where $n$ is the number of qubits, and the expectation values are taken over the output state of the circuit. 

This classical vector is then fed to a classical ridge regression (see Eq.~\ref{eq:ridge}), just as in classical RC. Notice that since the linear model has to extrapolate to unseen data (unseen values of $\vec{R}$), it is necessary to add regularization parameter $\gamma$ to the learning algorithm to prevent overfitting the training data. The regularization parameter $\gamma=10^{-7}$ is chosen to simultaneously prevent overfitting and provide accurate predictions. Although any other classical ML algorithm could indeed have been used instead of this linear model, at this point, the QR can extract valuable information from the quantum state. Thus, a simple linear model is plenty enough to predict the excited properties of the system.

\subsection{Performance of the quantum reservoirs}
\label{sect:performance_QRC}

\begin{figure*}[!ht]
 \includegraphics[width=1.0\textwidth]{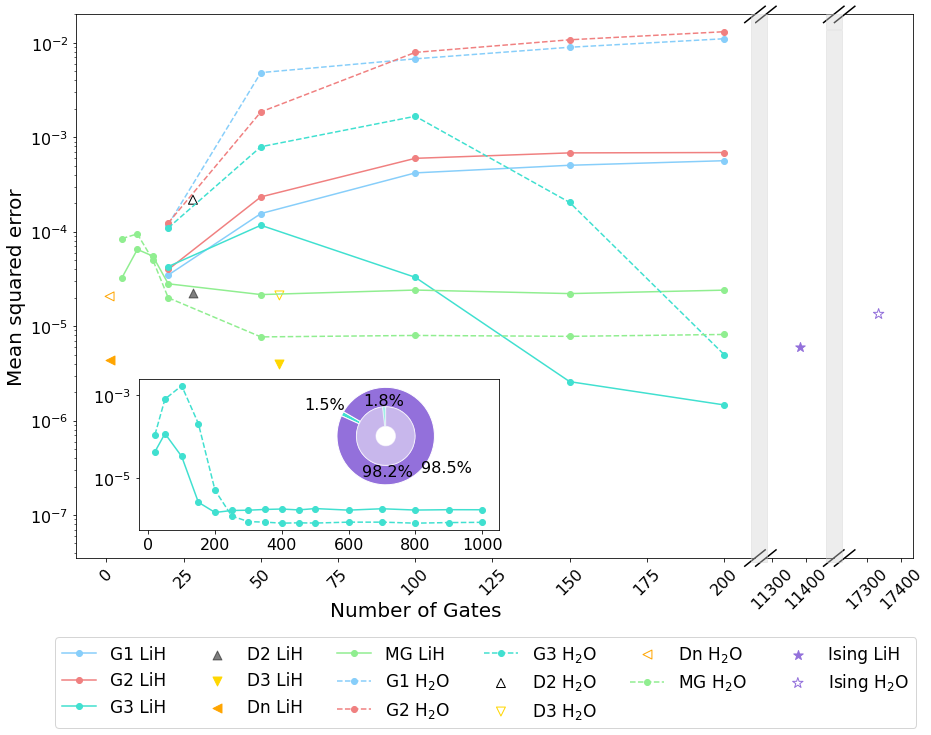}
  \caption[Average mean squared error for the seven families of quantum reservoirs 
  as a function of the number of gates of the circuit. ]{Average mean squared error of $\Delta E_1$ for the seven families of quantum reservoirs 
  as a function of the number of gates of the circuit. 
  Averages are made over 400 simulations. 
  The solid and dashed lines represent the results for the LiH and H$_2$O molecules respectively. 
  The plot also displays, for comparison, the average performance of the Ising model. 
  The inset contains the MSE of the G3 circuits for a larger number of gates. 
  The pie charts represent the proportion of gates needed to obtain optimal performance 
  for the G3 family and the Ising model. }
 \label{fig:MSE_QRC}
\end{figure*}

Figure~\ref{fig:MSE_QRC} shows the performance in terms of MSE (Mean Squared Error, introduced in Eq.~\ref{eq:MSE}) of QRC for the seven families 
of quantum circuits as a function of the number of gates of the circuits. Solid lines correspond to the LiH molecule and dashed lines to H$_2$O. As can be seen, the performance of the different circuits is qualitatively the same in both cases. As we explained in Sects.~\ref{sect:majorization} and~\ref{sect:families_gates}, in Ref.~\citep{majorization} the authors show how the fluctuations of the Lorentz curves differentiate the various families of random circuits, with the families with lower fluctuations corresponding to more complex random quantum circuits. The results in Fig~\ref{fig:MSE_QRC} agree with the classification given in Ref.~\citep{majorization}. Circuits with higher complexity have a better performance in the QML task. 

%
%---------------------------------------------------------------------
%Figure: Mean squared error for each family of gates.
%---------------------------------------------------------------------
%----------------------------------------------------------------------
%----------------------------------------------------------------------------

%-----------------------------------------------------------------
%Figure: MSE second excited state
%-----------------------------------------------------------------
\begin{figure}[!ht]
\includegraphics[width=1.0\columnwidth]{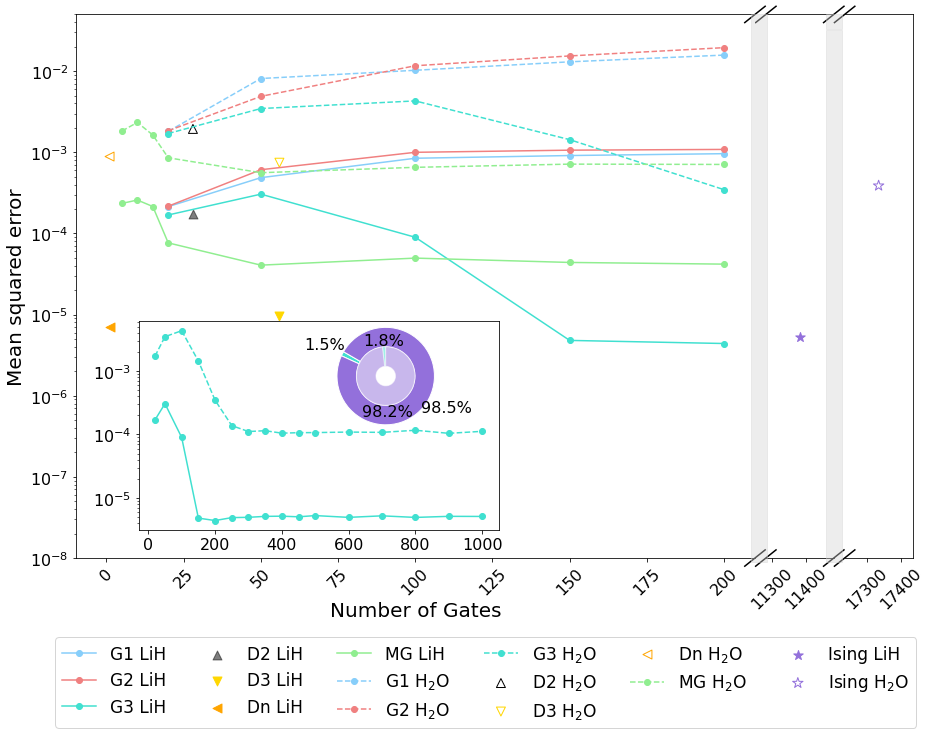}
\caption{Same as Fig.~\ref{fig:MSE_QRC} for the second excited electronic eigenenergy $\Delta E_2$.}
\label{fig:MSE_QRC2}
\end{figure}
%----------------------------------------------------------------------

In particular, the G1 and G2 families are the less complex families according to the majorization indicator, and they also give worse results in the QML task. On the other hand, the circuits in the G3 family give the lowest error in the predictions, which agrees with the G3 family having the highest complexity. The MG circuits have a slightly worse performance than the G3 family, which also agrees with the majorization criterion. Finally, the performances of the $D_3$ and $D_n$ circuits are very similar to each other, although the performance of the $D_2$ circuits is significantly worse. This difference between the diagonal circuits is also seen when using the majorization indicator.  

Figure~\ref{fig:MSE_QRC} also shows how the performance of the QRC changes with the number of gates of the circuit. The G1 and G2 circuits give worse results as the number of gates increases. On the contrary, the G3 circuits improve their performance with the number of gates, and this performance stabilizes around 200 gates for LiH and around 250 for H$_2$O. The MG circuits also improve their performance with the number of gates, but in this case, the optimal performance is achieved with only 20 gates (LiH) or 50 gates (H$_2$O),
even though this optimal error is higher than the optimal error of G3. 

Notice that the H$_2$O system is larger (10 qubits) than the LiH system (8 qubits). Therefore, predicting the excited energy, $\Delta E_1(\vec{R})$, for the H$_2$O is a harder task, and it is expected that the optimal circuit requires more gates. The same analysis has been performed to predict the second excited energy, $\Delta E_2(\vec{R}) = E_2(\vec{R}) - E_0(\vec{R})$, for the two molecules under study. The results are qualitatively equivalent to the results for the first excited energy and are shown in Fig.~\ref{fig:MSE_QRC2}.

In conclusion, the results of this analysis show that the families of quantum circuits with higher complexity according to the majorization criterion are the ones that give better performance in QRC. Therefore, the majorization criterion is an excellent indicator of both the complexity of random quantum circuits and performance in QML.

%-----------------------------------------------------------------
%Distribution in the Pauli space
%-----------------------------------------------------------------

\subsection{Distribution in the Pauli space}
\label{sect:distribution_QRC}
The quantum circuits used as QRs perform a unitary transformation to the input quantum state (the ground state of the molecule). This unitary transformation is followed by local measurements of Pauli operators which provide 
the features extracted from the input. Therefore, the choice of the unitary transformation determines the quality of the extracted features. 
In QRC, the quantum circuits are randomly sampled from a family of generator gates. The unitary transformation derived from each family of quantum circuits spans a subspace of all possible quantum operations. 
If a family of quantum circuits spans a larger space of quantum operations, it will potentially be able to extract more valuable information from the data. On the contrary, if a family of quantum circuits only allows the creation of concrete quantum operations, 
these circuits will have access to a limited amount of information about the input state. This will, in general, lead to poorer feature extraction. 

For this reason, inspecting how each of the random circuits spans the space of operators allows us to further understand the difference in the performance of the circuits presented in Sect.~\ref{sect:families_gates}. This subsection provides a visualization of the distribution of the different families of quantum circuits in the space of quantum operators. We also inspect the effect of changing the number of gates of the circuits, to explain how the performance of QRC varies with the number of gates.

For simplicity and easier visualisation, we create a toy model of two qubits and apply random circuits from each of the seven families. 
Each of the circuits is a unitary operator $U$, constructed by the successive application of the gates of the circuit. This operator can be written as a linear combination of the elements of the Pauli space 
$\{\mathds{1} \otimes \mathds{1}, \mathds{1} \otimes X, \mathds{1} \otimes Y, \mathds{1} \otimes Z, \cdots, X\otimes Z, Y \otimes Z, Z \otimes Z \}$. That is, the unitary $U$ can be written as
\begin{equation}
    U = \sum_i a_i P_i, \quad P_i \in \{\mathds{1} \otimes \mathds{1}, \ \mathds{1} \otimes X, \ \mathds{1} \otimes Y, \ \mathds{1} \otimes Z, \ \cdots, X\otimes Z, \ Y \otimes Z, \ Z \otimes Z \}.
    \label{eq:unitary_dec}
\end{equation}
For each of the seven families of quantum gates, we design 4000 random circuits, 
and see how their unitaries fill the Pauli space, i.e. we study the distribution of the coefficients $\{a_i\}$ in Eq.~\ref{eq:unitary_dec} compared to the uniform distribution. This experiment is repeated for circuits with 5, 10 and 20 gates respectively. For the 2-qubit model, there are 16 basis operators of the Pauli space. Therefore, to obtain a 2D visualization, we use a dimensionality reduction algorithm called \gls{UMAP}~\citep{UMAP2}. For more details on this algorithm, the reader can refer to Refs.~\citep{UMAP2} and~\citep{UMAP1}.

%-----------------------------------------------------------------
%Figure 2: Distribution in the Pauli space
%-----------------------------------------------------------------
\begin{figure*}
\centering
\includegraphics[width=1.05\textwidth]{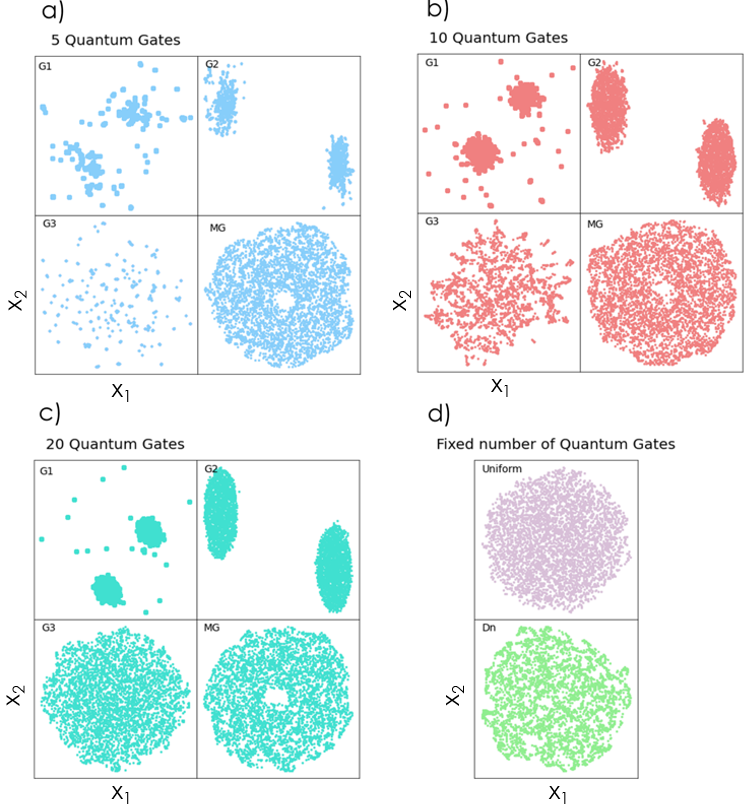}
\caption[2D representation of the distribution in the Pauli space for the G1, G2, G3, MG, and diagonal circuits, compared with the uniform sampling. ]{ 2D representation of the distribution in the Pauli space for the G1, G2, G3, MG, and diagonal circuits, compared with the uniform sampling. Panels a), b) and c) show the distribution for circuits containing 5, 10 and 20 gates respectively for the G1, G2, G3 and MG families. Panel d) shows the distribution of the uniform sampling and the distribution of the diagonal circuit, which has a fixed number of quantum gates. Notice that the $x_1$ and $x_2$ axes are chosen to display an appropriate 2D representation of the Pauli space using the UMAP algorithm ~\citep{UMAP1}, and thus lack physical meaning.}
\label{fig:Pauli_space}
\end{figure*}

The distribution of the families of random circuits in the Pauli space is shown in Fig.~\ref{fig:Pauli_space}, where panels a), b) and c) show the results for circuits containing 5, 10 and 20 quantum gates, respectively, for the G1, G2, G3 and MG families. Panel d) contains the uniform sampling of quantum operations and the results for the diagonal circuit. Notice that the uniform sampling is not obtained by creating a quantum circuit, but by selecting random linear combinations of the basis operators in the Pauli space (a uniform distribution of the coefficients $\{a_i\}$ from Eq.~\ref{eq:unitary_dec}). For this reason, uniform sampling does not depend on the number of gates. The diagonal circuit $D_n$ contains a fixed number of gates. For the 2-qubit system, the three families of diagonal gates $D_2$, $D_3$ and $D_n$ coincide and contain only one gate.

Figure~\ref{fig:Pauli_space} shows that the G1 and G2 families only fill a small subset of quantum operators when compared to the uniform sampling. For this reason, the reservoirs belonging to these families provide worse feature extractions, which leads to worse results in the QML task. Also, the subspace spanned by the G2 family is slightly larger than the subspace spanned by G1, a result which agrees with the fact that G2 has slightly higher complexity than G1. Moreover, we see that when the number of gates increases, the distribution concentrates even more in a region of the Pauli space. For this reason, increasing the number of gates for these circuits leads to worse results in performance. 

The opposite case is the G3 family. As can be seen, for quantum circuits with 20 gates, the distribution in the Pauli space resembles uniform sampling. This means that this family of circuits can perform every quantum operation possible, and thus they can effectively extract all the quantum information from the quantum input. Thus, the G3 family provides optimal results for predicting the input-output relationship. Also, it is seen that as the number of quantum gates increases, the size of the subspace spanned by the G3 family also increases, converging to the uniform sampling for 20-gates circuits. For this reason, the performance of the QRC improves when we increase the depth of the circuit with the G3 family. Notice that when the operator's distribution has converged to uniform sampling, increasing the number of gates will no longer improve the performance of the model. 

The MG circuits nearly span all the space of quantum operators, except for a small subset. Thus, the performance of these circuits is slightly worse than the performance of the G3 family. Also, the distribution of the quantum operators does not change with the number of gates of the circuit, maintaining the performance of the models constant with the number of gates. 

Finally, the diagonal circuits also fill the whole Pauli space. However, there are regions with higher density than others. This also coincides with the slightly higher errors in the QML task.

In this subsection, we have illustrated the relationship between the distribution of different families of quantum circuits in the space of quantum operators and their impact on the performance of the QML task.  We have seen that the families capable of offering a wide range of quantum operators exhibit superior performance in QRC. This phenomenon is analogous to classical RC, as discussed in Sect.~\ref{sect:echo_states}, where we saw how classical reservoirs had to be sparse and inhomogeneous to produce a diverse set of trajectories that could be used to predict the output. Similarly, in the quantum variant of the algorithm, diverse sets of quantum operators must be created by the QRs to ensure precise output prediction.

 \subsection{Comparison with the Ising model}
 \label{sect:Ising}
%==================================================================
After analysing the performance of the seven families of circuits, 
we can compare them with the results of the Ising model, which is a standard model widely used for QRC. Figure~\ref{fig:MSE_QRC} shows that the MSE of the Ising model is similar, slightly higher than the MSE for the G3 family. Therefore, the results do not show a big difference between the QRC performance of the G3 family and the Ising model.
 
However, apart from comparing the performance of the QRs, it is worth analyzing here how the Ising operator can be implemented in a gate-based quantum circuit. Let us show how the Ising model can be realized using only the G3 quantum gates, and prove that the number of gates required to implement this circuit is much higher than the optimal number of gates required to optimize the G3 quantum circuits. As a result, we will conclude that it is not necessary to use the Ising model as a QR, since a much shallow circuit of the same family can already provide optimal performance. 

\begin{figure}[!ht]
\centering
\includegraphics[width=0.4\columnwidth]{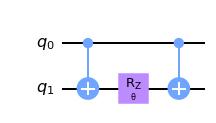}
\caption{Quantum circuit used to implement the $e^{i Z \otimes Z t}$ operator, where $\theta = 2t$.  }
\label{fig:gate_Ising}
\end{figure}

The Hamiltonian for the Ising model is given in Eq.~\ref{eq:Ising}. The time evolution operator $e^{-i\hat{H}_{\text{Ising}}T}$ provides the quantum unitary that will be applied to the initial state. To implement this time evolution in a gate-based quantum computer, this operator needs to be decomposed into a product of quantum gates. The Ising Hamiltonian is already a linear combination of Pauli tensor products as it can be written as $\hat{H} = \sum_k c_k P_k$, where $P_k$ are the Pauli product terms 
(the terms $Z_iZ_j$ and $X_i$), and $c_k$ are the coefficients of the Hamiltonian ($J_{ij}$ and $h_i$). Since the Ising Hamiltonian can be written as a combination of Pauli tensor products, this time evolution operator can be simulated by first-order Trotter decomposition~\citep{Trotter1}, resulting in 
\begin{equation}
    e^{-i\hat{H}_{\text{Ising}}T} = \Big(\prod_k e^{-ic_kP_kT/N} \Big)^N,
\end{equation}
where $T$ is the time, and $N$ is the Trotter number. By increasing $N$, it is possible to decrease the error of the Trotter decomposition as much as desired~\citep{Trotter1}. Therefore, to implement the time evolution, we only need to know how to implement individual operations $e^{iP_kt}$ (with $t=c_kT/N$) in a quantum computer. The Ising Hamiltonian Eq.~\ref{eq:Ising} contains terms of the form $e^{i Z \otimes Z t}$ and $e^{i X t}$.
Noting that $R_z(2t) = e^{i Z t}$, the $e^{i Z \otimes Z t}$ operator can be implemented with the circuit shown in Fig.~\ref{fig:gate_Ising}, where $\theta=2t$. To prove this equivalence, notice that 
\begin{equation}
    \text{CNOT} = \ketbra{0}{0} \otimes \mathds{1} + \ketbra{1}{1} \otimes Z, \text{ and } e^{i Z \otimes Z t} = \cos(t) \mathds{1} \otimes \mathds{1} + i \sin(t) Z \otimes Z.
\end{equation}
Therefore,
\begin{eqnarray}
    & & \text{CNOT} \times R_z(2t) \times \text{CNOT} =\Big(\ketbra{0}{0} \otimes \mathds{1} + \ketbra{1}{1} \otimes Z \Big) \nonumber \\
    & & \Big(\cos(t) \mathds{1} \otimes \mathds{1} + i \sin(t) \mathds{1} \otimes Z \Big) \Big(\ketbra{0}{0} \otimes \mathds{1} + \ketbra{1}{1} \otimes Z \Big)  \nonumber \\
    &  & = \cos(t) \mathds{1} \otimes \mathds{1} + i \sin(t) Z \otimes Z = e^{i Z \otimes Z t}.
\end{eqnarray}
%
%-----------------------------------------------------------------
%Figure: Gate Ising model
%-----------------------------------------------------------------
The other term in Eq.~\ref{eq:Ising} consists of a rotation around the $X$ axis $R_x(2t) = e^{i X t}$. Recall that $X = HZH$, and $e^{i t A} = \cos(t) \mathds{1} + i \sin(t) A$ for $A \in \{X,Y,Z\}$. Therefore, 
\begin{eqnarray}
    R_X(t) = \cos(t/2) \mathds{1} + i \sin(t/2) X  = \cos(t/2) \mathds{1} + i \sin(t/2) HZH = \\ \nonumber \cos(t/2) HH + i \sin(t/2) HZH =
    H(\cos(t/2) \mathds{1} + i \sin(t/2) Z)H =  HR_Z(t)H,
\end{eqnarray}
and the rotation around the X-axis can be implemented as a rotation around the Z-axis together with Hadamard gates.

\begin{figure}%[b]
\centering
\includegraphics[width=0.85\columnwidth]{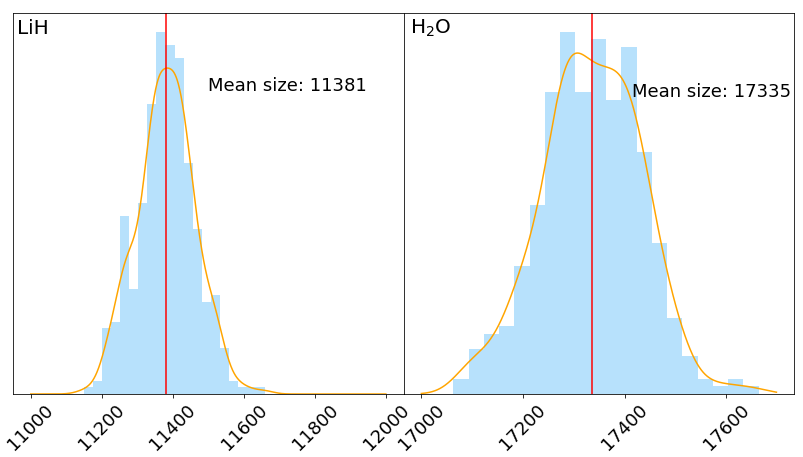}
\caption[Distribution of the number of gates required to implement an Ising model in a 
quantum circuit using the gates from the G3 family. ]{Distribution of the number of gates required to implement an Ising model in a 
quantum circuit using the gates from the G3 family. The orange line represents the empirical distribution, whose mean value is drawn with a red vertical line.}
\label{fig:Ising_decomposition}
\end{figure}

All in all, following the previous procedure one can write the time evolution operator as a product of gates in the set $\{\text{CNOT}, H, R_Z(\theta)\}_\theta$. In NISQ computation, it is necessary to design the quantum circuits in a way that they are protected from decoherence. Unfortunately, it is not possible to implement rotations $R_Z(\theta)$ with perfect accuracy. In current quantum computers, the rotations are not implemented directly but using multiple applications of simpler gates, such as the $T$ and $H$ gates. The problem of translating rotations to $T$ and $H$ gates was solved recently by Ross and Selinger~\citep{HT}, with a procedure based on number theory, where finding the decomposition in the G3 basis is based on solving a specific Diophantine equation. The authors also provided a command-line tool generating the $H-T$ decomposition~\citep{HT}. This software was used to obtain the sequence of gates necessary to implement the Ising model in a quantum circuit. 
Then, we performed 400 simulations varying the parameters of the Ising model. The distribution of the number of gates needed for each of the circuits is shown in Fig.~\ref{fig:Ising_decomposition}. For the LiH molecule, 11381 gates are needed on average, while for the H$_2$O molecule, 17335 gates are required. 
The G3 family of circuits achieved its optimal performance with only 200 gates for LiH, and 250 for H$_2$O. Thus, implementing a time evolution operator under the Ising model would require using far more gates than the ones needed to obtain an optimal QR.

%----------------------------------------------------------------------
%Figure: Ising circuit sizes
%-----------------------------------------------------------------
%----------------------------------------------------------------------
All in all, the results of this section provide a criterion for designing optimal reservoirs for QRC. This criterion is also easy to implement in NISQ devices, since it requires using circuits with only a few quantum gates, compared to the widely used Ising model. We have demonstrated that the optimal circuits generated from the G3 family of gates need significantly fewer quantum gates and provide a similar, slightly better performance than the commonly used Ising model. This is of uttermost importance for optimal implementations in current NISQ devices, since only circuits with few and simple gates can guarantee the required accuracy. The key point of this work is the use of the majorization principle~\citep{majorization} as an indicator for the complexity of the quantum circuits. The QRs with higher complexity according to the majorization criterion provide better results in QML tasks. In the next section, we study how the noise present in the current quantum devices affects the performance of the QRs.

 %%%%%%%%%%%%%%%%%%%%%%%%%%%%%%%%%%%%%%%%%%%%%%%%%%%%%%%%%%
 % NOISY QUANTUM RESERVOIRS
 %%%%%%%%%%%%%%%%%%%%%%%%%%%%%%%%%%%%%%%%%%%%%%%%%%%%%%%%%%
\section{Noisy quantum reservoirs}
\label{sect:noise_QRC}
One of the biggest challenges of the current quantum devices is the presence of noise. Such devices perform noisy quantum operations with limited coherence time, which affects the performance of quantum algorithms running on them. To overcome this limitation, great effort has been devoted to designing error-correcting methods~\citep{errorCorrecting, Domingo_RL, ErrorCorrectingAutores, google2023suppressing}, which correct the errors in the quantum hardware as the algorithm goes on, and also error-mitigation techniques~\citep{errorMitigation, errorMitigation2}, which aim to reduce the noise of the outputs after the algorithm has been executed. Even though these methods can sometimes successfully reduce quantum noise, a fundamental question remains open:
Can the presence of noise in quantum devices be beneficial for QML algorithms? 

The goal of this section is to address this issue in the QRC algorithm. We have already seen that the design of the QR is crucial for the performance of the model. However, all our results were obtained by \emph{noiseless} quantum simulations. Therefore, it is worth exploring the performance of noisy, real implementations of this algorithm. Surprisingly enough, in the next subsections, we will show that certain types of noise can actually provide better performance for QRC than noiseless reservoirs, while others should be prioritized for correction~\citep{Domingo_QRCNoise}. Our numerical experiments will be further supported by a theoretical mathematical demonstration.

\subsection{Noise models}
In this thesis, the effect of three noise models on the performance of the 
QML task is studied, considering different values of error probabilities and number of quantum gates. These noise models embody the majority of noise types to which modern hardware is subjected to, so the conclusions of this study are general to all quantum devices. 

The first noise model that we consider is the \textit{amplitude damping channel}, which reproduces the effect of energy dissipation, that is, the loss of energy of a quantum state to its environment. The second noise model is described by the \textit{phase damping channel}, which models the loss of quantum information taking place without loss of energy. The last noise model is described by the \textit{depolarizing channel}. In this case, a Pauli error $X$, $Y$ or $Z$ occurs with the same probability $p/3$. For more information about these noise models see Sect.~\ref{sect:noise_models}. 

The effect of the different noise channels in the algorithm performance is studied by varying the error probability $p$. We perform 100 simulations for probabilities ${p=0.0001}, 0.0005,\newline 0.001, 0.003$ for each quantum channel, and compare the performance of the model with the noiseless simulation ($p=0$). We also study how the number of quantum gates affects the performance of the reservoirs. We design circuits varying the number of gates from 25 to 215 in increments of 10 gates. Finally, we study the QRC performance for a large number of quantum gates, using 300, 500, 700 and 900 of them. The QML task is kept the same as in the previous study. However, in this case, we study only the LiH molecule for the first excited energy, since we have already seen that the results for the H$_2$O molecule and the second excited energy are qualitatively similar.

%%%%%%%%%%%%%%%%%%%%%%%%%%%%%%%%%%%%%%%%%%%%%%%%%%%%%%%%%%%%%%%%
% PERFORMANCE NOISY RESERVOIRS
%%%%%%%%%%%%%%%%%%%%%%%%%%%%%%%%%%%%%%%%%%%%%%%%%%%%%%%%%%%%%%%

\subsection{Performance of the quantum reservoirs}
%---------------------------------------------------------------------
%Figure: Mean squared error for each noise model
%---------------------------------------------------------------------
\begin{figure}[!ht]
\centering
 \includegraphics[width=0.8\columnwidth]{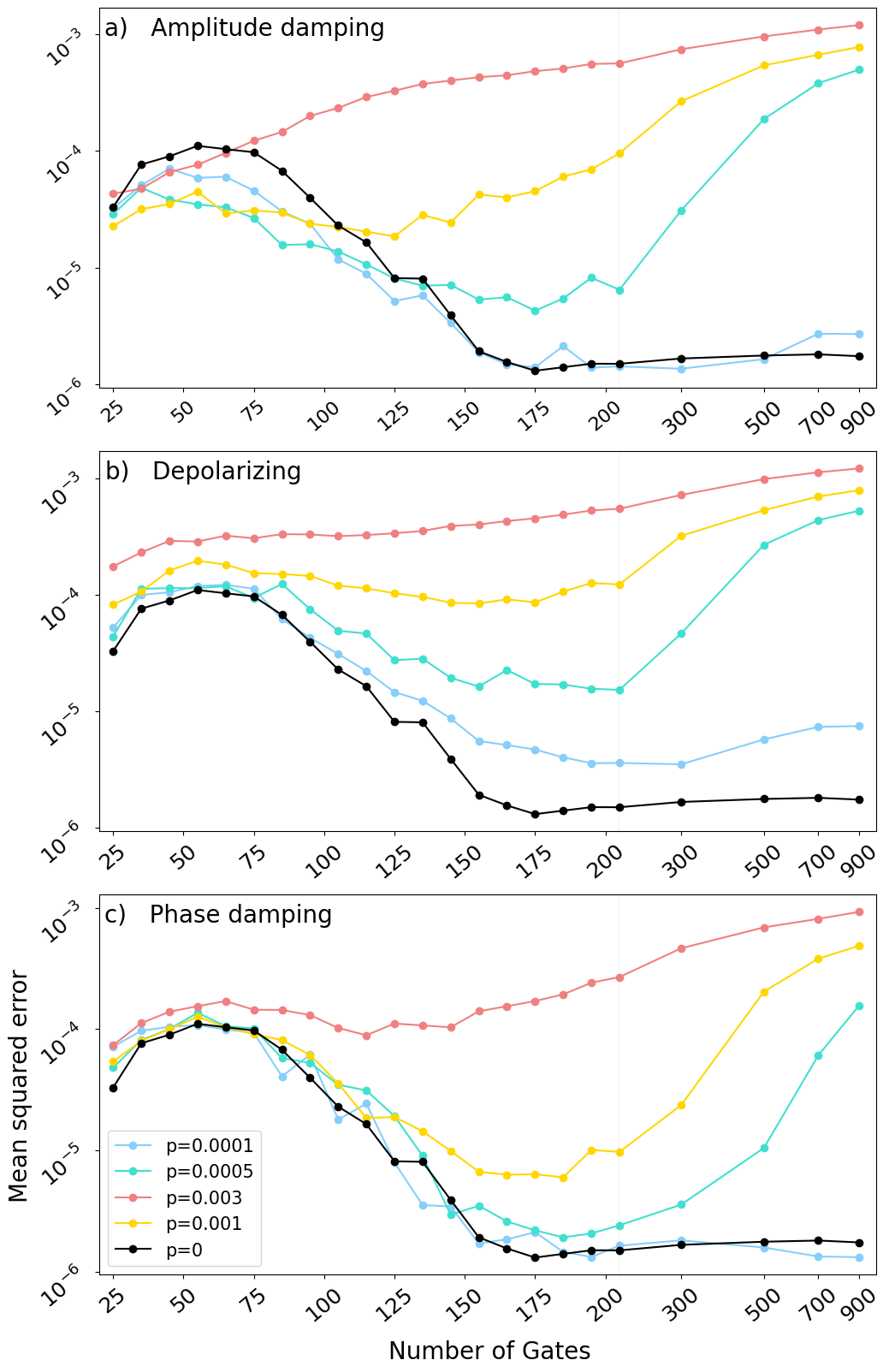}
  \caption{(Averaged) Mean squared error of the first excited energy $\Delta E_1$ for the quantum reservoirs with  a)
  amplitude damping noise, b) depolarizing noise and c) phase damping noise, 
  as a function of the number of gates of the circuit. 
  Averages are made over 100 simulations. }
 \label{fig:mse_noise}
\end{figure}
%----------------------------------------------------------------------
%----------------------------------------------------------------------------
Figure~\ref{fig:mse_noise} shows the MSE in $\Delta E_1$ predicted with our QRs as a function of the number of gates, for different values of the error probability $p$ (colored curves) and noise models (panels), together with the results for the corresponding noiseless reservoir (in black). As expected, the general tendency of the MSEs is to grow with the noise probability $p$. 

However, a careful comparison of the three plots in Fig.~\ref{fig:mse_noise} surprisingly demonstrates that the amplitude damping noise renders results that are significantly different from those obtained in the other two cases. Indeed, if the number of gates and the error probability are small enough, the QRs with amplitude damping noise render better results than the noiseless QR. The same conclusion applies to the higher values of $p$, although in those cases the threshold number of gates for better performance decreases. This is a very significant result since it means that, contrary to the commonly accepted belief, the presence of noise is here \textit{beneficial} for the performance of the quantum algorithm, and, more importantly, it takes place within the limitations of the NISQ era. As an example, notice that for $p=0.0005$ (green curve), all noisy reservoirs render better performance than the noiseless counterpart when the number of gates is smaller than $135$.

A practical criterion to decide when noise can be used to improve the performance of QRC is provided in Table~\ref{tab:optimal_fidelity}, which shows the average fidelity between the output noisy state $\rho$ and the noiseless state $\ket{\psi}$ for the circuits subjected to an amplitude damping noise with different values of $p$. Recall that the fidelity (introduced in Sect.~\ref{sect:fidelity}) measures the distance between two quantum states, which in this case are the noisy and noiseless QRs. In Table~\ref{tab:optimal_fidelity}, the number of gates has been chosen to be as large as possible provided that the noisy reservoirs outperform the noiseless ones. These results imply that when the fidelity is greater than $0.96$, the noisy reservoirs outperform the noiseless ones at QML tasks, and accordingly the noise does \textit{not need} to be corrected. Moreover, also notice that for $p=0.0001$ the fidelity is always higher than $0.96$, and thus the performance of the noisy QRs is always higher or equal than their noiseless counterparts.

%Table 
\begin{table}%[t]
    \centering
    \begin{tabular}{c|cc}
        \hline \hline \\[-2.5ex]
        Error prob. $p$  & Optimal of gates     & Fidelity (averaged)\\ [1ex]
        \hline
        0.0001      &  215           &  0.990\\
        0.0005      &  135          &  0.965 \\
        0.0010      &  105           &  0.956\\
        0.0030      &  65           &   0.962\\
        \hline
    \end{tabular}
    \caption[Fidelity between the noisy and noiseless final quantum states 
    for the circuits with amplitude damping noise.]{(Averaged) Fidelity between the noisy and noiseless final quantum states 
    for the circuits with amplitude damping noise (see text for details). 
    The number of quantum gates is chosen so that the performance of the noisy 
    reservoirs outperforms that of the noiseless reservoirs.}
    \label{tab:optimal_fidelity}
\end{table}

A second conclusion from the comparison among plots in Fig.~\ref{fig:mse_noise} is that the behavior for depolarizing and the phase damping channels are significantly different than that of the amplitude damping one. In the former cases, the performance of the noisy reservoirs is always worse than that of the noiseless one, even for small error probabilities. 

A third result that can be extracted from our calculations is that the tendency of the algorithm performance when the reservoirs have a large number of gates is the same for the three noise models considered (except for the smallest value of $p=0.0001$). While the performance of the noiseless reservoirs stabilizes to a constant value as the number of gates increases, the noisy reservoirs decrease their performance, seemingly going to the same growth behavior.  This is because the quantum channels are applied after each gate, and thus circuits with a large number of gates have larger noise rates, which highly decreases the fidelity of the output state. For this reason, even though increasing the number of gates has no effect in the noiseless simulations, it highly affects the performance of the noisy circuits, and thus the number of gates should be optimized. 

In Sect.~\ref{sect:Ising} we showed that a gate-based implementation of the Ising model would require around 11000 quantum gates. Even though the performance of both the Ising and G3 circuits were similar in the noiseless simulations, we now see that the presence of noise would highly decrease the performance of the Ising circuits because of the large number of gates needed to implement such circuits. Thus, the use of alternative circuits like the ones provided in this work, adapted to the NISQ era, appears to be highly important for current implementations of quantum algorithms. 
 
\subsection{Distribution in the Pauli space}
Having analyzed the performance of the QRs, we next provide a theoretical explanation for the different behavior of the three noisy reservoirs. That is, we will now provide a mathematical demonstration explaining why the amplitude damping channel is beneficial for QRC, while the depolarizing and phase damping channels are not. 

In the first place, recall that the depolarizing and phase damping channels give similar results, except that the performance of the former decreases faster than that of the latter. This effect can be explained with the aid of Table~\ref{tab:average_fidelity}, 
where the averaged fidelity of each error model over the first 200 gates is given. 
%Table 
\begin{table}[!t]
    \centering
    \begin{tabular}{c|ccc}
        \hline \hline \\[-0.3cm] 
        Error prob. $p$ &  Amplitude damping  &  Depolarizing  &  Phase damping \\
        \hline
        0.0001      & 0.995       & 0.994          &  0.998  \\
        0.0005      & 0.975       & 0.971          &  0.988  \\
        0.0010      & 0.951       & 0.944          &  0.976  \\
        0.0030      & 0.862       & 0.842          &  0.931  \\
        \hline
    \end{tabular}
    \caption[(Averaged) Fidelity between the noisy and noiseless final quantum states 
    for the circuits with the three noise models.]{(Averaged) Fidelity between the noisy and noiseless final quantum states 
    for the circuits with the three noise models. 
    Fidelity is averaged over all the quantum reservoirs with less than 200 gates, 
    with the same noise model.}
    \label{tab:average_fidelity}
\end{table}
As can be seen, the depolarizing channel decreases the fidelity of the output much faster than the phase damping, which explains the different tendencies in the corresponding ML performances. 

On the other hand, the amplitude damping channel is the only one that can improve the performance of the noiseless reservoirs in the case of few gates and small error rates. The main difference between amplitude damping and the other channels is that the former is not unital, i.e.~it does not preserve the identity operator. Let us consider now how this fact affects the distribution of noisy states in the Pauli space. For this purpose, let $\rho'$ be the $n$-qubit density matrix obtained after applying $N-1$ noisy gates, (with the noise described by the quantum channel $\epsilon$), and then apply the $N$-th noisy gate $U$. The state becomes $\epsilon(\rho)$, defined as
\begin{equation}
  \epsilon(\rho) = \sum_{m=1} M_m \rho M_m^\dagger, \quad \rho = U \, \rho' \, U^\dag,
\end{equation}
where $\rho$ is the state after applying gate $U$ \emph{without} noise. Now, both $\rho$ and $\epsilon(\rho)$ can be written as linear combinations of Pauli basis operators $\{P_i\}_i$, where each one of them is the tensor product of the Pauli operators $\{ X, Y, Z,\mathbb{I}\}$ as
\begin{eqnarray}
     \rho = \sum_i a_i P_i, \quad \text{with } \ \ a_i = \frac{1}{2^n} \tr (P_i \rho), \\\nonumber
    \epsilon(\rho) =\sum_i b_i P_i, \quad \text{with } \ \ b_i = \frac{1}{2^n}\tr [P_i \epsilon(\rho)].
\end{eqnarray}
Notice here that some of the coefficients $b_i$ will be used to feed the ML model after applying all the gates of the circuit. Thus, expanding the final quantum states on this basis is suitable for understanding the behavior of the QRC algorithm. 

Next, we study the relation between coefficients $\{a_i\}$ and $\{b_i\}$. Since the operators $P_i$ are tensor products of Pauli operators, it is sufficient to study how each of the noise models $\epsilon$ maps the four Pauli operators. The results are shown in Table~\ref{tab:quantum_channel}, where we see that $\epsilon(P_i)$ is always proportional to $P_i$, except for $\epsilon(\mathbb{I})$ with the amplitude damping channel. 

Indeed, it is for this reason that, with depolarizing or phase damping noises, the quantum channel only mitigates coefficients in the Pauli space. Moreover, this table also explains why the phase damping channel provides states with higher fidelity than the depolarizing channel. The phase damping channel leaves the $Z$ operator invariant and also produces lower mitigation of the $X$ and $Y$ coefficients compared to the depolarizing channel. For this reason, even though both the depolarizing and phase damping channels are unital, the depolarizing channel decreases the ML performance faster, and its correction should be prioritized.
%Table 
 \begin{table}
     \centering
     \begin{tabular}{c|ccc}
     \hline \hline                                     \\[-2.5ex]
          &  Amplitude damping  & Depolarizing  & Phase damping\\    
          \hline 
       $\epsilon(X)$   & $\sqrt{1-p}\;X$ & $(1-\frac{4}{3}p)X$ & $(1-p)\;X$ \\
       $\epsilon(Y)$   & $\sqrt{1-p}\;Y$ & $(1-\frac{4}{3}p)Y$ & $(1-p)\;Y$ \\
       $\epsilon(Z)$   & $(1-p)\;Z$      & $(1-\frac{4}{3}p)Z$ & $Z$ \\
       $\epsilon(\mathbb{I})$   & $\mathbb{I} + pZ$ & $\mathbb{I}$ & $\mathbb{I}$ \\
       \hline
     \end{tabular}
     \caption{Expressions for the error channel $\epsilon$ when applied 
     to the four basis Pauli operators.}
     \label{tab:quantum_channel}
 \end{table}
On the other hand, the amplitude damping channel can introduce additional non-zero terms to the Pauli decomposition. Also, this explains why, for low noise rates, the shapes of the MSE curves for depolarizing and phase damping are similar to that for the noiseless scenario, but not for the amplitude damping one. 

Let us provide a mathematical demonstration of this fact. For any Pauli operator $P_i$, the coefficient in the Pauli space with the depolarizing and phase damping channels is
 \begin{equation}
     b_i = \frac{1}{2^n}\tr[P_i\;\epsilon(\rho) ] = \frac{1}{2^n}\alpha_i \;tr(P_i  \rho ) = \alpha_i \; a_i, 
        \quad 0 \leq \alpha_i \leq 1,
 \end{equation}
 and therefore the noisy channel mitigates coefficient $a_i$. 
 However, let us consider what happens if we take a gate with amplitude damping noise. 
 Suppose channel $\epsilon$ acts non-trivially on qubit $j$, that is, the Kraus operators for $\epsilon$  take the form  $\tilde{M}_m = \mathbb{I} \otimes \cdots \otimes M_m \otimes \mathbb{I} \otimes \cdots \mathbb{I}$, with $M_m$ in the $j$-th position. Suppose now that we measure $P_i$ (the $i$-th operator in the Pauli basis  associated to coefficient $a_i$),  where $P_i$ acts as a $Z$ operator on the $j$-th qubit  ($P_i = P^0 \otimes \cdots P^{j-1} \otimes Z \otimes P^{j+1} \otimes \cdots P^n$). Let us also take $P_k=P^0 \otimes \cdots \otimes P^{j-1} \otimes \mathbb{I} \otimes P^{j+1} \otimes \cdots \otimes P^n$, with $a_k$ associated to $P_k$. Then, coefficient $b_i$ is
 \begin{eqnarray}
     b_i  = &\displaystyle\frac{1}{2^n} \tr[P_i\epsilon(\rho)] = 
     \frac{1}{2^n} \sum_l a_l \tr[P_i \epsilon(P_l)] =& \displaystyle\frac{1}{2^n} \Big(a_i \tr[P_i \epsilon(P_i)] 
                   +  a_k \tr[P_i \epsilon(P_k)]\Big) \nonumber \\
    =&\displaystyle\frac{1}{2^n}\Big(a_i (1-p) \tr[P_i^2] + a_k\tr[P_i(P_k + pP_i)]\Big)=& (1-p)a_i + p a_k.
\end{eqnarray}
When $a_i=0$ but $a_k \neq 0$, coefficient $b_i$ is different from 0, and thus the amplitude damping noise introduces an extra coefficient in the Pauli space. Therefore, we can conclude that the amplitude damping channel allows the introduction of additional non-zero coefficients in the Pauli space, instead of only mitigating them.
For this reason, for $p$ small enough, the amplitude channel can introduce new non-zero terms in the Pauli space without mitigating too much the rest of them. 

%-------------------------------------------------------------------
%Figure:  Example of coefficients in the Pauli space for AD, DP and PD noise
%-------------------------------------------------------------------
\begin{figure}[!ht]
\centering
  \includegraphics[width=0.85\columnwidth]{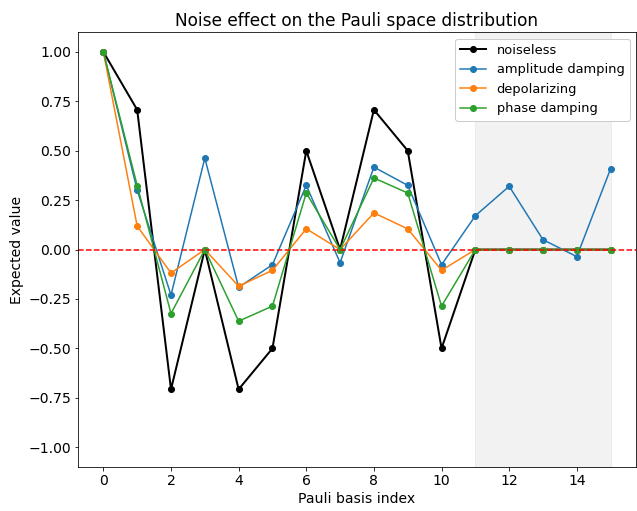}
  \caption[Coefficients in the Pauli space of a 2-qubits toy model.]{Coefficients in the Pauli space of a 2-qubits toy model
  consisting of a random quantum circuit with 10 gates from the G3 family and error probability $p=0.2$, for the three noise models studied in this work together with the noiseless coefficients in black.} 
\label{fig:example_coefficients}
\end{figure}
%-------------------------------------------------------------------    
The previous theorem can be further illustrated with a two qubits toy model example, just as we did in Sect.~\ref{sect:distribution_QRC}. We design a QR sampled from the G3 family with the three different quantum noise models and calculate the distribution of the Pauli coefficients $\{b_i\}$ at the end of the circuit. Figure~\ref{fig:example_coefficients} shows the outcomes of the measurements for a random circuit with 10 gates and an error rate of $p=0.2$. We see that all noise models mitigate the non-zero coefficients. 

However, the shadowed area shows a region where the noiseless simulation (as well as the depolarizing and phase damping simulations) gives zero expectation values. More importantly, the amplitude damping circuit has non-zero expectation  
  values for the same operators, which means that this quantum channel has introduced non-zero terms in the Pauli distribution.
  For small error rates, the noisy QRs provide a better performance, since having amplitude damping noise produces a similar effect as having more quantum gates in the circuit.
  
  %-------------------------------------------------------------------
%Figure:  Filling of the Pauli space for each noise model
%-------------------------------------------------------------------
\begin{figure}[!ht]
\centering
  \includegraphics[width=0.8\columnwidth]{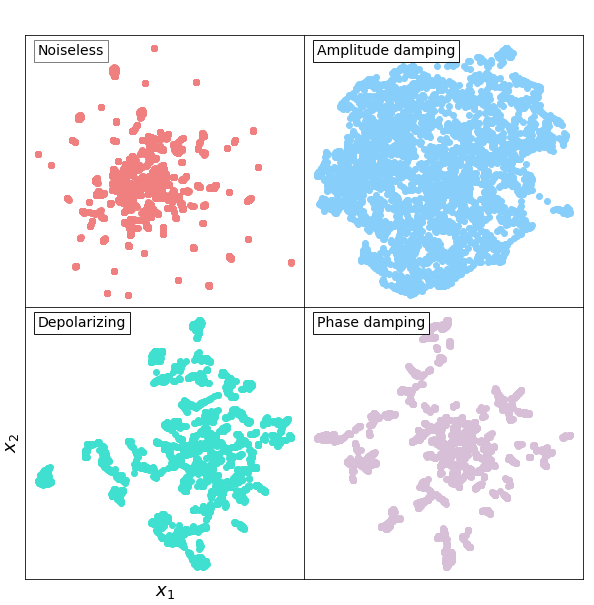}
  \caption[2D representation of the distribution in the Pauli space of 4000 simulations of the toy model of Fig.~\ref{fig:example_coefficients}.]{Reduced 2D (from 16D) representation of the distribution in the Pauli space of 4000 simulations of the toy model of Fig.~\ref{fig:example_coefficients}. Variables $x_1$ and $x_2$ are selected using the UMAP algorithm of Ref.~\citep{UMAP2}.}
\label{fig:pauli_space_noise}
\end{figure}
%-------------------------------------------------------------------
  To better visualize this effect, we design 4000 random circuits and see how the final state $\rho$ fills the Pauli space. Since the Pauli space in the 2-qubit system is a 16-dimensional space, we again use UMAP to visualize the distribution in 2D. The results are shown in Fig.~\ref{fig:pauli_space_noise}. We see that the amplitude damping channel fills the Pauli space faster than the other circuits, including the noiseless QR. Comparing this to  Fig.~\ref{fig:Pauli_space}, we confirm the hypothesis that the amplitude damping channel acts equivalently as having more quantum gates. 

In conclusion, we have studied the effect on the QRC performance of three different paradigmatic noise models, effectively covering all current possibilities affecting quantum devices. Contrary to common belief, our results demonstrate that, under certain circumstances, 
noise, which constitutes the biggest challenge for quantum computing and QML, is beneficial to quantum algorithms. Remarkably, the results show that for error rates $p \lesssim 0.0005$ or state fidelities of at least $0.96$, the presence of an amplitude damping channel renders better performance than noiseless reservoirs for QML tasks. On the other hand, the depolarizing and phase damping channels only reduce the amplitude of the coefficients in the Pauli space, thus producing poorer results. The depolarizing channel is the one that mitigates fastest these values, so its correction should be a priority.

\section{Quantum reservoir computing for time series forecasting}
\label{sect:QRC_forecasting}
In the preceding sections, we demonstrated that the majorization principle can serve as a criterion for optimizing QRs in QML applications. Our previous experiments focused on non-temporal tasks, specifically predicting the excited energies of electronic Hamiltonians for two molecules, which also served to give a theoretical explanation for the optimality of the QR designs. This is a significant task for quantum algorithms due to the exponential scaling of Hamiltonians with the number of molecules in a system, which can limit the computation with classical computers. 
However, these results only correspond to the QRC performance in non-temporal tasks. 

In contrast, in this section, we examine the QRC performance in predicting the time evolution of a price series, specifically the zucchini time series presented in Chapter~\ref{chapter4}. This task presents also a more applied characterization of a real-life problem. As we demonstrated in the previous chapter, forecasting this series is particularly challenging due to its high volatility, small training set, and heavy reliance on external factors. This setting is more challenging for the reservoir than the one studied in the previous sections since the extracted features not only need to capture information from the input quantum state encoding the dataset but also need to use the memory from previous states in an efficient way. The results of this study show that the majorization criterion also serves as a good performance indicator for QRC of time series forecasting, even for challenging, real-life time series. 

The next subsections are organized as follows. Section~\ref{sect:dataset_zucchini} introduces the time series and preprocessing used for this study. Then, Sect.~\ref{sect:QRC_zucchini} provides the training details for the QRC algorithm described in Sect.~\ref{sect:QRC}. Finally, in Sect.~\ref{sect:results_QRC_zucchini} we analyze the results and draw conclusions. 

\subsection{Dataset}
\label{sect:dataset_zucchini}
This study focuses on the weekly prices of zucchini, a relevant vegetable for the Spanish and European markets, which was already presented in Sect.~\ref{paper:RC_calabacines}. The dataset is divided into the same training, validation and test sets as in Sect.~\ref{paper:RC_calabacines}. The time series is scaled to the $[0, 1]$ range using a linear scale so that they are suitable inputs to the QR. In addition to the time series, the QRC algorithm makes use of nine regressor variables, that gather information on production volumes and international trade, thus providing supplementary data for the price prediction process. The optimal number of regressor variables for QRC is determined using the validation set.

\subsection{Training details}
\label{sect:QRC_zucchini}

In this subsection, we summarize the QRC algorithm (see Sect.~\ref{sect:QRC} for more details). The goal of QRC for temporal tasks is to predict the value of the time series at time $t+\Delta t$, $y(t + \Delta t)$, given the past values of the series $\{y(T)\}_{T\leq t}$. In contrast to the case of non-temporal tasks, the input series here has to be fed to the QR sequentially, so that the extracted features contain information from both the current input and the memory from past inputs. This requires performing an initialization process during computation, which causes the quantum state of the system to become a mixed state. Given a one-dimensional time series ($n=1$), consider a quantum system with $N>n$ qubits. Then, for each time step of the training process, the initial state of the system is initialized as
\begin{equation}
    \rho(t) = \ketbra{y(t)} \otimes \Tr_n\Big(\rho(t-\Delta t)\Big), \quad \ket{y(t)} = \sqrt{1-y(t)}\ket{0} + \sqrt{y(t)}\ket{1},
\end{equation}
where $y(t)$ is the value of the time series at time $t$, scaled in the $[0,1]$ domain, and the partial trace $\Tr_n(\cdot)$ is done on the first qubit. In our case, we set $N=7$ (see the reason for this choice in the discussion associated with Fig.~\ref{fig:num_qubits} below). After encoding the quantum state at time $t$, the system evolves under a unitary transformation $U$ until time $t + \Delta t$.
\begin{equation}
    \rho(t + \Delta t) = U \rho(t) U^\dag.
\end{equation}
Finally,  after a sufficiently long training time, the extracted features $\newline {\hat{x}(t) = \{\expval{P_i}(t)\} = (\expval{X_1}(t), \expval{Z_1}(t), \cdots \expval{X_N}(t), \expval{Z_N}(t))}$ are fed to a ridge regression (see Eq.~\ref{eq:ridge}) which predicts the target $y(t + \Delta t)$.  Moreover, one can gather the readout signals from intermediate times between $[t, t+ \Delta t]$ to increase the size of the feature vector. In this case, the input timescale $\Delta t$ is divided into $N_v = 2$ time steps, such that the intermediate steps are given by $t^k = t + k \Delta t/2$, $k \in \{0,1,2\}$. Then, the feature vector $\hat{x}(t)$ contains the 2$N$ expected values from the $N$ qubits and the intermediate times $t^k$, leading to a total dimension of $2N\times~N_v =~28$. Additionally, we can expand the features by adding extra regressor variables that contain useful external information. In this case, the feature vector will be $\hat{x}_{\text{extended}}(t) = (\hat{x}(t), x_{\text{extra}}(t))$. 

%-----------------------------------------------------------------
%Figure Setting 
%-----------------------------------------------------------------
\begin{figure}[!ht]
\includegraphics[width=1.0\textwidth]{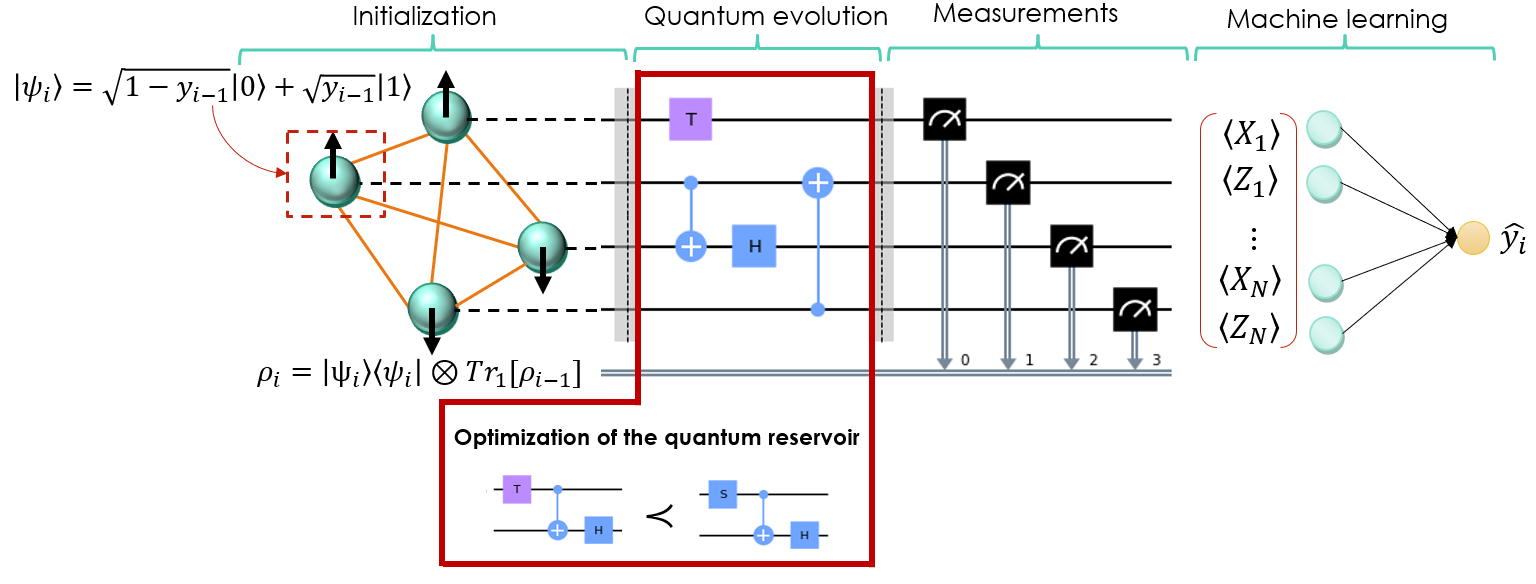}
\caption[Pipeline used to train the quantum reservoir computing model for time series forecasting.]{Pipeline used to train the quantum reservoir computing model. 
The time series at step $t-1$ is given as an input. Then, the reservoir evolves by applying a random quantum circuit sampled from one of the eight families studied in this work.
Local Pauli operators are then measured and fed to the classical machine learning algorithm, a linear regression model. The choice of the quantum reservoir is optimized according to the majorization principle introduced in Ref.~\citep{majorization}.}
\label{fig:QRC_zucchini}
\end{figure}
%----------------------------------------------------------------------
%

The final feature vector will be given to a ridge regression model which will learn to forecast the series. In this section, we will consider two scenarios. In the first one, only the time series will be given to the reservoir (i.e. no external variables). In the second scenario, apart from the time series, the additional regressor variables will also be given to the reservoir. The regularization parameter for the ridge regression in Eq.~\ref{eq:ridge} is set to $\gamma = 1 \times 10^{-10}$ when the feature vector \emph{does not} include the external factors $\hat{x}(t)$ (first scenario), and $\gamma = 0.1$ when the feature vector \emph{does} include the external factors  $\hat{x}_{\text{extended}}(t)$ (second scenario). 

The goal of this study is to identify the families of quantum circuits that provide optimal performance in forecasting the time series and are suitable for NISQ devices. We consider the same seven families of quantum circuits described in Sect.~\ref{sect:families_gates} together with the Ising model with the optimal parameters provided in Ref.~\citep{DynamicalIsing}, which were also the ones used in Sect.~\ref{sect:optimalQRC}. For a given family, the quantum circuit is built by adding a fixed number of random quantum gates from such family. For each family of QRs, we perform 100 simulations. The setting of the QRC algorithm is illustrated in Fig.~\ref{fig:QRC_zucchini}.

\subsection{Results}

\begin{figure*}[t!]
    \centering
    \includegraphics[width=0.87\textwidth]{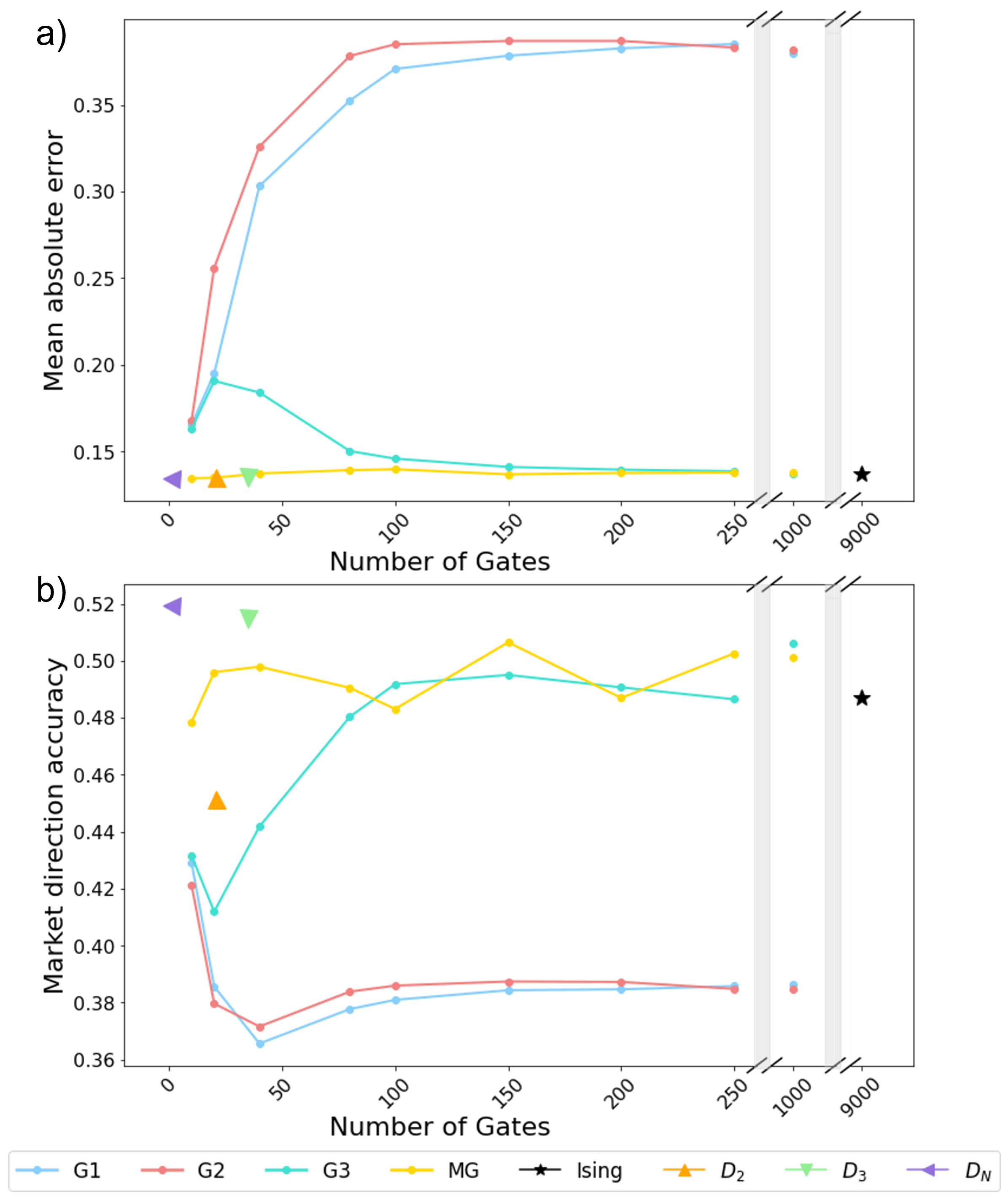}
    \caption[a) Mean absolute error and b) market direction accuracy of the price prediction for the eight families 
    of quantum reservoirs as a function of the number of gates of the circuit.]{a) Mean absolute error and b) market direction accuracy of the eight families 
    of quantum reservoirs as a function of the number of gates of the circuit. 
  Averages are made over 100 simulations. 
  The machine learning task consists of forecasting the zucchini prices \textit{without} 
  using external regressor variables.}
  \label{fig:no_vars}
\end{figure*}

\label{sect:results_QRC_zucchini}
%Figure 3
\begin{figure*}[!ht]
    \centering
    \includegraphics[width=0.95\textwidth]{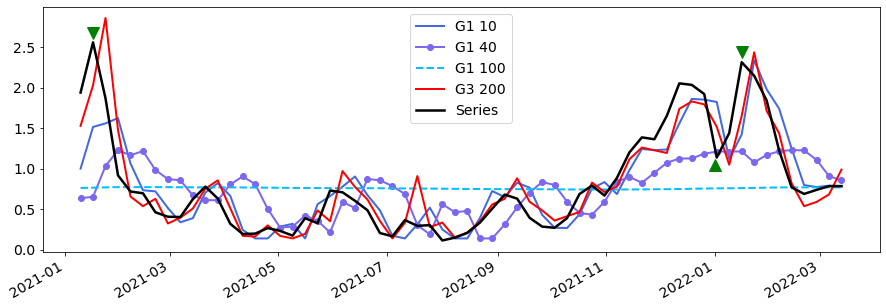}
    \caption[Prediction of the zucchini price (in €/kg) time evolution in the 
    test set 2021-2022 campaign for the G1 and G3 reservoirs.]{Prediction of the zucchini price (in €/kg) time evolution in the 
    test set 2021-2022 campaign for the G1 family with 10, 40 and 100 gates 
    \textit{without} regressor variables, compared with that for the G3 family 
    with 200 gates \textit{with} regressor variables, 
    and the actual price series in black.}
  \label{fig:example_pred}
\end{figure*}

%%%% First scenario
%%%%
\subsubsection{Predicting without regressor variables} \label{sec:scenario1}

Let us start by considering the first scenario, in which only the time series is provided to the QR, and no further external factors are considered for the prediction. Figure~\ref{fig:no_vars} shows the performance in predicting the series of zucchini prices of the seven families of QRs described in Sect.~\ref{sect:families_gates} as a function of the number of gates in the circuits, together with the results of the Ising model. The performance of the method is evaluated using two metrics: the standard MAE (Mean Absolute Error, introduced in Sect.~\ref{paper:RC_calabacines})
-in Fig.~\ref{fig:no_vars} a)- and the MDA (Market Direction Accuracy, see again Sect.~\ref{paper:RC_calabacines}) -in Fig.~\ref{fig:no_vars} b)-. Recall that low values of the MAE or high values of MDA correspond to good predictions. 

In Fig.~\ref{fig:no_vars}, we also show the asymptotic value of the MAE and MDA values, evaluated with a large number of gates, i.e.~1000, and the results obtained with the Ising model, which requires implementing a circuit with 9000 gates. As can be seen, the results clearly indicate that there are significant differences in the behavior computed with the different families. In particular, the worse results, both in terms of MDA and MAE, are obtained with the G1 and G2 gates. This is consistent with the fact that these gates are less complex according to the majorization criterion since they are not universal and generate unitaries within the Clifford group. Furthermore, as the number of quantum gates increases, the performance of the QRC algorithm worsens.
When the number of gates exceeds approximately 100, the MDA drops to around 38\%, which is equivalent to random guessing, and the MAE reaches up to a value equivalent to predicting the mean value of the series. 

This trend is also evident in Fig.~\ref{fig:example_pred}, which depicts the prediction in the test set of the G1 family for 10, 40, and 100 quantum gates. As the number of gates increases, the variability of the predictions decreases, eventually converging to the extreme case where the predictions are just a constant value similar to the average value of the time series (see the dashed line in Fig.~\ref{fig:example_pred}). This implies that the QRs generated with the G1 (and also the G2) families, instead of extracting useful features from the input data, slowly suppress the information encoded in the quantum states. This behavior worsens as the number of gates from these families increases, reaching the worst-case scenario at around 100 gates, where the information about the input series is completely lost.

The opposite behavior is observed in Fig.~\ref{fig:no_vars} for the G3 family. In this case, the performance of the QRC algorithm improves with the number of quantum gates, converging to the optimal value of MAE=0.13 and MDA=0.5 for around 150 quantum gates. The optimality of the G3 family performance agrees with the fact that it has the highest complexity in terms of the majorization criterion, as we discussed in the previous sections. The predictions obtained with the G3 family, in this case with the aid of external regressor variables, are also shown in Fig.~\ref{fig:example_pred} for comparison. These results are presented in the discussion regarding the second scenario. 

The performance of the MG circuits is very similar to that of the G3 family, both in terms of MAE and MDA. However, the optimal error is achieved with only 40 quantum gates, a value that is significantly lower than the number of gates required by the G3 family. Despite this advantage, the implementation of the MG quantum gates is much more complicated than that of the elementary quantum operations required for the G3 family. Therefore, although both the MG and G3 circuits perform equally well, the G3 circuits are more suitable for NISQ devices. Furthermore, the MG family is slightly less complex than the G3 family in terms of the majorization criterion, but this does not seem to have any significant impact on the performance of the QRC algorithm in this case.

Regarding the diagonal circuits, their performance is very similar to the optimal performance, as discussed in the previous sections. However, the $D_2$ circuits exhibit significantly lower values of MDA, which also agrees with the fact that they are less complex according to the majorization criterion, and with the analysis we performed with the quantum chemistry task.

\begin{figure}[t]
    \centering
    \includegraphics[width=0.7\columnwidth]{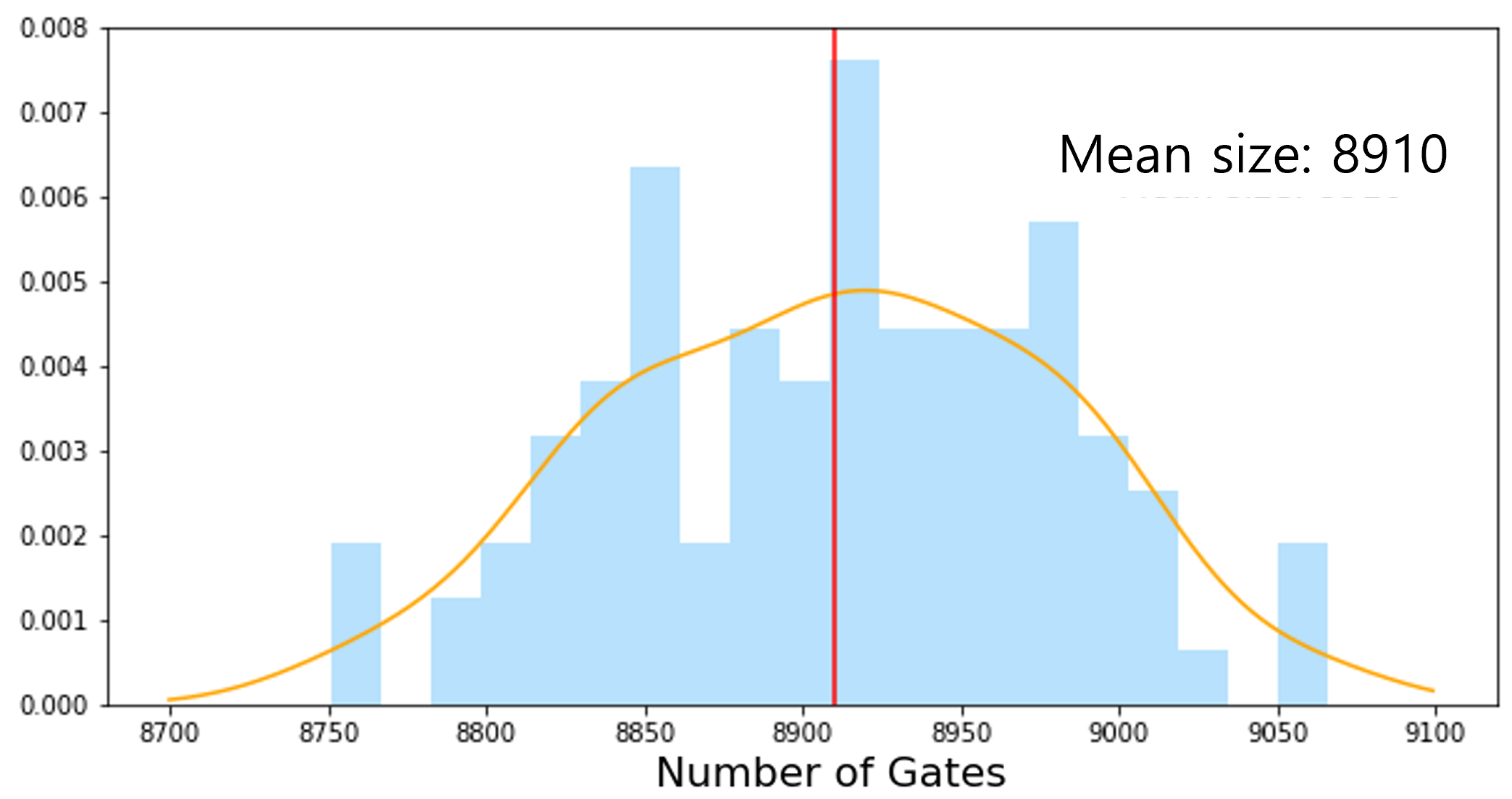}
    \caption[Probability distribution of the number of gates required to implement an Ising model in a quantum circuit using the gates from the G3 family.]{Probability distribution of the number of gates required to implement an Ising model in a quantum circuit using the gates from the G3 family. The y-axis shows the probability density as a function of the number of gates of the circuit. The orange curve represents the best fit of the empirical probability density function, whose average is marked with the red line.}
    \label{fig:Ising}
\end{figure}

Another important point is that the performance of the QR generated by the time evolution under the Ising model is also shown in Fig.~\ref{fig:no_vars}. As can be seen, this performance is also optimal both for MAE and MDA. In relation to this, we discussed, in Sect.~\ref{sect:Ising}, a method to estimate the number of gates required to implement the Ising model on a gate-based quantum computer with gates sampled only from the G3 family. To determine the number of G3 gates required for the Ising model implementation with different parameters, we conducted 100 simulations. The results are presented in Fig.~\ref{fig:Ising}, which shows the probability distribution of the number of G3 gates needed to implement each of the 100 simulations. The orange curve shows the best fit of the empirical distribution, where the average number of gates (vertical red line) appears to be very close to 9000. This value is significantly larger than the number of gates required for an optimal reservoir using the G3 family. Therefore, we conclude that implementing a time evolution operator based on the Ising model using the gates from the G3 family requires far more gates than those required to obtain an optimal reservoir. All in all, the results in this scenario coincide with those obtained in the static task in the quantum chemistry setting. 

%Figure 4
%Figure 
\begin{figure}[!ht]
    \centering
    \includegraphics[width=0.7\columnwidth]{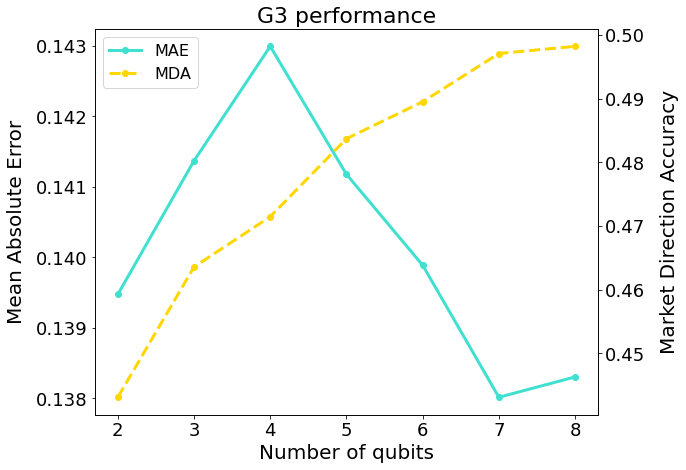}
    \caption[Mean squared error of the prices for the G3 family, evaluated in the test set, at forecasting the zucchini prices \textit{without} external regressor variables.]{Performance of the G3 family, evaluated in the test set, at forecasting the zucchini prices \textit{without} using external regressor variables with 150 quantum gates as a function of the number of qubits $N$. Averages are made over 100 simulations.
    }
  \label{fig:num_qubits}
\end{figure}

To conclude this part, we study the effect of the number of qubits $N$ used in the calculations in the performance of QRC. For this reason, we have performed 100 simulations using the optimal QRs, the G3 reservoirs with 150 gates, for $N=2,3,\cdots, 8$ qubits. The results are shown in Fig.~\ref{fig:num_qubits}. As can be seen, the performance of QRC improves in general with the number of qubits, until it converges to an optimal performance at $N=7$. For larger values of $N$, more information from the previous values of the time series is given to the quantum reservoir, thus yielding better results. However, increasing the number of qubits also increases the computational complexity of the task. Our experiments show that, for this task, $N=7$ qubits is the appropriate balance between computational complexity and performance in the machine learning task.
%%%% Second scenario
%%%%
\subsubsection{Getting help from regressor variables} \label{sec:scenario2}

Let us consider the second scenario, in which external regressor variables are also used to forecast the prices. These variables are known to provide highly relevant information to the behavior of the price time series. For this purpose, we have repeated the previous analysis by expanding the feature vector with nine external variables. The performance of the best model, namely the G3 family with 200 quantum gates, is depicted in Fig.~\ref{fig:example_pred}, where it is compared with the worst model, the G1 family with 100 gates and  \textit{no} regressor variables. As can be seen, the new model is better at predicting changes of tendency in the data, even though some abrupt changes, marked with green triangles, are not properly predicted, due to the high complexity of the task.

%Figure 
\begin{figure*}[!ht]
    \centering
    \includegraphics[width=0.85\textwidth]{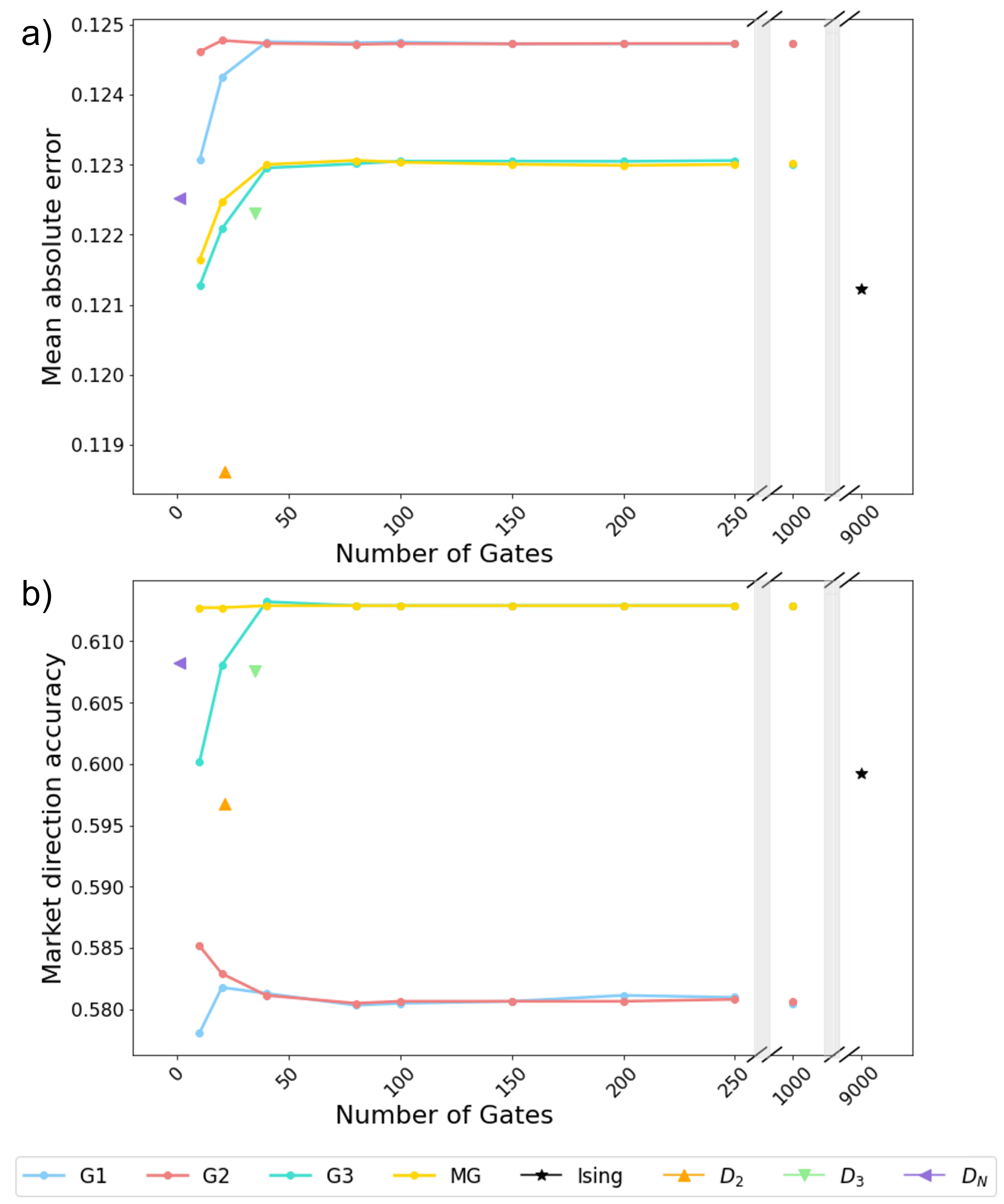}
    \caption{Same as Fig.~\ref{fig:no_vars} 
    using external regressor variables.}
    \label{fig:vars}
\end{figure*}

The performance of the seven families of QRs described in Sect.~\ref{sect:families_gates} are shown in Fig.~\ref{fig:vars}, together with the Ising model. The first thing to notice is that the differences in the performance of the models with gates G1 \textit{vs.~}G3 are now significantly smaller. Indeed, in the first scenario, the difference in MAE between the G1 and G3 family was around 0.23~€/kg, while the difference in MAE considering the external variables is now just 0.003~€/kg. In terms of MDA, the difference between the G3 and G1 family without considering the regressor variables was $13\%$, while this difference drops to $3\%$ when the influence of the external factors is included. 
This result indicates that the external factors contain critical information for the predictive model, which is not related to the design of the reservoir. Despite the strong influence of external variables, the difference in performance based on the majorization criterion is still apparent, which confirms the significant dependency of the QRC algorithm on the reservoir design. 

Regarding the other families of QRs, the MG families still provide optimal performance, while converging with slightly fewer gates than the G3 family. The $D_3$ and $D_n$ circuits provide similar MAE and slightly worse MDA than the G3 family, which agrees with the fact that they have somewhat smaller complexity according to the majorization criterion. The $D_2$ family presents a significantly worse MDA, which lies in between the G1 and G3 MDA values. This effect is also seen in the first scenario and also agrees with the majorization indicator. However, the $D_2$ circuit presents the lowest MAE, even though it is just 0.004~€/kg lower than the G3 MAE. This can be explained by noting that a model whose predictions copy the previous value of the series will present lower MAE but high MDA, thus providing worse overall performance (as was discussed in Sect.~\ref{sect:results_MDA}). For this reason, we believe that the G3 family produces better results than the $D_2$. Finally, the Ising model exhibits similar behavior as the $D_2$ family, where the MAE is smaller than that of the G3 family but the MDA is also smaller than the optimal one. As a result, in this case, we conclude that, when regressor variables are considered, the G3 family presents an optimal performance in forecasting the price series, which is obtained with only 60 quantum gates.

The results of this study show that the majorization criterion is a good performance indicator in time series forecasting with QRC. That is, the QRs with high complexity according to the majorization criterion also exhibit better performance in forecasting tasks. Due to this fact, our conclusion is that the majorization criterion is an optimal performance indicator in non-temporal tasks, like the one considered in Sect.~\ref{sect:optimalQRC}, as well as tasks containing time evolution, like the one discussed here.

%%%%%%%%%%%%%%%%%%%%%%%%%%%%%%%%%%%%%%%%%%%%%%%%%%%%
% HYBRID QNN FOR BINDING AFFINITY PREDICTIONS
%%%%%%%%%%%%%%%%%%%%%%%%%%%%%%%%%%%%%%%%%%%%%%%%%%%%%

\section{Hybrid quantum-classical neural networks for drug discovery}
\label{sect:hybrid_CNN}
The goal of this section is to use the knowledge gained in the previous sections from analyzing the optimal design and noise impact of QRs to tackle a significant and applicable challenge in the pharmaceutical industry; predicting the binding affinity of drug compounds. 

One of the main challenges in drug design is to find molecules that bind specifically and strongly to their target protein while having minimal binding to other proteins. By predicting binding affinity, it is possible to identify the most promising candidates from a large pool of compounds, reducing the number of compounds that need to be tested experimentally. Experimental determination of the binding affinities for a large number of small molecules and their targets is time-consuming and expensive. As a result, computational methods that can make accurate predictions have been greatly welcomed, and then widely used in this field.

Traditional methods are physics-based, meaning that they rely on biophysical models of the proteins-ligand structure to estimate their binding affinity. Various strategies exist within the realm of physics-based methods. For instance, molecular dynamics methods~\citep{molecularsim1, molecularsim4} simulate the temporal behavior of drug-protein complexes to estimate binding affinities.  Quantum mechanical calculations, encompassing semiempirical, density-functional theory, and coupled-cluster approaches have also been employed for binding affinity predictions~\citep{quantummechanical}. Finally, force-field scoring functions are also used to evaluate the energy associated with the complex formed by the ligand and protein~\citep{forcefield2, forcefield1,forcefield3}, considering bonding, electrostatic and van der Waals interactions. Although these traditional methods tend to offer high accuracy, their applicability is hindered by the size and complexity of protein-ligand structures, making it impractical to study larger molecules. Moreover, physics-based methods need to be adapted to every single protein-ligand pair, requiring long computations and domain expertise for every single binding affinity prediction. With the recently available large datasets, generating individual predictions is no longer practical.

On the other hand, data-based methods such as ML require only a one-time training process, enabling them to provide predictions indefinitely with minimal computational cost. Moreover, ML methods do not usually require extensive domain expertise since they are designed to uncover hidden patterns in the data. For these reasons, ML techniques, and especially deep learning methods, have recently gained attention for their ability to improve upon traditional physics-based methods.

A commonly used deep learning approach for binding affinity prediction is the three-dimensional convolutional neural network (3D CNN)~\citep{AtomNet,DeepSite, ScoringCNN, Wave, ATOM}. Recall that the architecture of the CNN was introduced in  Sect.~\ref{sect:CNN}. These networks represent atoms and their properties in a 3D space and take into account the local 3D structure of molecules, the bond lengths and the mutual orientation of atoms in space. The 3D representations used as input of the 3D CNN are high-dimensional matrices since millions of numbers are required to describe only one data sample. For example, in this work, we use a grid of the discretized 3D space to describe each molecular property of the protein-ligand pair. Such grid has size $48\times 48 \times 48$, and 19 features are used to describe the data sample (see Sect.~\ref{sect:data_drug} and Fig.~\ref{fig:protein} for the details on the data processing method). Therefore, the resulting four-dimensional matrix, which will be the input of the 3D CNN, contains more than 2 million values.

Because of the high dimensionality of the data, a complex deep learning model is required to uncover all the hidden patterns that can help predict the target value. Training such a model means finding the optimal value for parameters that minimize a loss function, such as the MSE. Complex deep learning models have more training parameters, requiring longer execution times which limits the exploration of different architectures, hyperparameters and data processing techniques. This training process can be heavily accelerated using powerful \gls{GPU}s. However, GPUs are expensive computational resources and may not always be available. Moreover, as the size of datasets continues to grow, it is crucial to not only scale computational resources, like GPUs but also enhance the efficiency of our algorithms to meet the demands of larger datasets.

The number of training parameters of a ML model also affects its generalisation capacity, that is, the chances of overfitting the training data. The number of training samples should be at least comparable to the complexity of the model to guarantee low errors in the predictions of new data (see Eq.~\ref{Hoeffding}). In some cases, when the test data is similar enough to the training data, a smaller training set can still allow for good performance. Nonetheless, increasing the complexity of a model always increases its chances of producing overfitting, and thus it is convenient to find simpler ML models. Because of the capacity of quantum devices to deal with larger datasets with few computational resources, combining quantum algorithms with ML allows to reduce the complexity of the classical ML methods, while maintaining its accuracy. 

In this last section of this thesis, a hybrid quantum-classical 3D CNN is proposed, where the first convolutional layer is replaced by a quantum circuit (a QR), effectively reducing the number of training parameters of the model. The results of this work show that as long as the QR is properly designed, the hybrid CNN maintains the performance of the classical CNN. Moreover, the hybrid CNN has 20\% fewer training parameters, so the training times are reduced by 20-40\%, depending on the hardware used to train the models. All the quantum circuits used in this work have been executed using quantum simulation due to the current limitations of quantum hardware. However, performance benchmarks considering different noise models and error probabilities are provided. The results show that with error probabilities lower than $p=0.01$ and circuits with 300 gates, a common error mitigation algorithm can accurately mitigate the errors produced by the quantum hardware.

 The next subsections are organized as follows. In Sect.~\ref{sect:data_drug} we present the protein-ligand data used for this study, and the processing techniques used to prepare the data for the CNN models. The details of the classic CNN are given in Sect.~\ref{sect:CNN_drug}. The hybrid CNN is described in Sect.~\ref{sect:Hybrid_drug}, where details of the data encoding process (Sect.~\ref{sect:encoding_drug}) and QRs (Sect.~\ref{sect:QR_drug}) are provided. The error mitigation strategy used to reduce the errors induced by the quantum hardware is explained in Sect.~\ref{sect:Error_mitig}. Finally, the results and comparison of the classical and hybrid CNNs are presented in Sect.~\ref{sect:results_drug}.

\subsection{Data}
\label{sect:data_drug}
 The data used for this study is sourced from the PDBBind database~\citep{PDBBind} -a curated subset from the Protein Data Bank- which contains a collection of protein-ligand biomolecular complexes, manually collected from the associated publications. For each protein-ligand complex, the data files contain information about the 3D morphology, types of bonds between their constituent atoms, together with the protein-ligand binding affinity.
 
 Binding affinities were experimentally obtained by measuring the equilibrium dissociation constant ($k_d$) between protein-ligand and the inhibition constant ($k_I$). Mathematically, the $k_d$ constant is the concentration of the dissociated ligand divided by the concentration of the target-ligand complex at equilibrium. A lower $k_d$ value indicates a stronger binding affinity, meaning that the ligand has a higher tendency to bind to the target and form the complex. On the other hand, the $k_I$ constant represents the concentration at which the ligand occupies 50\% of the receptor sites of the protein when no competing ligand is present. Lower values of $k_I$ indicate greater binding affinity since a smaller amount of ligand is needed to inhibit the protein's activity. Then, the binding affinity, in the PDBBind dataset, is defined as $-\log(\frac{k_d}{k_I})$.
 
  Because of its completeness and extension, the PDBBind dataset has recently become a popular benchmark for binding affinity prediction with both physics-based and ML methods~\citep{3DCNNBA, DeepLearningBA, Fingerprints}. The PDBBind dataset is already split into two non-overlapping sets, the general set and the refined set. The refined set is compiled to contain complexes with better quality out of the
general set. A subset from the refined set, called \emph{core set}, is separated to provide a small, high-quality set for testing purposes. In this study, we use the 2020 version of the PDBBind dataset. The general set (excluding the refined set) contains 14127 complexes, while the refined set contains 5316 complexes. The core set is significantly smaller, with only 290 data samples.

\subsubsection{Data processing}
In order to train classical and hybrid CNNs, the raw PDBBind data has to be transformed into an appropriate input format for the convolutional layers. The pre-processing used for this project was the same as the one from Refs.~\citep{Pafnucy, ATOM} to support a reproducible and comparable pipeline. This method uses 3D volume grids to represent the atomic relationships in a voxelized space. That is, each data sample has size $(C,N,N,N)$, where $N$ is the size of each dimension in space, and $C$ is the number of features extracted from the protein-ligand pair. For this project, we set $N=48$, so that each side of the volume space has a size of 48\r{A} and a voxel size of 1\r{A}. This size allows covering all the binding regions without having too large input sizes for the CNN models. Having set the dimension of the space, the following 19 features were extracted from each protein-ligand pair ($C=19$):

\begin{itemize}
    \item \textbf{Atom type:} One-hot encoding of the elements boron, carbon, nitrogen, oxygen, phosphorus, sulfur, selenium, halogen, or metal.
    \item \textbf{Atom hybridization:} Gives information about the number of $\sigma$ and $\pi$ bonds (i.e. the geometry of the bonds) connecting a particular atom to a neighboring one. Takes values 1, 2 and 3 for sp, sp$^2$ and sp$^3$ hybridizations, respectively.
    \item \textbf{Number of heavy atom bonds:} Heavy atoms are all atoms except for hydrogen.
    \item \textbf{Number of bonds with other heteroatoms:} Heteroatoms are those atoms different from hydrogen or carbon.
    \item \textbf{Structural properties:} one-hot encoding of hydrophobic, aromatic, acceptor, donor, and ring properties.
    \item \textbf{Partial charge:} Distribution of charge of an atom as a result of its chemical environment.
    \item \textbf{Molecule type:} Indicates whether it is a protein atom or a ligand atom (-1 for protein, 1 for ligand).
\end{itemize}

The feature extraction process was done with the OpenBabel tool (version 3.1.1.1)~\citep{openbabel}, following the steps in Ref.~\citep{ATOM}. OpenBabel is an open-source software program for cheminformatics, that allows one to read and write multiple chemical file formats and extract molecular properties from the raw data. 

%Figure 
\begin{figure*}[!ht]
\centering
\includegraphics
   [width=0.98\textwidth]{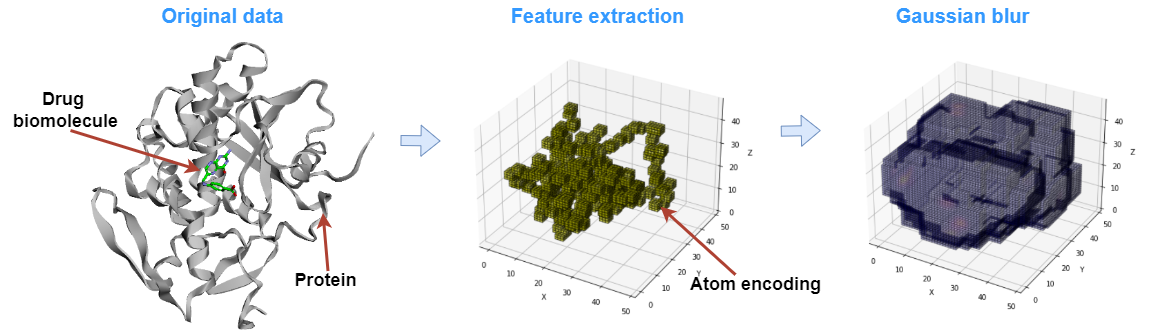}
\caption[Example of a data sample and further processing from the PDBBind dataset.]{Example of a data sample and further processing from the PDBBind dataset. The protein-ligand pair corresponds to the 1br6 sample from the refined set. The first step of the processing is feature extraction, where 19 features are voxelized in a 3D space. Then, a Gaussian blur is applied to produce a dense representation of the data.}
\label{fig:protein}
\end{figure*}

The Van-der-Waals radius was used to determine the size of each atom in the voxelized space. In this way, an atom could occupy one or more voxels depending on its Van-der-Waals radius. For atom overlaps, the features were added element-wise. The resulting 3D representations of the features resulted in sparse 3D matrices because most of the values in the 3D array were zero. Sparse data samples may make the training of NNs harder since the input samples are too similar to each other, and to a matrix with all zero entries. Therefore, NNs can have difficulties differentiating useful information from noise. For this reason, a Gaussian blur is applied to the voxelized features, populating the neighbouring atoms and thus reducing the number of zero-values voxels~\citep{Wave}. That is, given a data sample $X(x,y,z;c)$, where $x$, $y$, $z$ are the coordinates in the voxelized space and $c$ is the feature (i.e. one of the 19 features described above), the Gaussian blurring process consists of convolving the original image with a Gaussian kernel defined as
\begin{equation}
    K_\text{Gauss}(x,y,z) = \exp\Big(- \frac{x^2 + y^2 + z^2}{\sigma^2}\Big),
\end{equation}
with $\sigma=1$, and the convolution operation is defined as
\begin{equation}
    X_\text{Gauss}(x,y,z;c) = \sum_{\delta_x=-h}^h \sum_{\delta_y=-h}^h \sum_{\delta_z=-h}^h X(x + \delta_x,y + \delta_y,z + \delta_z;c) K_\text{Gauss}(\delta_x,\delta_y,\delta_z),
\end{equation}
where $h$ are the limits of the window where we define the convolution, in this case, $h=~4\sigma$. This leads to an exponential decay for every atom that fills multiple
voxels and blurs the picture, reducing sparsity and improving
the results of a CNN. Figure~\ref{fig:protein} shows a representation of the initial protein-ligand pair and the two main processing steps described above. Note that these 3D volume representations are very high-dimensional since more than 2 million real numbers are needed to represent only one data sample, as we discussed before. For this reason, large amounts of data samples and complex NN models are needed to make accurate predictions without overfitting the training data.

Apart from the general, refined and core sets, we further partitioned the general and refined sets into training and validation sets. This split is done to maintain the probability distribution of the binding affinities in both training and validation sets. In this way, we obtained training and validation sets for both the general and refined sets. Notice that the data processing is done independently for each one of the datasets used in this study.

\subsection{Classical neural network}
\label{sect:CNN_drug}
\begin{figure*}
\includegraphics[width=0.99\textwidth]{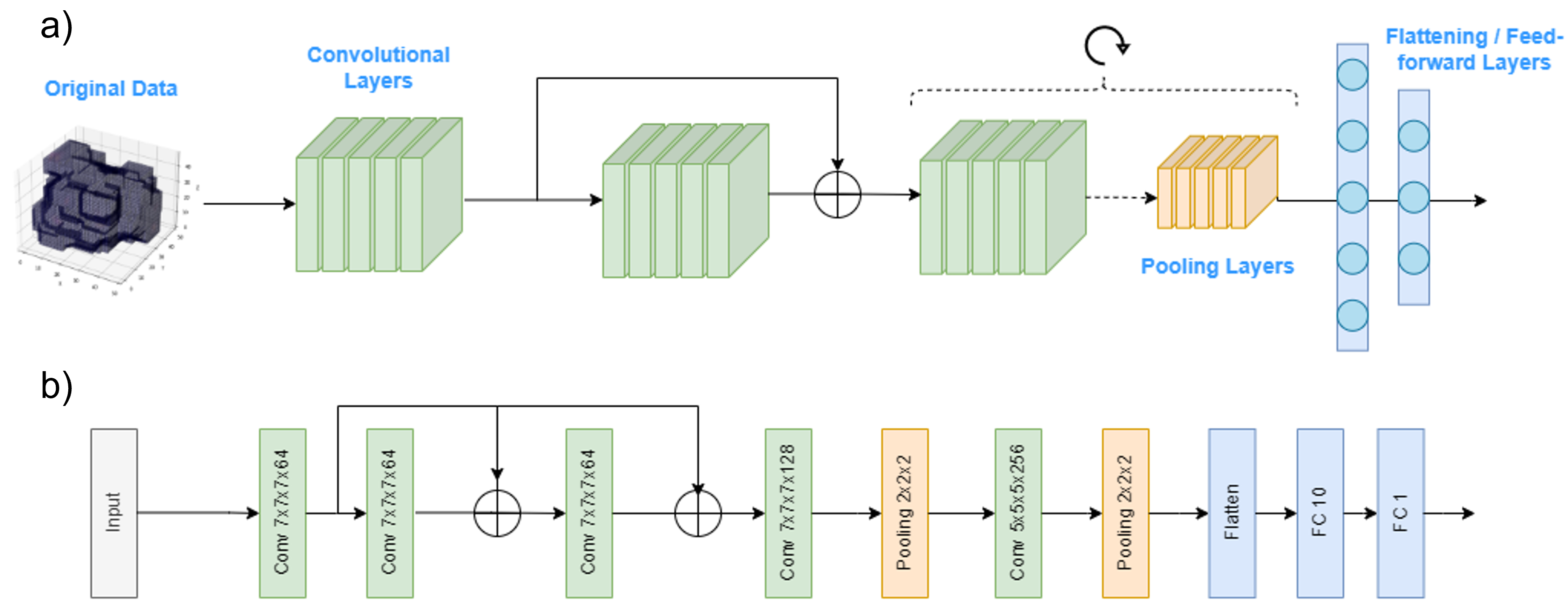}
\caption[a) Schematic representation of the components of a 3D convolutional neural network. b) Architecture of the 3D CNN used for this study.]{a) Schematic representation of the components of a 3D convolutional neural network. b) Architecture of the 3D CNN used for this study, proposed in Ref.~\citep{ATOM}.  }
\label{fig:CNN}
\end{figure*}

First, a classical 3D CNN is used to later compare the performance of a hybrid CNN with that of its classical analogue. Therefore, the architecture of both networks is the same, except for the quantum layer that replaces one of the classical convolutional layers. A representation of the layers of a 3D CNN is shown in Fig.~\ref{fig:CNN} a). 3D CNNs have been used for multiple applications such as volume image segmentation~\citep{ImageSegm}, medical imaging classification~\citep{medicalImages} and human action recognition~\citep{HumanAction}. A diagram of the 3D classical CNN used in this work is shown in Fig.~\ref{fig:CNN} b). The architecture is the same as the one proposed in Ref.~\citep{ATOM}, again for comparison purposes. The network contains five 3D convolutional layers, with 64, 64, 64, 128 and 256 filters respectively. The kernel size is 7 for all the layers except for the last one, which has kernel 5. The CNN contains two residual connections, as proposed in ResNet~\citep{ResNet}, which allow passing gradients to the next layers without a nonlinear activation function. Batch normalization is used after each convolutional layer, and we use ReLU as an activation function. The network contains two pooling layers and two fully-connected layers with 10 and 1 neurons respectively.

%Figure 

 %%%%%%%%%%%%%%%%%%%%%%%%%%%%%%%%%%%%%%%%%%%%%%%%%%%%%%%%%%
 % HYBRID NEURAL NETWORK
 %%%%%%%%%%%%%%%%%%%%%%%%%%%%%%%%%%%%%%%%%%%%%%%%%%%%%%%%%%
\subsection{Hybrid neural network}
\label{sect:Hybrid_drug}
Our hybrid (quantum-classical 3D) CNN is designed to reduce the complexity of the classical 3D CNN, described in the previous subsection, while maintaining its prediction performance. The hybrid CNN replaces the first convolutional layer with a quantum convolutional layer~\citep{Quanvolutional, tutorialQCNN, QCNN}. That is, each classical convolutional filter is replaced by a quantum circuit, which acts as a quantum filter. These quantum circuits should have significantly fewer training parameters than the classical convolutional layer, to reduce the overall complexity of the network. 

\begin{figure}
\centering
\includegraphics[width=0.90\textwidth]{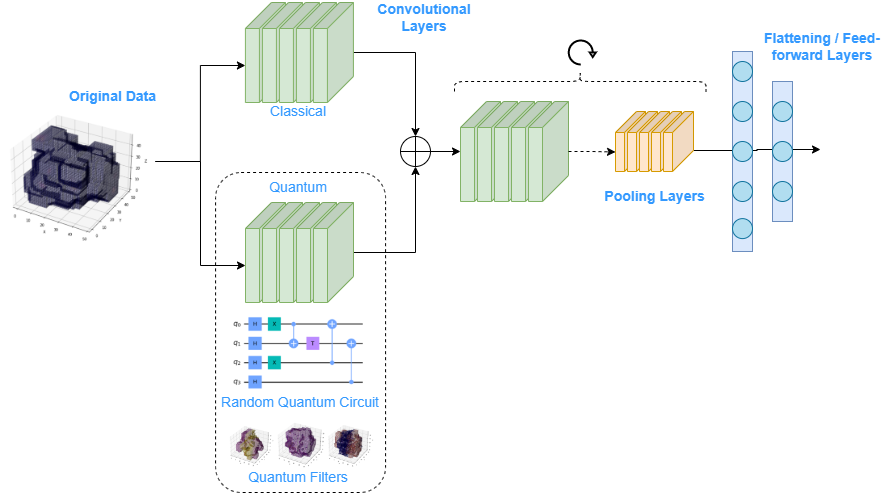}
\caption[Schematic representation of the hybrid quantum-classical 3D CNN.]{Schematic representation of the hybrid quantum-classical 3D CNN. The original data is processed by both a classical convolutional layer and a quantum convolutional layer. The outputs of both layers are then aggregated. The result is then fed to a set of convolutional and pooling layers, following the same architecture as the classical 3D CNN depicted in Fig.~\ref{fig:CNN}.}
\label{fig:QCNN}
\end{figure}

In our design, each quantum circuit is divided into two blocks: the \emph{data encoding}, which maps the input data into a quantum circuit, and the \emph{quantum reservoir}, where quantum operations are applied to retrieve information from the encoded data. The final architecture of the hybrid CNN is depicted in Fig.~\ref{fig:QCNN}. The processed protein-ligand data is fed to both a classical and a quantum convolutional layer. The outputs are aggregated by using a residual connection and then fed to the subsequent classical convolutional and pooling layers. The rest of the network is the same as its classical version. With this architecture, the first convolutional layer has been replaced by its quantum counterpart, while leaving the rest of the network unchanged. 
 
%Figure 

 \subsubsection{Data encoding}
 \label{sect:encoding_drug}

 The quantum convolutional layer aims to extract local features from the input data, just as the classical convolutional layer would do. For this reason, we split the input data into $(n\times n \times n)$ blocks and process each block individually. Given a block $B$, the data encoding process converts $B$ to a quantum state $\ket{B}$. Because of the high dimensionality of our data, we need to find a data encoding method that minimizes the number of qubits of the resulting quantum circuit. A suitable encoding should scale logarithmically with the dimension of the blocks. A popular data encoding mechanism that fulfills this property is called amplitude encoding~\citep{amplitudeEncoding}, which requires  [$\log_2(n^3)$] qubits to encode a block, which was introduced in Sect.~\ref{sect:QRC}. Given a data block with values $\vec{x} = (x_1, \cdots, x_{2^{n}-1})$, the amplitude encoding method would generate the quantum state
\begin{equation}
    \ket{B} = \frac{1}{||\vec{x}||} \sum_{i=0}^{2^{n}-1} x_i \ket{i}.
\end{equation}
 However, notice that the amplitude encoding scheme normalizes each block independently to produce a normalized quantum state. Therefore, the different blocks of the data would have different normalization constants and would not be comparable with each other.  
 
 For this reason, the Flexible Representation of Quantum Images (\gls{FRQI})~\citep{FRQI} method is chosen here, which normalizes the \emph{whole} data sample before the encoding, avoiding this problem, and uses only $\lceil \log_2(n^3) \rceil + 1$ qubits. FRQI was proposed to provide a normalized quantum state for 2D color images, which encodes both the value (colour) of a pixel and its position in an image. Given an image with $\theta = (\theta_0, \theta_1, \cdots, \theta_{2^{n-1}})$ pixels, where the pixels have been normalized such that $\theta_i \in [0, 2\pi), \forall i$, the encoded state is given by the following equation:
\begin{equation}
    \ket{I(\theta)} = \frac{1}{2^n} \sum_{i=0}^{2^{2n}-1} (\cos\theta_i \ket{0} + \sin \theta_i \ket{1}) \otimes \ket{i},
    \label{FRQI}
\end{equation}
where $\ket{i}, i=0,1,\cdots 2^{{2n}-1}$ are the basis computational states. Notice that this state is normalized since $||\ket{I(\theta)} ||= \frac{1}{2^n} \sqrt{\sum_{i=0}^{2^{2n}-1} (\cos^2(\theta_i) + \sin^2(\theta_i))} = 1$. 

For each $\theta_i$, the FRQI is composed of two parts: $\cos\theta_i \ket{0} + \sin \theta_i \ket{1}$ encodes the color of the pixel, and $\ket{i}$ encodes the position of the pixel in the image. As a simple example, a $(2\times 2)$ image and its representation are displayed in the following equation:
\begin{equation}
\begin{aligned}
    &\begin{array}{|c|c|}
        \hline
        \theta_{0} & \theta_{1}   \\
        \hline
        \theta_{2} & \theta_{3} \\
        \hline
    \end{array}, \qquad
    \begin{aligned}
        \ket{I}=\frac{1}{2}[ \; & \phantom{+} \left(\cos\theta_{0}\ket{0}+\sin\theta_{0}\ket{1} \right)\otimes\ket{00}&\\
        & + \left(\cos\theta_{1}\ket{0}+\sin\theta_{1}\ket{1} \right)\otimes\ket{01} \\
        & + \left(\cos\theta_{2}\ket{0}+\sin\theta_{2}\ket{1} \right)\otimes\ket{10}\\
        & + \left(\cos\theta_{3}\ket{0}+\sin\theta_{3}\ket{1} \right)\otimes\ket{11} \;].
    \end{aligned}
\end{aligned}
\label{FRQI_ex}
\end{equation}
The number of qubits needed to construct the FRQI state increases logarithmically with the number of pixels (angles) of the image since the dimension of the computational basis increases exponentially with the number of qubits of the Hilbert space. Even though the FRQI was designed for 2D colour images, the generalization to 3D blocks is straightforward. Let $B$ be a $(n \times n \times n)$ block, with normalized values $(\theta_0, \theta_1, \cdots, \theta_{n^3 - 1}), \theta_i \in [0, 2\pi), \forall i$.   The FRQI state would then be given by Eq.~\ref{FRQI3D}.

\begin{equation}
    \ket{B} = \frac{1}{n^3} \sum_{i=0}^{n^3 - 1} (\cos\theta_i \ket{0} + \sin \theta_i \ket{1}) \otimes \ket{i}.
    \label{FRQI3D}
\end{equation}

Notice that the only difference between Eq.~\ref{FRQI} and Eq.~\ref{FRQI3D} is the number of angles of the quantum state (the number of terms in the sum). When $n^3$ is a power of 2 (i.e. $n^3 = 2^l, l \in \mathbb{N}$), the state in Eq.~\ref{FRQI3D} has non-zero components in all the states of the computational basis. Therefore, choosing $n^3$ as a power of 2 mostly exploits the use of the Hilbert space, since we are creating entanglement through all the basis states. For this reason, $n=4$ is set for the experiments, since $n^3 = 4^3 = 2^6$. A higher value of $n$ could also be used, such as $n=8$. However, since we will be running the experiments on classical hardware (CPUs and GPUs), simulating larger quantum systems would require larger memory capacities and longer execution times. Therefore, we will for now keep a small value of $n$ for our simulations, and leave it as future work considering larger blocks when using actual quantum hardware. 

\begin{figure}[!ht]
\centering
\includegraphics[width=0.75\textwidth]{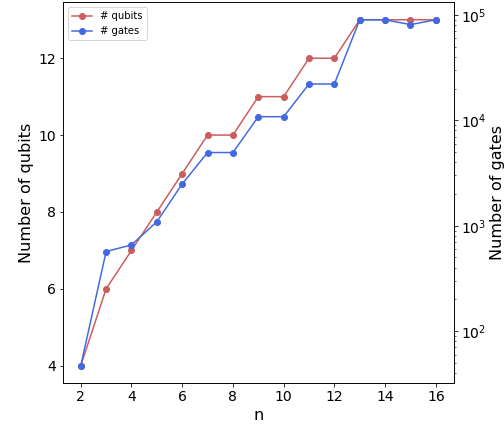}
\caption[Example of scaling of the flexible representation of quantum images.]{Example of scaling of the flexible representation of quantum images. The number of qubits scales logarithmically with the dimension of the block $n^3$. The number of gates of the quantum circuit scales linearly with the dimension of the block $n^3$.  }
\label{fig:FRQI}
\end{figure}

Figure~\ref{fig:FRQI} shows an example of the scaling of the number of qubits and the number of gates with the block size $n$. The number of qubits needed for the FRQI encoding is $\lceil \log_2(n^3) \rceil + 1$, so it scales logarithmically with the dimension of the block.

On the other hand, we have also calculated the number of gates needed to implement the FRQI on a real quantum device. The number of gates depends on the values of the block~$\theta_i$. If there are some angles with the same value, the quantum circuit can be compressed to reduce the number of gates. In Ref.~\citep{FRQI}, the authors showed how the FRQI quantum circuits can be simplified by minimizing boolean expressions. As an example of how the number of gates can scale with the block size, we have considered blocks from our data with the highest mean absolute sum, to ensure that we chose blocks with highly different angles. Figure~\ref{fig:FRQI} shows that the number of gates of a general FRQI encoding scales linearly with the dimension of the block, $n^3$.

% Figure 4

 \subsubsection{Quantum reservoirs}
 \label{sect:QR_drug}
 After the data has been encoded in a quantum circuit, a set of quantum gates is applied to perform the quantum transformation, followed by a set of measurements that convert the data back to a classical representation. In this work, we decided to use QRs as the quantum transformation of the data, since we have already tested its performance in other QML tasks and found the optimal design for the quantum circuits. Moreover, the main advantage of using QRs is the low complexity of the model, and thus, its easy training strategy. 

 %Figure 
\begin{figure}[!ht]
\includegraphics[width=0.98\textwidth]{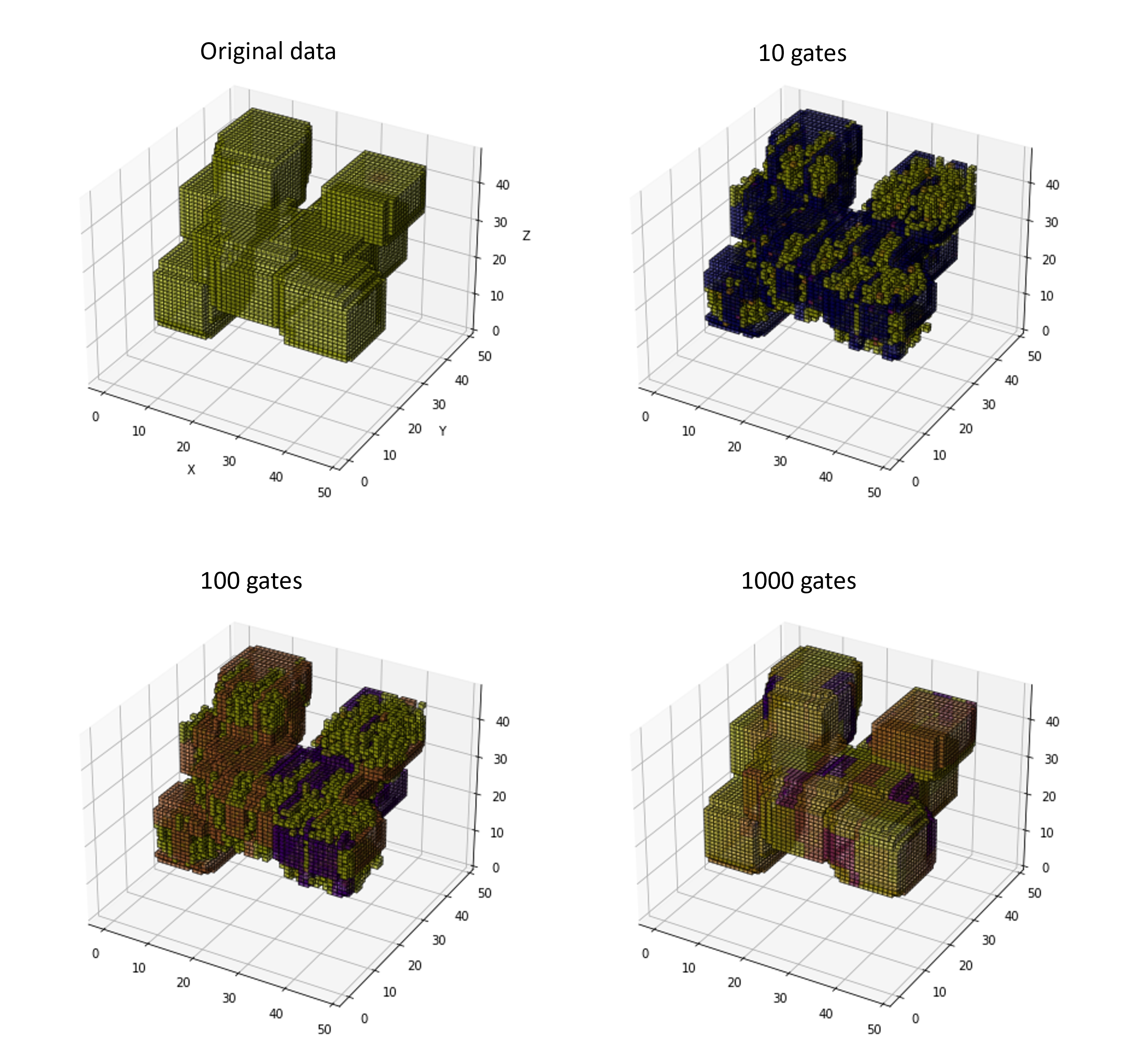}
\caption[Example of an output of the quantum convolutional layer, together with its input.]{Example of an output of the quantum convolutional layer, together with its input. The quantum convolutional layer is composed of a FRQI encoding layer followed by a quantum transformation generated with a random quantum circuit from the G3 family. Examples of the output are given for circuits with 10, 100 and 1000 quantum gates.}
\label{fig:quantum_filters}
\end{figure}

 As a result, the QR consists of a quantum circuit randomly generated with gates from the G3 family. Then, the qubits are measured on the computational basis, providing the output of the quantum convolutional layer. The hybrid CNN is trained with QRs having 20, 50, 100, 200, 300, 400, 500 and 600 quantum gates. In this way, we can evaluate how the depth of the QR influences the performance of the model for this task, and compare the results with the ones in Sect.~\ref{sect:performance_QRC}, where we found that the optimal design of QRs was obtained with the G3 family and 200-250 gates.  Figure~\ref{fig:quantum_filters} shows an example of the output of the quantum convolutional layer. We see that with a low number of gates, the quantum layer extracts simpler quantum features than with a higher number of gates. 
 
 Apart from considering circuits generated from the G3 family, we also compare the performance of the models with QRs generated from the Ising model from Eq.~\ref{eq:Ising}. The parameters are chosen to be the same as in Ref.~\citep{DynamicalIsing}, just as in the previous sections. Since the current quantum computers have limited availability and high access queue times, which limit the number of iterative runs we can do for training, the hybrid CNNs are run using quantum simulation on classical hardware. The code has been optimized using Qiskit~\citep{Qiskit} and PyTorch~\citep{pytorch}, and adapted so that it could be trained on GPUs, just like the classical CNN.

%%%%%%%%%%%%%%%%%%%%%%%%%%%%%%%%%%%%%%%%%%%%%%%%%%%%%%%%%%%%%%%%%%
% ERROR MITIGATION
%%%%%%%%%%%%%%%%%%%%%%%%%%%%%%%%%%%%%%%%%%%%%%%%%%%%%%%%%%%%%%%%%%%

\subsection{Error mitigation}
\label{sect:Error_mitig}
One of the biggest challenges that the current quantum devices face is the presence of noise, as we broadly discussed in Sect.~\ref{sect:noise_QRC}. Even though the quantum circuits used in this study are run using quantum simulation, we have also evaluated the performance of the noisy quantum circuits using the three noise models in Sect.~\ref{sect:noise_QRC} for a small set of samples. That is, we performed noisy simulations with the amplitude damping channel, the depolarizing channel and the phase damping channel, for error probabilities $p=0.03, 0.01, 0.008, 0.005, 0.003, 0.001$. In Sect.~\ref{sect:noise_QRC} we showed that the amplitude damping model can actually improve the performance of QRs. In this section, however, we will not be able to evaluate if the amplitude damping noise also improves the performance of the hybrid CNN, because that would require performing noisy simulations with the whole dataset, which would require more computational resources than the ones available to us at the moment. Nonetheless, we will take a different approach, by studying whether the errors induced by the noisy simulations can be minimized using an error mitigation method.

Error mitigation methods aim to reduce the noise of the outputs after the quantum algorithm has been executed. In this work, the \emph{data regression error mitigation} (\gls{DREM})~\citep{DREM} algorithm is used to mitigate the noise of the quantum circuits. The DREM algorithm trains a ML model to correct the errors of noisy quantum circuits. To obtain the training set, random quantum circuits with 300 gates sampled from the G3 family are executed with both noisy and noiseless simulations. Thus, the training set consists of pairs $(X_i,y_i)$ where $X_i$ contains the counts of the noisy distribution and $y_i$ contains the counts of the noiseless distribution. In our case, the ML model is again a ridge regression model. The DREM is trained with 1000 samples. Then, the performance is tested with 500 noisy quantum circuits used in the quantum convolutional layer. In this case, the 3D volumetric space is divided into blocks of size $n=8$, leading to quantum circuits of 10 qubits and 300 gates. The DREM algorithm is suitable for this task since, once the ML model is trained, it can be used to mitigate multiple quantum circuits requiring very few classical computational resources. This makes it practical for large datasets.

%%%%%%%%%%%%%%%%%%%%%%%%%%%%%%%%%%%%%%%%%%%%%%%%%%%%%%%%%%%%%%%%%%%
% RESULTS
%%%%%%%%%%%%%%%%%%%%%%%%%%%%%%%%%%%%%%%%%%%%%%%%%%%%%%%%%%%%%%%%%%%
\subsection{Results}
\label{sect:results_drug}
Let us study now the performance of the classical CNN and the multiple variations of our hybrid CNN. In this way, we will see if the hybrid CNN keeps the same prediction performance as the classical CNN. Moreover, we will evaluate different designs of the hybrid CNN (using QRs with different number of quantum gates), and show how these designs affect the performance of the final model.

The performance of the models is evaluated against the core set of the 2020 PDBBind dataset. The training and validation steps are done with the refined set (separated into training and validation sets, see Sect.~\ref{sect:data_drug}) for all the models. To reduce overfitting of the training data, we use an early stopping procedure, finishing the training step when the performance in the validation set has converged. To evaluate the convergence of the training process, we assess five error metrics:

\begin{itemize}
    \item \textbf{Root mean squared error (\gls{RMSE})}
    \item \textbf{Mean absolute error (\gls{MAE})}
    \item \textbf{Coefficient of determination R squared (\gls{R2})}: proportion of the variation of the dependent variable (binding affinity) that is predictable from the independent variable (prediction of the model).
    \item \textbf{Pearson correlation coefficient (Pearson)}: Linear correlation between two variables (binding affinity and prediction of the model). It ranges between $-1$ (anticorrelation) and $+1$ (full correlation).
    \item \textbf{Spearman coefficient}: Monotonic correlation coefficient. It ranges between $-1$ and $+1$. A Spearman correlation of $+1$ or $-1$ occurs when a variable is a perfect monotone function of the other.
\end{itemize}

\subsubsection{Performance on the validation set}

During the training of a NN, the training parameters of the network are modified to minimize a loss function (the MSE in our case). However, if we train the network for too many epochs, the model may stop learning general patterns of the data, and start overfitting the training set. For this reason, it is important to stop training the NNs when the loss function in the validation set has stopped decreasing. At this point, the NN will have learned all the useful information from the data and will be ready to be used for prediction purposes. This standard procedure is called \emph{early stopping}.

%Figure 6
\begin{figure}[!ht]
\centering
\includegraphics[width=1.0\textwidth]{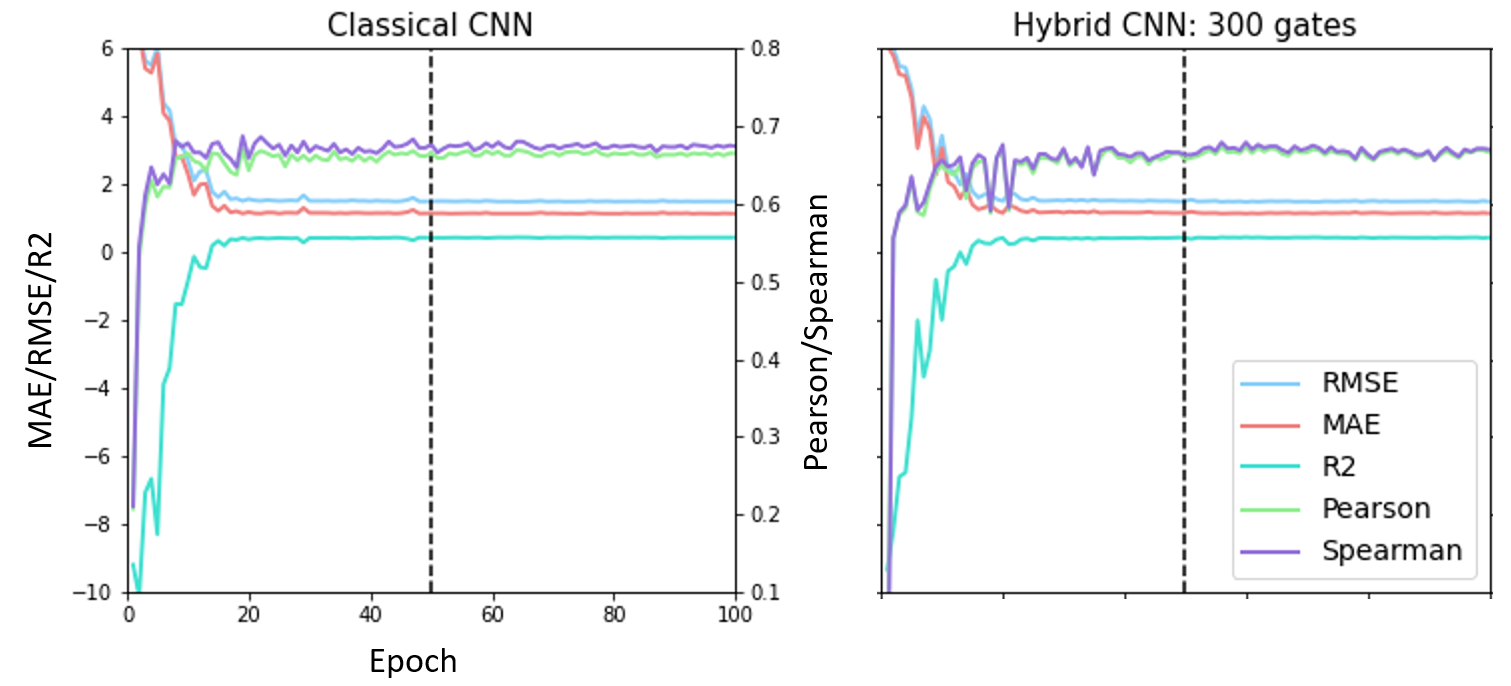}
\caption[Validation error metrics for the binding affinity predictions evaluated in the validation set as a function of the training epochs.]{Validation error metrics (RMSE, MAE, R2, Pearson and Spearman) for the binding affinity predictions evaluated in the validation set as a function of the training epochs. Panel a) shows the training of the classical CNN and panel b) shows the hybrid CNN with 300 quantum gates. Both models converged after 50 epochs.}
\label{fig:validation}
\end{figure}

Figure~\ref{fig:validation} shows the evolution with the number of training epochs of the five error metrics in the validation set, for the classical CNN and the hybrid CNN with 300 quantum gates. We see that for both cases, all the error metrics have stabilized after 50 epochs. Further training the models with the same data could lead to overfitting of the training data, decreasing the generalization capacity. For this reason, the training of all the models is stopped at 50 epochs. Figure~\ref{fig:validation} also shows that the Pearson and Spearman coefficients oscillate more than the other error metrics, even when the training has converged. For this reason, we conclude that, in this case, the RMSE, MAE and R2 are better measurements of convergence of the models.

\subsubsection{Performance on the test set: classical vs hybrid CNN}
\begin{figure}[!ht]
\centering
\includegraphics[width=0.95\textwidth]{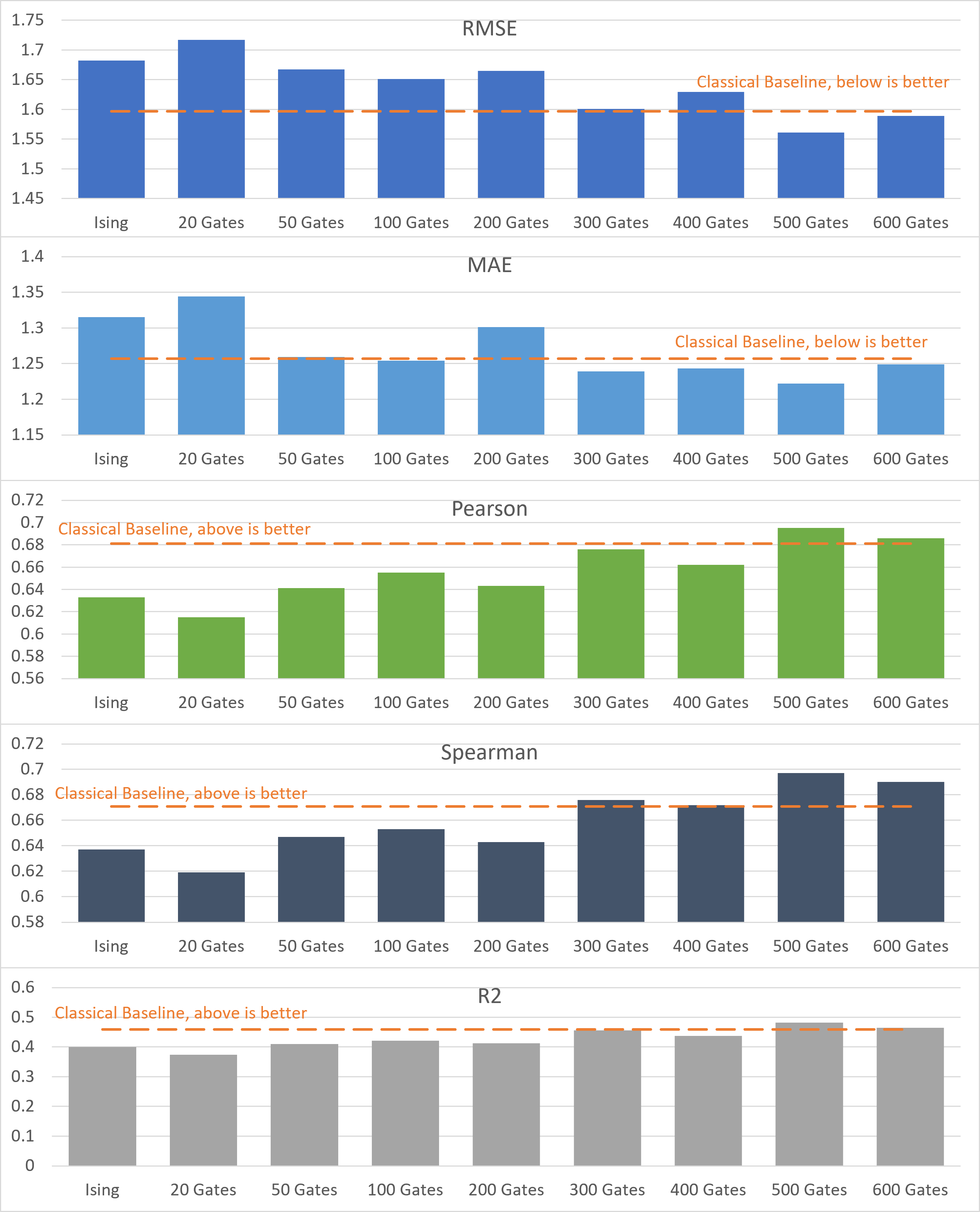}
\caption[Evaluation of the five error metrics (RMSE, MAE, R2, Pearson and Spearman) in the core set.]{Evaluation of the five error metrics (RMSE, MAE, R2, Pearson and Spearman) for the binding affinity prediction in the core set. Comparison of the hybrid CNN models constructed from the Ising model and the G3 family with 20,50,100,200,300,400,500 and 600 gates, together with the classical CNN. All the models have been trained with the refined set.}
\label{fig:test}
\end{figure}

Once the classical and hybrid models have been trained, we evaluate their performance on the test set. Figure~\ref{fig:test} shows the five error metrics presented before, evaluated on the test set (the core set from the PDBBind), for all the studied CNN models. The performance of the hybrid models with 20 - 600 quantum gates is compared with the performance of the classical CNN. 

%Figure 

The results show that, in general, the performance of the hybrid CNN models increases with the number of quantum gates until it reaches roughly the same performance as the classical CNN (the dashed orange line), at 300 quantum gates. From that point, the performance of the models with 400, 500 and 600 gates oscillates around the classical performance and does not improve significantly with the number of quantum gates. Therefore, we conclude that the number of quantum gates does affect the performance of the model, for shallow quantum circuits, and it stabilizes when the quantum circuits achieve a certain depth. The minimal number of gates needed to achieve classical performance, in this case, would be around 300 quantum gates. Thus, for a certain choice of quantum circuits, hybrid CNNs with lower complexity than the classical CNN have the same performance as the classical CNN.

Figure~\ref{fig:test} also shows the performance of the hybrid model constructed from the Ising model. We see that the performance is a bit worse than the optimal G3 hybrid model. Notice that these results completely agree with the conclusions of our previous studies (see Sects.~\ref{sect:optimalQRC} and~\ref{sect:QRC_forecasting}), where the ML tasks were significantly different. Therefore, we can conclude that this analysis is versatile and holds for different problems and ML algorithms.

%Table 
\begin{table}[!t]
    \centering
    \begin{tabular}{c|ccc}
    \hline \hline \\[-0.3cm]
        Hardware & Hybrid CNN & Classical CNN & Difference\\
        \hline\\
        Azure CPU & 10.7 days & 18.3days & 42\%\\
        Azure GPU & 24h & 39h & 38\%\\
        Purdue Anvil GPU & 16.3h & 22.1h & 26\%\\
        \hline \hline\\
        Training parameters & & & \\
        \hline\\
        All hardware & 8088499 & 10137129 & 20\%\\
         \hline
    \end{tabular}
    \caption[Training times and training parameters for the classical and hybrid 3D CNN trained with the refined set of the PDBBind dataset.]{Training times and training parameters for the classical and hybrid 3D CNN trained with the refined set of the PDBBind dataset. The details of the hardware are given in Table~\ref{tab:hardware}}
    \label{tab:time}
\end{table}

%Table 
\begin{table}[!t]
    \centering
    \begin{tabular}{c|ccc}
    \hline \hline \\[-0.3cm]
         & Azure CPU & Azure GPU & Purdue Anvil GPU\\
        \hline\\
        CPU &	Intel Xeon E5-2673  &	Intel Xeon E5-2690  & 2 x 3rd Gen AMD EPYC 7763\\
        Cores	& 4	& 6	& 128\\
        GPU	& -	& 1 x NVIDIA Tesla K80	& 4 x NVIDIA A100\\
        Memory	& 14GB	& 56GB	& 512GB\\
         \hline
    \end{tabular}
    \caption{Hardware specification for the different devices used to train the classical and hybrid 3D CNNs.}
    \label{tab:hardware}
\end{table}

\subsubsection{Training times}
The main motivation for designing a hybrid CNN model was to reduce the complexity and thus the training time of the NN. A measure of complexity that does not depend on the hardware where the model is trained is the number of training parameters. Table~\ref{tab:time} shows the training parameters of the classical CNN and all the hybrid CNN models. The classical CNN has around 10 million parameters, while the hybrid CNNs have around 8 million parameters, providing a 20\% reduction in model complexity. As can be seen, the training times of the models depend on the hardware where the training is executed. Training the CNNs with only CPUs requires much longer execution times. Using GPUs highly accelerates the training process, reducing the training time from many days to hours. In our experiments, the models have been trained using only CPUs and with two types of GPUs. The details of the used hardware are shown in Table~\ref{tab:hardware}. The improvement in training times with the hybrid model over the classical varies from 26\% to 42\%. Using more powerful GPUs reduces the difference in training times, but of course, the hardware is also more expensive. The difference in training times in all cases is limited by the difference in training parameters, which is a hardware-agnostic measure of complexity. 

% Figure 
\begin{figure}[!ht]
\centering
\includegraphics[width=0.95\textwidth]{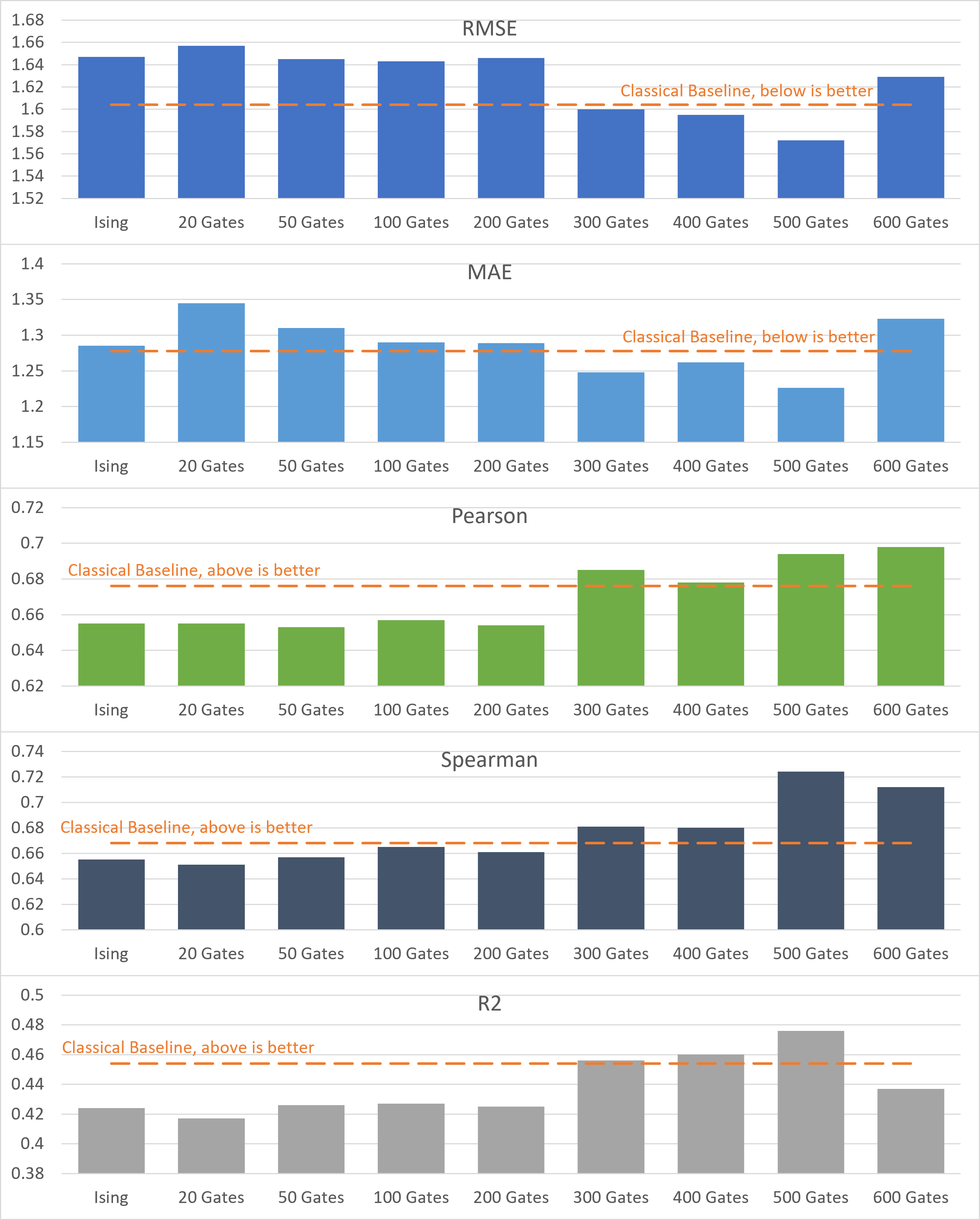}
\caption{Same as Fig.~\ref{fig:test}, where all the models have been trained with the general set.}
\label{fig:test_general}
\end{figure}

\subsubsection{Performance of the models trained with the general set}
After analyzing the results from the models trained with the refined set (see subsection \textbf{Performance on the test set} above), we repeated the experiments training the models using the general set. The general set has almost three times more data than the refined set, and thus the training takes more time and computational resources. Our experiments showed that the models required more epochs for the performance to converge in the validation set. The corresponding performance metrics evaluated in the test set are displayed in Fig.~\ref{fig:test_general}. The results are equivalent to the ones from the models trained with the refined set (see Fig.~\ref{fig:test}). The performance of the hybrid G3 models increases with the number of quantum gates until it converges at around 300 gates. Then, the performance oscillates around the classical performance. The Ising model has suboptimal performance compared to the classical CNN or the hybrid G3 CNN with 300 gates. 

We conclude that training the models with the general set leads to equivalent results to training the models with the refined set, but it requires longer training times and more computational resources. 

\subsubsection{Scaling of the hybrid CNN complexity with the dimensionality of the data}
CNNs are widely used models to learn from data such as time series, images or volumetric representations. Their goal is to unravel hidden patterns from the input data which are used to predict the target. Thus, the complexity of a CNN model highly depends on the complexity of the data. Hybrid quantum-classical CNN models can help reduce the number of parameters of the NN while maintaining prediction accuracy. 

One natural question to ask is how this reduction of training parameters scales with the size of the data. Let us consider that each sample has size $(C,N,N,N)$, where $C$ is the number of features and $N$ is the size of the volume side. The reduction of model complexity corresponds to the number of parameters of the first layer of the network. Therefore, the reduction of training parameters scales linearly with the number of features $C$. The number of training parameters does not explicitly depend on $N$, because each filter is applied locally to a portion of the data, as many times as needed to cover the whole sample. However, when the dimensionality of the data increases, usually more filters are needed for the CNN to converge. As the data complexity increases, more complex models are needed to learn useful information from it.

\begin{figure}[!ht]
    \centering
    \includegraphics[width=1.0\textwidth]{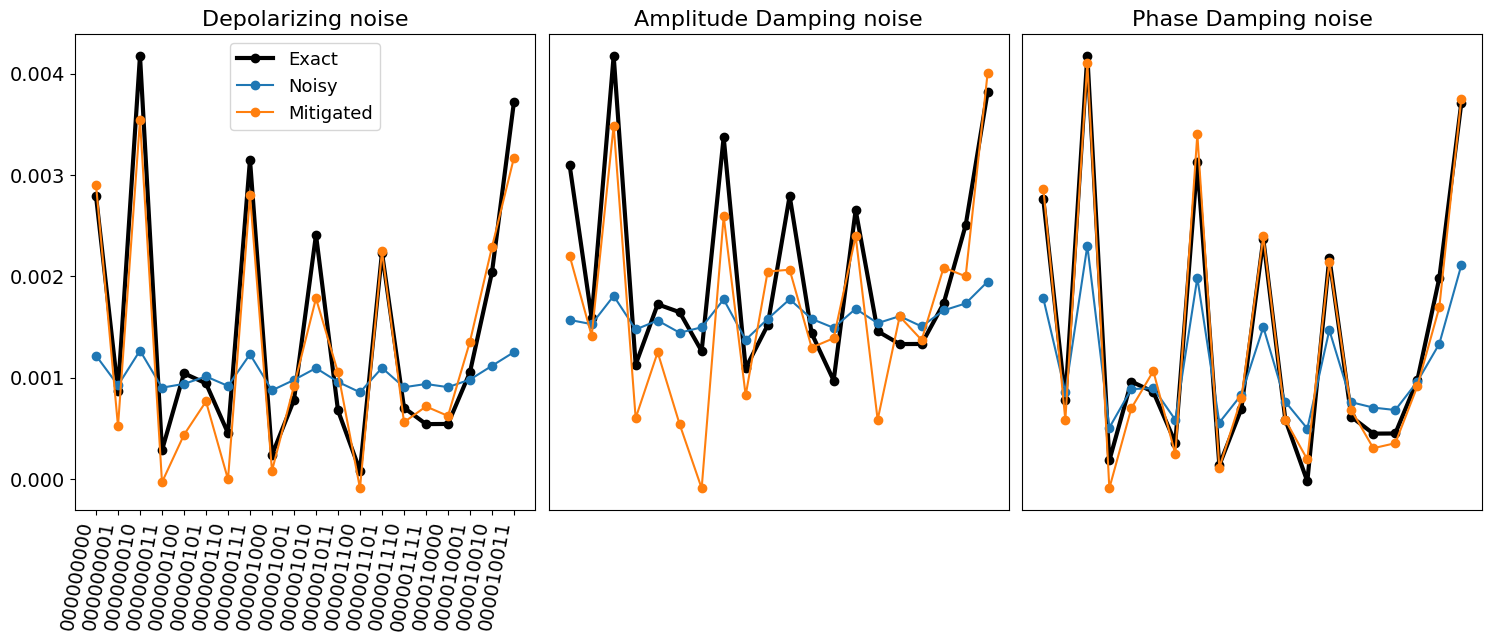}
    \caption[Output of the quantum circuit (measured in the computational basis) used in the quantum convolutional layer for three noise models and different error rates.]{Example of an output of the quantum circuit used in the quantum convolutional layer for three noise models and different error rates. For easier visualization, only the first 20 outputs are displayed in the figure.}
    \label{fig:counts_noise}
\end{figure}

\begin{figure}[!ht]
    \centering
    \includegraphics[width=1.0\textwidth]{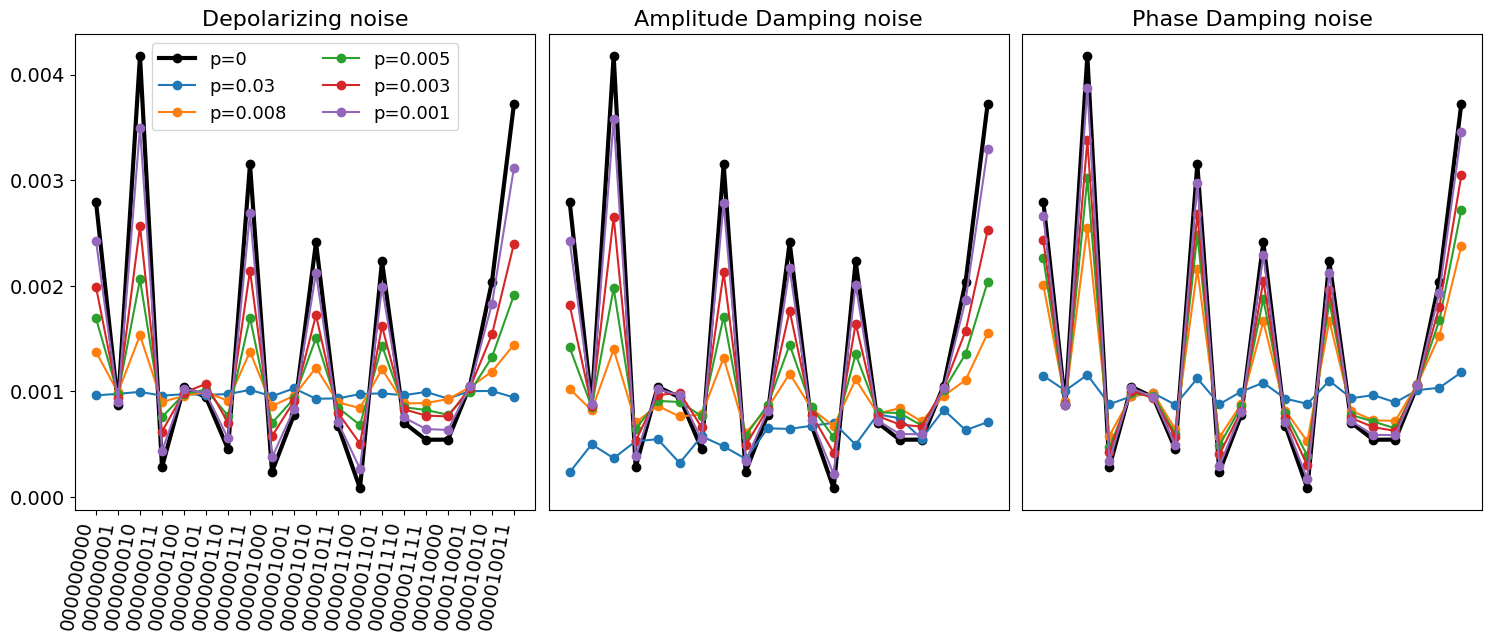}
    \caption[Output of the quantum circuit (measured in the computational basis) for three noise models, together with the output of the error mitigation algorithm.]{Example of an output of the quantum circuit used in the quantum convolutional layer for three noise models and error rate $p=0.01$, together with the output of the error mitigation algorithm. For easier visualization, only the first 20 outputs are displayed in the figure.}
    \label{fig:counts_mitig}
\end{figure}

\begin{table}[!ht]
    \centering
    \begin{tabular}{c|ccccc}
    \hline
    \hline \\[-0.3cm]
         Error model & Error rate & Regularization  & MSE Noisy  & MSE Mitigated  & Tendency \\
          & $p$ & $\alpha$ & circuits & circuits & accuracy \\
         \hline \\[-0.2cm]
         \multirow{6}{*}{Depolarizing} & 0.03 & 0.1 & $9.7 \times 10^{-7}$ &  $9.7 \times 10^{-7}$ & 0.59\\
          & 0.01 & 0.01 & $7.7 \times 10^{-7}$ &  $2.1 \times 10^{-7}$ & 0.75\\
          & 0.008 & $1 \times 10^{-5}$ & $6.7 \times 10^{-7}$ &  $1.0 \times 10^{-7}$ & 0.80\\
          & 0.005 & $1 \times 10^{-5}$ & $4.3 \times 10^{-7}$ &  $3.4 \times 10^{-8}$ & 0.84\\
          & 0.003 & $1 \times 10^{-5}$ & $2.3 \times 10^{-7}$ &  $1.1 \times 10^{-8}$ & 0.87\\
          & 0.001 & $1 \times 10^{-6}$ & $4.0 \times 10^{-8}$ &  $1.4 \times 10^{-9}$ & 0.89\\
         \hline\\[-0.2cm]
         \multirowcell{6}{Amplitude\\ Damping} & 0.03 & 0.1 & $1.1 \times 10^{-6}$ &  $9.9 \times 10^{-7}$ & 0.54\\
          & 0.01 & $1 \times 10^{-4}$ & $7.3 \times 10^{-7}$ &  $5.3 \times 10^{-7}$ & 0.62\\
          & 0.008 & $1 \times 10^{-4}$ & $6.2 \times 10^{-7}$ &  $3.4 \times 10^{-7}$ & 0.66\\
          & 0.005 & $1 \times 10^{-5}$ & $3.9 \times 10^{-7}$ &  $6.8 \times 10^{-8}$ & 0.75\\
          & 0.003 & $1 \times 10^{-5}$ & $2.0 \times 10^{-7}$ &  $2.1 \times 10^{-8}$ & 0.80\\
          & 0.001 & $1 \times 10^{-5}$ & $3.2 \times 10^{-8}$ &  $2.4 \times 10^{-9}$ & 0.84\\
         \hline\\[-0.2cm]
         \multirowcell{6}{Phase\\ Damping} & 0.03 & $1 \times 10^{-5}$ & $8.2 \times 10^{-7}$ &  $4.0 \times 10^{-7}$ & 0.67\\
          & 0.01 & $1 \times 10^{-5}$ & $3.2 \times 10^{-7}$ &  $3.7 \times 10^{-8}$ & 0.81\\
          & 0.008 & $1 \times 10^{-5}$ & $2.4 \times 10^{-7}$ &  $2.4 \times 10^{-8}$ & 0.82\\
          & 0.005 & $1 \times 10^{-5}$ & $1.2 \times 10^{-7}$ &  $9.4 \times 10^{-9}$ & 0.85\\
          & 0.003 & $1 \times 10^{-5}$ & $5.1 \times 10^{-8}$ &  $3.6 \times 10^{-9}$ & 0.85\\
          & 0.001 & $1 \times 10^{-6}$ & $6.9 \times 10^{-9}$ &  $5.0 \times 10^{-10}$ & 0.85\\
         \hline
    \end{tabular}
    \caption{Performance of the error mitigation algorithm by means of the mean squared error and tendency accuracy for different noise models and error rates.}
    \label{tab:error_mitig}
\end{table}
\vspace{0.5cm}

\subsubsection{Noisy simulations and error mitigation}
After analyzing the results of the noiseless quantum circuits, we now perform noisy simulations for three different quantum channels (amplitude damping, depolarizing and phase damping) and analyze the difference between the noisy and noiseless outputs. An example of the output of a quantum circuit used for the quantum layer for different noise models and different error probabilities is shown in Fig.~\ref{fig:counts_noise}. We see that the three noise models reduce the probability amplitude of the circuit outputs. As we discussed in Sect.~\ref{sect:noise_QRC}, the main difference between the phase damping channel and the depolarizing channel is that the former reduces the probability amplitude slower than the latter. In addition, the amplitude damping channel introduces additional non-zero coefficients in the outputs, apart from mitigating the amplitudes. 

In all cases shown in Fig.~\ref{fig:counts_noise}, when the error probability reaches $p=0.03$, the quantum information is lost, since the amplitude peaks can no longer be distinguished. On the other hand, when the error rate is smaller than $p=0.03$, the DREM algorithm can successfully mitigate the noisy outputs. An example of the performance of the DREM algorithm for $p=0.01$ is shown in Fig.~\ref{fig:counts_mitig}. Even though the noise of the quantum device significantly reduces the amplitudes of the distribution, the DREM algorithm can recover the original amplitudes with significant accuracy. 

For every noise model and error rate $p$, we performed a hyperparameter optimization to obtain the best linear model to mitigate the quantum errors. The results are shown in Table~\ref{tab:error_mitig}. Apart from evaluating the MSE, we also evaluate the tendency accuracy, that is, the proportion of times the DREM algorithm modifies the output in the correct direction. Let $\vec{y}_\text{noisy}$, $\vec{y}_\text{noiseless}$, $\vec{y}_\text{mitigated}$ be the noisy, noiseless and mitigated counts respectively. Then, the tendency accuracy measures the proportion of times $|\vec{y}_\text{mitigated} - \vec{y}_\text{noiseless}| < |\vec{y}_\text{noisy} - \vec{y}_\text{noiseless}|$. 

The results in Table~\ref{tab:error_mitig} show that when $p \leq 0.01$, the MSE of the mitigated circuits is smaller than the MSE of the noisy circuits. On the other hand, for $p=0.03$, the MSE of the mitigated circuits is similar or even larger than the MSE of the noisy circuits, and the tendency accuracy is barely better than random guessing. This result agrees with the results in Fig.~\ref{fig:counts_noise} since the noisy simulations with $p=0.03$ basically render a constant value. Table~\ref{tab:error_mitig} also shows that the tendency accuracy increases when the error probability $p$ decreases. For the depolarizing quantum channel, the tendency accuracy reaches the value $0.8$ when $p=0.008$ and increases to $0.89$ with $p=0.001$. 

The amplitude damping noise seems to be the hardest to mitigate since the tendency accuracy increases slower than the tendency accuracy of the other noise models. This is because the amplitude damping channel introduces non-zero counts apart from mitigating the amplitudes of the noiseless simulation. To mitigate this kind of error, a more complex ML algorithm could be used to improve the mitigation capacity. Therefore, even though this noise type can improve the performance of QRC~\citep{Domingo_QRCNoise}, it is also the hardest to mitigate.

On the other hand, the tendency accuracy increases faster with the phase damping channel, where it reaches the value $0.81$ with $p=0.01$. All in all, these results show that as long as the error rates are smaller than $p=0.01$, the DREM algorithm can successfully mitigate the errors introduced by the quantum device on the quantum convolutional layer with 300 gates.

In conclusion, the study reported here shows how QML offers the potential to reduce the complexity and long training times of classical NNs by leveraging the advantages of quantum computing to handle large and high-dimensional datasets and speed up ML algorithms.

\chapter{Conclusions}
\label{chapter6}
\begin{fquote}[Richard P. Feynman][1918-1988]
    If you want to learn something, read about it. If you want to understand something, write about it. If you want to master something, teach it.
\end{fquote}

The main objective of this thesis was to thoroughly investigate the principles of reservoir computing, a novel machine learning algorithm used to propagate dynamical systems, and develop effective architectures that provide applications in diverse domains. The results presented here unequivocally showcase the immense power and robustness of the algorithm. Indeed, the same reservoir computing framework has been tailored to tackle fundamentally different challenges, ranging from forecasting financial time series to studying complex Hamiltonians and enhancing drug design methods. Moreover, due to the model-agnostic nature of the algorithm, which relies solely on data,  reservoir computing can be applied across various domains without needing extensive expertise in each area. Additionally,  the algorithm's versatility has enabled us to explore not only its classical version but also its quantum computing counterpart. Leveraging the principles of quantum mechanics, quantum reservoir computing offers an enhanced feature representation and the ability to process larger datasets with fewer quantum computational resources. As the frontier of machine learning continues to expand, reservoir computing stands as an exciting avenue for future research and advancements in solving cutting-edge challenges in multiple domains. Let us now thoroughly 
 present the conclusions of this PhD thesis:

\begin{enumerate}
     \item Forecasting agricultural time series is challenging because of the small datasets, high volatility and strong dependence on external factors of the time series. To properly evaluate the performance of a forecasting model, it is not only necessary to look at typical error metrics, such as the mean absolute error, but also to consider their accuracy in anticipating the direction of the series. In this respect, the market direction accuracy is an excellent metric to evaluate whether a model can accurately predict a change in the series trend.
    
    \item We have compared five different reservoir computing variants for predicting the prices of zucchini, aubergine and tomato from 2013 to 2022. The optimal model consists of decomposing the time series into trend, seasonality and residuals and predicting each component with reservoir computing separately. With this approach, the method could accurately anticipate market falls and regions of high volatility. Moreover, the reservoir computing approach clearly outperforms traditional forecasting models such as the long short term memory networks and seasonal auto-regressive integrated moving average. These conclusions are not limited to the agri-food market but can be extended to many other domains where the time series have similar properties.

    \item Estimating the confidence intervals of the forecasts provides a better understanding of the uncertainty of the predictions and variability of the agricultural market. While such confidence intervals can be estimated by modeling the error distribution of the reservoir computing model, complex models that consider the intricate patterns of the time series and incorporate the relevant additional variables are required to achieve a precise estimation of the confidence intervals. The development of such models will be explored in future works. 

    \item Reservoir computing models that incorporate the correlation between prices of related products in a multivariate forecasting framework exhibit superior performance compared to reservoir computing models that solely consider individual product prices.
    
    \item We have designed and trained a fully-connected neural network (1D case) and a convolutional neural network (2D case) to solve the time-independent Schrödinger equation, using randomly generated polynomial potentials and their associated eigenfunctions. The network can successfully generalize to unseen, non-polynomial potentials, modeled by Morse oscillators, predicting with high accuracy the associated ground and excited eigenstates. 
    
    \item Using a convolutional neural network, we can recover the eigenfunctions of a Hamiltonian $\hat{H}$ by training on the eigenfunctions of a simpler Hamiltonian $\hat{H}_0$ that approximates $\hat{H}$. This approach was applied successfully to find the eigenfunctions of the 2D coupled Morse oscillator. The network was trained on analytical solutions of decoupled Morse Hamiltonians and a customized loss function was used to ensure optimal performance. The technique allows for the reproduction of high-energy states with large distortions compared to the training data.
    
    \item The neural network framework allows obtaining high-energy states without calculating lower-energy states, which improves the scalability of the method, compared to traditional numerical methods. However, obtaining each eigenstate requires training a separate neural network. 
    
    \item To overcome this limitation, we proposed a reservoir computing-based method to obtain all eigenenergies and eigenstates within a certain energy range. To integrate the time-dependent Schrödinger equation with reservoir computing, we introduced a complex-valued activation function and extended ridge regression to the complex domain. Additionally, we used a multi-step learning strategy to improve the generalization capacity of reservoir computing.

    \item The previous method was evaluated using four quantum systems: 1D harmonic oscillator, 1D Morse oscillator, 1D polynomial potential and 2D harmonic oscillator. After propagating an initial wavepacket with time, Fourier Transform was used to recover eigenenergies and eigenfunctions at similar energies. The multi-step method showed slower error growth in wavefunction predictions, which is critical for high-dimensional systems to avoid overfitting. Additionally, the multi-step method was 4-6 times faster than the traditional numerical integrator.

    \item Reservoir computing was used to solve the time-independent Schrödinger equation for a 2D coupled Morse oscillator. To reduce dimensionality, wavefunctions were represented using the coefficients of their decomposition in the decoupled Morse Hamiltonian basis. Predicting time evolution became harder for higher-energy states, but all predicted eigenenergies had errors below $1\cdot10^{-3}$ and eigenfunctions had absolute errors below $1\cdot10^{-4}$, showing successful propagation. Reservoir computing is 7-9 times faster than the traditional numerical integrator and less prone to numerical error propagation.

    \item The multi-step reservoir computing has been successfully applied to obtain the eigenstates and scar functions at different energy levels of the coupled quartic oscillator. The multi-step reservoir computing method is 8-15 times faster than the numerical integrator for the eigenstates calculation, and 9-15 times faster for the scars calculation. 
    
    \item We have analyzed the performance of seven families of quantum reservoirs (G1, G2, G3, MG, $D_2$, $D_3$ and $D_n$), with different complexity according to the majorization criterion, in predicting excited energy levels for LiH and H$_2$O molecules. The results show that the families with higher complexity according to the majorization criterion are the ones that perform better in quantum reservoir computing.
    \item The G3 family provide the optimal designs of quantum reservoirs. Its performance increases with the number of gates until it converges at around 200 gates (8-qubits) system and 250 gates (10-qubits). This is because the G3 family spans a larger subspace of the Pauli space as the number of gates increases, which allows for extracting more quantum information from the input data.
    \item The G1 and G2 families provide the worst performance in quantum reservoir computing, which decreases as the number of gates increases. These families span a smaller subset of the Pauli space as the number of gates increases.
    \item The MG and diagonal circuits provide slightly worse performance than the G3 family, which also agrees with the majorization criterion. The MG performance increases with the number of gates but converges faster than the G3 family. Nonetheless, the MG circuits are harder to implement in quantum devices.
    \item The performance of the $D_3$ and $D_n$ circuits are very similar to each other, although the performance of the $D_2$ circuits is significantly worse. 
    \item The commonly used Ising model can be implemented using gates only from the G3 family. The resulting circuits have 30-70 times more gates than the optimal G3 family. For this reason, the G3 family is more suitable for implementation in current quantum devices.
    \item The previous results also hold for quantum reservoir computing in temporal tasks. The design of quantum reservoirs is particularly relevant when no external regressor variables are used to forecast the time series. 
    \item We have studied the effect of quantum noise in the performance of quantum reservoir computing for three different error models: depolarizing, amplitude damping and phase damping noise. When the fidelity of the final state is larger than $0.96$, the quantum reservoirs with amplitude damping noise provide better results than the noiseless ones.  
    \item The depolarizing and phase damping channels always worsen the performance of the quantum reservoirs, even for small error probabilities. The depolarizing channel decreases the fidelity of the final state faster than the phase damping channel. For this reason, the depolarizing noise should be prioritized for correction.
    \item We provide a theoretical explanation for the differences between the three quantum channels. The amplitude damping channel, apart from mitigating the amplitude of the coefficients in the Pauli space, also introduces extra non-zero coefficients. For this reason, when the error rate is small, it provides a similar effect as increasing the number of gates in the noiseless reservoir. On the other hand, the depolarizing and phase damping channels only mitigate the amplitude of the extracted measurements, reducing the information provided by the classical machine learning model.
    \item Quantum reservoirs have been used to design a hybrid quantum-classical neural network to predict the binding affinity between potential drugs and their target proteins. The first layer of the classical convolutional neural network has been replaced by a layer of quantum reservoirs sampled from the G3 family. The performance of the hybrid network improves with the number of gates of the quantum reservoir until it stabilizes at around 300 quantum gates. This behavior agrees with the analysis performed with both temporal and non-temporal machine learning tasks.
    \item The hybrid network contains $20\%$ less training parameters than the classical network. The training times are reduced by $20-40\%$, depending on the hardware where they are trained.
    \item Noisy quantum reservoirs have been tested with the three error models and different error probabilities, for a subset of the validation set. For error probabilities lower than $p=0.01$ and circuits with 300 gates, the data regression error mitigation algorithm can accurately mitigate the errors produced by the quantum hardware.
\end{enumerate}

% ********************************** Back Matter *******************************
% Backmatter should be commented out, if you are using appendices after References
%\backmatter

% ********************************** Bibliography ******************************
\begin{spacing}{0.9}

% To use the conventional natbib style referencing
% Bibliography style previews: http://nodonn.tipido.net/bibstyle.php
% Reference styles: http://sites.stat.psu.edu/~surajit/present/bib.htm
%% Imports the package natbib

%% Sets the bibliography style
%\bibliographystyle{unsrtnat}

%\bibliographystyle{plainnat} % use this to have URLs listed in References
%\bibliographystyle{myabbrvnat} % Use for unsorted references  

\cleardoublepage
\bibliography{References/references} % Path to your References.bib file

% If you would like to use BibLaTeX for your references, pass `custombib' as
% an option in the document class. The location of 'reference.bib' should be
% specified in the preamble.tex file in the custombib section.
% Comment out the lines related to natbib above and uncomment the following line.

%\printbibliography[heading=bibintoc, title={References}]

\end{spacing}

% ********************************** Appendices ********************************

\begin{appendices} % Using appendices environment for more functunality

%!TEX root = ../thesis.tex
% ******************************* Thesis Appendix A ****************************
%!TEX root = ../thesis.tex
% ******************************* Thesis Appendix B ********************************

\chapter{The Variational method for a harmonic oscillator basis set}
\label{appendix1}

%=============================================================
Given the 1D Hamiltonian
\begin{equation}
\hat{H} = \frac{\hat{p}^2}{2m} + V(\hat{x}),
\end{equation}
the mean energy corresponding to the normalized wavefunction $\phi(x)$ is given by
\begin{equation}
\expval{\hat{H}} = \expval{\hat{H}}{\phi} = \int_{-\infty}^\infty \phi^*(x) H(x) \phi(x)dx.
\end{equation}
For simplicity, we will use from now on $m=1$ and $\hbar=1$ 
(which are the values that have been used throughout this work) 
and thus, these parameters will be omitted in all expressions and calculations.

\subsection*{Variational principle}
The variational principle states that the mean energy under a Hamiltonian $\hat{H}$ for a wavefunction is always greater or equal to the exact ground state energy of such Hamiltonian. That is
\begin{equation}
E_0 \leq \expval{\hat{H}} = \expval{\hat{H}}{\phi} \quad \forall \ket{\phi} \in \mathcal{H}.
\end{equation}
This principle can be extended to higher eigenenergies by imposing that the state $\psi$ is orthogonal to the 
previous eigenstates.

\subsection*{The harmonic oscillator basis set}

We choose as a basis set for $\mathcal{H}$ consisting of the eigenfunctions of the harmonic oscillator (HO) 
with $\omega=1$, i.e., 
\begin{equation}
\chi_n(x) = \frac{1}{\sqrt{2^n n! \sqrt{\pi}}} e^{-x^2/2} H_n(x)\, ,
\end{equation}
where $H_n(x)$ is the $n$-th Hermite polynomial, defined by the recurrence
\begin{equation}
H_n(x) = 2xH_{n-1}(x) - 2n H_{n-2}(x),\qquad \mbox{and} \quad H_0(x) = 1, \ H_1(x) = x^2.
\end{equation}
Since $\{\phi_n(x)\}_n$ form a complete basis for $\mathcal{H}$, we can write any wavefunction $\psi(x)$ as a 
linear combination of the eigenfunctions or basis set elements $\{\chi_n\}$.
\begin{equation}
\psi(x) = \sum_{n=0}^\infty a_n \chi_n(x),
\end{equation}
and the associated mean energy of $\psi$ under the effect of the Hamiltonian $H$ is
\begin{multline}
\expval{\hat{H}} = \expval{\hat{H}}{\psi} = \int_{-\infty}^\infty \Big(\sum_{n=0}^\infty a_n \chi_n(x)\Big) \hat{H} 
  \Big(\sum_{m=0}^\infty a_m \chi_m(x)\Big) dx =\\
  \sum_{n=0}^\infty \sum_{m=0}^\infty a_n a_m \int_{-\infty}^\infty \chi_n(x) H(x) \chi_m(x)dx =
  \sum_{n=0}^\infty \sum_{m=0}^\infty a_n a_m C_{nm},
\end{multline}
where the coefficients $C_{nm}$ are given by
\begin{equation}
C_{nm} = \int_{-\infty}^\infty A_n e^{-x^2/2} H_n(x) \Big(-\frac{1}{2} \frac{\partial^2}{\partial x^2} 
   + V(x) \Big) A_m e^{-x^2/2} H_m(x) dx, \qquad \mbox{being} \quad A_n = \frac{1}{\sqrt{n!2^n \sqrt{\pi}}}.
\end{equation}
Finally, to find a good estimation of the ground state, one uses a finite basis of harmonic oscillator eigenstates 
$\{\chi_n\}_{n=0}^N$ and then find the coefficients $\{a_n\}$ which minimize the mean energy $\expval{H}$. 

\subsection*{Finding the coefficients $\{a_n\}$}

To find the coefficients $\{a_n\}$ which minimize the energy $\expval{\hat{H}}$ we use the Lagrange Multipliers theorem, 
which state that the local minimum of a function $F$, under a constraint $G$ is the solution of
\begin{equation}
\nabla F = \lambda \nabla G, \quad \lambda \in \mathbb{R}.
\end{equation}
In this case $F$ is the mean energy $F(\{a_n\}) = \expval{\hat{H}}$, and $G$ is the normalization constraint. 
Since $\{\phi_n\}$ is a basis set of of $\mathcal{H}$, then $G(\{a_n\}) = \sum_{n=0}^N a_n^2 = 1$. Next, we calculate the partial derivative of $F$
\begin{multline}
\frac{\partial F}{\partial a_i} = \frac{\partial}{\partial a_i} \Big( \sum_{n=0}^N \sum_{m=0}^N a_n a_m C_{nm} \Big)
=  \frac{\partial}{\partial a_i} \Big( \sum_{n=0}^N a_n \Big) \Big(\sum_{m=0}^N a_m C_{nm}\Big) 
+ \frac{\partial}{\partial a_i} \Big( \sum_{m=0}^N a_m \Big) \Big(\sum_{n=0}^N a_n C_{nm}\Big) =\\
\sum_{m=0}^N a_m C_{im} + \sum_{n=0}^N a_n C_{ni} = \sum_{n=0}^N a_n (C_{in} + C_{ni}),
\end{multline}
so that the gradient of $F$ is linear equations system
\begin{equation}
\nabla F(\vec{a}) = D \vec{a}, \quad \vec{a} = 
\begin{pmatrix} 
a_1\\
\vdots\\
a_N
\end{pmatrix}, \quad D \in \mathcal{M}_{N}(\mathbb{R}), \ [D]_{ij} = C_{ij} + C_{ji}.
\end{equation}
Moreover, the partial derivative of $G$ is
\begin{equation}
\frac{\partial G}{\partial a_i} = \frac{\partial }{\partial a_i} \Big(\sum_{n=0}^N a_n^2\Big) = 2a_i ,
\end{equation}
and the Lagrange Multiplier equation becomes
\begin{equation}
\nabla F(\vec{a}) = \lambda \nabla G(\vec{a}) \Longleftrightarrow D \vec{a} = 2\lambda \vec{a},
 \label{eq:41}
\end{equation}
which is an eigenvalue problem. 

The solution will then be found by solving the eigenvalue problem (\ref{eq:41}), and then selecting the vector $\vec{a_0}$ 
which minimizes $\expval{\hat{H}}$. 
Since the basis $\chi_n(x)$ is finite (we take up to $N$ functions), the solution will be an approximation of the exact eigenvector. 
When $N \rightarrow \infty$ the solution $\psi(x)$ will converge to the ground state of $H$. 
Moreover, since the eigenvectors of $D$ are orthonormal, the vector with the $n$-th lowest energy will be an 
approximation to the $n$-th excited state of the Hamiltonian.  

\subsection*{Integrals involving Hermite polynomials}

In order to generate a basis of the Hilbert space that we will use to approximate the ground state wavefunctions,
we need to perform some integrals involving Hermite polynomials, i.e.,
\begin{equation}
I(n,m,r) = \int_{-\infty}^\infty x^r e^{-x^2} H_n(x) H_m(x) dx
\end{equation}
which can be obtained by the recurrence relation
\begin{multline}
I(n,m,r) = \int_{-\infty}^\infty x^r e^{-x^2} H_n(x) H_m(x) dx = 
  \int_{-\infty}^\infty x^r e^{-x^2}\frac{1}{2x}\Big(H_{n+1}(x) + 2nH_{n-1}(x)\Big)H_m(x)dx = \\
  \frac{1}{2}I(n+1,m,r-1) + nI(n-1,m,r-1), 
\end{multline}
and taking into account that
\begin{equation}
I(n,m,0) = \sqrt{\pi} 2^n n! \delta_{n,m} 
\end{equation}

\subsection*{Calculating $C_{nm}$}

In order to compute matrix $D$ in the eigenproblem expression (\ref{eq:41}), we need to calculate the coefficients $C_{nm}$
\begin{equation}
C_{nm} = A_nA_m \int_{-\infty}^\infty e^{-x^2/2} H_n(x) (-\frac{1}{2} \frac{\partial^2}{\partial x^2} 
+ V(x)) H_m(x) e^{-x^2/2} dx,
 \qquad \mbox{with} \quad A_n = \frac{1}{\sqrt{n! 2^n \sqrt{\pi}}}.
\end{equation}
In order to do so we need to calculate
\begin{equation}
\frac{\partial^2}{\partial x^2}(H_m(x) e^{-x^2/2} ) = e^{-x^2/2}\Big((x^2-1) H_m(x) - 4mx H_{m-1}(x) 
+ 4m(m-1)H_{m-2}(x)\Big) := e^{-x^2/2} P(x)
\end{equation}
\begin{multline}
C_{nm} = A_n A_m \Big( - \frac{1}{2} \int_{-\infty}^\infty  H_n(x) P(x) e^{-x^2} dx 
+ \int_{-\infty}^\infty e^{-x^2} H_n(x)H_m(x)V(x) dx\Big) = \\
A_nA_m\Big(- \frac{1}{2} I(n,m,2) + 1/2 I(n,m,0) + 2mI(n,m-1,1) - 2m(m-1)I(n, m-2, 0) + I_V\Big),
\label{eq:Cnm1D}
\end{multline}
where $I_V$ is the integral corresponding to the potential $V(x)$. If this potential is a polynomial
\begin{equation}
V(x) = \sum_{i=1}^N \alpha_i x^i ,
\end{equation}
then
\begin{equation}
I_V = \sum_{i=1}^N \alpha_i I(n,m,i) 
\end{equation}
\section{Variational method in 2D}
\label{sec:supp2}
The previous variational method can be extended to 2D as described below (we will only focus on the differences between 
the 1D and 2D problems). 

\subsection*{The harmonic oscillator basis set for 2D}

We choose as a basis of $\mathcal{H}$ the eigenfunctions of the harmonic oscillator for 
$\omega_x$, $\omega_y$ with $\omega_x/\omega_y \in \mathbb{R}-\mathbb{Q}$ in 2D, so that
\begin{equation}
\hat{H} = \frac{\hat{p}^2}{2m} + \frac{1}{2}m(\omega_x^2x^2 + \omega_y^2y^2).
\end{equation}
Notice that the frequency is different for the two dimensions. 
Since this Schr\"{o}dinger equation is separable, the eigenfunctions come as the product of 1D eigenfunctions in both coordinates
\begin{eqnarray}
\nonumber \chi_{n_x, n_y}(x,y) = \frac{1}{\sqrt{n_x!2^{n_x} \sqrt{\pi/\omega_x}}}\frac{1}{\sqrt{n_y!2^{n_y} \sqrt{\pi/\omega_y}}} 
  e^{-\frac{\omega_x x^2}{2}}e^{-\frac{\omega_y y^2}{2}} H_{n_x}(\sqrt{\omega_x}x)H_{n_y}(\sqrt{\omega_y}y) \\ 
 = \chi_{n_x}(x)\chi_{n_y}(y).
\end{eqnarray}
and the corresponding eigenenergies come in terms of $n_x$ and $n_y$ as
\begin{equation}
E_{n_x, n_y} = \hbar (\omega_x n_x + \omega_y n_y +1) .
\end{equation}
Choosing $\omega_x/\omega_y \in \mathbb{R}-\mathbb{Q}$, there is no degeneracy in the energy levels, which can then be ordered in an ascending mode. Therefore, there exists a unique bijective order in $(n_x,n_y)$ which sorts the energy levels. Hence, we can write $\chi_{n_x,n_y}(x,y) = \phi_n (x,y)$ where $n = n (n_x,n_y)$. 
Since $\{\chi_n(x,y)\}_n$ form a complete basis set for $\mathcal{H}$, 
we can write any wavefunction $\psi(x,y)$ as the following linear combination
\begin{equation}
\psi(x,y) = \sum_{n=0}^\infty a_n \chi_n(x,y),
\end{equation}
and the associated mean energy is
\begin{multline}
\expval{\hat{H}}= \expval{H}{\psi} = \int_{-\infty}^\infty\int_{-\infty}^\infty \Big(\sum_{n=0}^\infty a_n \chi_n(x,y)\Big) 
\hat{H} \Big(\sum_{m=0}^\infty a_m \chi_{m}(x,y)\Big) dx dy=\\
\sum_{n=0}^\infty \sum_{m=0}^\infty a_n a_m\int_{-\infty}^\infty \int_{-\infty}^\infty \chi_n(x,y) H(x,y) \chi_{m}(x,y)dx 
= \sum_{n=0}^\infty \sum_{m=0}^\infty a_n a_m C_{nm},
\end{multline}
where
\begin{multline}
C_{nm} = \int_{-\infty}^\infty \int_{-\infty}^\infty A_{n_x} A_{n_y}  e^{-x \omega_x^2/2} H_{n_x}(\sqrt{\omega_x}x)e^{-y^2\omega_y/2} H_{n_y}(\sqrt{\omega_y}y) 
\Big(-\frac{1}{2} \frac{\partial^2}{\partial x^2} -\frac{1}{2} \frac{\partial^2}{\partial y^2} 
+ V(x,y) \Big)\times \\ A_{m_x}A_{m_y} e^{-x^2\omega_x/2}e^{-y^2\omega/2} H_{m_x}(\sqrt{\omega_x}x) H_{m_y}(\sqrt{\omega_y}y) dx dy, 
\qquad \mbox{with} \quad A_{n_x} = \frac{1}{\sqrt{n_x!2^{n_x} \sqrt{\pi/\omega_x}}}
\end{multline}
\subsection*{Calculating $C_{nm}$}

Now, for any given potential of the form
\begin{equation}
V(x,y) = \sum_{j+j\leq k} \alpha_{ij} x^i y^j ,
\end{equation}
and taking into account
\begin{equation}
\frac{\partial^2}{\partial x^2}(H_m(x) e^{-x^2/2} ) = e^{-x^2/2}\Big((x^2-1) H_m(x) - 4mx H_{m-1}(x) 
+ 4m(m-1)H_{m-2}(x)\Big) := e^{-x^2/2} P_m(x),
\end{equation}
and defining the new variable $\tilde{y} = \sqrt{\omega}y$, the expression for $C_{nm}$ becomes for 2D
\begin{multline}
C_{nm} = \int_{-\infty}^\infty \int_{-\infty}^\infty A_{n_x} A_{n_y} e^{-\tilde{x}^2/2} H_{n_x}(\tilde{x})e^{-\tilde{y}^2/2} H_{n_y}(\tilde{y}) 
\Big(-\frac{\omega_x}{2} \frac{\partial^2}{\partial \tilde{x}^2} -\frac{\omega_y}{2} \frac{\partial^2}{\partial \tilde{y}^2} 
+ V(\tilde{x}/\sqrt{\omega_x},\tilde{y}/\sqrt{\omega}) \Big)\times \\ A_{m_x}A_{m_y} e^{-x^2/2}e^{-\tilde{y}^2/2} H_{m_x}(x) H_{m_y}(\tilde{y}) d\tilde{x} d\tilde{y}.
\end{multline}
Or alternatively
\begin{multline}
C_{nm} = \Big( - \frac{\omega_x}{2} \int_{-\infty}^\infty e^{-\tilde{y}^2} H_{n_y}(\tilde{y}) H_{m_y}(\tilde{y}) d\tilde{y}
\int_{-\infty}^\infty  H_{n_x}(\tilde{x}) P_{m_x}(\tilde{x}) e^{-\tilde{x}^2} d\tilde{x} - \\
\frac{\omega_y}{2} \int_{-\infty}^\infty e^{-\tilde{x}^2} H_{n_x}(\tilde{x}) H_{m_x}(\tilde{x}) d\tilde{x}\int_{-\infty}^\infty  
H_{n_y}(\tilde{y}) P_{m_y}(\tilde{y}) e^{-\tilde{y}^2} d\tilde{y} \ + \\
 \int_{-\infty}^\infty \int_{-\infty}^\infty e^{-\tilde{x}^2-\tilde{y}^2} H_{n_x}(\tilde{x})H_{n_y}(\tilde{y})H_{m_x}(\tilde{x})
 H_{m_y}(\tilde{y})V(\tilde{x}/\sqrt{\omega_x},\tilde{y}/\sqrt{\omega_y}) d\tilde{x}d\tilde{y}\Big) = \\
A\Big( \omega_x\sqrt{\pi}2^{n_y}n_y! \delta_{n_ym_y}I_P(n_x,m_x) 
+ \omega_y\sqrt{\pi}2^{n_x}n_x! \delta_{n_xm_x}I_P(n_y,m_y) 
+ \\
\sum_{i+j\leq k} \alpha_{ij} \omega_x^{-i/2}\omega_y^{-j/2} I(n_x, m_x, i)I(n_y,m_y,j) \Big), \\ A= A_{n_x} A_{n_y} A_{m_x}A_{m_y},
\label{eq:Cnm2D}
\end{multline}
where $I_P(n,m)$ is:
\begin{equation}
I_P(n,m) = - \frac{1}{2} I(n,m,2) + 1/2 I(n,m,0) + 2mI(n,m-1,1) - 2m(m-1)I(n, m-2, 0),
\end{equation}
being
\begin{equation}
I(n,m,r) = \int_{-\infty}^\infty x^r e^{-x^2} H_n(x) H_m(x) dx.
\end{equation}
\section{Variational method for the coupled Morse Hamiltonian}

In this section we adapt the variational method to integrate the coupled Morse Hamiltonian, given by the expression 
\begin{equation}   
H(x_1,x_2) = \frac{1}{2}G_{11}p_1^2 + \frac{1}{2}G_{22}p_2^2  +  G_{12}p_1 p_2 + U(x_1) + U(x_2), 
\end{equation}
where $U(x)=D_e(e^{-2a(x-x_0)} - 2e^{-a(x-x_0)})$, $\ G_{11}=\frac{M_H + M_O}{M_HM_O}$ and $G_{12}=\frac{\cos(\alpha)}{M_O} < 0$.

Just as in the previous sections, the 2D harmonic oscillator basis is chosen as a basis of the Hilbert space. The problem reduces to finding the coefficients of the eigenfunctions in the harmonic oscillator basis $\{a_n\}_n$. The only difference from the previous derivation is the calculation of the coefficients $C_{nm}$. In this case, we have:
\begin{eqnarray}
     &C_{nm} =A \int_{-\infty}^\infty \int_{-\infty}^\infty e^{-\tilde{x_1}^2/2}e^{-\tilde{x_2}^2/2} H_{n_1}(\tilde{x_1}) H_{n_2}(\tilde{x_2}) \times \\ \nonumber
    &\Biggr[-\frac{\omega_1}{2}G_{11} \frac{\partial^2}{\partial \tilde{x_1}^2} 
    -\frac{\omega_2}{2}G_{11} \frac{\partial^2}{\partial \tilde{x_2}^2} 
    -\sqrt{\omega_1\omega_2}G_{12}\frac{\partial^2}{\partial \tilde{x_1}\partial \tilde{x_2}}+ \\ \nonumber
    &D_e \left(e^{-\frac{2a}{\sqrt{\omega_1}}(\tilde{x_1}-\tilde{x}_{1,0})} - 2e^{-\frac{a}{\sqrt{\omega_1}}(\tilde{x_1}-\tilde{x}_{1,0})}+e^{-\frac{2a}{\sqrt{\omega_2}}(\tilde{x_2}-\tilde{x}_{2,0})} - 2e^{-\frac{a}{\sqrt{\omega_2}}(\tilde{x_2}-\tilde{x}_{2,0})} \right)\Biggr] \times \\ \nonumber
    &e^{-\tilde{x_1}^2/2}e^{-\tilde{x_2}^2/2} H_{m_1}(\tilde{x_1}) H_{m_2}(\tilde{x_2})  d\tilde{x_1} d\tilde{x_2} = \\ \nonumber
    &A\Biggr[\sqrt{\pi}G_{11}\left(\omega_12^{n_2}n_2!I_P(n_1,m_1)\delta_{n_2,m_2}+\omega_22^{n_1}n_1!I_P(n_2,m_2)\delta_{n_1,m_1}\right)-\\ \nonumber
    &\sqrt{\omega_1\omega_2}G_{12}K_P(n_1,m_1,n_2,m_2)+  D_e\sqrt{\pi}(2^{n_2}n_2!\delta_{n_2,m_2}J_P(n_1,m_1,\omega_1,a)+\\ \nonumber
    &2^{n_1}n_1!\delta_{n_1,m_1}J_P(n_2,m_2,\omega_2,a))\Biggr].
\end{eqnarray}
In the previous equation the following new integrals are defined:
\begin{itemize}
    \item \textbf{Coupled kinetic energy integrals:} The integrals involving the coupling kinetic term ($G_{12}p_1p_2$) is the following:
    \begin{eqnarray}
    K_P(n_1,m_1,n_2,m_2)= & 4m_1m_22^{n_1+n_2}n_1!n_2!\pi\delta_{n_1,m_1-1}\delta_{n_2,m_2-1} \\ \nonumber &-2m_12^{n_1}n_1!\sqrt{\pi}I(n_2,m_2,1)\delta_{n_1,m_1-1}\\ \nonumber
    &-2m_22^{n_2}n_2!\sqrt{\pi}I(n_1,m_1,1)\delta_{n_2,m_2-1}+I(n_1,m_1,1)I(n_2,m_2,1).
    \end{eqnarray}
    \item \textbf{Potential energy integrals:} The last integral that we need to calculate is the following:
    \begin{equation}
        J_P(n,m,\omega,a)=J(n,m,\frac{2a}{\sqrt{\omega}})-2J(n,m,\frac{a}{\sqrt{\omega}}), \quad J(n,m,a)=\int_{-\infty}^{\infty}e^{-x^2}e^{-ax}H_n(x)H_m(x)dx.
    \end{equation}
    Using the recurrence relation $H_m=2xH_{m-1}-2(m-1)H_{m-2}$ we can write 
    \begin{equation}
        J(n,m,a)=-2(m-1)J(n,m-2,a)-\int_{-\infty}^{\infty}(-2xe^{-x^2})e^{-ax}H_n(x)H_{m-1}(x)dx.
    \end{equation}
    Now we integrate by parts the second term, taking into account that 
    \begin{eqnarray}
        &\frac{d}{dx}\left(e^{-ax}H_n(x)H_{m-1}(x)\right)= \\ \nonumber
        &e^{-ax}\Big(-aH_n(x)H_{m-1}+2nH_{n-1}(x)H_{m-1}(x)+2(m-1)H_n(x)H_{m-2}(x)\Big).
    \end{eqnarray}
    Putting everything together we obtain the recurrence relation 
    \begin{equation}
        J(n,m,a)=-aJ(n,m-1,a)+2nJ(n-1,m-1,a).
    \end{equation}
    The starting values for $m$ can be easily computed 
    \begin{eqnarray}
        &J(n,0,a)=\int_{-\infty}^{\infty}e^{-x^2}e^{-ax}H_n(x)dx= \\ \nonumber
        &\int_{-\infty}^{\infty}e^{-x^2}e^{-ax}\Big(2xH_{n-1}(x)-2(n-1)H_{n-2}\Big)dx=-aJ(n-1,0,a),
    \end{eqnarray}
    where we applied integration by parts for the first term.
    The initial values for $n$ are $J(0,0,a)=\sqrt{\pi}e^{\frac{a^2}{4}}$, $J(1,0,a)=-a\sqrt{\pi}e^{\frac{a^2}{4}}$, $J(2,0,a)=a^2\sqrt{\pi}e^{\frac{a^2}{4}}$. By induction, we conclude that
    \begin{equation}
        J(n,0,a)=\sqrt{\pi}(-1)^na^ne^{\frac{a^2}{4}}.
    \end{equation}
\end{itemize}

\subsection{Convergence of the variational method}
\begin{table}[!ht]
    \centering
    \begin{tabular}{rccc}
    \hline\hline \\[-0.3cm]
    n &  $E_n$ &  Absolute Error &  Squared Error \\ \\[-0.3cm]
    \hline \\
    0 & -0.388129 & 1.7$\times 10^{-14}$ & 1.3$\times 10^{-31}$ \\
    1 & -0.371424 & 1.1$\times 10^{-14}$ & 2.3$\times 10^{-31}$ \\
    2 & -0.371180 & 1.4$\times 10^{-14}$ & 3.7$\times 10^{-31}$ \\
    4 & -0.355232 & 6.6$\times 10^{-15}$ & 1.9$\times 10^{-31}$ \\
    5 & -0.354408 & 4.8$\times 10^{-15}$ & 1.2$\times 10^{-31}$ \\
    17 & -0.312634 & 2.2$\times 10^{-15}$ & 8.2$\times 10^{-32}$ \\
    19 & -0.311284 & 2.0$\times 10^{-15}$ & 6.0$\times 10^{-32}$ \\
    25 & -0.297732 & 2.1$\times 10^{-15}$ & 8.2$\times 10^{-32}$ \\
    40 & -0.260768 & 1.4$\times 10^{-15}$ & 9.0$\times 10^{-32}$ \\
    42 & -0.259161 & 1.5$\times 10^{-15}$ & 9.0$\times 10^{-32}$ \\
    63 & -0.233235 & 8.8$\times 10^{-16}$ & 3.8$\times 10^{-32}$ \\
    53 & -0.245563 & 1.3$\times 10^{-15}$ & 7.6$\times 10^{-32}$ \\
    69 & -0.230218 & 9.2$\times 10^{-16}$ & 3.1$\times 10^{-32}$ \\
    82 & -0.216359 & 1.5$\times 10^{-15}$ & 1.0$\times 10^{-31}$ \\
    96 & -0.206224 & 2.2$\times 10^{-09}$ & 2.6$\times 10^{-19}$ \\
    119 & -0.192591 & 6.6$\times 10^{-12}$ & 2.7$\times 10^{-24}$ \\
    143 & -0.179755 & 1.3$\times 10^{-13}$ & 9.1$\times 10^{-28}$ \\
    158 & -0.171747 & 7.4$\times 10^{-05}$ & 8.6$\times 10^{-10}$ \\
    164 & -0.174886 & 2.4$\times 10^{-04}$ & 1.5$\times 10^{-08}$ \\
    209 & -0.155549 & 6.6$\times 10^{-04}$ & 4.5$\times 10^{-08}$ \\
    214 & -0.152665 & 7.1$\times 10^{-05}$ & 3.4$\times 10^{-09}$ \\
    267 & -0.134658 & 1.0$\times 10^{-03}$ & 1.2$\times 10^{-07}$ \\
    272 & -0.132021 & 9.7$\times 10^{-04}$ & 1.1$\times 10^{-07}$ \\
    321 & -0.112876 & 8.4$\times 10^{-04}$ & 1.1$\times 10^{-07}$ \\
    322 & -0.112577 & 8.5$\times 10^{-04}$ & 7.6$\times 10^{-08}$ \\
    383 & -0.092889 & 1.0$\times 10^{-03}$ & 1.6$\times 10^{-07}$ \\
    396 & -0.087234 & 1.0$\times 10^{-03}$ & 1.7$\times 10^{-07}$ \\
    \hline
\end{tabular}
    \caption[Eigenenergies obtained with variational method.]{Eigenenergies obtained with variational method for states at different energies with $N=6500$, $\omega_x = 20$, $\omega_y=20.06225$, together with the average relative error and the average mean squared error.}
    \label{tab:convergence_variational}
\end{table}

\begin{figure}[!ht]
    \centering
    \includegraphics[width=1.0\textwidth]{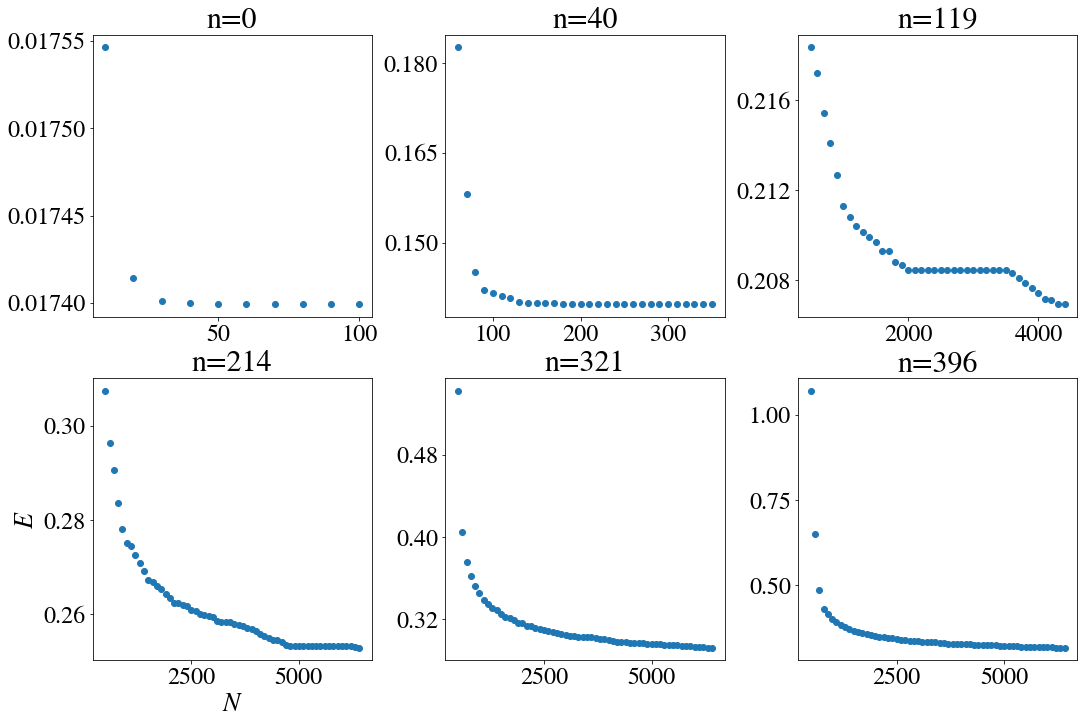}
    \caption{Convergence of the variational method for different energy levels.}
    \label{fig:convergence_vari}
\end{figure}

In this section, we analyze the convergence of the variational method for the coupled Morse potential. To do so, we calculate the eigenenergies for increasing the number of basis states $N$. We consider that the method has converged when the energies stabilize to a constant value. Figure \ref{fig:convergence_vari} shows the energy convergence for different energy levels. For lower energies, the variational method converges faster, requiring a lower number of basis states. As can be seen, at around $N=5500$ all the eigenenergies have stabilized. In order to estimate the accuracy of this method, we calculate the average relative error, defined as

\begin{equation}
    E_\text{rel} = \frac{1}{1000} \sum_{N=5500}^{6500} \frac{|E(N) - E(N-1)|}{E(N)},
\end{equation}

where $E(N)$ corresponds to the eigenenergy calculated by using $N$ basis elements. We also report the average mean squared error, defined as

\begin{equation}
    E_\text{MSE}= \frac{1}{1000} \sum_{N=5500}^{6500} (E(N) - E(N-1))^2.
\end{equation}

The eigenenergies, together with the relative and squared errors are provided in Table \ref{tab:convergence_variational}. We see that for the highest energy levels, the relative error is around 0.1\%, while the mean squared error is of around $1 \times 10^{-7}$. This illustrates the scalability problem present in the variational method. As the energy of the system increases, this method requires cubically more computational resources to obtain results with similar accuracy.

\chapter{Canonical transformations of the coupled Morse Hamiltonian}
\label{appendix2}

Providing only the potential function to the neural network, instead of the whole Hamiltonian, allows having an easy representation of the Hamiltonian as a grid containing the values of the potential energy on a rectangular spatial domain. The coupled Morse Hamiltonian contains a coupling term in the momentum coordinates instead of in the spatial coordinates. 
Therefore, to be able to provide only the potential energy to the neural network, we have to make a change of coordinates in the Hamiltonian in such a way that the coupling appears only in the spatial part, i.e.,~ the potential. To this end, we can rewrite the Hamiltonian in generalized coordinates so that the coupling appears in the spatial coordinates instead of in the momentum coordinates.

In order to do so, we apply a canonical transformation to the Hamiltonian $H(x_1,x_2,p_1,p_2)$ 
obtaining the Hamiltonian $H'(x_1', x'_2, p_1', p_2')$. 
We use a generating function of the form $F_2(x_1, x_2, p_1', p_2')$~\citep{Goldstein}, so that
\begin{equation}
\begin{array}{l}
p_i = \displaystyle\frac{\partial F_2}{\partial x_i}, \qquad i=1,2\vspace{0.07in}\\
x_i' = \displaystyle\frac{\partial F_2}{\partial p_i'}, \qquad i=1,2\vspace{0.07in}\\
H' = H + \displaystyle\frac{\partial F_2}{\partial t}.
\end{array} 
\end{equation}
In particular, by choosing $F_2(x_1,x_2, p_1', p_2')$ of the form

\begin{equation}
F_2(x_1,x_2, p_1', p_2') = f_1(x_1,x_2)\,p_1' + f_2(x_1,x_2)\,p_2'
\end{equation}
then, the generalized coordinates fulfill the following equations

\begin{equation}
\begin{array}{ll}
p_1 = \displaystyle\frac{\partial F_2}{\partial x_1} = \frac{\partial f_1}{\partial x_1}p_1' 
  +\frac{\partial f_2}{\partial x_1}p_2'\vspace{0.07in} \\
p_2 = \displaystyle\frac{\partial F_2}{\partial x_2} = \frac{\partial f_1}{\partial x_2}p_1' 
  +\frac{\partial f_2}{\partial x_2}p_2'\vspace{0.07in}\\
x_1' = \displaystyle\frac{\partial F_2}{\partial p_1'} = f_1(x_1,x_2)\vspace{0.07in}\\
x_2' = \displaystyle\frac{\partial F_2}{\partial p_2'} = f_2(x_1,x_2).
\end{array} 
\label{eq:16}
\end{equation}

The kinetic energy can be written in matrix form as

\begin{equation}
T(p_1,p_2) = \frac{1}{2} \begin{pmatrix}
p1 & p2\\
\end{pmatrix} 
\begin{pmatrix}
G_{11} & G_{12}\\
G_{12} & G_{11}
\end{pmatrix}
\begin{pmatrix}
p1\\
p2\\
\end{pmatrix} := \begin{pmatrix}
p1 & p2\\
\end{pmatrix} 
M
\begin{pmatrix}
p1\\
p2\\
\end{pmatrix},
\end{equation}
so that, diagonalizing $M$

\begin{equation}
M = S D  S^T \qquad \mbox{with} \quad
S = \frac{1}{\sqrt{2}}
\begin{pmatrix}
-1 & 1\\
1 & 1\\
\end{pmatrix}, \qquad \mbox{then} \quad
D = 1/2 \begin{pmatrix}
G_{11} - G_{12} & 0\\
0 & G_{11} + G_{12}
\end{pmatrix}.
\end{equation}
Accordingly, defining new coordinates $\vec{p'} = S^T \vec{p}$, the new kinetic energy only consists of by diagonal terms

\begin{equation}
T'(p_1', p_2') = \frac{1}{2} \left(G_{11}-G_{12}\right) \, p_1'^2 + \frac{1}{2} \left(G_{11}+G_{12}\right) \, p_2'^2.
\end{equation}

Now, the transformation defined by Eqs.~(\ref{eq:16}) can be used to compute the new set of coordinates $(x_1', x_2')$

\begin{equation}
x_1' = f_1(x_1,x_2) = \frac{1}{\sqrt{2}}(-x_1 + x_2), \qquad x_2' = f_2(x_1,x_2) = \frac{1}{\sqrt{2}}(x_1 + x_2),
\end{equation}

to obtain the new Hamiltonian

\begin{equation}
H'(x_1', x_2', p_1', p_2') = \frac{1}{2} \left(G_{11}-G_{12}\right)\,p_1'^2 + 
                                     \frac{1}{2} \left(G_{11}+G_{12}\right)\,p_2'^2 + 
                                     U_M\left(\frac{1}{\sqrt{2}}\left[x_2' - x_1'\right]\right) + 
                                     U_M\left(\frac{1}{\sqrt{2}}\left[x_2' + x_1'\right]\right) .
\end{equation}
Finally, scaling the coordinates so that both particles have the same mass, we define

\begin{equation}
x = \frac{1}{\sqrt{G_{11} - G_{12}}}x_1', \quad y = \frac{1}{\sqrt{G_{11} + G_{12}}}x_2', \quad 
      p_x = \dot{x}, \quad p_y = \dot{y},
\end{equation}
obtaining the Hamiltonian with the coupling in the spatial coordinates as

\begin{multline}
K(x,y,p_x,p_y) = \frac{1}{2}(p_x^2 + p_y^2) + 
   U_M\left(\frac{1}{\sqrt{2}}\left[\sqrt{G_{11} + G_{12}}y- \sqrt{G_{11} - G_{12}}x\right]\right) + \\
   U_M\left(\frac{1}{\sqrt{2}}\left[\sqrt{G_{11} + G_{12}}y+ \sqrt{G_{11} - G_{12}}x\right]\right).
 \label{eq:23}
\end{multline}
Let us remark, that this equivalent Hamiltonian has no kinetic coupling in the momenta $p_x$ and $p_y$, 
but it has been moved to the potential term, between spatial coordinates $x$ and $y$,
which is more adequate for our computational purposes when using neural networks. To further illustrate the effect of the above transformation, we compute the Taylor expansion of the new Hamiltonian,
thus obtaining

\begin{multline}
K(x,y,p_x,p_y) = \frac{1}{2}(p_x^2 + p_y^2) -2D_e + D_ea^2(G_{11}-G_{12})x^2 + D_ea^2(G_{11}+G_{12})y^2 - \\ 
 \frac{3}{\sqrt{2}}D_ea^3(G_{11}-G_{12})\sqrt{G_{11} + G_{12}}yx^2 - \frac{1}{\sqrt{2}}D_ea^3 (G_{11}+G_{12})^{3/2} y^3 + 
 {\cal O}(4)
\label{eq:Tayor_morse}
\end{multline}

where we see reappearing the two Morse parameters $D_e$ and $a$. 
We observe that the expansion contains a coupling term $yx^2$, and an anharmonicity in $y^3$, 
thus indicating that up to order 3 this Hamiltonian is identical to that proposed by H\'enon and Heiles to study the stability of some galaxies, 
at the dawn of nonlinear science~\citep{HH}.
\chapter{Author Publications}
\label{appendix3}
The research publications associated to this doctoral thesis are the following:

\begin{itemize}
    \item \textbf{L. Domingo} and F. Borondo. Deep learning methods for the computation of vibrational wavefunctions. \emph{Communications in Nonlinear Science and Numerical Simulation}, 103:105989, 2021. doi:   \href{https://doi.org/10.1016/j.cnsns.2021.105989}{10.1016/j.cnsns.2021.105989}.
    \item \textbf{L. Domingo}, J. Borondo, and  F. Borondo. Adapting reservoir computing to solve the Schrödinger equation. \emph{Chaos}, 32(6):063111, 2022. doi: \href{https://doi.org/10.1063/5.0087785}{10.1063/5.0087785}.
    \item \textbf{L. Domingo}, G. Carlo, and F. Borondo. Optimal quantum reservoir computing for the noisy intermediate-scale quantum era. \emph{Physical Review E}, 106(4):L043301, 2022. doi: \href{https://doi.org/10.1103/PhysRevE.106.L043301}{10.1103/PhysRevE.106.L043301}.
    \item \textbf{L. Domingo}, G. Carlo, and F. Borondo. Taking advantage of noise in quantum reservoir computing. \emph{Scientific Reports}, 13(1):8790, 2023. doi: \href{https://doi.org/10.1038/s41598-023-35461-5}{10.1038/s41598-023-35461-5}. 
    \item \textbf{L. Domingo}, M. Grande, F. Borondo and J. Borondo. Anticipating food price crises by reservoir computing. \emph{Chaos Solitons and Fractals}, 174:113854, 2023. doi: \href{https://doi.org/10.1016/j.chaos.2023.113854}{10.1016/j.chaos.2023.113854}. 
    \item \textbf{L. Domingo}, M. Djukic, C. Johnson and F. Borondo. Binding affinity predictions with hybrid quantum-classical convolutional neural networks. \emph{Submitted}, 2023.
    \item \textbf{L. Domingo}, M. Grande, G. Carlo, F. Borondo and J. Borondo. Optimizing quantum reservoir computing for time series prediction. \emph{Submitted}, 2023.
    \item \textbf{L. Domingo}, J. Rifà,  J. Borondo and F. Borondo. Vibrational eigenstates calculated using reservoir computing. \emph{Submitted}, 2023.
    \item \textbf{L. Domingo} and F. Borondo. Using reservoir computing to construct scar functions. \emph{Submitted}, 2023.
\end{itemize}
\end{appendices}

% *************************************** Index ********************************
\printthesisindex % If index is present

\thispagestyle{empty}
\end{document}